\documentclass[preprint,sort&compress]{elsarticle}
\usepackage[T1]{fontenc}
\usepackage[latin1]{inputenc}
\usepackage{indentfirst}
\usepackage{graphicx}
\usepackage{amsmath}
\usepackage{mathtools}
\usepackage{amssymb}
\usepackage{paralist}
\usepackage[section]{placeins}
\usepackage{pstricks,pst-node}
\usepackage{psfrag}
\usepackage[colorlinks=false,pdfborder={0 0 0}]{hyperref}
\DeclareMathOperator{\sech}{sech}

\DeclareMathOperator{\Li}{Li}
\DeclareMathOperator{\sgn}{sgn}

\begin{document}
\title{Nonequilibrium Statistical Mechanics of Systems with Long-Range Interactions: Ubiquity of Core-Halo Distributions}
\author[ifufrgs]{Yan Levin\corref{cor1}}
\ead{levin@if.ufrgs.br}
\author[ifufrgs]{Renato Pakter}
\ead{pakter@if.ufrgs.br}
\author[ifufrgs]{Felipe B. Rizzato}
\ead{rizzato@if.ufrgs.br}
\author[ifufrgs]{Tarc\'isio N. Teles\fnref{fn2}}
\ead{tarcisio.teles@fi.infn.it}
\author[ifufrgs]{Fernanda P. da C. Benetti}
\ead{fbenetti@if.ufrgs.br}

\cortext[cor1]{Corresponding author}
\fntext[fn2]{Present address: Istituto Nazionale Di Fisica Nucleare, Viale delle Idee 1, Zona Osmanoro, 50019 - Sesto
Fiorentino, Italy}
\address[ifufrgs]{Instituto de F\'isica, Universidade Federal do Rio Grande do Sul,
Caixa Postal 15051, CEP 91501-970, Porto Alegre, RS, Brazil}

\begin{abstract}

Systems with long-range (LR) forces, for which the interaction 
potential decays with the interparticle distance with an exponent 
smaller than the dimensionality of the embedding space, remain an 
outstanding challenge to statistical physics.  
The internal energy of such systems lacks extensivity and additivity. Although the extensivity 
can be restored by scaling the interaction potential with the number of particles, 
the non-additivity still remains. Lack of additivity leads to inequivalence of statistical ensembles. 
Before relaxing to thermodynamic equilibrium, isolated systems 
with LR forces become trapped in out-of-equilibrium quasi-stationary
state (qSS), the lifetime of which diverges with the number of particles.  Therefore, 
in thermodynamic limit LR systems will not relax to equilibrium.
The qSSs are attained through the process of collisionless relaxation. 
Density oscillations lead to particle-wave interactions and excitation of 
parametric resonances. The resonant particles escape 
from the main cluster to form a tenuous halo.  Simultaneously, this cools down the core of 
the distribution
and dampens out the oscillations.  When all the oscillations die out the ergodicity is 
broken and a qSS is born. 
In this report, we will review a theory which allows
us to quantitatively predict the particle distribution in the qSS.  The theory is applied to 
various LR interacting systems, ranging from plasmas to self-gravitating clusters and 
kinetic spin models.

\end{abstract}


\maketitle

\tableofcontents
\section{Introduction}

A long time ago Einstein expressed his belief that thermodynamics 
is ``the only physical theory of universal content concerning which I am convinced
that, within the framework of applicability of its basic concepts, it will never be overthrown'' \cite{Kle1967}.
One can, however,  wonder about the extent of the ``applicability'' to which Einstein was referring.  
For example, can thermodynamics in any form 
be applied to study non-neutral plasmas
or  galaxies in which ``particles'' interact by long-range (LR) forces?

The difficulty of studying systems with LR interactions
was already well appreciated by Gibbs, who 
noted the inapplicability of statistical mechanics when interparticle
potentials decay with exponents smaller than the dimensionality of the embedding space~\cite{Fis1964,Rue1970}. For such
systems energy is not extensive and traditional thermodynamics fails.  
One way to correct  the lack of extensivity
is to scale the interaction energy with the inverse of the number of particles.  This is the so-called Kac prescription designed to restore extensivity to the free energy~\cite{Kac1959,KacUhl1963,Bak1963}.  The problem, however,
remains  --- although the energy is now extensive, it is still non-additive.  On the other hand, 
it is a fundamental 
postulate  of thermodynamics  that entropy and energy must be additive over the subsystems  ---  that is, the 
interfacial contributions should be negligibly small.  For systems with short-range forces
this condition is clearly satisfied ---  in the thermodynamic limit the interfacial energy is much smaller than the energy of the bulk.  This, however, is not true for systems with LR forces
for which the interfacial region cannot be clearly defined~\cite{CamDau2009} ---  every
particle interacts with every other particle of the system,  so that no clear separation into
bulk and interface exists.  

One can still hope that although the additivity of energy breaks down,  it might still be possible to
use  equilibrium statistical mechanics to describe stationary states of  systems with LR
interactions.  Very quickly, however, one runs into difficulties.  For example,
depending on the ensemble used, one finds that a system can remain either in one phase or
undergo a phase transition~\cite{BarMuk2001}.  One also finds that in the microcanonical ensemble
such systems can have negative specific heat~\cite{Thi1970,LynLyn1977,El-1998,Lyn1999,KieNeu2003,ThiNar2003}, contrary to the laws of usual thermodynamics.

There is, however, an even more profound problem with applying classical statistical mechanics
to systems with LR forces.  The underlying assumption of Boltzmann-Gibbs (BG) statistics
is the existence of ergodicity and mixing~\cite{Rei1998}. For a closed system of particles (in a microcanonical ensemble)
the initial distribution should uniformly spread over the available phase space, so that
in equilibrium all microstates corresponding to a given thermodynamic macrostate should be equally probable.
Although there is no general proof of ergodicity and mixing, in practice it has been found to apply to most
nonintegrable systems with short-range forces.  There is, however, no indication that ergodicity and mixing exist 
for systems with LR interactions~\cite{TsuKon1994,BorCel2004,CheCro1996,CamDau2009,BenTel2012}.  In fact, one should expect precisely the opposite.  
Kac renormalization of the interaction potential kills off the correlations between particles.
Within the kinetic theory, it is precisely these correlations (collisions) that drive a 
system to thermodynamic equilibrium.  In the absence of correlations, the dynamical evolution of the
one-particle distribution function $f(\mathbf{r},\mathbf{p},t)$ is governed by the collisionless Boltzmann (Vlasov) equation~\cite{BraHep1977,GabJoy2010}.
Starting from an arbitrary initial condition, a solution of this equation 
does not evolve to a stationary state --- the spatiotemporal
evolution  continues  {\it ad infinitum} on smaller and smaller length scales. 
It is only in a coarse-grained sense that we can say that the system has reached an ``equilibrium''  --- 
a finite resolution imposed on us by an experiment or a computer simulation will  not allow us to see the full 
fine-grained 
evolution of the distribution function.
The coarse-grained stationary state will, 
in general, be very different from the normal thermodynamic equilibrium.  Unlike the state of thermodynamic equilibrium, 
it will explicitly depend on the initial
distribution of particle positions and velocities \cite{LuwSev1984}. In particular,  the velocity distribution in the stationary state (SS) 
will not have the characteristic Maxwell-Boltzmann form~\cite{LevPak2008a,TelPak2009,PakLev2011}.   
Indeed, observations and simulations of both gravitational clusters~\cite{Hen1964,Sal1965,Lec1966,HohCam1968,Mil1971,WriMil1982,LuwSev1985,Car1986,ReiMil1987,YawMil2003,DeWu2004,KliMil2004,LevPak2008,TelLev2010,JalTre2012} and 
confined non-neutral plasmas~\cite{Len1961,EdwLen1962,EldFei1963,Ste1963,Ran1965,KadPog1970,Rei1991,HuaDri1994,LunBar1995,AllCha2002,OkaIke1997,WanCra1998,LevPak2008a}, indicate presence of such nonequilibrium  
stationary states.  

It is, therefore,  clear
that in the thermodynamic limit, 
traditional methods of equilibrium statistical mechanics cannot be applied to  systems 
with LR forces.  A new theory is needed.  The goal of the
present Report is to show how such theory can be constructed.  Using the
properties of Vlasov dynamics and the
theory of parametric resonances,  we will derive coarse-grained
distribution functions for the nonequilibrium stationary states of 
systems with LR interactions, without explicitly solving the collisionless Boltzmann equation. 
Comparing the theory with the explicit $N$-body simulations, we will
show that it is able to quantitatively predict
both position and velocity distribution functions of self-gravitating clusters~\cite{LevPak2008,TelLev2010,TelLev2011}, 
magnetically confined plasmas~\cite{LevPak2008a,TelPak2009},
and of kinetic spin models~\cite{BarBou2002,AntCal2007,PakLev2011}, without any adjustable parameters. 
We will focus on statistical theory of nonequilibrium quasi-stationary states; 
only briefly shall we address the thermodynamic equilibrium, 
which has already been thoroughly covered by Campa \emph{et al} in Ref \cite{CamDau2009}.

The Report is organized as follows:  in section 1 we begin with an introduction to the principal properties of systems with LR interactions, followed by a review of the Vlasov dynamics. Sections 2, 3, and 4 present results for self-gravitating clusters in one, two, and three dimensions, respectively. In section 5  we address the non-equilibrium properties of magnetically confined plasmas, and  in sections 6 and 7 we discuss two different kinetic spin models. 
Section 8 concludes the Report, reviewing the theories and the results obtained so far and outlining the perspectives for future research.

\section{Systems with Long Range Forces}\label{sec:lr}

Among the physical systems, a significant fraction involves those whose particles interact by long-range potentials of the form $\psi(r)\sim 1/r^{\alpha}$, where $\alpha < d$ and $d$ is the dimensionality of the embedding space. Examples of such systems include galaxies and globular clusters \cite{Kap1922,Jea1922,Oor1932,Cam1950,Ogo1957,Pad1990,TreOst1999,Tre1999,BinTre2009}, two-dimensional and geophysical flows and vortex models \cite{CheCro1996,ChaSom1996,Cha2000,AndLim2007,BouBar2008,VenBou2009,BouVen2012}, quantum spin models \cite{Kas2011}, dipolar excitons \cite{BerKez2012}, cold atom models \cite{SlaKre2008}, as well as magnetically confined plasmas \cite{CheDav1994,DavQin2001,KagDav2010,LevPak2008a}.
In order to predict the behavior of systems with short-range forces we can rely on thermodynamics and statistical mechanics both of which, however,
fail for systems with LR interactions.

Thermodynamics requires  extensivity and additivity \cite{Cal1985}, neither of which is valid for LR systems \cite{CamDau2009}. A system of 
$N$ particles confined inside a volume $V$ is said to be extensive if, when the number of 
particles and the volume are scaled by  $\lambda$, the internal energy $U(\lambda N,\lambda V)$ of the system scales as $\lambda U(N,V)$. 
It is easy to see that systems with short-range forces are 
extensive.  If the interaction potential is short-range, each particle will interact
only with the particles which are within the range $\gamma$ of the interaction potential.  Suppose
that a system is homogeneous, the number of particles within the distance $\gamma$
of a given particle will then be proportional to $N \gamma^d/V$ and the internal
energy must have the form of $U(N,V)=N f(N/V)$, where $f(x)$ is a function
that depends on the microscopic interactions between the particles.  
This form of internal energy is clearly extensive. In fact, it is not necessary for 
the interaction potential to be strictly short-range --- bounded by $\gamma$ --- algebraically decaying potentials will lead to extensive thermodynamics as long
as they decay sufficiently rapidly, i.e. if  $\alpha > d$ \cite{FisRue1966}.  We shall call all
such systems ``finite range''.

Extensivity is important for the
existence of a nontrivial thermodynamic limit and the equivalence of different statistical ensembles.   
A thermodynamic system in contact with a thermal reservoir at
temperature $T$ --- canonical ensemble --- must be at the minimum of its Helmholtz free energy 
$F(N,V)=U(N,V)-T S(N,V)$,
where $S(N,V)$ is the entropy. The celebrated Boltzmann formula $S=k_B \ln W$
%, where $W$ is the number of microstates available to the system, 
relates the thermodynamics with dynamics by associating $W$, the number of microstates available to the system through its dynamics, to the concept of entropy of classical thermodynamics.   
The phase space volume of a confined Hamiltonian system, which is proportional to $W$, grows exponentially with the number
of particles so that $S \sim N$,  irrespective of the range of interactions.  Therefore, both the internal energy and entropy of a finite range system
scale linearly with the number of particles in the system, allowing for a
nontrivial thermodynamic equilibrium.

LR systems, however, are intrinsically different. 
The infinite range of the interaction potential
results in an internal energy that scales superlinearly with the number of particles in the system,  $U \sim N^2$.  Therefore if such system is put in contact with a thermal bath, for large $N$ the Helmholtz free energy will be dominated by the internal energy.  The equilibrium state will then correspond to the  minimum of the internal energy $U$.  The entropy will be irrelevant, unless the temperature of the reservoir is unrealistically large and scales with the number of particles in the system, $T \sim N$.

In practice, most LR systems are isolated from the environment. This is the case for galaxies and magnetically confined plasmas.
Gravity in three dimensions is particularly challenging because of the evaporation of particles \cite{Hen1964,Hoh1978,YawMil1997}; however,
one and two dimensional gravitational systems and magnetically confined plasmas can be studied straightforwardly using molecular
dynamics simulations (MD). Unlike systems with short-range forces  --- which must be confined to a box in order to have a nontrivial thermodynamics --- one and two dimensional
gravitational systems are self-confining and 
can exist in an infinite volume,  $V \rightarrow \infty$. Once again, however, one runs
into a difficulty with the long-range nature of the interaction potential.  The superextensive
interaction energy leads to strong forces and velocities which rapidly exceed that of the speed of light.  To avoid this problem and
to obtain a well defined thermodynamic limit it is necessary to rescale the gravitational
coupling constant by a factor $1/N$. This is the so-called Kac prescription \cite{Kac1959}. For a gravitational system of $N$ particles in
an infinite volume, the Kac prescription is equivalent to the requirement that the
mass of each particle $m \rightarrow 0$, while $mN$ remains finite, $mN=M$.  
One can show that this leads to a well defined thermodynamic limit as $N \rightarrow \infty$.

Although the rescaled gravity and plasmas are extensive, they remain nonadditive. 
For a $d$-dimensional system of particles interacting by a finite-range potential, the interfacial energy scales with the number of particles as $N^{\frac{d-1}{d}}$,
while the  bulk energy grows as $N$. Thus, the total energy of a finite-range system in the thermodynamic limit is equivalent to the sum of the energies of its macroscopic 
subsystems.  This is not true for LR systems.  As the interaction range grows, the concept of interface loses its meaning.  One can no longer consider a total system as a sum of smaller subsystems, since the LR nature of the potential leads to a nontrivial interaction between all the subsystems. The lack of additivity can result in a negative specific heat for an isolated LR system \cite{Thi1970,LynLyn1977,Lyn1999,CamDau2009}. On the other hand, if a LR system is in contact with a thermal bath, its specific heat must be positive. Contrary to what happens with finite-range systems the predictions of microcanonical and canonical ensembles may, therefore, be inequivalent for systems with LR interactions \cite{HerThi1971,BouBar2005,Cha2006a,FilAma2009}. Similarly, the canonical and the grand-canonical ensembles may also become inequivalent \cite{CohMuk2012}. Besides inequivalence of ensembles, it has also been debated that negative specific heat may result in yet another abnormality: the violation of the zeroth law of thermodynamics \cite{RamLar2008,RamLar2008a,MicSan2009,RamLar2009}.

Another difficulty with the statistical treatment of LR systems is the lack of ergodicity.
The ergodic theorem allows us to replace the time averages by the ensemble averages \cite{Pen1979}. 
Consider a $2dN$ dimensional phase space of $N$ interacting particles. 
Each point ${\bf X}$ in this phase space represents a possible 
configuration (microstate) of the system. For a given thermodynamic macrostate
there is a huge number of possible microstates. This allows us to define a statistical
ensemble of microstates with a probability density $\rho({\bf X},t)$. The dynamics of $\rho({\bf X},t)$ is
governed by the Liouville equation.
For equilibrium statistical mechanics to work, the initial probability density should uniformly spread
over the energy surface --- producing a, so-called, mixing flow \cite{Rei1998}.

The fundamental problem of ergodic theory is to understand under what conditions
a nonstationary phase space density will converge to a stationary one \cite{LebPen1973,Leb1999}. Note that
for a time reversible system one can not have a ``fine-grained'' equilibrium,
a thermodynamic equilibrium exists only in a coarse-grained sense. On a fine-grained scale, the dynamical evolution of the probability density will never stop, so that if at some point during the dynamical evolution the velocities of all the particles are reversed, the system will diverge from the equilibrium. 
Although ergodicity and mixing
have been verified for many different systems with finite-range forces, both seem to fail for systems with LR interactions \cite{TsuKon1994,BorCel2004,MukRuf2005,BenTel2012}.

%A system, initially at one point, should evolve in accordance with the Hamiltonian laws of motion along a surface of constant energy in this phase space and should be able to explore all of the . Alternatively, if the system is ergodic, this situation may be represented as an infinite number of systems, each in a different point along this surface. On the other hand, if the system is not ergodic, there are regions of phase space which may be inaccessible to certain initial conditions. In this case, the two situations are not equivalent. This collection of infinite copies of the system along the constant energy surface constitutes the microcanonical ensemble, and its validity depends on ergodicity. While it is generally observed in systems with short-range interactions, there are indications of breaking of ergodicity in long-range systems \cite.

%The Boltzmann's hypothesis requires that all the microstates be  ergodic. Only then can a purely statistical description of the equilibrium becomes possible and the time averages can be replaced by the ensemble averages.  

The relaxation to a stationary state (SS) of systems with LR interactions is fundamentally different from the relaxation to equilibrium of systems with short-range forces. For the latter, the relaxation is collisional and the reduced probability
densities are governed by the BBGKY (Born, Bogoliubov, Green, Kirkwood, Yvon) hierarchy of equations \cite{Bal1997}. At the leading order of this hierarchy is the Boltzmann equation
$Df/Dt=(\partial f/\partial t)_{\mathrm{col}}$, where $Df/Dt\equiv\partial f/\partial t+(\mathbf{p}/m)\cdot\nabla_r f+\mathbf{F}\cdot\nabla_p f$ is the convective derivative of $f(\mathbf{r},\mathbf{p},t)$ and $\mathbf{F}=\dot{\mathbf{p}}$. This equation describes the evolution of 
the one-particle distribution function 
$f(\mathbf{r},\mathbf{p},t)$ \cite{Hua1987}. The right hand side of the Boltzmann equation is the
collision term that drives the system towards thermodynamic equilibrium \cite{Hua1987}. The distribution functions in thermodynamic equilibrium do not depend on the initial condition, but only on the global
conserved quantities, and are described by the Boltzmann-Gibbs statistical mechanics \cite{Gib1928}.

The situation is very different for systems with LR forces. In the thermodynamic limit $N \rightarrow \infty$ the dynamics of these systems 
is completely dominated by the mean-field and the collisions (correlations) 
are negligible.  To see why this is so, let us consider, for example, a one dimensional gravitational system of particles of mass $m$, interacting by $\varphi(x)=G m^2 |x|$, where $G$ is the gravitational constant.  
As was discussed above, to have
a well defined thermodynamic limit we need to require that $m \rightarrow 0$, while
the total mass of the system remains fixed, $mN=M$.  Although the interaction between any two
particles is vanishingly small, the infinite range of the potential results in a finite total force
acting on each particle. To quantify the discreteness (correlations) effects \cite{Lev2002}
we can define a plasma parameter --- corresponding to the ratio of 
the characteristic two-body interaction energy 
and the average kinetic energy --- $\Gamma \equiv 2 G m^2 a/m \langle v^2 \rangle$, where $\langle v^2 \rangle$ is the average particle velocity and $a$ is a characteristic separation between the particles. 
$\Gamma$ measures the degree to which the dynamics of a system is dominated by the correlations ---
if $\Gamma>1$ the correlations (collisions) are important and if $\Gamma<1$ the dynamics is governed purely by the mean-field.  Starting from an initial particle distribution, a one dimensional gravitational cluster
will relax to a stationary state, with a characteristic velocity $\langle v^2 \rangle \sim O(1)$.  It will be shown in the following sections that the
extent of the mass distribution is controlled by the parametric resonances, so that starting
from an initial particle distribution with a compact support, the final distribution
will be restricted to a finite "volume" or radius $r_h$, so that $a \sim r_h/N$. We then come to the conclusion that 
$\Gamma \sim 1/N^2$, in the thermodynamic limit the correlations vanish and the dynamics of a LR system is determined purely by the mean-field.  

The argument above suggests that for LR systems the (collisional) right-hand side
of the Boltzmann equation should vanish and the one-particle distribution function
should satisfy the collisionless Boltzmann equation $Df/Dt=0$.  This equation is also known
as the Vlasov equation \cite{BraHep1977}. While the stationary solution to the Boltzmann equation is the Maxwell-Boltzmann distribution, the Vlasov equation has an infinite number of stationary states, depending on the initial particle distribution. The one-particle distribution function evolves on ever-decreasing length scales. Eventually, the dynamical scale becomes so small that the evolution of $f(\mathbf{r},\mathbf{p},t)$ can no longer be observed at any resolution available to us. It is only in this coarse-grained sense that a LR system achieves a stationary state (SS). 

For a finite number of particles, the correlations --- although very small --- remain finite.  The  cumulative effect of weak correlations will drive a LR system from a quasi-stationary state (qSS) towards the true thermodynamic equilibrium. 
The relaxation time $t_\times$, however, is very slow, diverging with the number of particles as $N^{\gamma}$ \cite{YamBar2004,JaiBou2007,TelLev2011}.  The value of the exponent $\gamma$ depends on each system \cite{TelLev2010}, but is usually $\gamma \ge 1$. We expect that $t_\times \sim  1/\Gamma$,
so that for 1D gravity $t_\times \sim N^2$.  For 2D gravitational clusters 
the interaction potential is logarithmic, 
so that the crossover time should scale as $t_\times \sim N/\ln N$.  In the following sections we will see if 
these simple estimates of the relaxation time agree with the results of $N$-body simulations.

Although interesting theoretically, the strong divergence of $t_\times$ precludes the equilibrium state from
ever being reached by most physically relevant systems, such as galaxies and plasmas.  To achieve equilibrium these systems would
require a span of time longer than
the age of the universe \cite{BinTre2009,Pad1990,SakGou1991}.

%Solving the Vlasov equation for arbitrary initial conditions is not a trivial process. However, the dynamics governed by this equation is essential for describing the quasi-stationary states achieved by systems with long-range interactions. Therefore, there exists a demand of statistical theories based on Vlasov dynamics, able to predict and describe the quasi-stationary states without explicitly solving the Vlasov equation nor performing full $N$-body molecular dynamics simulations. Such theories will be the main topic of this review.

%One motivation for the development of a general long-range theory is its application in real systems. In many astrophysical systems, for example, the lifetime of the quasi-stationary states is greater than the age of the universe itself \cite{Pad1990}. Regarding the Coulomb potential, long-range systems include non-neutral, magnetically confined plasmas \cite{LevPak2008,TelPak2009}. Other examples include models in hydrodynamics \cite{VenBou2009} and atomic physics \cite{SlaKre2008}.

\section{Vlasov dynamics}

In the thermodynamic limit $N \to \infty$, the correlations between the particles of a LR system
vanish and the dynamics of the one-particle distribution function $f(\mathbf{q},\mathbf{p},t)$ is governed exactly \cite{BraHep1977} by the Vlasov equation,

%%%%%%%%%%%%%%%%%%%%%
\begin{equation}\label{eq:vlasov}
\left(\frac{\partial}{\partial t}+\mathbf{p}\cdot\frac{\partial}{\partial \mathbf{q}}-\frac{\partial \psi}{\partial \mathbf{q}}\cdot
\frac{\partial}{\partial \mathbf{p}}\right)f(\mathbf{q},\mathbf{p},t)=0.
\end{equation}
%%%%%%%%%%%%%%%%%%%%%

%For finite systems, the collisional contribution of the Boltzmann equation remains finite, as it scales with $1/N$. In this case, eventually the correlations will drive the system to thermodynamic equilibrium, but the time it takes to do so grows with the size of the system. Until then, the system remains in a quasi-stationary state of Vlasov dynamics, achieved through the initial collisionless relaxation. In order to describe these states, theories which take into account the collisionless nature of the dynamics are required.

The one-particle distribution function evolves in the phase space as the density of an incompressible fluid  --- its local value remains constant along the flow.  
The $\psi(\mathbf{q})$ represents the potential felt by a ``fluid element'' located at 
$(\mathbf{q}, \mathbf{p})$. It can be shown that the Vlasov dynamics has an infinite number of
conserved quantities called Casimir invariants \cite{ChaBou2005,FilFig2005}. Any local functional
of the distribution function is a Casimir invariant,
%%%%%%%%%%%%%%%%%%%%%
\begin{equation}\label{eq:casimir}
C[f]=\int g(f)d\mathbf{q} d\mathbf{p},
\end{equation}
%%%%%%%%%%%%%%%%%%%%%
In particular, the fine-grained Boltzmann entropy 
%%%%%%%%%%%%%%%%%%%%%
\begin{equation}\label{eq:ent}
S(f)=- \int f(\mathbf{q}, \mathbf{p},t) \ln f(\mathbf{q}, \mathbf{p},t) d\mathbf{q} d\mathbf{p}.
\end{equation}
%%%%%%%%%%%%%%%%%%%%%
is a Casimir invariant and is conserved by the Vlasov flow.
The entropy can increase only in a coarse-grained sense \cite{TreHen1986}.  To see this let us define a coarse-grained
distribution function
%%%%%%%%%%%%%%%%%%%%%
\begin{equation}\label{eq:fbar}
\bar f(\mathbf{q}, \mathbf{p},t)=\frac{1}{(\Delta p \Delta q)^d} \int_{\Delta p, \Delta q} f(\mathbf{q'}, \mathbf{p'},t)d\mathbf{q'} d\mathbf{p'}.
\end{equation}
%%%%%%%%%%%%%%%%%%%%%
Consider the evolution of the coarse-grained entropy
%%%%%%%%%%%%%%%%%%%%%
\begin{equation}\label{eq:sbar}
\Delta \bar S=\bar S(t_1)- S(t_0)=\int \left[s(\bar f,t_1)-s(f,t_0)\right] d\mathbf{q} d\mathbf{p},
\end{equation}
%%%%%%%%%%%%%%%%%%%%%
where we have defined the Boltzmann entropy density $s(f,t)=- f(\mathbf{q}, \mathbf{p},t) \ln f(\mathbf{q}, \mathbf{p},t)$.
We have also supposed that at $t=t_0$ the exact particle distribution is known. Since the fine-grained entropy is conserved, we can rewrite Eq \eqref{eq:sbar} as
%%%%%%%%%%%%%%%%%%%%%
\begin{equation}\label{eq:sbar2}
\Delta \bar S=\int \left[s(\bar f,t_1)-s(f,t_1)\right] d\mathbf{q} d\mathbf{p},
\end{equation}
%%%%%%%%%%%%%%%%%%%%%
To perform the coarse-graining, we divide the macrocells of volume $(\Delta p \Delta q)^d$
into $K$ microcells, with the local value of the distribution function inside the microcell $i$
given by $f_i$. Now, consider the variation of the coarse-grained entropy inside  the
{\it macrocell} $j$,
%%%%%%%%%%%%%%%%%%%%%
\begin{equation}\label{eq:coarse}
\Delta \bar S_j= (\Delta p \Delta q)^d \sum_{i}^{K} \left[s\left(\frac{\sum_{i}^{K} f_i}{K}\right) -s(f_i)\right]=(\Delta p \Delta q)^d \left[ K s\left(\frac{\sum_{i}^{K} f_i}{K}\right)-\sum_{i}^{K} s(f_i)\right].
\end{equation}
%%%%%%%%%%%%%%%%%%%%%
Since the entropy density $s(x)$ is a concave function it must satisfy Jensen's inequality
%%%%%%%%%%%%%%%%%%%%%
\begin{equation}\label{eq:convex}
\frac{1}{K}\sum_{i}^{K} s(f_i) \le s\left(\frac{\sum_{i}^{K} f_i}{K}\right)\,,
\end{equation}
%%%%%%%%%%%%%%%%%%%%%
from which we conclude that the coarse-grained entropy of the system should increase with time, 
$\Delta \bar S \ge 0$.  The Boltzmann entropy will be maximum in equilibrium, this however, does not
mean that the equilibrium can always be reached.  As we shall see, in the thermodynamic limit,
systems with LR interactions can become trapped in a non-ergodic stationary state.

\begin{figure}[!htb]
\begin{center}
\includegraphics[width=0.8\textwidth]{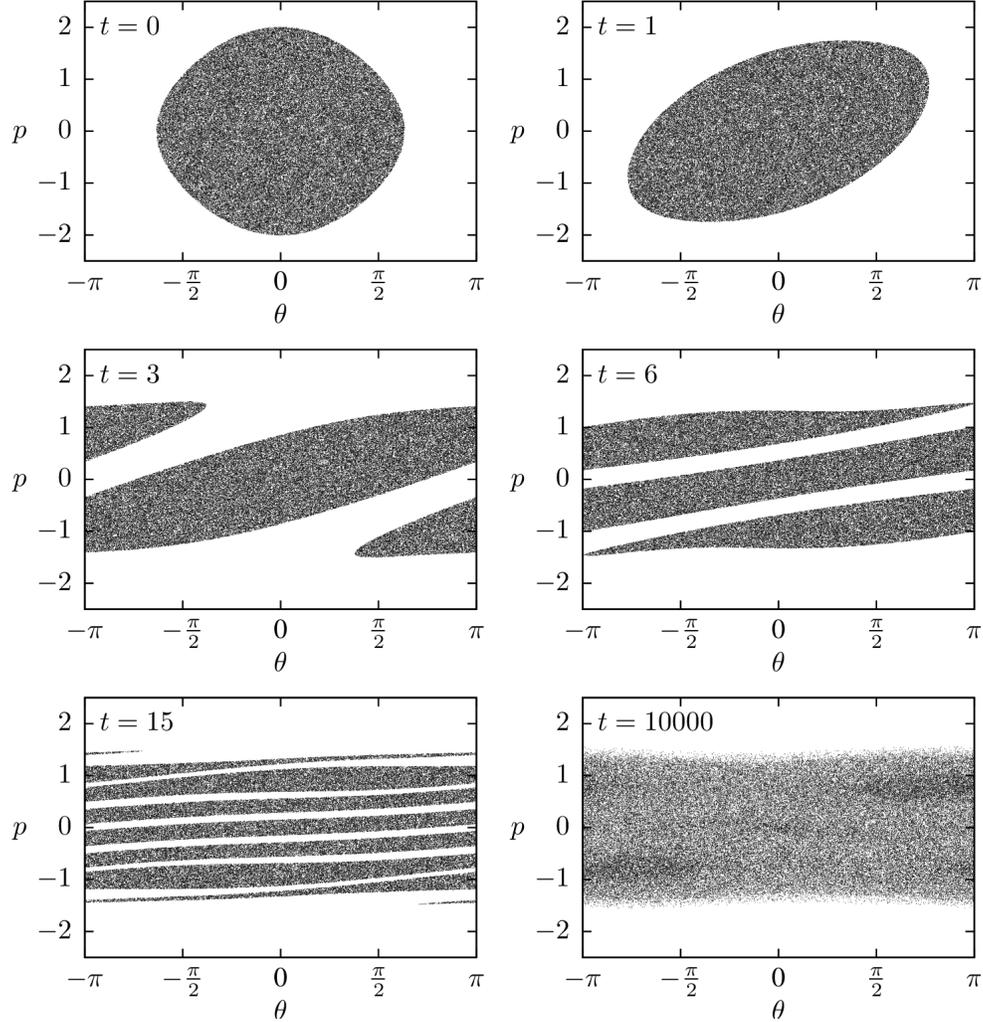}
\end{center}
\caption{Evolution of particle distribution in the phase space of the Hamiltonian mean field (HMF) model.\label{fig:hmfmix}}
\end{figure}

If the initial fine-grained distribution function $f_0(\mathbf{q},\mathbf{p})$ is divided into $p$ levels of phase space density $\eta_j$, Vlasov dynamics will preserve the hypervolume of each level, $C(\eta_j)=\int\delta[f(\mathbf{q},\mathbf{p},t)-\eta_j]d\mathbf{q}d\mathbf{p}$. In this review, we will concentrate on one-level (waterbag) initial distributions of the form
%%%%%%%%%%%%%%%%%%%%%
\begin{equation}
f_0(\mathbf{q},\mathbf{p})=\eta\Theta(q_m-|\mathbf{q}|)\Theta(p_m-|\mathbf{p}|),
\end{equation}
%%%%%%%%%%%%%%%%%%%%%
where $\Theta(x)$ is the Heaviside step function and $q_m$ and $p_m$ represent the maximum values for the generalized coordinates and momentum, $\eta$ is the phase space density of the
initial particle distribution.  Starting from this initial condition, 
the fine-grained distribution function $f(\mathbf{q},\mathbf{p},t)$ will evolve 
in phase space through the process of filamentation,  developing 
structure on smaller and smaller length scales, see Fig \ref{fig:hmfmix}. 
%%%%%%%%%%%%%%%% Figure%%%%%%%%%%%%%%%%%%%%%
%%%%%%%%%%%%% End of figure%%%%%%%%%%%%%%%%%
Eventually, the length scale of the
dynamical evolution will become so small, that to an observer it will appear
that the dynamics has ceased.  
At this stage, we may say that the coarse-grained distribution, $\bar{f}$, has achieved a stationary state, even though the fine-grained distribution $f$ is still evolving. 
For a practical purpose  of describing the results of 
molecular dynamics simulations --- which, of course, have finite precision  --- we only need to have the knowledge of $\bar{f}(\mathbf{q},\mathbf{p})$.

\subsection{Lynden-Bell Statistics}

In a seminal work, Lynden-Bell (LB) proposed a statistical approach for calculating $\bar{f}(\mathbf{q},\mathbf{p})$ for the final stationary state \cite{Lyn1967}. 
%Physically, this granulation of phase space implies that, in the quasi-stationary state, fluctuations in position and velocity of a particle are not relevant in determining the macroscopic, coarse-grained behavior of the system. This allows for the fine-grained distribution $f(\mathbf{q},\mathbf{p},t)$ to be substituted by the coarse-grained distribution $\bar{f}(\mathbf{q},\mathbf{p},t)$, which is the average of $f(\mathbf{q},\mathbf{p},t)$ over a macrocell of size $d\mathbf{q}\,d\mathbf{p}$. The coarse-grained distribution $\bar{f}(\mathbf{q},\mathbf{p},t)$ should relax to a final distribution $\bar{f}(\mathbf{q},\mathbf{p})$. For gravitational systems, this process was called violent relaxation, for it occurs in a time scale of the order of the system's typical dynamical time.
LB theory is similar in its construction to the usual Boltzmann statistics, but instead
of working with the particles, Lynden-Bell studied the distribution of the phase space
density levels, $\eta$. It is important to keep in mind that, similar to the usual equilibrium 
statistical mechanics, the LB approach requires the existence of ergodicity and mixing \cite{SakGou1991,CamDau2009}. 

\begin{figure}[!h]
\begin{center}
\includegraphics[width=0.8\textwidth]{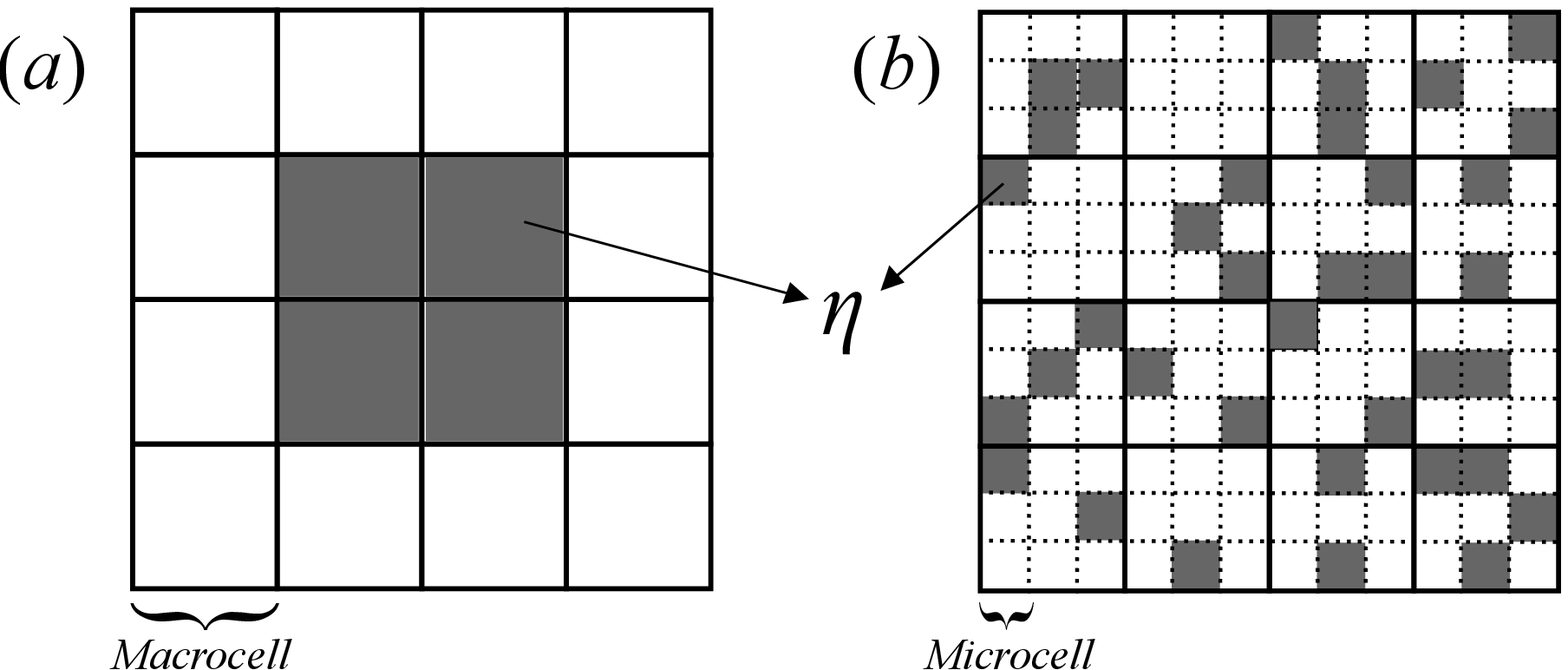}
\end{center}
\caption{Schematic of phase-space evolution described by the Vlasov dynamics: (a) initial and (b) final 
stationary state for a
distribution with initial phase-space density $\eta$. In this example, $\nu=9$.\label{fig:cells}}
\end{figure}

The phase space is divided into $P$ macrocells which are in turn subdivided into $\nu$ microcells of volume $h^d$. As the dynamics progresses, the distribution function spreads over the phase space, occupying more macrocells than it did initially. This process is illustrated in Fig \ref{fig:cells}. The volume fraction occupied by the level $\eta$ inside the macrocell $i$ is
%%%%%%%%%%%%%%%%%%%%%
\begin{equation}
\rho(\mathbf{q},\mathbf{p})=\frac{n_i}{\nu},
\end{equation}
%%%%%%%%%%%%%%%%%%%%%
where $n_i$ is the number of microcells inside a macrocell $i$ 
occupied by the level $\eta$.
The volume fraction is related to the distribution function by $\rho(\mathbf{q},\mathbf{p})=\bar{f}(\mathbf{q},\mathbf{p})/\eta$.  
Due to the incompressibility of Vlasov dynamics, each microcell 
can be occupied by at most one level $\eta$, so that the density must satisfy
%%%%%%%%%%%%%%%%%%%%%
\begin{equation}
\rho(\mathbf{q},\mathbf{p})\leq 1.
\end{equation}
%%%%%%%%%%%%%%%%%%%%%

%
%%%%%%%%%%%%%%%% figure %%%%%%%%%%%%%%%%%%%%%
%
%
%%%%%%%%%%%%%%%% end of figure %%%%%%%%%%%%%%%%%%%%%

LB supposed that in a stationary state the dynamics of the density levels is ergodic  ---  $\eta$'s have an equal probability of occupying any of the microcells.  He then 
applied the usual Boltzmann counting to calculate the most probable distribution of the density level over the phase space.
 
The total number of occupied microcells, 
%%%%%%%%%%%%%%%%%%%%%
\begin{equation}
N=\sum_i n_i.
\end{equation}
%%%%%%%%%%%%%%%%%%%%%
remains constant throughout the dynamics.
The number of ways in which these $N$ microcells can be divided among the $P$ macrocells is given by
%%%%%%%%%%%%%%%%%%%%%
\begin{equation}\label{eq:micmac}
\frac{N!}{\prod_i^P n_i!}.
\end{equation}
%%%%%%%%%%%%%%%%%%%%%
Now consider a macrocell.  The number of ways in which $n_i$ of its $\nu$ microcells can be occupied by a density level is 
%%%%%%%%%%%%%%%%%%%%%
\begin{equation}\label{eq:ocumic}
\frac{\nu !}{(\nu-n_i)!}.
\end{equation}
%%%%%%%%%%%%%%%%%%%%%
Note that the density levels are treated as distinguishable. Multiplying expressions \eqref{eq:ocumic} and \eqref{eq:micmac} we obtain the total number of possible microstates, 
%%%%%%%%%%%%%%%%%%%%%
\begin{equation}
W(n_i)=\frac{N!}{\prod_i^P n_i!}\prod_i\frac{\nu !}{(\nu-n_i)!}.
\end{equation}
%%%%%%%%%%%%%%%%%%%%%
The coarse-grained entropy of the system is defined as $S_{lb}\equiv -k_B \ln W(n_i)$ where $k_B$ is the Boltzmann constant. In the limit in which the variations of $\rho(\mathbf{q},\mathbf{p})$ between the macrocells are infinitesimal, the entropy can be written as
%%%%%%%%%%%%%%%%%%%%%
\begin{equation}\label{eq:lbentropy}
S_{lb}=-k_B \int \frac{d\mathbf{q} d\mathbf{p}}{h^d}\{ \rho(\mathbf{q},\mathbf{p})\ln[\rho(\mathbf{q},\mathbf{p})]+[1-\rho(\mathbf{q},\mathbf{p})]
\ln[1-\rho(\mathbf{q},\mathbf{p})]\}.
\end{equation}
%%%%%%%%%%%%%%%%%%%%%

Similar to the usual thermodynamic
equilibrium, LB proposed that the SS of a LR system corresponds to the 
most probable distribution of the density levels among the macrocells.
To find this distribution, we must 
maximize the LB entropy under the constraints of energy 
%%%%%%%%%%%%%%%%%%%%%
\begin{equation}\label{eq:encons}
\int \left(\frac{p^2}{2m}+\frac{\psi(\mathbf{q})}{2} \right) \bar{f}(\mathbf{q},\mathbf{p})d\mathbf{q} d\mathbf{p}=\mathcal{E}_0
\end{equation}
%%%%%%%%%%%%%%%%%%%%%
and particle
%%%%%%%%%%%%%%%%%%%%%
\begin{equation}\label{eq:normcons}
\int \bar{f}(\mathbf{q},\mathbf{p})d\mathbf{q} d\mathbf{p}=1,
\end{equation}
%%%%%%%%%%%%%%%%%%%%%
conservation.  In the above equations $\mathcal{E}_0$ is the average particle energy in the initial distribution and  $\psi(\mathbf{q})$ is the
potential at position $\mathbf{q}$ in the stationary state. Maximizing the entropy Eq \eqref{eq:lbentropy}, under the constraints given by Eqs. \eqref{eq:encons} and \eqref{eq:normcons}, we find the coarse-grained 
distribution function $\bar{f}(\mathbf{q},\mathbf{p})=\eta \rho(\mathbf{q},\mathbf{p})$ for the SS,
%%%%%%%%%%%%%%%%%%%%%
\begin{equation}\label{eq:flb}
f_{lb}(\mathbf{q},\mathbf{p})=\bar{f}(\mathbf{q},\mathbf{p})=\frac{\eta}{1+e^{\beta[\epsilon(\mathbf{q},\mathbf{p})-\mu]}}
\end{equation}
%%%%%%%%%%%%%%%%%%%%%
where $\epsilon(\mathbf{q},\mathbf{p})=\frac{p^2}{2m}+\psi(\mathbf{q})$ is the one-particle energy. The Lagrange multipliers $\beta$ and $\mu$ are the inverse temperature and the chemical potential of the stationary state. The expression \eqref{eq:flb} is similar to the distribution function of fermions in an equilibrium system.

Besides Lynden-Bell's theory, other statistical approaches have also been proposed to study 
qSSs which arise in the process of collisionless relaxation.  Example include,
statistics based on particles instead of the distribution function \cite{Shu1978} and an information-theoretical approach \cite{Nak2000,FilFig2005}. 
Just like LB theory these approaches require
existence of ergodicity and good mixing \cite{AraLyn2005,BinSec2008} which, in general, are not valid for systems with LR forces. In this Report, we will only focus on LB theory. In the following sections we will see how well it compares with the simulations.

\section{Gravitation in one dimension}\label{sec:grav1d}

Due in part to complications of 3D gravitational systems, which will be addressed later on, 
many studies of self-gravitating systems have focused
on one and two dimensions \cite{MilPre1968,MilPre1970,Mil1971,SevLuw1984,WriMil1984,SevLuw1986,ReiMil1987,YanGou1998,YouMil2000,MilRou2002,Val2006,JoySic2011}. The reduced dimensionality makes the study of these systems much simpler. The fact that
the gravitational potential in one and two dimensions is unbounded from above prevents particle evaporation
which makes theoretical and simulation work on 3D systems very difficult.  In spite of their
greater simplicity, 1D and 2D gravitational systems share many characteristics of 3D gravity. For example, 
the global structure of disk-like galaxies, found using 3D numerical simulation, are also reproduced by 2D simulations \cite{Hoh1978}.
One-dimensional self-gravitating systems have also been used to study the stellar dynamics of 
galaxy clusters and of cosmological models \cite{Oor1932,Cam1950,Lec1966,HohFei1967,Ryb1971,WriMil1982,Mat1990,JoyWor2010,JoyWor2011,MilRou2006,MilRou2007,
MilRou2010,JoySic2011,SchDeh2013}. 

A 1D self-gravitating system consists of $N$ sheets of mass density $m$ uniformly distributed in the $y$--$z$ plane, free to move along the $x$ axis. 
%In order to simplify the calculations, the sheets are assumed to have the same mass density, $m=M/N$, where $M$ is the total mass per unit area of the system. 
The dynamics of the sheets is the same as the dynamics of point particles of mass $m$ 
interacting by a linear potential.   The particles are free to cross one another.  
The thermodynamic limit, $\lim_{N\to\infty} mN=M=\mathrm{constant}$, is equivalent to the Kac prescription necessary to guarantee the extensivity of the energy.

The Poisson equation for this system is
%%%%%%%%%%%%%%%%%%%%%
\begin{equation}
\nabla^2\psi(x,t)=4\pi G \lambda(x,t)
\end{equation}
%%%%%%%%%%%%%%%%%%%%%
where $G$ is the gravitational constant and $\lambda(x,t)$ is the mass density. In order to simplify the expressions, we will work with dimensionless variables. We shall rescale the mass, length, velocity, potential, and energy\footnote{A system's energy takes into account the total work necessary to bring a particle from infinity (or from a position where the potential is zero) to a position $\mathbf{q}$, i.e. $\int{[\psi(\mathbf{q})-\psi(\infty)]}d\mathbf{q}$. For 3D self-gravitating systems the potential at infinity is zero, and for plasmas it is zero at the conducting wall. However, it is important to note that for 1D and 2D self-gravitating systems, the potential diverges at infinity. Since this divergent term appears in both the initial and the final state, the problem is avoided by using a renormalized energy, see Ref \cite{TelLev2010} for more details.} by $M$, $L_0$ (an arbitrary length scale), $V_0=\sqrt{2\pi G M L_0}$, $\psi_0=2\pi G M L_0$ and $E_0=M V_0^2=2\pi G M^2 L_0$, respectively. This is equivalent to considering $G=M=1$ and to defining a dynamical time scale
%%%%%%%%%%%%%%%%%%%%%
\begin{equation}
\tau_D=(4\pi G \rho_0)^{-1/2}.
\end{equation}
%%%%%%%%%%%%%%%%%%%%%
Thus, the Poisson equation becomes
%%%%%%%%%%%%%%%%%%%%%
\begin{equation}\label{eq:poisson}
\nabla^2\psi(x,t)=2 \rho(x,t).
\end{equation}
%%%%%%%%%%%%%%%%%%%%%
For a particle (sheet) of (reduced) mass density located at $x'$, the density is $\rho(x,x')=\delta(x-x')$, and the long-range potential is given by the Green's function,
%%%%%%%%%%%%%%%%%%%%%
\begin{equation}
G(x,x')=|x-x'|.
\end{equation}
%%%%%%%%%%%%%%%%%%%%%
A particularly interesting aspect of the one-dimensional gravity is that the interaction potential 
does not have any singularities, which simplifies significantly molecular dynamics (MD) simulations,
allowing us to explore in great detail the relaxation of this model to the qSS.

\subsection{Molecular Dynamics}

The reduced Hamiltonian for a system of $N$ particles interacting by a one-dimensional gravitational potential
is
%%%%%%%%%%%%%%%%%%%%%
\begin{equation}
\mathcal{H}(x,v)=\sum_{i=1}^N\frac{v_i^2}{2}+\frac{1}{2N}\sum_{i,j}^N|x_i-x_j|,
\end{equation}
%%%%%%%%%%%%%%%%%%%%%
This Hamiltonian, along with Hamilton's equations of motion, completely determines the dynamics
of the system. The acceleration of a particle at position $x$, due to its interaction with the other $N-1$ particles, is given by
%%%%%%%%%%%%%%%%%%%%%
\begin{equation}\label{eq:g1dac1}
\ddot{x}=-\frac{1}{N}\sum_{i=1}^N\frac{x-x_i}{|x-x_i|},
\end{equation}
%%%%%%%%%%%%%%%%%%%%%
which may be expressed as
%%%%%%%%%%%%%%%%%%%%%
\begin{equation}\label{eq:g1dac2}
\ddot{x}=\frac{N_{>}(x)-N_{<}(x)}{N},
\end{equation}
%%%%%%%%%%%%%%%%%%%%%
where $N_{>}(x)$ and $N_{<}(x)$ represent the number of particles to the right and to the left of $x$, respectively. To simulate the system according to equation \eqref{eq:g1dac2} requires time that scales with $N^2$. However, the simulation may be simplified by using a vector containing the indices of each particle, and reordering it according to each particle's position at each new calculation. The expression in equation \eqref{eq:g1dac2} then may be written as
%%%%%%%%%%%%%%%%%%%%%
\begin{equation}
\ddot{x}=\frac{(N-i)-(i-1)}{N}=\frac{N-2i+1}{N},
\end{equation}
%%%%%%%%%%%%%%%%%%%%%
where $i$ is the index of the particle at position $x$. This simplification involves no approximation; the advantage is purely computational, for the simulations become more efficient regarding the computational time \cite{NouFan2003}---the typical time required to order a vector of size $N$ varies at most with $N\ln N$ \cite{PreTeu1992}. Using this method, the trajectories may be obtained exactly, that is, at machine precision \cite{NouFan2003}. However, for the exact procedure, the trajectories must be calculated at each collision, and the number of collisions grows as $N^2$. Therefore, in our simulations, we used a fourth-order symplectic integrator, reordering the index vector at each time step and maintaining the relative error in energy at $10^{-5}$.

We simulate numerically the evolution of a system of particles that are initially distributed uniformly with positions $x_i$ where $x_i\,\in \,[-x_m,x_m]$ and velocities $v_i\, \in \, [-v_m,v_m]$, so that the initial distribution function is given by
%%%%%%%%%%%%%%%%%%%%%
\begin{equation}
f_0(x,v)=\eta\Theta(x_m-|x|)\Theta(v_m-|v|)
\end{equation}
%%%%%%%%%%%%%%%%%%%%%
where $\eta=(4 x_m v_m)^{-1}$. In order to calculate the initial energy, we must find the potential that is the solution of the Poisson equation \eqref{eq:poisson} at $t=0$,
%%%%%%%%%%%%%%%%%%%%%
\begin{equation}
\frac{d^2}{dx^2}\psi(x)=
  \begin{cases}
    \frac{1}{x_m} & \text{for } |x| \leq x_m \\
    0 & \text{for } |x| \geq x_m
  \end{cases}
\end{equation}
%%%%%%%%%%%%%%%%%%%%%
with boundary conditions $\lim_{|x| \rightarrow \infty} \psi(x)=|x|$ and $\psi'(0)=0$. The solution is given by
%%%%%%%%%%%%%%%%%%%%%
\begin{equation}\label{eq:g1dmdpot}
\psi(x)=
  \begin{cases}
    \frac{x^2}{2x_m}+\frac{x_m}{2} & \text{for } |x| \leq x_m \\
    |x| & \text{for } |x| \geq x_m.
  \end{cases}
\end{equation}
%%%%%%%%%%%%%%%%%%%%%
Using the definition of the mean energy, equation \eqref{eq:encons}, the initial energy of the system is found to be
%%%%%%%%%%%%%%%%%%%%%
\begin{equation}\label{eq:g1de0}
\mathcal{E}_0=\frac{v_m^2}{6}+\frac{1}{3}
\end{equation}
%%%%%%%%%%%%%%%%%%%%%
where without loss of generality we have set $x_m=1$.

\subsection{Equilibrium}\label{subsec:g1deq}

\begin{figure}[!htb]
\vspace{5mm}
\begin{center}
\includegraphics[width=0.8\textwidth]{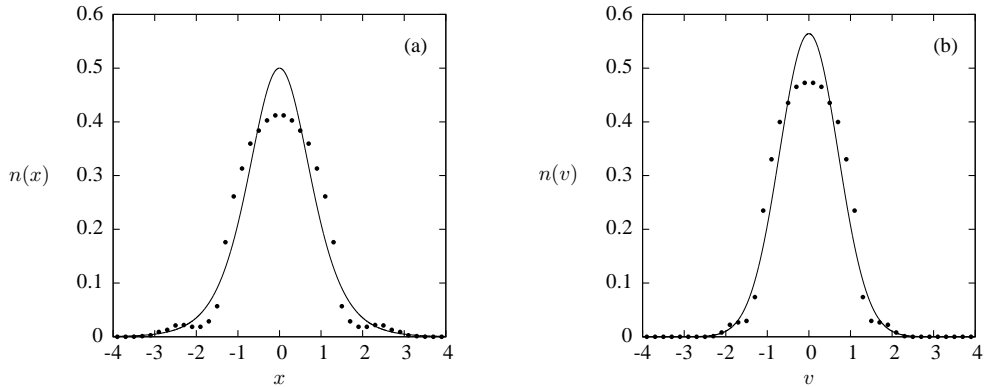}%{figg1deqqss}
\end{center}
\caption{Distributions in (a) position and (b) velocity for a 1D gravitational system with $\mathcal{E}_0=0.75$, obtained using MD simulations (points), averaged over times $t=1000\tau_D$ to $t=1100\tau_D$, compared with the equilibrium distributions (lines), given by Eqs \eqref{eq:g1dmbpos} and \eqref{eq:g1dmbvel}. Repeating the MD simulation for the same initial energy but different initial conditions and taking the average value of the resulting distributions, error bars showing the standard error are smaller than the symbol size. \label{fig:g1deqqss}}
\end{figure}

If the system relaxes to equilibrium the gravitational potential must satisfy the
Poisson equation 
%%%%%%%%%%%%%%%%%%%%%
\begin{equation}\label{eq:g1dpoi1}
\nabla^2\psi(x)=2  n(x)
\end{equation}
%%%%%%%%%%%%%%%%%%%%%
where $n(x)$ is the equilibrium density distribution. Using the Maxwell-Boltzmann distribution, $f_{mb}(x,v)=Ce^{-\beta(v^2/2+w(x))}$, the equilibrium density distribution is given by
%%%%%%%%%%%%%%%%%%%%%
\begin{equation}\label{eq:g1dbolt1}
n(x)=\int f_{mb}(x,v)\,dv=\sqrt{\frac{2\pi}{\beta}}C\mathrm{e}^{-\beta \omega(x)}\,,
\end{equation}
%%%%%%%%%%%%%%%%%%%%%
where $\beta$ is the Lagrange multiplier used to conserve total energy, $C$ is the normalization constant and $\omega(x)$ is the potential of mean force \cite{Lev2002}. As $N \rightarrow \infty$, interparticle correlations vanish and 
$\omega(x)\sim\psi(x)$. Substituting equation \eqref{eq:g1dbolt1} into equation \eqref{eq:g1dpoi1}, we obtain the Poisson-Boltzmann equation in its dimensionless form
%%%%%%%%%%%%%%%%%%%%%
\begin{equation}
\nabla^2 \psi_{eq}(x)=\sqrt{\frac{8\pi}{\beta}}C e^{-\beta \psi_{eq}(x)} \,.
\end{equation}
%%%%%%%%%%%%%%%%%%%%%
Solving this equation using the boundary conditions $\lim_{|x|\to\infty}\psi_{eq}(x)=|x|$ and ${\psi'}_{eq}(0)=0$ %\cite{JoyWor2011}
, the potential is found to be \cite{Ryb1971}
%%%%%%%%%%%%%%%%%%%%%
\begin{equation}\label{eq:g1deqpot}
\psi_{eq}(x)=-\frac{1}{\beta}\ln\left[\frac{1}{4}\sech^2\left(\frac{\beta x}{2}\right)\right],
\end{equation}
%%%%%%%%%%%%%%%%%%%%%
and the distribution function is given by
%%%%%%%%%%%%%%%%%%%%%
\begin{equation}\label{eq:g1deqdist}
f_{eq}(x,v)=\sqrt{\frac{\beta^3}{32\pi}}\mathrm{e}^{-\frac{\beta v^2}{2}}\sech^2 \left[\frac{\beta x}{2}\right].
\end{equation}
%%%%%%%%%%%%%%%%%%%%%
The value of $\beta$ is determined by the conservation of energy, Eq. \eqref{eq:encons} with 
$\bar{f}(x,v)=f_{eq}(x,v)$,  yielding
%%%%%%%%%%%%%%%%%%%%%
\begin{equation}\label{eq:g1deqencons}
\beta=\frac{3}{2\mathcal{E}}.
\end{equation}
%%%%%%%%%%%%%%%%%%%%%
The equilibrium density and velocity distributions are given by
%%%%%%%%%%%%%%%%%%%%%
\begin{equation}\label{eq:g1dmbpos}
n(x)=\frac{\beta}{4}\sech^2\left(\frac{\beta x}{2}\right)
\end{equation}
%%%%%%%%%%%%%%%%%%%%%
and
%%%%%%%%%%%%%%%%%%%%%
\begin{equation}\label{eq:g1dmbvel}
n(v)=\sqrt{\frac{\beta}{2\pi}}e^{-\beta v^2/2}.
\end{equation}
%%%%%%%%%%%%%%%%%%%%%
In  Fig \ref{fig:g1deqqss} we compare the equilibrium distributions, Eqs \eqref{eq:g1dmbpos} and \eqref{eq:g1dmbvel}, with the results
of MD simulations.  As can be seen, the predictions of equilibrium statistical mechanics are
very different from those of MD simulations.  This clearly shows that the ergodicity
required by the Boltzmann-Gibbs statistical mechanics is violated.  

%%%%%%%%%%%%%%%% Figure%%%%%%%%%%%%%%%%%%%%%
%%%%%%%%%%%%% End of figure%%%%%%%%%%%%%%%%%

In the next section
we will compare the predictions of Lynden-Bell statistics with the results of MD simulations.

\subsection{Lynden-Bell theory for one-dimensional gravity}

The application of Lynden-Bell statistics to one-dimensional gravitational systems has spanned various decades, with divergent results. While early studies have suggested some correspondence between numerical simulations and the predictions of LB statistics, especially for low-energies, they have also shown the occurrence of high-energy tails in the distribution, which LB statistics could not describe \cite{HohCam1968,CupGol1969,GolCup1969,LecCoh1971,AarLec1975}.  
The more recent works demonstrated that although for some very specific 
initial conditions LB theory agrees well with
MD simulations, in general it fails to describe the qSS \cite{MinFei1990,BinSec2008,Yam2008,JoyWor2011}. In this section we will examine the predictions of LB statistics and compare them with the results of MD simulations for various initial conditions.

\begin{figure}[!htb]
\vspace{5mm}
\begin{center}
\includegraphics[width=0.8\textwidth]{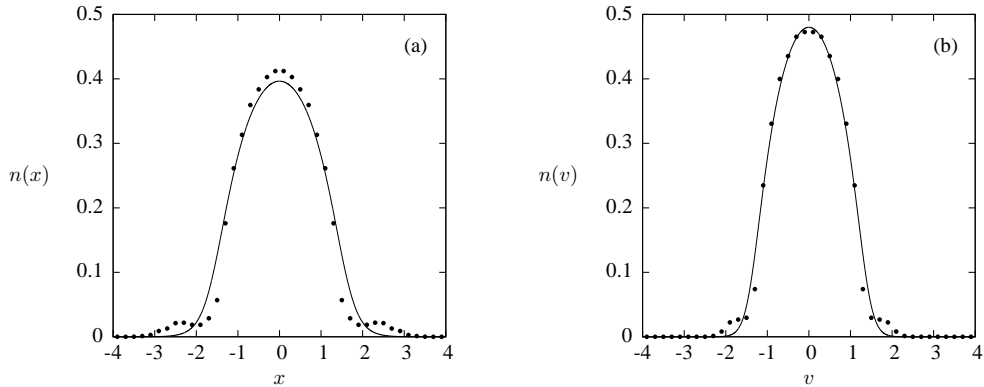}%{figg1dlynqss}
\end{center}
\caption{Distributions in (a) position and (b) velocity for a 1D gravitational system with $\mathcal{E}_0=0.75$, obtained using MD simulations (points), averaged over $t=1000\tau_D$ and $t=1100\tau_D$, compared with the LB distributions (lines), $n(x)=\int f_{lb}(x,v)dv$ and $n(v)=\int f_{lb}(x,v)dx$ with $f_{lb}(x,v)$ given by Eq \eqref{eq:flb}. Error bars are smaller than the symbol size.\label{fig:g1dlynqss}}
\end{figure}

In order to determine $f_{lb}(x,v)$, Eq \eqref{eq:flb}, for a one-dimensional gravitational system, we 
need to calculate the gravitational potential $\psi_{lb}(x)$.  To do this we must solve the Poisson equation (equation \eqref{eq:poisson}) with $f(x,v)=f_{lb}(x,v)$ and the one-particle energy given by
$\epsilon(x,v)=v^2/2+\psi_{lb}(x)$.  Integrating the LB distribution over
momentum, we obtain the Poisson equation
%%%%%%%%%%%%%%%%%%%%%
%\begin{equation}\label{eq:g1dpoisson1}
%\frac{d^2\psi_{lb}(x)}{dx^2}=2\int dv \frac{\eta}{1+e^{\beta[v^2/2+\psi_{lb}(x)-\mu]}},
%\end{equation}
%%%%%%%%%%%%%%%%%%%%%
%or
%%%%%%%%%%%%%%%%%%%%%
\begin{equation}\label{eq:g1dpoisson2}
\frac{d^2\psi_{lb}(x)}{dx^2}=-\sqrt{\frac{8\pi}{\beta}}\eta \Li_{1/2}\left[-e^{-\beta(\psi_{lb}(x)-\mu)}\right],
\end{equation}
%%%%%%%%%%%%%%%%%%%%%
with boundary conditions $\lim_{|x|\to\infty}\psi_{lb}(x)=|x|$ and ${\psi'}_{lb}(0)=0$, where $\Li_n(x)$ is the polylogarithm function of order $n$ \cite{Duf2001}. The solution to this equation is obtained numerically. We see that the predictions of LB statistics are in general quite different 
from the results of MD simulations, as exemplified in Fig \ref{fig:g1dlynqss}, which compares the position and the velocity distributions $n(x)=\int f_{lb}(x,v)dv$ and $n(v)=\int f_{lb}(x,v)dx$ with the results of MD simulations.

%%%%%%%%%%%%%%%% Figure%%%%%%%%%%%%%%%%%%%%%
%%%%%%%%%%%%% End of figure%%%%%%%%%%%%%%%%%

%In cases where the mixing process of the distribution function in phase space is not complete, violent relaxation theory becomes inadequate in describing the quasi-stationary states. 

The problem, common to both BG and LB statistics, is that in thermodynamic limit, systems with LR forces are intrinsically non-ergodic, invalidating the basic assumptions that underlie both theories.  For systems with a finite number of particles, however, ergodicity is
restored on a sufficiently long time scale.  Such systems
will eventually relax to the BG equilibrium (if it exists, and the BG entropy has
a maximum), after being trapped in a qSS for a time proportional to the number
of particles in the system.

The Kac scaling required by the LR nature of the interaction potential
destroys the correlations (collisions) between the particles~\cite{Lev2002}.  Therefore, in  thermodynamic limit, LR systems are
intrinsically collisionless  -- particles move under the action of the mean-field potential produced by all the other particles. In general, the mean-field potential has a complex
dynamics, characterized by quasi-periodic oscillations \cite{Mat1990}. It is possible, therefore, for some particle to enter in resonance with the oscillations and gain large amounts of energy at the expense of the collective motion \cite{Wu1962,KadPog1970,Sag1994}. This process is known as Landau damping~\cite{Lan1946}.   
The Landau damping diminishes the amplitude of the oscillations
and leads to the formation of a tenuous halo of highly energetic particles which surround the high density core \cite{Glu1994}.  After all the oscillations have died out, a SS state is established.   The phase space
distribution of particles in the qSS has a characteristic core-halo structure, very different from the predictions of either BG or LB statistics. Once the stationary state is established, there
is no longer a mechanism through which highly energetic particles of the halo can equilibrate with the particles of the core, and the ergodicity is broken.

\subsection{The virial condition}\label{subsec:virial}

If the system is in a stationary state, it must satisfy the virial theorem. 
Consider a system with a Hamiltonian given by
%%%%%%%%%%%%%%%%%%%%%
\begin{equation}\label{eq:hamgeral}
\mathcal{H}=\sum_i\frac{\mathbf{p}_i^2}{2m_i}+\frac{1}{2}\sum_{i, j}V(\mathbf{r}_i-\mathbf{r}_j)+\frac{\kappa}{2}\sum_{i=1}^N |\mathbf{r}_i|^{\gamma}
\end{equation}
%%%%%%%%%%%%%%%%%%%%%
where $(\mathbf{r}_i,\mathbf{p}_i)$ are respectively the coordinates of position and momentum of the $i$th particle, $V(\mathbf{r}_i-\mathbf{r}_j)$ is the interaction potential and $(\kappa/2)\sum_{i=1}^N(\mathbf{r}_i)^{\gamma}$ is a generic confining potential. The virial function $I$ is defined as
%%%%%%%%%%%%%%%%%%%%%
\begin{equation}\label{eq:virial}
I=\left\langle \sum_i \mathbf{r}_i\cdot \mathbf{p}_i\right\rangle,
\end{equation}
%%%%%%%%%%%%%%%%%%%%%
where $\langle x \rangle$ represents a time average. Differentiating the virial function with respect to time and using Hamilton's equations \cite{GolPoo2001}, we find
%%%%%%%%%%%%%%%%%%%%%
\begin{equation}\label{eq:didt}
\frac{d}{dt}I=\left\langle\sum_i \frac{\mathbf{p}_i^2}{m_i}\right\rangle-\left\langle\sum_i \mathbf{r}_i\cdot \frac{\partial}{\partial \mathbf{r}_i}\left(\tilde{V}+\frac{\kappa}{2}\sum_{j}|\mathbf{r}_j|^{\gamma}\right)\right\rangle,
\end{equation}
%%%%%%%%%%%%%%%%%%%%%
where
%%%%%%%%%%%%%%%%%%%%%
\begin{equation}\label{eq:vdef}
\tilde{V}=\frac{1}{2}\sum_{j, k}V(\mathbf{r}_j-\mathbf{r}_k).
\end{equation}
%%%%%%%%%%%%%%%%%%%%%
If $\tilde{V}$ is a homogeneous function of order $p$, that is, $\tilde{V}(\mathbf{r})=\lambda^{-p}\tilde{V}(\lambda \mathbf{r})$, then by Euler's theorem,
%%%%%%%%%%%%%%%%%%%%%
\begin{equation*}
p\tilde{V}=\sum_i\mathbf{r}_i\cdot\frac{\partial}{\partial \mathbf{r}_i}\tilde{V}.
\end{equation*}
%%%%%%%%%%%%%%%%%%%%%
For a stationary state, $dI/dt=0$, which determines the virial condition 
%%%%%%%%%%%%%%%%%%%%%
\begin{equation}\label{eq:virialcondition}
2 K-p U-\frac{\gamma\kappa}{2} r_m^{\gamma}=0
\end{equation}
%%%%%%%%%%%%%%%%%%%%%
where $K=\frac{1}{N}\langle \sum_i^N \mathbf{p}^2_i/2m_i\rangle$ is the average kinetic energy
per particle in a SS, $U=\frac{1}{N}\langle \tilde{V}\rangle$ is the average potential energy per particle in a SS, and $r_m^\gamma=\frac{1}{N}\langle \sum_i^N |\mathbf{r}_i |^\gamma \rangle$. In the case of two-dimensional gravity\footnote{The specific case of two-dimensional gravity is addressed in Ref \cite{ChaSir2006}, which presents a study of the virial theorem in the general case of $d$ dimensions and includes terms for friction and noise.}, which will be discussed in section \ref{sec:grav2d}, the interaction potential is logarithmic, $V=2Gm^2\ln(|\mathbf{r}_i-\mathbf{r}_j|)$, and is not a homogeneous function.    However, writing the logarithm as $\ln x=\lim_{p\to 0}\left(\frac{x^p}{p}-\frac{1}{p}\right)$, after some manipulation (see \cite{TelLev2010}), we find
%%%%%%%%%%%%%%%%%%%%%
%\begin{equation}\label{eq:g2dv1}
%V\sim\lim_{p\to 0}\frac{|\mathbf{r}_i-\mathbf{r}_j|^p}{p}
%\end{equation}
%%%%%%%%%%%%%%%%%%%%%
%which is the logarithm plus an infinite constant. Therefore,
%%%%%%%%%%%%%%%%%%%%%
%\begin{equation}\label{eq:g2dvtil}
%\tilde{V}=2Gm^2 \lim_{p\to 0}\frac{1}{2}\sum_{i,j}^N\frac{|\mathbf{r}_i-\mathbf{r}_j|^p}{p},
%\end{equation}
%%%%%%%%%%%%%%%%%%%%%
%and
%%%%%%%%%%%%%%%%%%%%%
%\begin{equation}\label{eq:g2deuler}
%p\tilde{V}=Gm^2\lim_{p\to 0}\sum_{i,j}^N|\mathbf{r}_i-\mathbf{r}_j|^p.
%\end{equation}
%%%%%%%%%%%%%%%%%%%%%
%Therefore, according to the Euler theorem,
%%%%%%%%%%%%%%%%%%%%%
\begin{equation}\label{eq:g2deulerfinal}
GM^2\frac{(N-1)}{N}=\sum_i\mathbf{r}_i\cdot\frac{\partial}{\partial \mathbf{r}_i}\tilde{V}.
\end{equation}
%%%%%%%%%%%%%%%%%%%%%
Using equation \eqref{eq:g2deulerfinal} in equation \eqref{eq:didt}, the virial condition for a 2D gravitational system is found to be
%%%%%%%%%%%%%%%%%%%%%
\begin{equation}\label{eq:g2dvirial_general}
\langle v^2 \rangle=GM\frac{N-1}{N}
\end{equation}
%%%%%%%%%%%%%%%%%%%%%
where we have set $\kappa=0$ in equation \eqref{eq:hamgeral}.

In 1D the gravitational potential is a homogeneous function of order $p=1$, so that the virial condition reduces to
%%%%%%%%%%%%%%%%%%%%%
\begin{equation}\label{eq:g1dvirial}
2K=U.
\end{equation}
%%%%%%%%%%%%%%%%%%%%%

If at $t=0$ the initial distribution function is not a stationary solution of the Vlasov equation,
the system will undergo oscillations.  When the relaxation is completed and a qSS is established, equation \eqref{eq:g1dvirial} must be satisfied.  However, even if the initial distribution function does
not satisfy the stationary Vlasov equation --- as is the case for the waterbag distributions considered above  --- we can significantly diminish the amplitude of oscillations during
the relaxation process if the initial distribution is forced to 
satisfy the virial condition, Eq \eqref{eq:g1dvirial}.  For such
distributions, even though the initial state is not stationary, it is not ``too far'' from a qSS.    
To quantify this, we define the virial number for 1D gravity as $\mathcal{R}=2K/U$. When $\mathcal{R}=1$, the virial condition is satisfied and the oscillations should be suppressed; on the other hand, if $\mathcal{R}\neq 1$, the system will experience strong density oscillations due to the imbalance between the kinetic and the potential energies.  We expect that the process of relaxation to the qSS should be quite different for these two cases. 
%%%%%%%%%%%%%%%% Figure%%%%%%%%%%%%%%%%%%%%%
\begin{figure}[!htb]
\vspace{5mm}
\begin{center}
\includegraphics[width=0.75\textwidth]{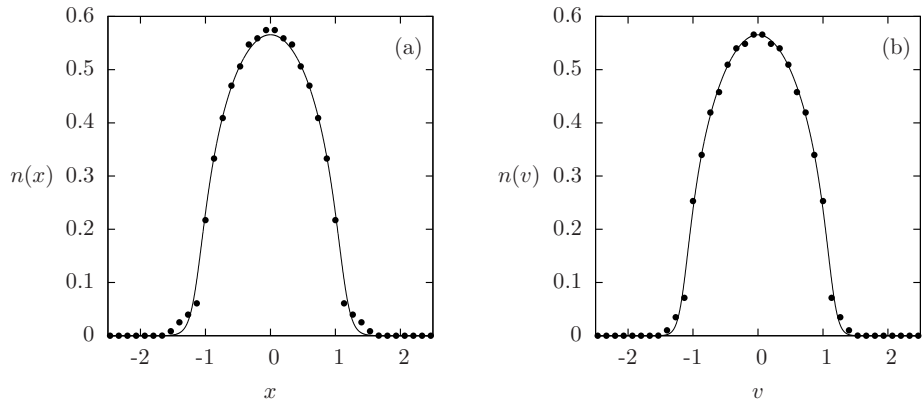}%{figg1d_lbdist}
\end{center}
\caption{Distributions in (a) position and (b) velocity of a system that initially was in a waterbag distribution with $\mathcal{R}_0=1$. The solid line represents the predictions of LB theory, Eq. \eqref{eq:flb},
while the points are results of MD simulation, averaged over $t=1000\tau_D$ to $t=1100\tau_D$. For this case, LB theory provides a fairly accurate approximation for the qSS distribution, despite a small deviation in the distribution tails. Error bars in the distributions are comparable to the symbol size.\label{fig1dlyn}}
\end{figure}
%%%%%%%%%%%%% End of figure%%%%%%%%%%%%%%%%%
Indeed, we find that 
when $\mathcal{R}_0 = 1$, where $\mathcal{R}_0$ is the virial number at time $t=0$, the resulting qSS has
a compact structure, which is reasonably well captured by LB theory, see Fig. \ref{fig1dlyn}.  On the other hand
when $\mathcal{R}_0 \neq 1$, the system separates into a central core surrounded by a halo of highly energetic particles.  To understand the mechanism of the core-halo formation
we need to explore the parametric resonances which appear as a result of the density oscillations.

\subsection{Envelope equation}

To explore the density oscillations, we define the envelope $x_e(t)$ to be the initial size of the system, $x_e(t) \equiv \sqrt{3\langle x^2(t) \rangle}$. Note that at $t=0$, the envelope $x_e(t)$ coincides with 
the boundary of the initial waterbag distribution, $x_e(0)=1$. 
Differentiating $x_e(t)$ twice with respect to time, we have
%%%%%%%%%%%%%%%%%%%%%
\begin{equation}\label{eq:g1denvdd}
\ddot{x}_e(t)=\frac{3\langle x(t) \ddot{x}(t)\rangle}{x_e(t)}+\frac{3\langle \dot{x}^2(t)\rangle}{x_e(t)}-\frac{9\langle x(t) \dot{x}(t)\rangle^2}{x_e^3(t)}.
\end{equation}
%%%%%%%%%%%%%%%%%%%%%

\begin{figure}[!htb]
\vspace{5mm}
\begin{center}
\includegraphics[width=0.45\textwidth]{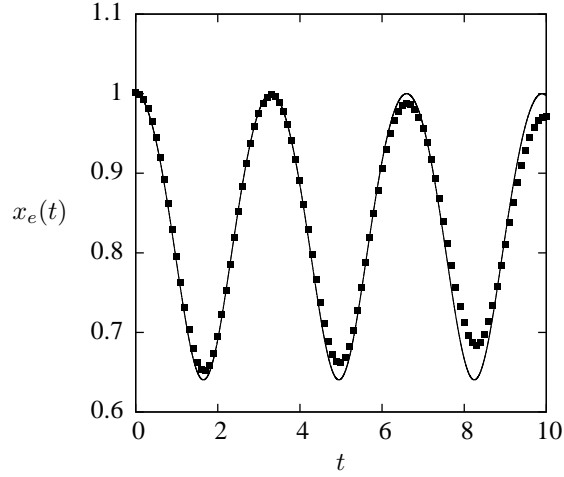}%{figg1denve.eps}
\end{center}
\caption{Oscillations of the envelope $x_e(t)$ determined by equation \eqref{eq:g1denvmove} (solid line) compared to results of MD simulation (squares). The virial number is $\mathcal{R}_0=0.5$.\label{fig:g1denv}}
\end{figure}

To simplify the first term, we suppose that the mass density oscillations are smooth, so that the particle distribution remains uniform. Under these conditions, the oscillating gravitational potential $\psi_e(x,t)$ maintains the functional form given by Eq. \eqref{eq:g1dmdpot}, but with $x_m \rightarrow x_e(t)$,
%%%%%%%%%%%%%%%%%%%%%
\begin{equation}\label{eq:g1denvpot}
\psi_e(x,t)=
  \begin{cases}
    \frac{x^2}{2x_e(t)}+\frac{x_e(t)}{2} & \text{for } |x| \leq x_e(t) \\
    |x| & \text{for } |x| \geq x_e(t).
  \end{cases}
\end{equation}
%%%%%%%%%%%%%%%%%%%%%
Similarly, the distribution function will be approximated by a waterbag
%%%%%%%%%%%%%%%%%%%%%
\begin{equation}\label{eq:wb}
f_e(x,v,t)=\eta_e \Theta(x_e(t)-|x|)\Theta(v_m-|v|)
\end{equation}
%%%%%%%%%%%%%%%%%%%%%
with $\eta_e=[4x_e(t)v_m]^{-1}$. The average $\langle x \ddot{x}\rangle$ can then be expressed as
%%%%%%%%%%%%%%%%%%%%%
\begin{align}
\langle x \ddot{x} \rangle &= -\left\langle x\frac{d}{dx}\psi_e(x,t)\right\rangle \nonumber \\
&= -\int x\frac{d}{dx}\psi_e(x,t)f_e(x,v,t)\,dx\,dv \nonumber \\
&= -\frac{1}{2x_e(t)}\int_{-x_e(t)}^{x_e(t)}\frac{x^2}{x_e(t)}\,dx,
\end{align}
%%%%%%%%%%%%%%%%%%%%%
resulting in
%%%%%%%%%%%%%%%%%%%%%
\begin{equation}
\langle x \ddot{x}\rangle=-\frac{x_e(t)}{3}.
\end{equation}
%%%%%%%%%%%%%%%%%%%%%
The second and the third terms of Eq. \eqref{eq:g1denvdd} are
%%%%%%%%%%%%%%%%%%%%%
\begin{equation}
\langle \dot{x}^2\rangle=\frac{1}{2v_m}\int_{-v_m}^{v_m}v^2\,dv=\frac{v_m^2}{3}
\end{equation}
%%%%%%%%%%%%%%%%%%%%%
and
%%%%%%%%%%%%%%%%%%%%%
\begin{equation}
\langle x \dot{x}\rangle=\frac{1}{4 x_e(t) v_m}\int_{-x_e(t)}^{x_e(t)}x\,dx\int_{-v_m}^{v_m}v\,dv=0,
\end{equation}
%%%%%%%%%%%%%%%%%%%%%
considering that at $t=0$ there is no correlation between position and velocity.
The envelope equation reduces to
%%%%%%%%%%%%%%%%%%%%%
\begin{equation}\label{eq:g1denvmove}
\ddot{x}_e(t)=\frac{\mathcal{R}_0}{x_e(t)}-1,
\end{equation}
%%%%%%%%%%%%%%%%%%%%%
where $\mathcal{R}_0=2 K(t=0)/U(t=0)=v_m^2$ and the initial conditions are $x_e(0)=1$ and
$\dot{x}_e(t)=0$. 
If $\mathcal{R}_0=1$, then $\ddot{x}_e(t)=0$, and the system does not develop oscillations.
Fig \ref{fig:g1denv} compares the oscillations of the envelope predicted by Eq \eqref{eq:g1denvmove} with
the results of MD simulation, showing a reasonable agreement for short times.

%%%%%%%%%%%%%%%% Figure%%%%%%%%%%%%%%%%%%%%%
%%%%%%%%%%%%% End of figure%%%%%%%%%%%%%%%%%

\subsection{The test particle model}

To understand the mechanism of halo formation, we first study the dynamics of noninteracting test particles initially located at positions $x_i^0 \in [-1,1]$ with velocities $v_i^0 \in [-v_m,v_m]$, where $v_m=\sqrt{\mathcal{R}_0}$. Each particle moves in a gravitational 
potential produced by the oscillating mass density
\begin{equation}
\rho(t)=\frac{1}{2 x_e(t)} \Theta(x_e(t)-|x|)
\end{equation}
%%%%%%%%%%%%%%%%%%%%%
where $x_e(t)$ is governed by the envelope equation, Eq \eqref{eq:g1denvmove}.  
The trajectory of a test particle is then determined by the equation of motion
\begin{equation}\label{eq:g1dtpmove}
\ddot{x}_i(t)=
  \begin{cases}
    -\frac{x_i(t)}{x_e(t)} & \text{for } |x_i(t)| \leq x_e(t) \\
    -\sgn [x_i(t)-x_e(t)] & \text{for } |x_i(t)| \geq x_e(t)
  \end{cases}
\end{equation}
%%%%%%%%%%%%%%%%%%%%%
where  $\sgn$ is the sign function \cite{ArfWeb2001}.

%%%%%%%%%%%%%%%% Figure%%%%%%%%%%%%%%%%%%%%%
\begin{figure}[!htb]
\vspace{5mm}
\begin{center}
\includegraphics[width=0.99\textwidth]{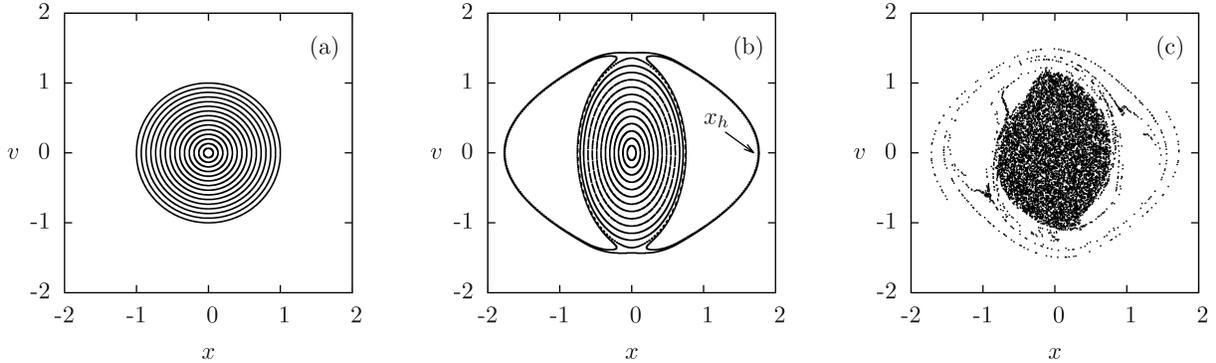}%{figg1dpoins.eps}
\end{center}
\caption{Poincar\'e sections of test particle dynamics (see Eq \eqref{eq:g1dtpmove}) for $\mathcal{R}_0\approx 1$ (a) and $\mathcal{R}_0=0.5$ (b). In (a) the dynamics is integrable while in (b), two resonance islands are formed. Panel (c) shows the phase space obtained using MD simulation of a 1D self-gravitating system with $\mathcal{R}_0=0.5$ at $t=7$. The test particle dynamics enables us to determine the maximum energy $\epsilon_h$ that a particle in the full $N$-body simulation can achieve. In the case of 1D gravitation, $\epsilon_h=|x_h|$, where $x_h$, indicated in panel (b), is the maximum position reached by a test particle.\label{fig:g1dpoin1}}
\end{figure}
%%%%%%%%%%%%% End of figure%%%%%%%%%%%%%%%%%

%%%%%%%%%%%%%%%% Figure%%%%%%%%%%%%%%%%%%%%%
%\begin{figure}[!htb]
%\begin{center}
%\includegraphics[width=12cm]{figpoincnb1d.eps}
%\end{center}
%\caption{Phase space at $t=7$ of a system with $\mathcal{R}_0=0.5$ (a) and the Poincar\'e section of the test particle dynamics for $%\mathcal{R}_0=0.5$ (b). The test particle method enables us to determine the maximum energy $\epsilon_h$ that a particle in the full $N$-body dynamics can obtain. In the case of 1D gravitation, $\epsilon_h=|x_h|$, where $x_h$ is the maximum position reached by a test particle.\label{fig:g1dpoin2}}
%\end{figure}
%%%%%%%%%%%%% End of figure%%%%%%%%%%%%%%%%%

In Fig \ref{fig:g1dpoin1} we show the 
Poincar\'e sections for test particle dynamics. When $\mathcal{R}_0=1$, the trajectories of the test particles correspond
to harmonic oscillators and the dynamics
is completely regular; on the other hand, when $\mathcal{R}_0\neq 1$, we see the appearance
of resonance islands. At short times a very similar structure of the phase space is also found in the complete MD simulation, as shown in panel (c) of the same figure. The formation of resonance islands is the result of the particle(density)-wave interactions \cite{SimRiz2006,RizPak2007,RizPak2009}. The parametric resonances
allow some particles to move into the regions of the phase space which are highly
improbable from the perspective of BG or LB statistics.  Once the oscillations die out,
these particles are trapped, becoming a part of a halo.   

\subsection{The core-halo distribution}

From the Jeans theorem, a steady-state solution of the Vlasov equation depends on the phase space coordinates only through the integrals of motion of the mean-field potential. Conversely, any function of the integrals of motion is a steady-state solution of the Vlasov equation \cite{BinTre2009}. In all the cases treated in this Report, the only integral of motion is the one-particle energy.
Thus, a Maxwell-Boltzmann distribution is only one of the infinite number of solutions of the Vlasov
equation.  In particular an arbitrary initial distribution will not converge to the Maxwell-Boltzmann distribution, as is the case for systems with finite-range forces.

Unlike gravitation in three dimensions, in 1D particles cannot escape to infinity. The test particle dynamics shows, however, 
that the resonant particles may gain a lot of energy from collective oscillations and
form a tenuous high-energy halo that surrounds the central core region.  
Since the Hamiltonian dynamics is conservative, the gain of energy of resonant particles must result in the loss of energy (cooling down) of the core particles.  In principle, the halo formation will
continue until the oscillations of the core have completely died down.  Once the SS state is established, the core particles should be in the "ground state".  The incompressibility constraint
imposed by the Vlasov dynamics, however, does not allow the core particles to collapse to the
minimum of the potential energy. Rather, these particle will arrange in such a way as to
occupy all of the low energy states up to the allowed maximum 
phase space density $\eta$,   
%%%%%%%%%%%%%%
\begin{equation}
\bar{f}_{core}(x,v)=\eta\Theta(\epsilon_F-\epsilon(x,v)),
\end{equation}
%%%%%%%%%%%%%%
where $\epsilon_F$ is the "Fermi energy" of the core.

\begin{figure}[!htb]
\vspace{5mm}
\begin{center}
\includegraphics[width=0.75\textwidth]{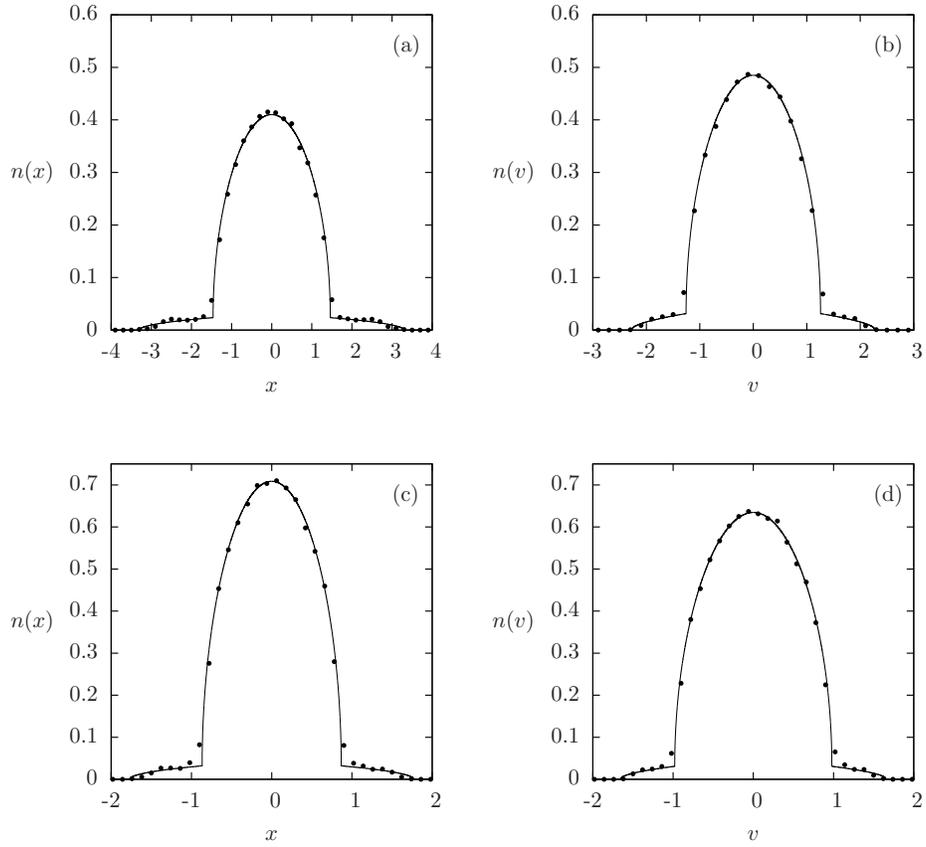}%{figg1dchdist.eps}
\end{center}
\caption{Distributions inside qSS in (a) position and (b) velocity, for a system with $\mathcal{R}_0=2.5$, and distributions in (c) position and (d) velocity for a system with $\mathcal{R}_0=0.5$. Points show the results of MD simulations averaged over $t=1000\tau_D$ to $t=1100\tau_D$, and the solid lines correspond to the marginal distributions predicted by the core-halo theory. Error bars in the distributions are comparable to the symbol size.\label{figg1dchdist}}
\end{figure}

The maximum energy that a halo particle can gain corresponds to the resonant orbit.  As the oscillations die down, the resonances shift toward the smaller energies, resulting in a quasi-homogeneous
population of the phase space between $\epsilon_F$ and the maximum halo energy, $\epsilon_h$.
We will, therefore,  suppose that in a qSS the halo particles are distributed according to
%%%%%%%%%%%%%%
\begin{equation}
\bar{f}_{halo}(x,v)=\chi\Theta(\epsilon(x,v)-\epsilon_F)\Theta(\epsilon_h-\epsilon(x,v)),
\end{equation}
%%%%%%%%%%%%%%
where $\chi$ is the phase space density of the halo particles and the maximum halo energy, $\epsilon_h$, 
can be calculated using the test particle dynamics and is given by $\epsilon_h=|x_h|$, see Fig \ref{fig:g1dpoin1}.
The complete core-halo distribution is then
%%%%%%%%%%%%%%
\begin{equation}\label{eq:g1dfch}
\bar{f}_{ch}(x,v)=\eta\Theta(\epsilon_F-\epsilon(x,v))+\chi\Theta(\epsilon(x,v)-\epsilon_F)\Theta(\epsilon_h-\epsilon(x,v)).
\end{equation}
%%%%%%%%%%%%%%
From now on, for simplicity we will write $f_{ch}$ instead of $\bar{f}_{ch}$. After determining $\epsilon_h$ using the test particle dynamics, two unknowns remain, $\epsilon_F$ and $\chi$, which are obtained using the conservation of the
total energy and the number of particles in the system.
Integrating the core-halo distribution function over velocities and substituting the resulting particle density into Poisson equation, 
the gravitational potential is found to satisfy
%%%%%%%%%%%%%%%%%%%%%
\begin{equation}\label{eq:g1dpoissonch}
\frac{d^2}{dx^2}\psi_{ch}(x)=2\sqrt{2}
  \begin{cases}
    (\eta-\chi)\sqrt{\epsilon_F-\psi_{ch}(x)}+\chi\sqrt{\epsilon_h-\psi_{ch}(x)} & \text{for } \psi_{ch}(x) \leq \epsilon_F, \\
    \chi\sqrt{\epsilon_h-\psi_{ch}(x)} & \text{for } \epsilon_F\leq \psi_{ch}(x) \leq \epsilon_h, \\
    0 & \text{for } \psi_{ch}(x)\geq \epsilon_h,
  \end{cases}
\end{equation}
%%%%%%%%%%%%%%%%%%%%%
with the boundary conditions given by $\lim_{|x| \to \infty} \psi_{ch}(x)= |x|$ and ${\psi'}_{ch}(0)=0$. The parameters $\chi$ and $\epsilon_F$ are determined self-consistently from the numerical solution of equation \eqref{eq:g1dpoissonch} and the conservation of the total energy and the number of particles (equations \eqref{eq:encons} and \eqref{eq:normcons}) in the system.  Once the potential 
is known we can easily calculate the distributions $n(x)=\int f_{ch}(x,v)\,dv$ and $n(v)=\int f_{ch}(x,v)\,dx$, see Fig. \ref{figg1dchdist}.

%%%%%%%%%%%%%%%% Figure%%%%%%%%%%%%%%%%%%%%%
%%%%%%%%%%%%% End of figure%%%%%%%%%%%%%%%%%

\subsection{Thermodynamic equilibrium}

\begin{figure}[!htb]
\vspace{5mm}
\begin{center}
\includegraphics[width=0.8\textwidth]{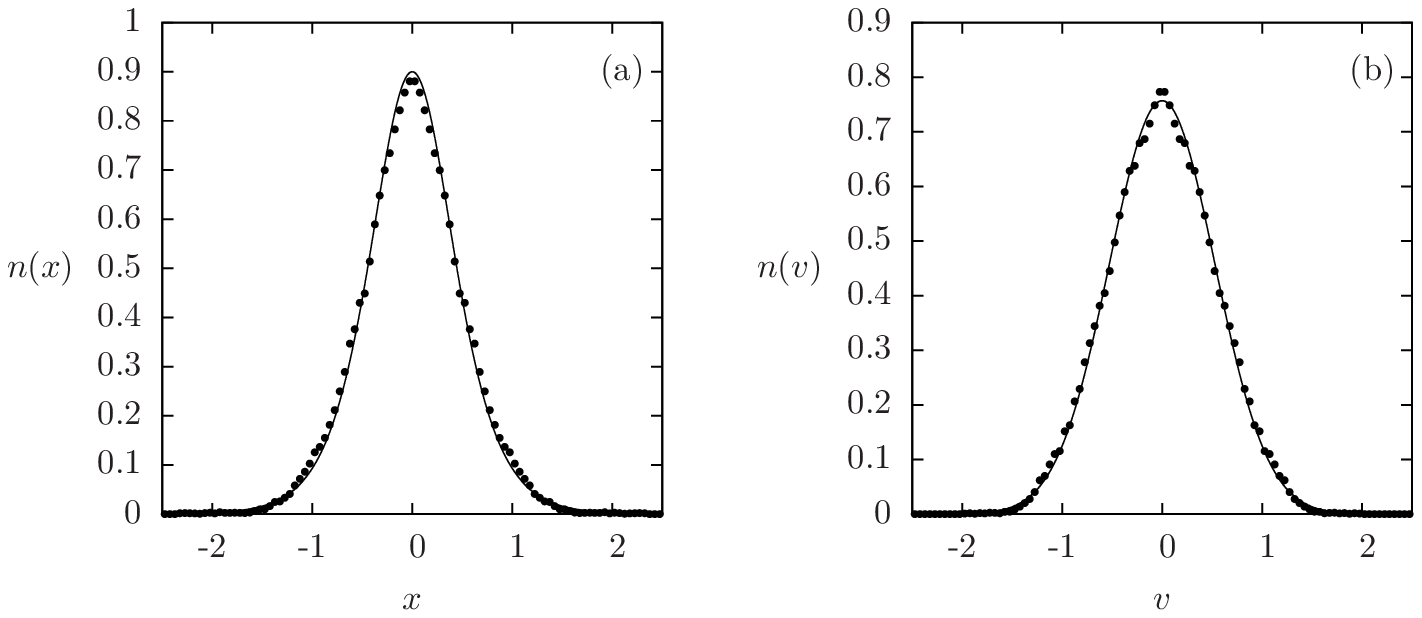}
\end{center}
\caption{Equilibrium distributions: (a) position and (b) velocity, for a system with $\mathcal{R}_0=2.5$ at time $t=3\times 10^6\tau_D$, obtained using MD simulation (points). Solid lines are the predictions of BG statistics, Eqs. \eqref{eq:g1dmbpos} and \eqref{eq:g1dmbvel}.\label{fig:g1deqreached}}
\end{figure}

For finite $N$, correlations are not completely negligible and eventually they will drive the system to thermodynamic equilibrium. The equilibrium state should be described by the MB distribution Eq \eqref{eq:g1deqdist}, discussed in subsection \ref{subsec:g1deq}. 
Therefore if the number of particles in the system is not too large and  the simulation is run for a sufficiently long time 
the thermodynamic equilibrium  should be observed. Figure \ref{fig:g1deqreached} shows the results of 
MD simulation for $t=3\times 10^6 \tau_D$. We see that after this time the system indeed relaxes to the thermodynamic equilibrium with the particle distribution given by Eq \eqref{eq:g1deqdist}.

%%%%%%%%%%%%%%%% Figure%%%%%%%%%%%%%%%%%%%%%
%%%%%%%%%%%%% End of figure%%%%%%%%%%%%%%%%%

%%%%%%%%%%%%%%%% Figure%%%%%%%%%%%%%%%%%%%%%
\begin{figure}[!htb]
\vspace{5mm}
\begin{center}
\includegraphics[width=0.85\textwidth]{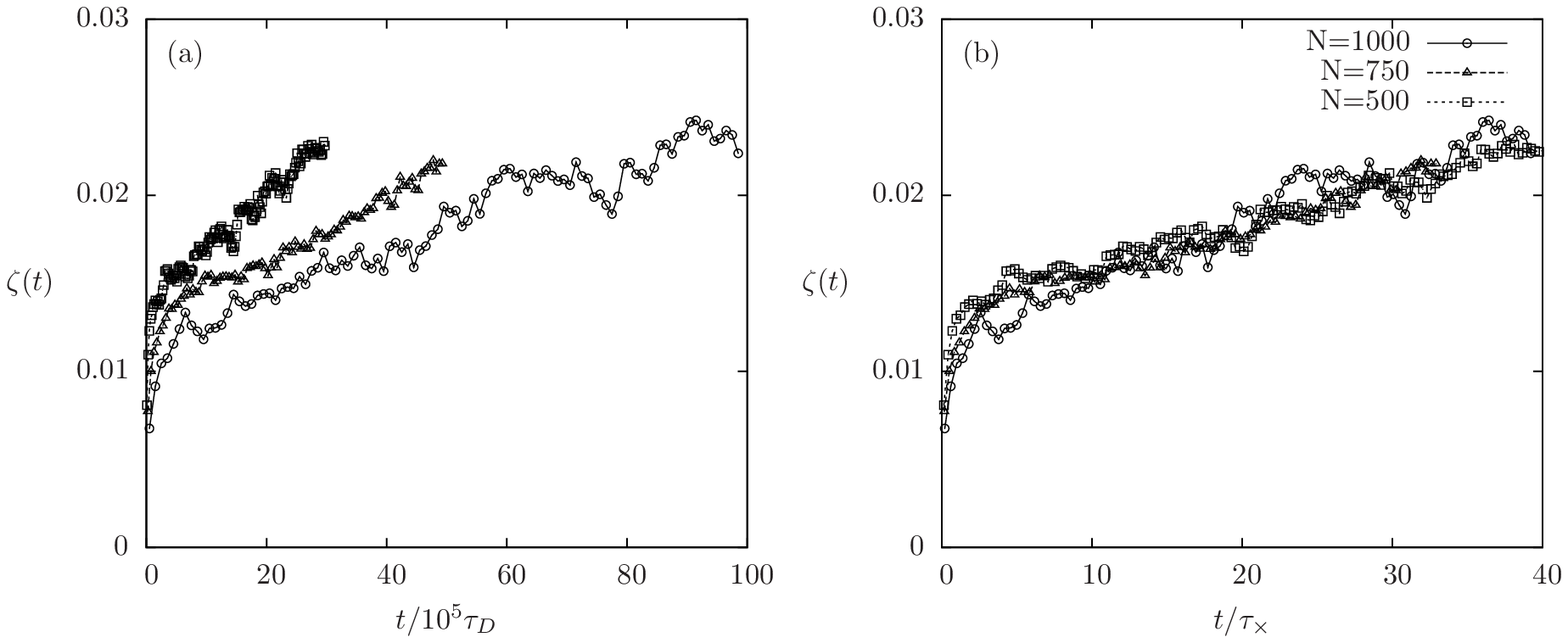}
\end{center}
\caption{Relaxation to equilibrium, shown by the crossover parameter $\zeta(t)$, Eq \eqref{eq:g1dzeta}, with time rescaled by $10^5\tau_D$ in (a) and by $\tau_{\times}=\tau_D N^{1.8}$ in (b). In this case, the equilibrium value $\zeta_{eq}$ is approximately $0.032$. The virial number is $R_0=0.5$.\label{fig:g1dzeta}}
\end{figure}
%%%%%%%%%%%%% End of figure%%%%%%%%%%%%%%%%%

The approach to equilibrium can be observed using a crossover parameter, $\zeta(t)$, which measures how well the system's density profile is described by the core-halo distribution $f_{ch}(x,v)$, Eq \eqref{eq:g1dfch} at each instant. We define
%%%%%%%%%%%%%%%%%
\begin{equation}\label{eq:g1dzeta}
\zeta(t)=\frac{1}{N^2}\int{[N(x,t)-N_{ch}(x)]^2{ d}x}
\end{equation}
%%%%%%%%%%%%%%%%%
where $ N(x,t)$ is the number of particles located between $x$ and $x+{ d}x$ at time $t$ and $N_{ch}(x)=N\int{f_{ch}(x,v){ d}v}$. The smaller the value of $\zeta(t)$, the better the agreement between the system's marginal distribution in position and the predicted distribution of the core-halo theory. When the system starts to cross over to equilibrium, $\zeta(t)$ begins to deviate from its minimum, growing until it reaches the equilibrium value, given by $\zeta_{eq}=\frac{1}{N^2}\int{[N_{mb}(x)-N_{ch}(x)]^2{ d}x}$ where $N_{mb}(x)=N n(x)$ with $n(x)$ given by Eq \eqref{eq:g1dmbpos}. In Fig \ref{fig:g1dzeta}, we show the evolution of $\zeta(t)$ for different values of $N$. After relaxing to the qSS,  $\zeta(t)$ rises and approaches the equilibrium value. Rescaling time with $\tau_{\times}=\tau_D N^{\gamma}$, with $\gamma=1.8$, all the curves collapse onto one universal curve. This value of $\gamma$ is approximate --- to find a precise value of $\gamma$, a very large number of particles must be used in MD simulations.  Nevertheless, the observed value of 
$\gamma$ agrees quite well with the exponent $\gamma=2$ predicted by the theoretical argument of Section \ref{sec:lr}. While our simulations find $\gamma=1.8$, other previous simulations with smaller number of particles find $\gamma=1$ \cite{SevLuw1984}, $\gamma=2$ \cite{HohBro1967} and greater \cite{WriMil1982,LuwSev1985,TsuGou1996}.

\section{Gravitation in two dimensions}\label{sec:grav2d}

We next consider self-gravitating systems in two dimensions. Such systems and their dynamics have been applied to study  topics ranging from the spiral structure of disk-like galaxies \cite{MilPre1968,MilPre1970,Hoh1971} to the large-scale structure of the universe \cite{DorKot1980}. They have also been analyzed in the context of equilibrium thermodynamics \cite{Aly1994,AlyPer1999}. 

The system consists of $N$ particles of mass $m$ in a two-dimensional space.  The total mass of the system is $M=m N$. It is convenient to define dimensionless variables by rescaling length, velocity, potential, and energy with respect to $L_0$ (an arbitrary length scale), $V_0=\sqrt{2GM}$, $\psi_0=2GM$ and $E_0=MV_0^2=2GM^2$, respectively, where $G$ is the gravitational constant. This process is equivalent to setting 
$M=G=1$ and to defining the dynamical time
%%%%%%%%%%%%%%%%
\begin{equation}\label{eq:g2ddynamictime}
\tau_D=\frac{L_0}{\sqrt{2GM}}.
\end{equation}
%%%%%%%%%%%%%%%%
In three-dimensional space, the system corresponds to rods of mass density $m$ \cite{Aly1994}.

Considering only systems with azimuthal symmetry, the corresponding gravitational potential $\psi$ satisfies the dimensionless Poisson equation,
%%%%%%%%%%%%%%%%
\begin{equation}\label{eq:g2dpoisson}
\nabla^2\psi(r,t)=2\pi \rho(r,t)\,,
\end{equation}
%%%%%%%%%%%%%%%%
where  $\rho(r,t)$ is the mass density of a self-gravitating system
which is obtained from the one particle distribution function,
$\rho(r,t)=\int f(r,\mathbf{v};t)\,d^2v$.
For an isolated particle the density is 
%%%%%%%%%%%%%%%%
\begin{equation}
\rho(\mathbf{r},\mathbf{r'})=\delta(|\mathbf{r}-\mathbf{r}'|),
\end{equation}
%%%%%%%%%%%%%%%%
so that the Green's function solution to Eq \eqref{eq:g2dpoisson} is
%%%%%%%%%%%%%%%%
\begin{equation}\label{eq:g2dgreen}
G(\mathbf{r},\mathbf{r'})=\ln |\mathbf{r}-\mathbf{r}'|.
\end{equation}
%%%%%%%%%%%%%%%%
The Hamiltonian for a $N$ particle gravitational system is then
%%%%%%%%%%%%%%%%
\begin{equation}\label{eq:2dham}
\mathcal{H}=\sum_{i=1}^N\frac{p_i^2}{2m}+\frac{m^2}{2}\sum_{i,j=1}^N\ln |\mathbf{r}_i-\mathbf{r}_j|.
\end{equation}
%%%%%%%%%%%%%%%%

\subsection{Molecular dynamics}

We will study 2D gravitational systems in the thermodynamic limit.  
In this limit, if the initial distribution is azimuthally symmetric,  
the mean-field potential will also retain this
symmetry, so that the angular momentum, $p_{\theta}=mr^2\dot{\theta}$, of each particle is conserved.  This allows us to use an effective Hamiltonian description based on the Gauss's law.
A particle at position $r_i$ is subject to an interaction potential produced by all the particles with $r\le r_i$, leading to an effective Hamiltonian
%%%%%%%%%%%%%%%%
\begin{equation}\label{eq:g2deffham}
\mathcal{H}_{eff}(r_i,\theta_i,p_{r_i},p_{\theta_i})=\sum_{i=1}^N \left(\frac{p_{r_i}^2}{2m}+\frac{p_{\theta_i}^2}{2mr_i^2}\right)+\sum_{i=1}^N m_{eff}(r_i)m\ln r_i,
\end{equation}
%%%%%%%%%%%%%%%%
where
%%%%%%%%%%%%%%%%
\begin{equation}\label{eq:g2deffmass}
m_{eff}(r_i)=m\sum_{j=1}^N\Theta(r_i-r_j),
\end{equation}
%%%%%%%%%%%%%%%%
is the mass of all the particles with the radial coordinates $r<r_i$.
The equation of motion for $r_i$ is then
%%%%%%%%%%%%%%%%
\begin{equation}\label{eq:g2deffaccel}
\ddot{r_i}=\frac{v_{\theta_i}^2}{r_i^3}-\frac{m_{eff}(r_i)}{r_i},
\end{equation}
%%%%%%%%%%%%%%%%
where $v_{\theta_i}=p_{\theta}/m$ is determined by the initial distribution.
The advantage of the effective Hamiltonian is that the simulation time of the system's dynamics depends exclusively on the time of sorting a vector composed of $N$ elements, similar to 1D gravity.

At the start of the simulation the $N$ point particles are distributed uniformly inside a circle of
radius $r_m$.  They are also assigned velocities from a uniform distribution with the maximum value $v_m$.  
This corresponds to a one-level initial distribution of the form
%%%%%%%%%%%%%%%%
\begin{equation}\label{eq:g2df0}
f_0(r,v)=\eta\Theta(r_m-r)\Theta(v_m-v)
\end{equation}
%%%%%%%%%%%%%%%%
where $\eta=(\pi^2 r_m^2 v_m^2)^{-1}$ is the normalization constant.

Since in the thermodynamic limit the mean-field potential is purely radial, the angular momentum of each particle will remain constant throughout the  simulation.  The radial dynamics of each particle is then determined by the Eq \eqref{eq:g2deffaccel}, while the  $\theta_i(t)$ dynamics
is controlled by the angular momentum conservation $v_{\theta_i}(t)=v_{\theta_i}(0)$.  The magnitude of the
velocity of the particle $i$ is  $v_i=\sqrt{v_{r_i}^2+(r_i\dot{\theta}_i)^2}$. 

The potential $\psi$ associated with the initial distribution satisfies the Poisson equation,
%%%%%%%%%%%%%%%%%%%%%
\begin{equation}\label{eq:g2dpois1}
\frac{d^2\psi(r)}{dr^2}+\frac{1}{r}\frac{d\psi(r)}{dr}=
  \begin{cases}
    \frac{2}{r_m^2} & \text{for } r \leq r_m \\
    0 & \text{for } r > r_m
  \end{cases}
\end{equation}
%%%%%%%%%%%%%%%%%%%%%
with the boundary conditions given by $\lim_{r \to \infty} \psi(r)=\ln(r)$ and ${\psi'}(0)=0$. The solution to this equation is
%%%%%%%%%%%%%%%%%%%%%
\begin{equation}\label{eq:g2dpoissol}
\psi(r)=
  \begin{cases}
    \frac{r^2-r_m^2}{2r_m^2}+\ln(r_m) & \text{for } r \leq r_m \\
    \ln(r) & \text{for } r > r_m.
  \end{cases}
\end{equation}
%%%%%%%%%%%%%%%%%%%%%
Using this potential, Eq \eqref{eq:g2dpoissol}, and the initial distribution function, Eq \eqref{eq:g2df0}, in the expression for conservation of energy, Eq \eqref{eq:encons}, the initial energy of the system is calculated to be
%%%%%%%%%%%%%%%%%%%%%
\begin{equation}\label{eq:g2de0}
\mathcal{E}_0=\frac{v_m^2}{4}-\frac{1}{8},
\end{equation}
%%%%%%%%%%%%%%%%%%%%%
where without loss of generality we have set $r_m=1$.

We now consider two cases: one in which the initial distribution 
obeys the virial condition (${\cal R}_0 = 1 $) and one in which it does not ($ {\cal R}_0 \ne 1 $).
In section \ref{subsec:virial}, we have shown that the virial condition for a two-dimensional gravitational system requires that $\langle v^2 \rangle=GM(N-1)/N$. In the thermodynamic limit, using the rescaled variables, the virial condition reduces to
%%%%%%%%%%%%%%%%%%%%%
\begin{equation}\label{eq:g2dvirial}
\langle v^2 \rangle =\frac{1}{2}.  
\end{equation}
%%%%%%%%%%%%%%%%%%%%%
We then define the virial number for a 2D gravitational system to be
%%%%%%%%%%%%%%%%%%%%%
\begin{equation}\label{eq:g2dvirialnumber}
{\cal R} = 2 \langle v^2\rangle.
\end{equation}
%%%%%%%%%%%%%%%%%%%%%

\subsection{Lynden-Bell theory for a 2D self-gravitating system}

\begin{figure}[!htb]
\vspace{5mm}
\begin{center}
\includegraphics[width=0.8\textwidth]{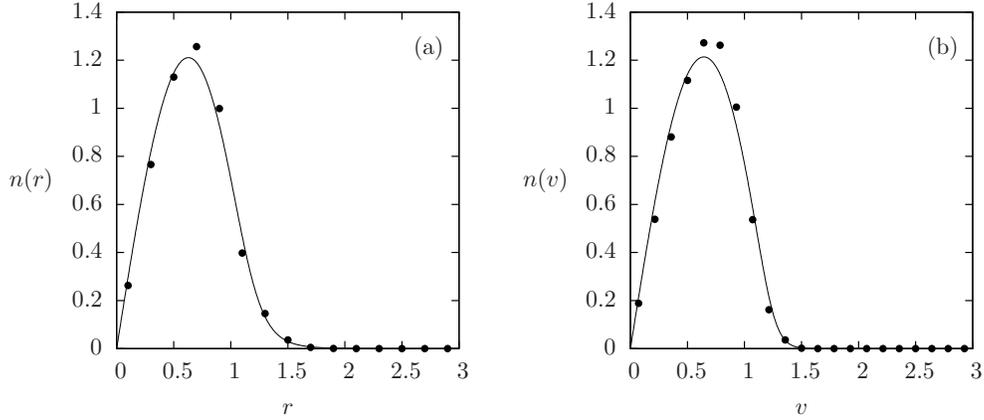}%{figg2dlb.eps}
\end{center}
\caption{Particle distributions in (a) position and (b) velocity of a 2D gravitational system that initially satisfied the virial condition. The solid lines represent the prediction of LB statistics, $n(r)=N(r)/N$ and $n(v)=N(v)/N$, with $N(r)$ and $N(v)$ given by Eqs \eqref{eqnr2dlb} and \eqref{eqnvlb}.
Points are results of MD simulation for $N=10000$ particles, averaged over times $t=1000$ to $t=1100$. Error bars in the distributions are comparable to the symbol size.}
\label{figdistlb2d}
\end{figure}

\begin{figure}[!ht]
\vspace{5mm}
\begin{center}
\includegraphics[width=0.4\textwidth]{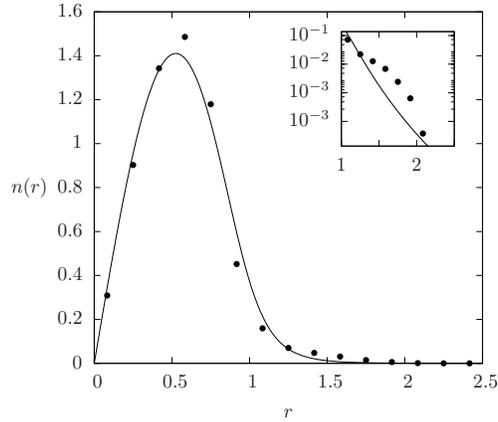}%{figg2dlbamp}
\end{center}
\caption{Distribution in position for a 2D self-gravitating system with ${\cal R}_0 = 0.694$. The solid line represents the prediction of LB theory, $n(r)=N(r)/N$ with $N(r)$ given by Eq \eqref{eqnr2dlb}, while the symbols are the results of MD simulation with $ N=10000$ particles, averaged over times $t=2000$ to $t=2100$. Error bars in the distributions are comparable to the symbol size.}
\label{fig2distlb2d}
\end{figure}

In analogy with 1D gravity,
if the initial distribution of a 2D self-gravitating system obeys the virial condition, 
we expect that the parametric resonances will not be excited and the
qSS should be well described by the Lynden-Bell statistics.  
The mean-field potential should then satisfy the Poisson equation
with the mass density given by the momentum integral of Eq \eqref{eq:flb},
%%%%%%%%%%%%%%%%%%%%%
\begin{equation}
  \frac{d^2\psi_{lb}(r)}{dr^2}+\frac{1}{r}\frac{d\psi_{lb}(r)}{dr}=\frac{4\pi^2 \eta}{\beta}\ln[1+e^{-\beta(\psi_{lb}(r)-\mu)}].
\end{equation}
%%%%%%%%%%%%%%%%%%%%%
The boundary conditions for this equation are $ \lim_{r \rightarrow \infty}\psi_{lb} (r) = \ln(r) $ and $ {\psi'}_{lb}(0) = 0$. The parameters $\beta$ and $\mu$ are determined self-consistently by the conservation of energy and norm of the distribution function, Eqs \eqref{eq:encons} and \eqref{eq:normcons}.  
Once $\psi_{lb}(r)$, $\beta$, and $\mu$ are calculated, we can compare the theoretical predictions with
the results of
the MD simulations.  
To do this we calculate the marginal distributions: the number of particles located between $[r,r+dr]$,
%%%%%%%%%%%%%%%%%%%%%
\begin{equation}
\label{eqnr2dlb}
N(r)=2 \pi N r \int{ d^2{\rm v}} f_{lb}({\bf r},{\bf v})=\frac{4 N r}{\beta v_m^2}\ln[1+e^{-\beta(\psi_{lb}(r)-\mu)}]\,;
\end{equation}
%%%%%%%%%%%%%%%%%%%%%
and the number of particles with velocities between $[v,v+dv]$,
%%%%%%%%%%%%%%%%%%%%%
\begin{equation}
\label{eqnvlb}
N(v)=2 \pi N v \int{ d^2{\rm r}} f_{lb}({\bf r},{\bf v})\,.
\end{equation}
%%%%%%%%%%%%%%%%%%%%%
Comparing the theory and the simulation, we see a reasonably good agreement between the LB statistics and the results
of MD simulations, Fig \ref{figdistlb2d}.
%%%%%%%%%%%%%%%% Figure%%%%%%%%%%%%%%%%%%%%%
%%%%%%%%%%%%% End of figure%%%%%%%%%%%%%%%%%
However, if the initial distribution does not satisfy the virial condition,  LB theory starts to deviate
from the results of MD simulations.  A 
tail in the marginal distribution functions emerges, showing the
formation of a core-halo structure, see Fig \ref{fig2distlb2d}.
%%%%%%%%%%%%%%%% Figure%%%%%%%%%%%%%%%%%%%%%
%%%%%%%%%%%%% End of figure%%%%%%%%%%%%%%%%%
%In this case, some particles gain great
%amounts of energy, thus reaching more energetic regions of phase space and inhibiting the mixing process.

\subsection{The envelope equation}\label{subsec:g2denveq}

The appearance of the core-halo structure is a consequence of the parametric resonances which arise from the
density oscillations.  To study these oscillations we 
define the  envelope of the particle distribution as $r_e(t)=\sqrt{2\langle{\bf r}\cdot{\bf r}\rangle}$.  Note that with this definition $r_e(0)=r_m$.
Differentiating $r_e(t)$ twice with respect to time, we find
%%%%%%%%%%%%%%%%%%%%%
\begin{equation}
\ddot{r}_e(t)=\frac{2\langle {\bf r}\cdot \ddot{{\bf r}}\rangle}{r_e(t)}+\frac{2\langle \dot{{\bf r}}\cdot \dot{{\bf r}}\rangle}{r_e(t)}-
\frac{4 \langle {\bf r} \cdot \dot{{\bf r}}\rangle^2}{{r_e}^3(t)},
\end{equation}
%%%%%%%%%%%%%%%%%%%%%
which can be rewritten as
%%%%%%%%%%%%%%%%%%%%%
\begin{equation}
\ddot{r}_e(t)=\frac{2<{\bf r}\cdot\ddot{{\bf r}}>}{r_e(t)}+\frac{\varepsilon^2(t)}{r_e^3(t)}
\end{equation}
%%%%%%%%%%%%%%%%%%%%%
where
%%%%%%%%%%%%%%%%%%%%%
\begin{equation}
\label{emit}
\varepsilon^2(t) \equiv 4\left(\langle  {\bf r}\cdot{\bf r}  \rangle\langle  {\dot{\bf r}}\cdot{\dot{\bf r}}  
\rangle - \langle  {\bf r}\cdot\dot{{\bf r}} \rangle^2\right)
\end{equation} 
%%%%%%%%%%%%%%%%%%%%%
is known as the ``emittance''. The emittance is an important parameter in the physics of charged particle beams, and is related to the area occupied by the particles in the phase space \cite{DavQin2001}. Unlike the one-dimensional case, in two dimensions the term $ \langle {\bf r}\cdot\ddot{\bf r}\rangle$ can be simplified using the Poisson equation \eqref{eq:g2dpoisson},
%%%%%%%%%%%%%%%%%%%%%
\begin{eqnarray}
\langle {\bf r}\cdot\ddot{{\bf r}} \rangle&=&\int{{\bf r}\cdot\ddot{{\bf r}}\ f_{e}({\bf r},{\bf v},t)
d^2{r}d^2{\rm v}}\nonumber\\
&=&\frac{1}{2\pi}\int{{\bf r}\cdot\ddot{{\bf r}}\ \nabla^2\psi_{e} d^2{r}}\nonumber\\
&=&-\int{r^2\frac{\partial\psi}{\partial r}\nabla^2\psi_{e} dr} \nonumber\\
&=&-\int{r\frac{\partial\psi_{e}}{\partial r}\frac{\partial}{\partial r}\left(r\frac{\partial\psi_{e}}{\partial r}
\right)dr}\nonumber\\
&=&-\frac{1}{2}\int_0^{r_e(t)}{dr \frac{\partial}{\partial r}\left[\left(r\frac{\partial\psi_{e}}{\partial r}
\right)^2\right]}.
\label{Aprforce}
\end{eqnarray}
%%%%%%%%%%%%%%%%%%%%%
The gradient of the potential at $r_e$ is $1/r_e$, and we obtain 
%%%%%%%%%%%%%%%%%%%%%
\begin{equation}
\langle {\bf r}\cdot\ddot{{\bf r}} \rangle=-{1/2} \,.
\end{equation}
%%%%%%%%%%%%%%%%%%%%%

We are interested to study the behavior of a 2D self-gravitating system when its initial distribution does
not deviate significantly from the virial condition.  In this case, we expect that the emittance
will remain close to its initial value,  ${\varepsilon}^2(0)=v_m^2={\cal R}_0$,
so that the envelope equation reduces to 
%%%%%%%%%%%%%%%%%%%%%
\begin{equation}
\ddot{r}_e(t)=\frac{{\cal R}_0}{r_e^3(t)}-\frac{1}{r_e(t)}\;. 
\label{eqenve2d}
\end{equation}
%%%%%%%%%%%%%%%%%%%%%
As expected, if $r_e(0) = 1 $ and $ {\cal R}_0 = 1 $,
$\ddot{r}_e = 0$, so that the envelope   does not develop oscillations.

\begin{figure}[!ht]
\vspace{5mm}
\begin{center}
\includegraphics[width=0.45\textwidth]{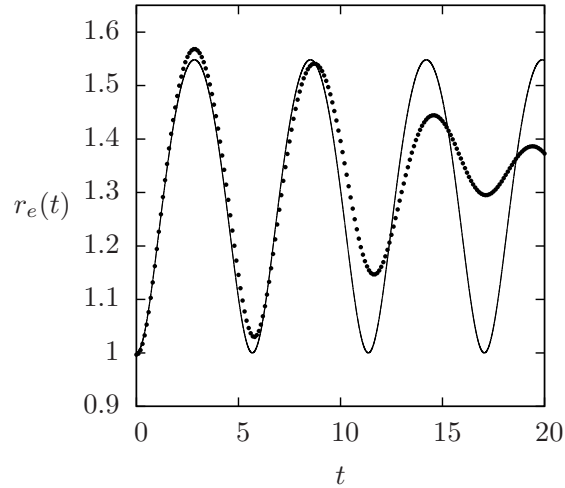}%{figg2denve.eps}
\caption{Evolution of the envelope $r_e$ according to Eq \eqref{eqenve2d} (solid line)
compared to MD simulation (points) for a 2D self-gravitating system with ${\cal R}_0=1.5$. A reasonably good agreement is seen for short times.\label{envenbody2d}}
\end{center}
\end{figure}

Comparing the temporal evolution of $r_e(t)$ with the data from MD simulations, we see that 
there is a reasonably good agreement between the two, especially for short times (Fig. \ref{envenbody2d}).
%****************************************************************************
%****************************************************************************

\subsection{The test particle model}

We now study the  behavior of test particles
subject to a gravitational potential $\psi_{e}(t)$ produced by an oscillating uniform mass distribution,  
%%%%%%%%%%%%%%%%%%%%
\begin{equation}
 \rho(t)=\frac{1}{\pi r^2_e(t)}\Theta(r_e(t)-r)\,.
\end{equation}
%%%%%%%%%%%%%%%%%%%%
Solving the Poisson equation we find
%%%%%%%%%%%%%%%%%%%%
\begin{eqnarray}
\label{pot2denve}
\psi_{e}(r,t)&=&\left\{
\begin{array}{l}
 \frac{r^2-r_e^2(t)}{2r_e(t)^2} \>\ \text {for} \>\ r \le r_e(t) \\
\\
 \ln(r) \>\ \text{for} \>\ r \ge r_e(t)
\end{array}
\right.
\end{eqnarray}
%%%%%%%%%%%%%%%%%%%%
This means that the dynamics of a test particle $i$ which at $t=0$ was at $r_i(0)$ and had an angular momentum $p_{\theta_i}$ 
will be governed by the equation of motion 
%%%%%%%%%%%%%%%%%%%%
\begin{eqnarray}
\label{eqpoinc2d}
\ddot{r}_i(t)-\frac{{v_{\theta_i}}^2}{{r_i}^3(t)}&=&
\left\{
\begin{array}{l}
-\frac{r_i(t)}{r_e^2(t)}\>\, \text{ for }\>\,r_i(t)\le r_e(t)
\\
\\
-\frac{1}{r_i(t)}\>\, \text{ for }\>\,r_i(t)\ge r_e(t)
\end{array}
\right.
\end{eqnarray}
%%%%%%%%%%%%%%%%%%%%
where $r_e(t)$ is the solution of Eq \eqref{eqenve2d}.

\begin{figure}[!ht]
\vspace{5mm}
\begin{center}
\includegraphics[width=0.8\textwidth]{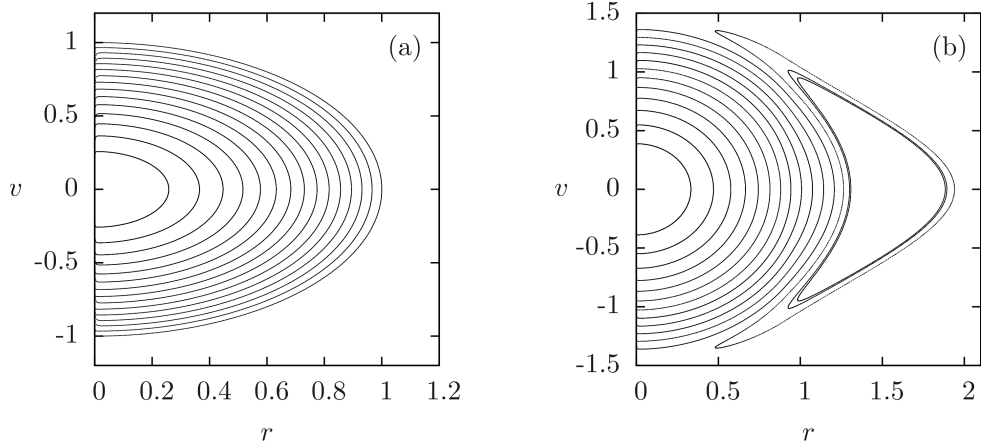}%{figg2dpoin.eps}
\caption{Poincar\'e sections for a 2D self-gravitating system with (a) ${\cal R}_0 \approx 1$ and (b) ${\cal R}_0 = 0.9$.
While in (a) the dynamics is completely regular in (b) we see the formation of a resonance island. We have considered $v_{\theta_i}=0$ so that only the radial velocity appears in the Poincar\'e sections. \label{figpoinctese2d}}
\end{center}
\end{figure}

We integrate the equations of motion \eqref{eqpoinc2d} for $15$ test particles, uniformly distributed
at $t=0$, $ r_i(0) \in [0,1] $ and $ v_i(0) \in [0,v_m] $, with $v_m =\sqrt{ {\cal R}_0} $. 
The  Poincar\'e section is constructed by plotting the position and velocity of each test particle when
the envelope $r_e(t) $  is at its minimum value, see Fig. \ref{figpoinctese2d}.
% ************************************************* ***************************
% ************************************************* ***************************

\begin{figure}[!ht]
\vspace{5mm}
\begin{center}
\includegraphics[width=0.8\textwidth]{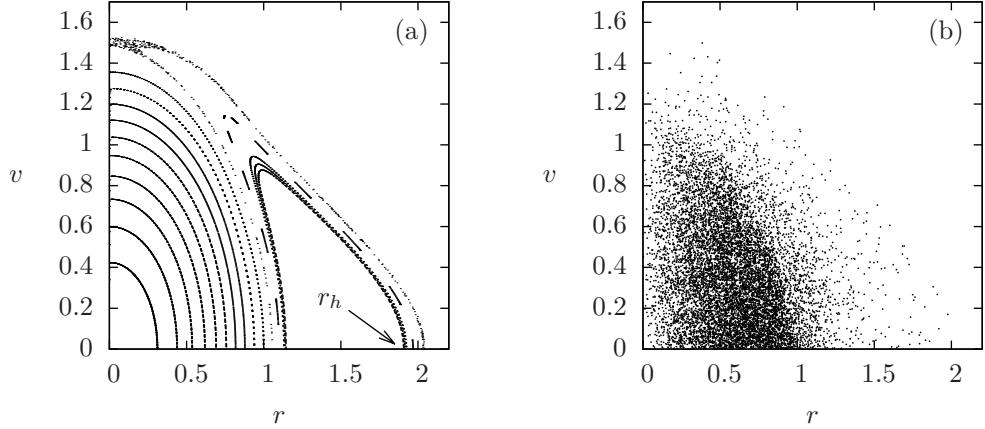}%{figg2dpoinphase.eps}
\caption{Poincar\'e section of test particles (a) moving in an effective potential given by Eq \eqref{pot2denve} and the phase space of MD simulation at $t = 2000$ with $N=20000$ (b). The virial number is
$ {\cal R}_0 = 0.694 $. Comparing the two phase spaces, we see that the test particle dynamics allows us
to accurately determine the maximum energy $ \epsilon_h$ that a particle of a 2D self-gravitating system
can gain from the density oscillations. In this particular case, $\epsilon_h = \ln(r_h)$, where $ r_h $ is the
maximum position reached by a test particle, as indicated in the panel (a).\label{figpoincnb2d}}
\end{center}
\end{figure}

In Fig. \ref{figpoincnb2d} we compare the phase space structure of the test particle dynamics
to a snapshot of the phase space obtained using MD simulation, after the qSS has been established.
% ************************************************* ***************************
% ************************************************* ***************************
We see that the test particle dynamics allows us to calculate the maximum energy that a particle of a self-gravitating system
can gain from the density oscillations.

\subsection{Core-Halo Distribution}

\begin{figure}[!hb]
\vspace{5mm}
\begin{center}
\includegraphics[width=0.8\textwidth]{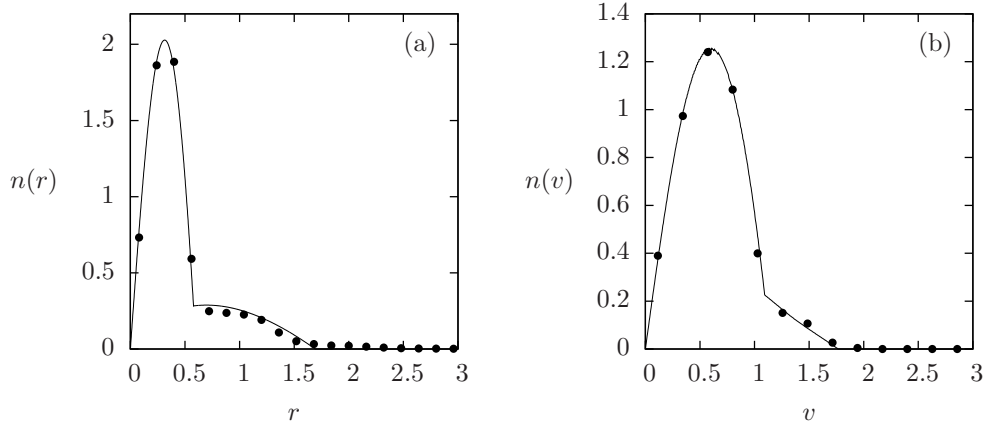}%{figg2dchdist.eps}
\end{center}
\caption{Distributions in (a) position and (b) velocity for a 2D self-gravitating system with ${\cal R}_0=0.25$.
The solid line corresponds to the prediction of the core-halo distribution function, Eq \eqref{eq:g2dfch},
and points are results of MD simulation with $N=10000$ particles averaged over times $t=2000$ to $t=2100$. Error bars in the distributions are comparable to the symbol size.\label{figdist2d}}
\end{figure}

The particles which enter in resonance with the core density oscillations escape from the central region producing
a tenuous halo.  The  halo formation progressively dampens the oscillations, bringing the resonances closer and closer to the core.  When the qSS is established, we expect that the particle distribution will, once again, correspond to the core-halo
distribution, given by
%%%%%%%%%%%%%%%%
\begin{equation}\label{eq:g2dfch}
f_{ch}(r,v)=\eta\Theta(\epsilon_F-\epsilon(r,v))+\chi\Theta(\epsilon(r,v)-\epsilon_F)\Theta(\epsilon_h-\epsilon(r,v)),
\end{equation}
%%%%%%%%%%%%%%%%
where $\epsilon_F$ and $\chi$ are calculated using conservation of energy and norm and $\epsilon_h$ is determined by the test particle dynamics, see Fig \ref{figpoincnb2d}.
%be given by $f_{ch}(r,v)$.
Integrating the core-halo distribution over $v$, we obtain the particle density in the qSS state.
Substituting this into Poisson equation \eqref{eq:g2dpoisson}, we obtain the equation for the gravitational potential of a 2D cluster 
%%%%%%%%%%%%%%%%%
\begin{eqnarray}\label{eqpoisson2dch}
\nabla^2\psi_{ch}(r)=4\pi^2\left\{
\begin{array}{l}
\eta(\epsilon_F - \psi_{ch}(r))+\chi(\epsilon_h-\epsilon_F)\>\, \text{ for } \psi_{ch}(r) <  \epsilon_F  \;,\\
\\
\chi(\epsilon_h - \psi_{ch}(r))\>\, \text{ for } \epsilon_F \le \psi_{ch}(r) \le \epsilon_h \;,\\
\\
0 \>\, \text{ for } \psi_{ch}(r) > \epsilon_h \,,
\end{array}
\right.
\end{eqnarray}
%%%%%%%%%%%%%%%%%
with boundary conditions $\lim_{r\to\infty}\psi_{ch}(r)=\ln(r)$ and ${\psi'}_{ch}(0)=0$. The system of equations \eqref{eqpoisson2dch} can be solved analytically, see Ref \cite{TelLev2010}.  Comparing 
the marginal distributions predicted by the core-halo theory to the results
of MD simulations (Fig. \ref{figdist2d}), an excellent agreement between the two is observed.
%%%%%%%%%%%%%%%% Figure%%%%%%%%%%%%%%%%%%%%%
%%%%%%%%%%%%% End of figure%%%%%%%%%%%%%%%%%

%------------------------------------------%
\subsection{Relaxation time}
Finally, it is interesting to explore how much time $\tau_\times(N)$ a finite system of $N$ particles remains in the qSS before relaxing to the true thermodynamic equilibrium. To this end, we use the crossover parameter $\zeta(t)$, defined as
%%%%%%%%%%%%%%%%%
\begin{equation}
\zeta(t)=\frac{1}{N^2}\int_0^{\infty}{[N(r,t)-N_{ch}(r)]^2{ d}r}
\end{equation}
%%%%%%%%%%%%%%%%%
where $ N(r,t)$ is the number of particles located inside shells between $r$ and $r+{ d}r$ at time $t$ and $N_{ch}(r)=2\pi Nr\int{f_{ch}\left({\bf r},{\bf v}\right){ d}^2{\rm v}}$, where $f_{ch}\left({\bf r},{\bf v}\right)$ is the core-halo distribution, Eq \eqref{eq:g2dfch}.
%%%%%%%%%%%%%%%% Figure%%%%%%%%%%%%%%%%%%%%%
\begin{figure}[!ht]
\vspace{5mm}
\begin{center}
\includegraphics[width=0.9\textwidth]{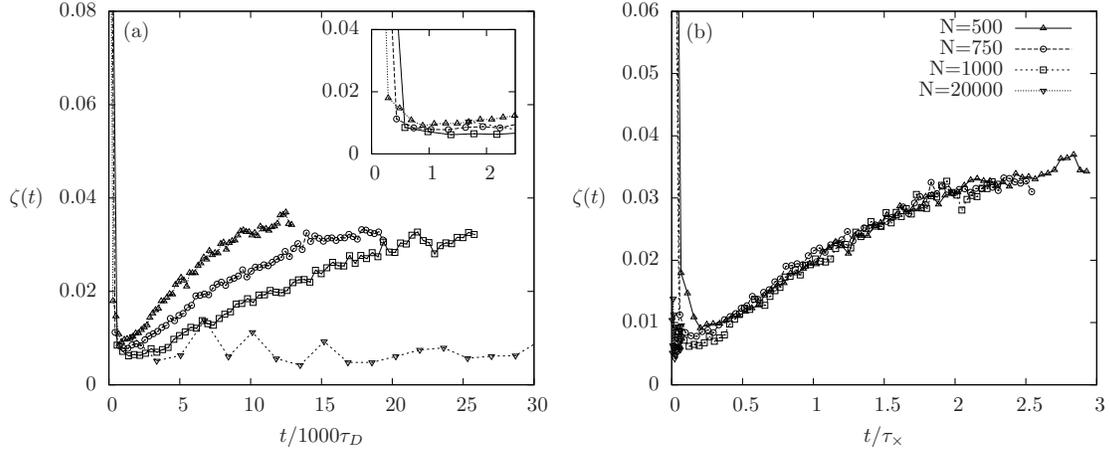}%{figg2dzeta.eps}
\end{center}
\caption{(a) $\zeta(t)$ for different numbers of particles in the system.
In the inset, we show the fast ($N$ independent) relaxation to the core-halo qSS
after a time $t\approx 2000\tau_D$.  The system remains in the qSS for a time interval that 
scales with the
number of particles.  When the time is rescaled by $\tau_{\times}(N)$
all the data in (a) fall on a universal curve (b).\label{figtime2d}}
\end{figure}
%%%%%%%%%%%%% End of figure%%%%%%%%%%%%%%%%%
Figure \ref{figtime2d} shows the value of $\zeta(t)$ for systems with different numbers of particles.
The panel \ref{figtime2d}b shows that if the time is rescaled by $\tau_{\times}=N^{\gamma}\tau_D$, where
$\gamma=1.35$ and $\tau_D$ is the dynamical time defined by Eq \eqref{eq:g2ddynamictime}, all the curves fall on a universal curve, indicating the divergence of the crossover time in the thermodynamic limit.  Thus, in the
limit $N \rightarrow \infty$ a self-gravitating system will remain forever trapped in a nonequilibrium stationary
state. Recent simulations performed with discrete particles instead of the concentric shells used in this Report have lead toexponent, $\gamma\approx 1$ \cite{Mar2012}, which is in good agreement with the scaling argument presented in Section \ref{sec:lr}, $t_\times N/\ln N$.

\subsection{Thermodynamic equilibrium}

For a finite number of particles, after a time $\tau_{\times}(N)$, we expect the system to relax to thermodynamic equilibrium, with 
%%%%%%%%%%%%%%%%
\begin{equation}\label{feq}
f_{mb}({\bf r},{\bf v})= C e^{-\beta\left( \frac{v^2}{2} +\psi_{eq}(r)\right)},
\end{equation}
%%%%%%%%%%%%%%%%
where $C$ is the normalizations constant. 
To see that this is the case, we calculate the gravitational potential 
and the marginal distributions and compare them to the results of MD simulations. 
The gravitational potential in equilibrium $\psi_{eq}$ will satisfy the Poisson-Boltzmann equation
%%%%%%%%%%%%%%%%
\begin{equation}\label{eq:g2deqpoisson}
\nabla^2\psi_{eq}(r)=\frac{d^2\psi_{eq}(r)}{dr^2}+\frac{1}{r}\frac{d\psi_{eq}}{dr}=\frac{4\pi^2
C}{\beta}e^{-\beta\psi_{eq}(r)},
\end{equation}
%%%%%%%%%%%%%%%%
where  $\beta=1/T$ is the Lagrange multiplier used to enforce the
conservation of the total energy. The solution of this equation is given in Ref. \cite{TelLev2010},
%%%%%%%%%%%%%%%%
\begin{equation}\label{eq:g2dpsieq}
\psi_{eq}(r)=\frac{1}{2}\ln\left(e^{2(2\mathcal{E}-1)}+r^2\right).
\end{equation}
%%%%%%%%%%%%%%%%
Curiously, an isolated 2D gravitational system can only exist at one temperature, $T=1/4$, independent of the
initial energy.  If such a system is put in contact with a thermal bath, it will either gain energy 
from the bath and grow without bound or lose energy and shrink, 
depending if the temperature of the bath is greater or smaller than $T=1/4$, respectively. 
%%%%%%%%%%%%%%% figure %%%%%%%%%%%%%%%%%%%%%
\begin{figure}[!ht]
\vspace{5mm}
\begin{center}
\includegraphics[width=0.8\textwidth]{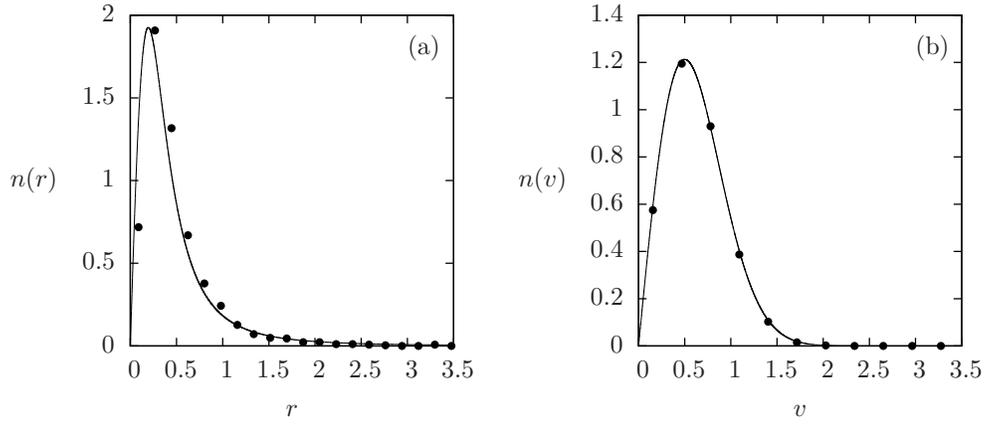}%{figg2deqdist.eps}
\end{center}
%\vspace{1cm}
\psfrag{r}{$r/\sqrt{K/\kappa_z}$}
\caption{Equilibrium distributions: (a) position and (b) velocity of a system with
${\cal E}_0=-0.0434$. The solid line corresponds to the equilibrium distributions, Eqs \eqref{eq:g2dnr} and \eqref{eq:g2dnv}, and the points are results of MD simulations with $N=10000$ particles  at $t=10^6$.\label{fig-1cap2}}
\end{figure}
%%%%%%%%%%%%% end of figure %%%%%%%%%%%%%%%%%%%

Figure \ref{fig-1cap2} compares the marginal distributions obtained using the MD simulations 
with the predictions of equilibrium statistical mechanics.  
The number density of particles located between $[r,r+dr]$ is
%%%%%%%%%%%%%%%%
\begin{equation}\label{eq:g2dnr}
N(r)=2\pi Nr\int d^2v\,f_{mb}(\mathbf{r},\mathbf{v})=\frac{2N e^{2(2\mathcal{E}-1)} r}{(e^{2(2\mathcal{E}-1)}+r^2)^2},
\end{equation}
%%%%%%%%%%%%%%%%
and the number density of particles with velocities between $[v,v+dv]$ is
%%%%%%%%%%%%%%%%
\begin{equation}\label{eq:g2dnv}
N(v)=2\pi Nv\int d^2r\,f_{mb}(\mathbf{r},\mathbf{v})=4Nve^{-2v^2}.
\end{equation}
%%%%%%%%%%%%%%%%
The figure shows a good agreement between the results of MD simulations and BG statistics.  
However, to reach thermodynamic equilibrium, it was necessary to run a simulation with $N=10000$ particles
for $t=10^6$ dynamical times.  Up to this time, the system remained trapped in a qSS state with the
particles distributed in accordance with the core-halo distribution function $f_{ch}(r,v)$, Eq \eqref{eq:g2dfch}.

\section{Gravitation in three dimensions}

The relaxation of 3D self-gravitating systems is extremely 
difficult to study.  There are two basic problems arising from the fact
that Newton's gravitational potential has no lower bound, but is bounded from above.
The consequence of the upper bound is that 
some particles of a self-gravitating system can gain enough kinetic energy to
escape the gravitational field of the cluster.  
In principle, there is no limit to the particle evaporation since the 
energy can be constantly 
supplied  by the two-body collisions ~\cite{Pad1990,ChaBou2005,LevPak2008} and
the gravitational collapse. As a consequence,
the Poisson-Boltzmann equation for a 3D open system has no solutions.
Based on cosmological simulations, however, it has been observed that 3D 
systems do relax to qSSs \cite{Bin2004,Hen2006,Sax2013}.  There have been
a number of  phenomenological models proposed  to describe
the observed density profiles in such qSS: ``de Vaucouleurs'', ``S\'{e}rsic'' and ``NFW'' models
\cite{Vau1948,Ser1963,HjoMad1995,NavFre1996,NavFre1997,WilHjo2010a,WilHjo2010b}.  These, phenomenological density distributions, however, lack the theoretical foundation.  

The fact that the Poisson-Boltzmann equation does not have a solution indicates that open 3D self-gravitating systems
are intrinsically unstable in the {\it infinite time limit}.  
This instability, is a consequence of the the binary collisions which lead to
a flux of evaporating particles.  On shorter
time scales, however,  it is possible for a system to relax to a collisionless qSS.   
Again, however, the situation in 3D is 
much more complex than in one and two dimensions \cite{MilPre1968,MilPre1970,Mil1971,SevLuw1984,WriMil1984,SevLuw1986,ReiMil1987,YanGou1998,MilRou2002,Val2006,JoySic2011}.  Significant evaporation of particles can happen even on very
short time scales, leading to a halo that extends all the way to infinity.  At the moment, there is no theory that  can account for the particle distribution inside a 3D halo.  The theory of parametric resonances,
which was so successful for treating 1D and 2D gravity, 
can not be applied in 3D since, in general, there are no bounded resonant orbits.

Although the particle distribution in a qSS can not be  predicted {\it a priori}, we expect
that it will have a core-halo structure.  Evaporation  should progressively 
cool down the core region.  Statistically only a completely degenerate core can remain stable in an infinite
space --- at finite temperature  the entropy gain will always favor particle evaporation.  
%While collisional relaxation, and the resulting particle evaporation leads to formation of a singularity and a diverging phase space density, 
Furthermore, since the collisionless relaxation is controlled by the Vlasov equation, 
the phase space density in the core can not exceed that of the initial waterbag distribution.  
We, therefore, expect that 
the core will be described by a fully degenerate Fermi-Dirac distribution \cite{LevPak2008} with the
"spin" degeneracy equal to the phase space density of the initial waterbag distribution.
The difficulty, however, is that without knowing the full particle distribution in the halo, we can not calculate the self-consistent gravitational potential and close all the equations of the theory.  

For a 3D gravitational system of total mass $M$, the gravitational potential in the qSS must 
satisfy the Poisson equation,
%%%%%%%%%%%%%%%%
\begin{equation}
\label{eqpoisson3d}
 \nabla^2\psi(r)=4 \pi G M \int{f({\bf r}, {\bf v}) { d^3{{\rm v}}}} \,,
\end{equation}
%%%%%%%%%%%%%%%%
where $f({\bf r}, {\bf v})$ is the one particle distribution function.
If the potential $\psi(r)$ has a radial symmetry, the particles 
can be represented as spherical shells of mass $m=M/N$. 
This approach greatly facilitates the numerical simulations, 
and becomes exact in the thermodynamic limit.

It is convenient to measure all the distances in an arbitrary length unit $r_0$,
the time in units of dynamical time,
%%%%%%%%%%%%%%%%
\begin{equation}
\tau_D=\sqrt{\frac{r_0^3}{GM}},
\end{equation}
%%%%%%%%%%%%%%%%
and the gravitational potential in units of $\psi_0=GM/r_0$.
The Poisson equation \eqref{eqpoisson3d} then reduces to
%%%%%%%%%%%%%%%%
\begin{equation}
 \nabla^2\psi(r,t)=4 \pi \int{f({\bf r}, {\bf v},t) {\rm d^3{v}}}\;.
\end{equation} 
%%%%%%%%%%%%%%%%
For a particle located at 
$\mathbf{r}'$, $\rho(\mathbf{r})=\delta(|\mathbf{r}-\mathbf{r}'|)$, the Green's function of  Poisson equation is the usual Newton's gravitational potential
%%%%%%%%%%%%%%%%
\begin{equation}\label{green3d}
 G(|\mathbf{r}-\mathbf{r}'|)=1/|\mathbf{r}-\mathbf{r}'|.
\end{equation}
%%%%%%%%%%%%%%%%
This potential diverges at small distances and is bounded from above.  We saw already that in 1D
and 2D some particles enter in resonance with the density oscillations
and gain a lot of energy.  The situation in 3D is even more complex --- the potential is  bounded
from above so that the resonant particles can gain enough energy to completely escape from the 
gravitational field of the cluster.

\subsection{Test particle dynamics} 

To get a better idea of the relaxation process which leads to the core-halo formation, we
study the dynamics of test particles moving under the action of an 
oscillating gravitational potential.  Once again we consider particles which at $t=0$ were distributed
uniformly in the phase space inside a sphere of radius $0<r \le r_m$ and  $0<v \le v_m$.
We define the "envelope radius" as $r_e(t)=\sqrt{\frac{5\langle r^2\rangle}{3}}$, which at $t=0$ satisfies
$r_e(t)=r_m$.  We will work in dimensionless units and set $r_0=r_m$.  Differentiating
twice with respect to time and performing manipulations similar to those for 1D and 2D gravitational 
systems, we obtain a differential equation that governs the envelope dynamics,
%%%%%%%%%%%%%%%%
\begin{equation}
{\ddot r}_{e}+{\frac{1}{r_{e}^2}}-\frac{{\cal R}_0}{ r_{e}^3}=0,
\label{rb3D}
\end{equation}
%%%%%%%%%%%%%%%%
where
%%%%%%%%%%%%%%%%
\begin{equation}
{\cal R}_0= -\frac{2K_0}{V_0}\;
\label{emit3d}
\end{equation}
%%%%%%%%%%%%%%%%
is the virial number, and 
$K_0$ and $V_0$ are the kinetic and the potential energy of the initial distribution.

We consider the dynamics of $10$ test particles, initially distributed uniformly
with positions $r_{i} \in [0,1]$ and velocities $v_{i}\in [0,v_m]$,
%%%%%%%%%%%%%%%%
\begin{eqnarray}
\label{eqpoinc3d}
\ddot{r}_{i}(t)-\frac{l_i^2}{{r_{i}}^3(t)}&=&
\left\{
\begin{array}{l}
-\frac{r_{i}(t)}{r_{e}^3(t)}\>\, \text{ for }\>\,r_{i}(t)\le r_{e}(t)
\\
\\
-\frac{1}{r_{i}^2(t)}\>\, \text{ for }\>\,r_{i}(t)\ge r_{e}(t)\;,
\end{array}
\right.
\end{eqnarray}
%%%%%%%%%%%%%%%%
where $l_i=|{\bf r}_i(0) \times {\bf v}_i(0)|$ and $r_{e}(t)$ evolves according to Eq \eqref{rb3D}.
%****************************************************************************
\begin{figure}[!ht]
\begin{center}
%  \vbox to 70mm{\vfil
\includegraphics[width=0.8\textwidth]{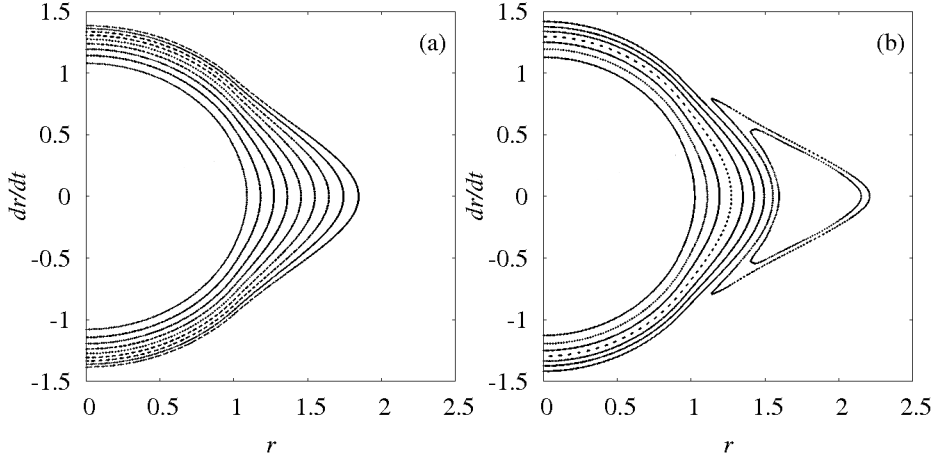}%{figpoincnb3gn.eps}
\caption{Poincar\'{e} sections of a 3D gravitational system, for (a) ${\cal R}_0\approx 1$  and (b) 
${\cal R}_0=0.97$. 
In (a) the orbits are completely integrable, whereas in (b), we see a resonance island.}\label{figpoincnb3gn}
%    \vfil}
\end{center}
\end{figure}
%****************************************************************************
Fig \ref{figpoincnb3gn} shows the the Poincar\'{e} sections for two systems with ${\cal R}_0 \approx 1$ 
and ${\cal R}_0=0.97$.  For  ${\cal R}_0 \approx 1$, the orbits remain integrable, while even a small deviation
from the virial condition results in the appearance of a resonance island.  For slightly larger or smaller 
${\cal R}_0$ the resonant orbit becomes unbounded.

\subsection{Lynden-Bell theory for a 3D self-gravitating system}

It is interesting to consider the predictions of the LB theory for a 3D self-gravitating system. In this case
the one-particle distribution function becomes
%--------------------------- 
\begin{eqnarray}
\label{e2}
f_{lb}({\bf r},{\bf v})=
\frac{\eta_1}{e^{\beta [\epsilon({\bf r},{\bf v})-\mu]}+1}\;,
\end{eqnarray}
%-----------------------------
where $\eta_1=9/16 v_m^3$ and $ \epsilon(r,v)= v^2/2 + \psi(r)$.  Integrating over the velocities we obtain
the density distribution corresponding the the LB stationary state.  Substituting this into the Poisson equation
allows us to write a self-consistent equation for the gravitational potential 
%%%%%%%%%%%%%%%%%%%%%%%%%%%%%%%%%%%%%%%
\begin{equation}
\label{e5}
\frac{1}{r^2}\frac{\partial}{\partial r} r^2 
\frac{\partial \psi}{\partial r} = - 16 \pi^2 \,\eta_1\,\sqrt{\frac{\pi}{2 \beta^3}}\,\Li_{3/2}(- e^{\beta\left[\mu - \psi(r)\right]}),
\end{equation}
%%%%%%%%%%%%%%%%%%%%%%%%%%%%%%%%%%%%%%%%%%
where $\Li_n(x)$ is the $n^{th}$ polylogarithm function of $x$. This equation has to be solved numerically and the two Lagrange  
multiplier $\beta$ and $\mu$ must be calculated to preserve the number of particles and the energy of the system.
The solution of Eq \eqref{e5} is complicated by the open boundary conditions.  In practice, we
will solve this equation by enclosing the system in a spherical box of radius $r_w$ and then take the limit $r_w \to \infty$.
%****************************************************************************
\begin{figure}[!ht]
\begin{center}
%  \vbox to 70mm{\vfil
\includegraphics[width=0.65\textwidth]{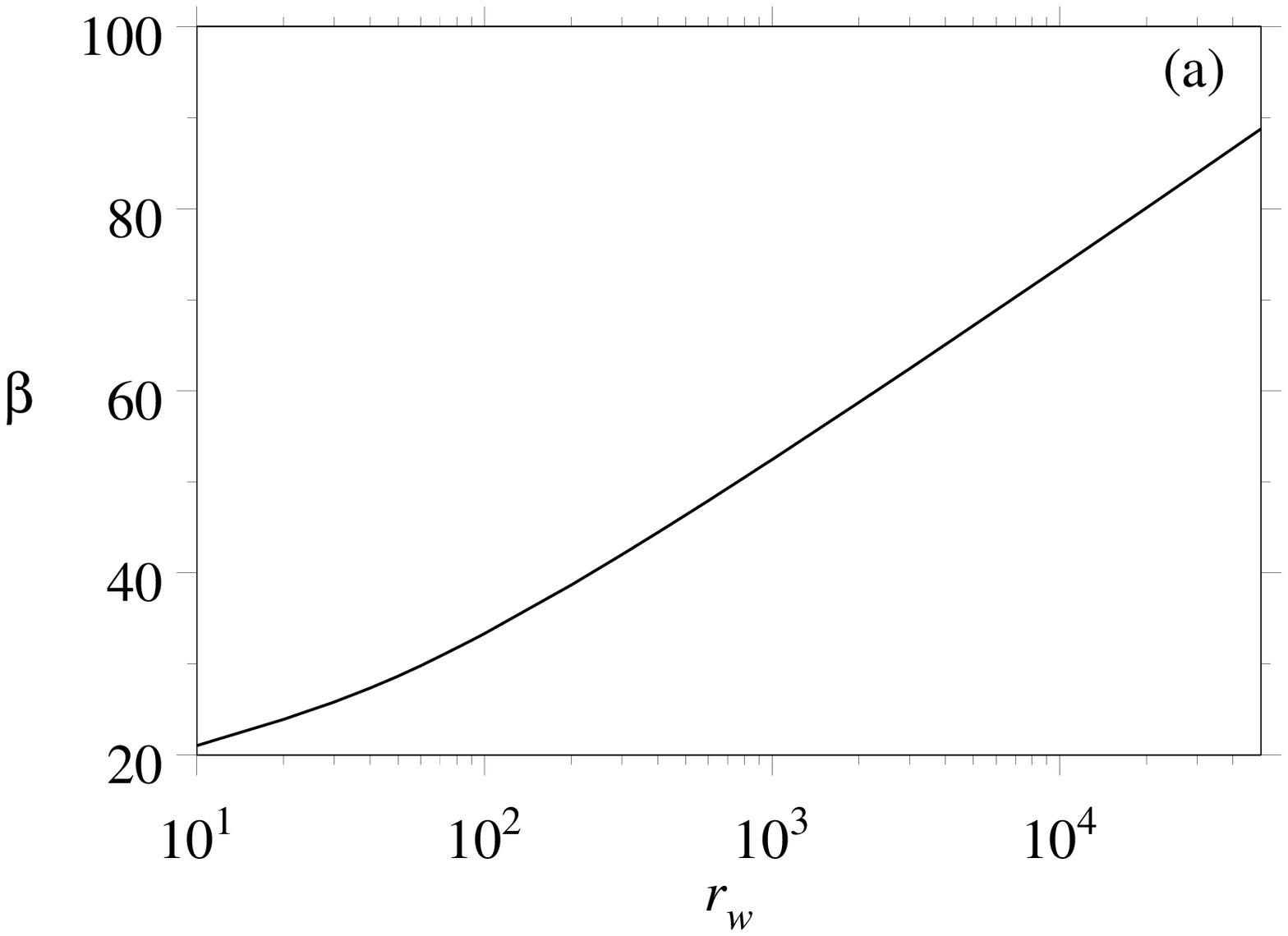}\\%{betavsrw.eps}
\includegraphics[width=0.7\textwidth]{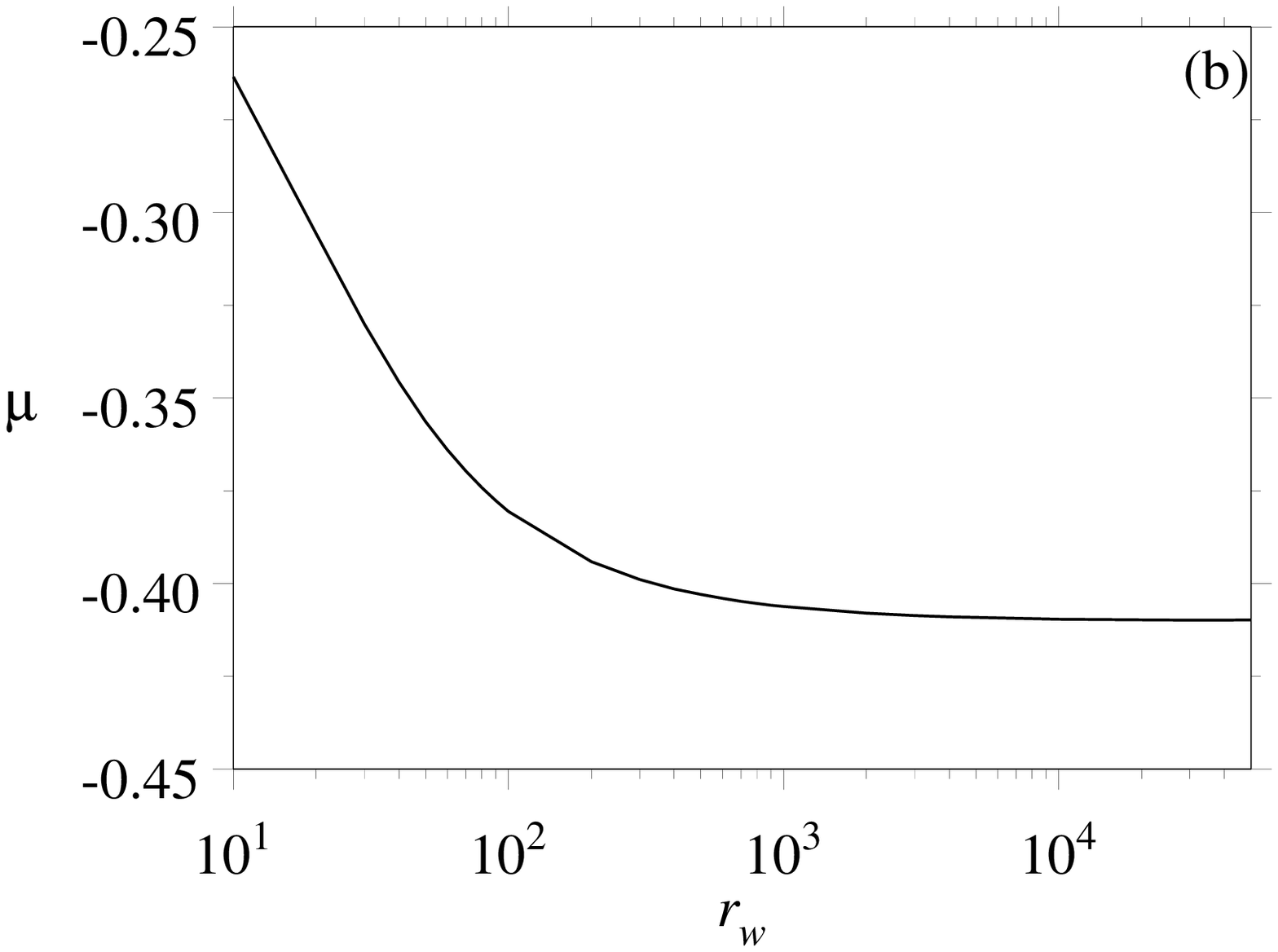}%{muvsrw.eps}
\caption{(a) $\beta$ and (b) $\mu$ as a function of $r_w$. While the inverse temperature parameter $\beta$ diverges in (a), the chemical potential $\mu$ in (b) asymptotically goes to a finite value $\mu \approx -0.41$, as $r_w$ increases.
The virial number is ${\cal R}_0=1.7$.}\label{betmu}
%    \vfil}
\end{center}
\end{figure}
%****************************************************************************
As expected, when $r_w \to \infty$, the LB distribution separates into a completely degenerate 
core and a very tenuous halo which extends all the way to $r_w$.  However, the particle distribution in the halo
is very different from the ones found in MD simulations, see Fig. \ref{3dist}, so that 
LB theory fails to correctly describe a 3D self-gravitating system.
%****************************************************************************
\begin{figure}[!ht]
\begin{center}
%  \vbox to 70mm{\vfil
\includegraphics[width=0.65\textwidth]{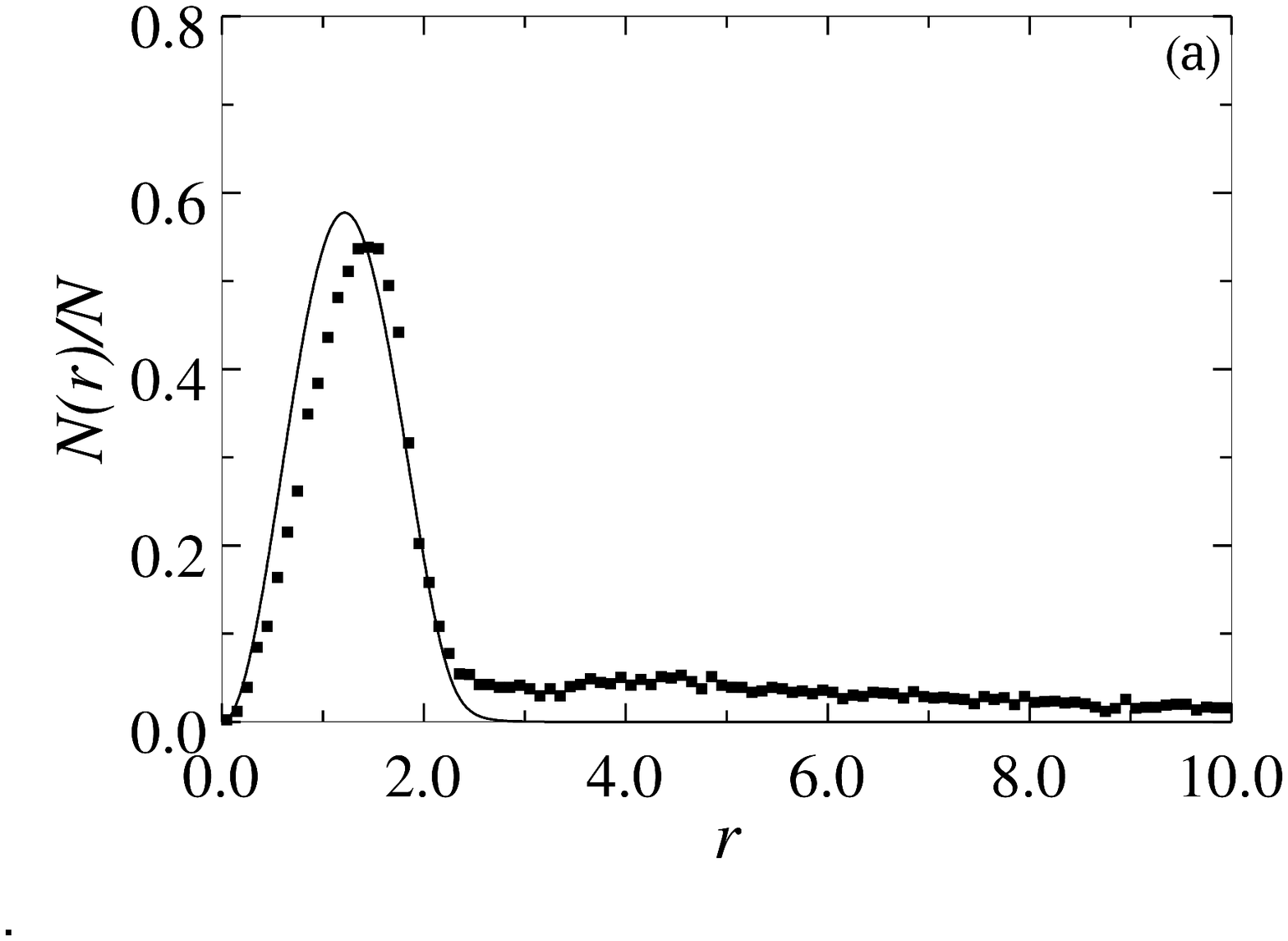}\\%{histR1p7.eps}
\includegraphics[width=0.65\textwidth]{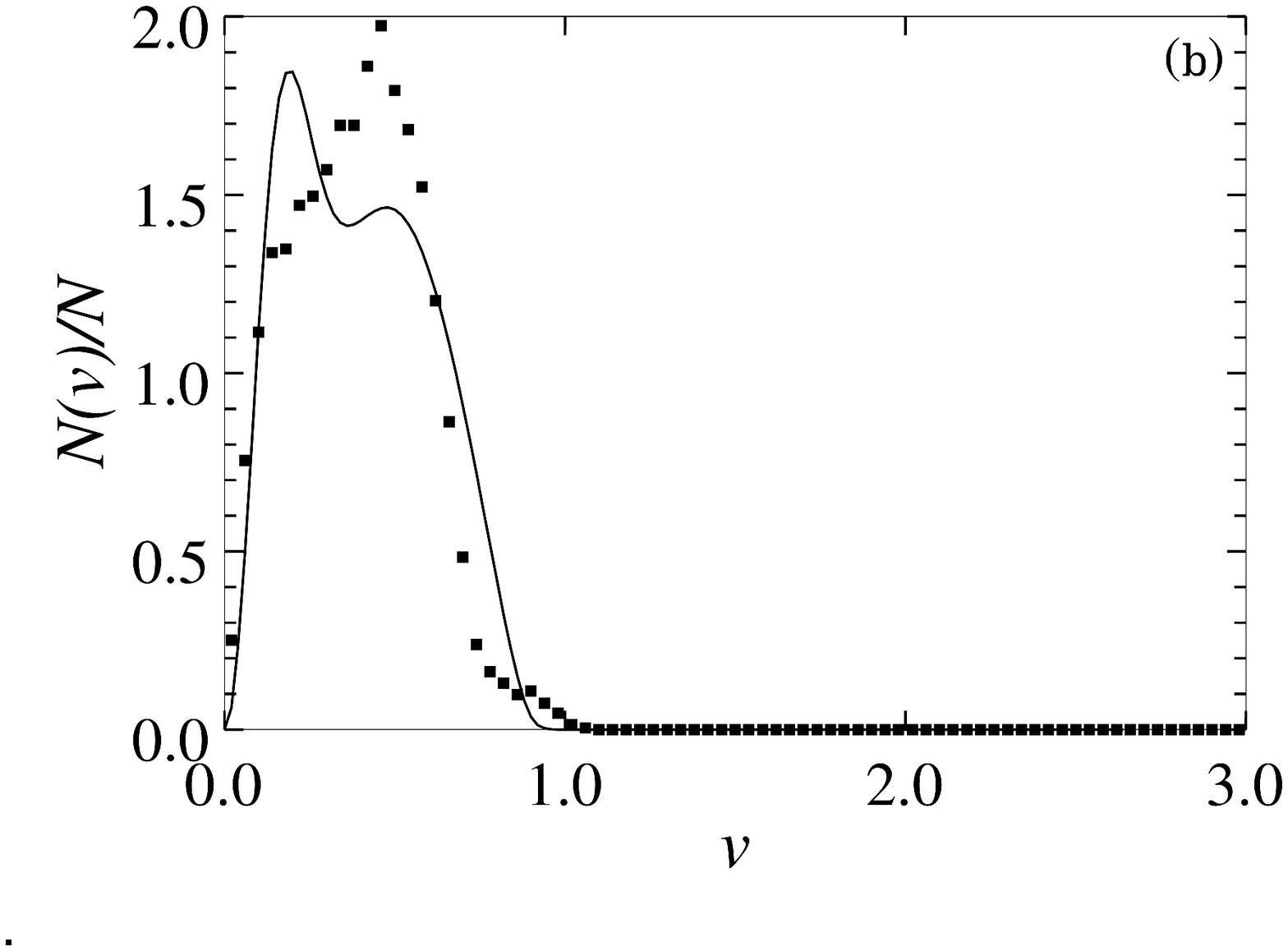}%{histR1p7v.eps}
\caption{The (a) mass and (b) velocity distributions in the qSS obtained by MD simulation (symbols) and the LB prediction (solid line). The wall radius is placed at $r_w=10^4$ and the virial number is ${\cal R}_0=1.7$}. \label{3dist}
%    \vfil}
\end{center}
\end{figure}
%**************************************************************************** 

\subsection{Systems with ${\cal R}_0=1$}

\begin{figure}[!ht]
\begin{center}
%  \vbox to 70mm{\vfil
\includegraphics[width=0.65\textwidth]{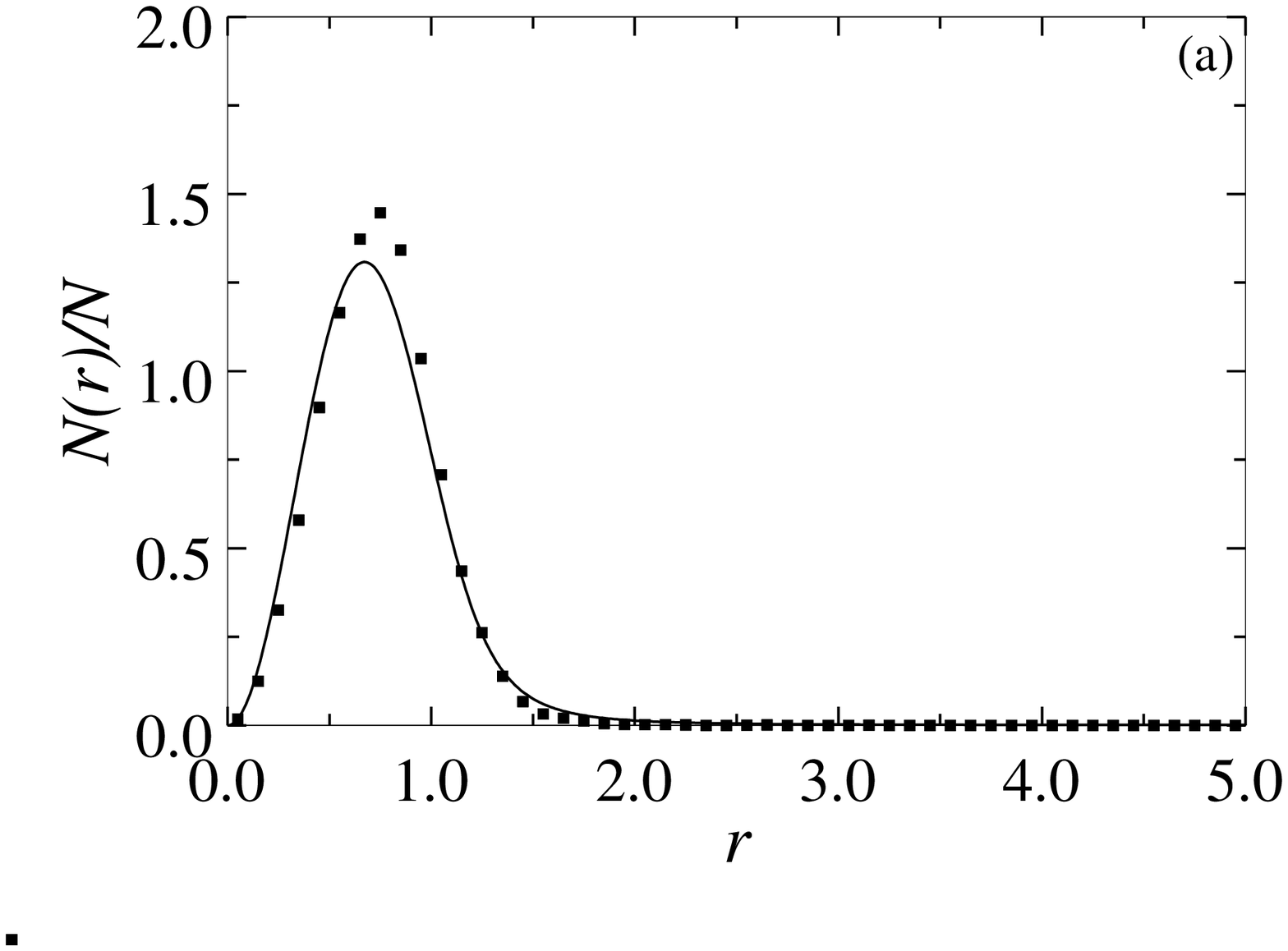}\\%{3dr1r.eps}
\includegraphics[width=0.65\textwidth]{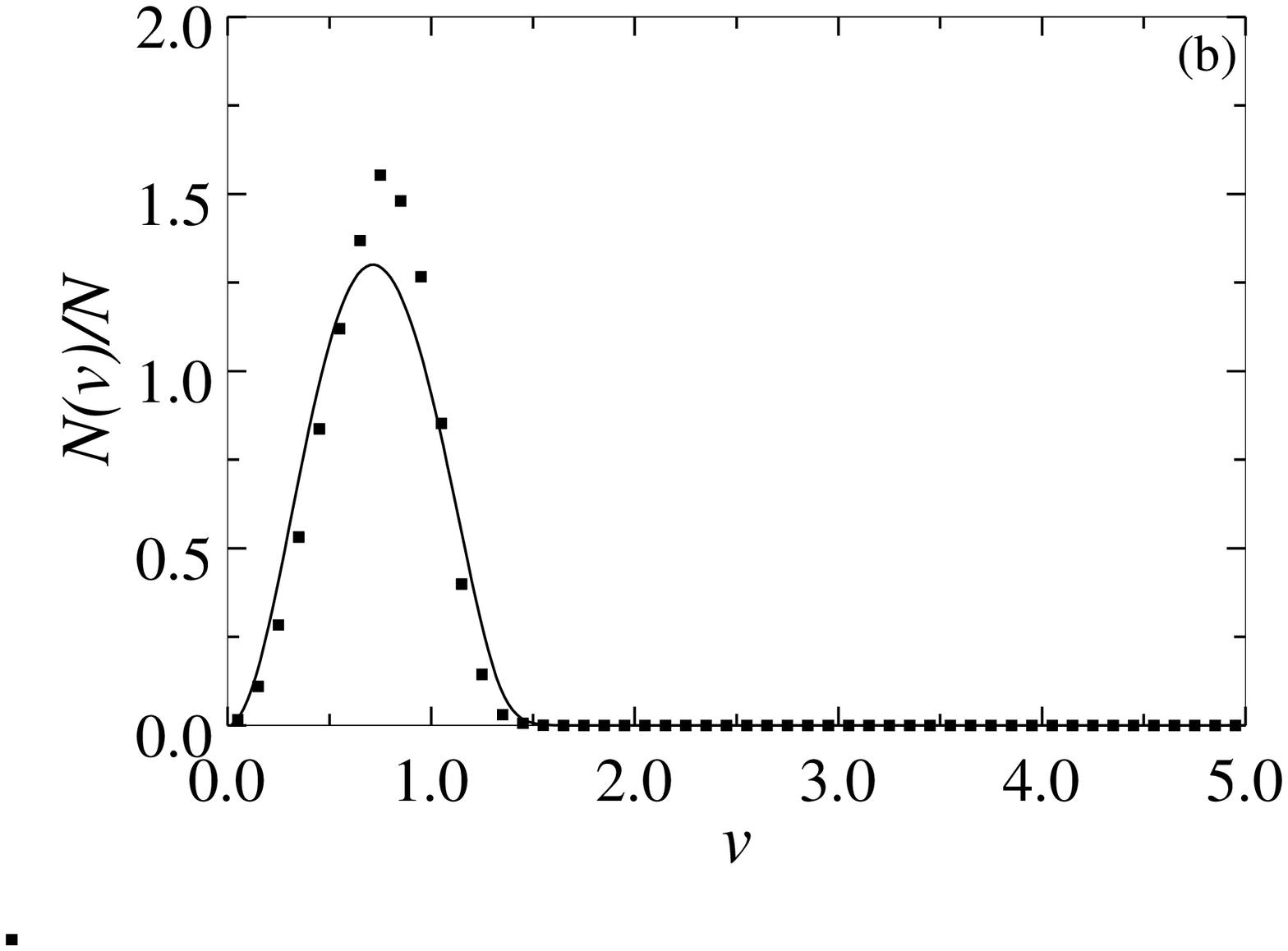}%{3dr1v.eps}
\caption{The mass (a) and velocity (b) distributions in the qSS obtained by MD simulation (symbols) and the
distributions obtained using LB theory with a cutoff at $r_w=10$ (solid line),
for an
initially virialized waterbag distribution, ${\cal R}_0=1$ .\label{3dr1}}
%    \vfil}
\end{center}
\end{figure}

If the initial particle distribution satisfies the virial condition ${\cal R}_0=1$, the macroscopic oscillations
will be suppressed and the parametric resonances will not be excited, see Fig. \ref{figpoincnb3gn}.  For such
initial distributions, we saw that LB theory worked reasonably well for 1D and 2D gravitational systems.
For 3D systems, however, LB theory fails even when ${\cal R}_0=1$.  
As  $r_w \to \infty$, the solution of Eq \eqref{e5}
requires that $\beta \to \infty$ (see Fig \ref{betmu}) and the distribution function approaches the degenerate limit  
$f_{core}({\bf r},{\bf v})=\eta_1\,\Theta(\mu-\epsilon)$ (plus halo particles at infinity).  Thus, for an open system, LB theory will always predict a fully degenerate core \cite{ChaSom1998}.  This conclusion,
however, is valid only in  the asymptotic $t \rightarrow \infty$ limit.  
In this limit, even small oscillations of the envelope will lead to particle evaporation
and result in formation of a cold core.  
In practice, however, for ${\cal R}_0=1$ the rate of evaporation is very low, 
so that the degenerate limit will not be reached in the time of simulation.  
To treat this ``short'' time limit, we can introduce an effective cutoff (a wall)  
at $r_w$.   The precise value of the cutoff is unimportant --- as long as it is not
too large  $5 \le r_w \le 100$.  The wall will prevent the particle evaporation and a complete cooling of the
core region.  Indeed, the cutoff-LB distribution (cLB) is found to describe reasonably the qSS state for ${\cal R}_0=1$ \cite{LevPak2008}, see Fig. \ref{3dr1}
%****************************************************************************
%**************************************************************************** 

\section{Non-neutral plasmas}

In this chapter we will analyze qSSs of magnetically confined non-neutral plasmas.  
The non-neutrality condition is crucial for the plasma to be a long-ranged interacting system --- 
for neutral two component plasmas, Debye screening leads to an effective 
short-range interaction potential \cite{DubGil1962, Bal1975, DavQin2001}.  Equilibrium state of neutral plasmas and electrolytes, therefore,  
can be studied using the
usual Boltzmann-Gibbs statistical mechanics \cite{Lev2002}.

Many different applications, such as heavy ion fusion, high-energy physics, communications, materials processing, and cancer therapy, depend on the physics of transport of intense charged-particle beams. 
The goal is to avoid the heavy particle losses produced by the 
parametric resonances \cite{Rei1991, NunPak2007}, which can lead to halo formation 
that is detrimental to the beam quality, and can result in damage to the accelerator walls. 
A theory which can quantitatively predict this effect is, therefore,  highly 
desirable for a better understanding of the physics of
beam transport \cite{Glu1994,BanSch2002,AllCha2002,CheGon2005,MugBlu2008}. 

In general, the dynamics of the beams is influenced by multiple effects, including
 the mismatched envelope (rms radius of the beam) 
\cite{Glu1994,OkaIke1997,WanCra1998,ChePak2000,AllCha2002}, movement outside the axis of symmetry \cite{HesChe2000,MorPak2004,MorPak2005,Hes2008,MarRiz2009,BabGos2012}, 
nonuniformities in the beam distribution \cite{BerKis1999,AndRos2000,LunGro2005,RizPak2007}, and the image 
forces due to the surrounding conducting walls \cite{QiaZho2003,ZhoQia2003,PakLev2007}. 
Of all these, the study of  parametric resonances 
resulting from the transverse beam oscillations has attracted the most attention. Envelope mismatch is believed to be the main cause of the halo formation in space-charge dominated beams \cite{OCWan1993}. In this section we will show that the mismatch of the beam envelope  is closely related to the virial condition --- similarly to the one found for self-gravitating systems --- and that the final qSS is, once again, described by the core-halo distribution function.

\subsection{The model}

Our system consists of a beam of charged point particles, confined by an external 
magnetic field ${\bf B}^{ext}({\bf r}) = B_0 \hat z$, propagating 
along the axial $\hat z$ direction, with velocity $V_b$. 
The beam has a characteristic radius $r_b$ and is surrounded by a conductive cylindrical wall of
radius $r_w$ \footnote{A conducting grounded wall requires that the electric potential at the wall vanishes $\phi^s(r_w) = 0 $.}. 
%%%%%%%%%%%%%%%%%%%%%%%%%%%%%%%%%%%%%%
\begin{figure}[!ht]
\vspace{5mm}
\begin{center}
\includegraphics[width=0.8\textwidth]{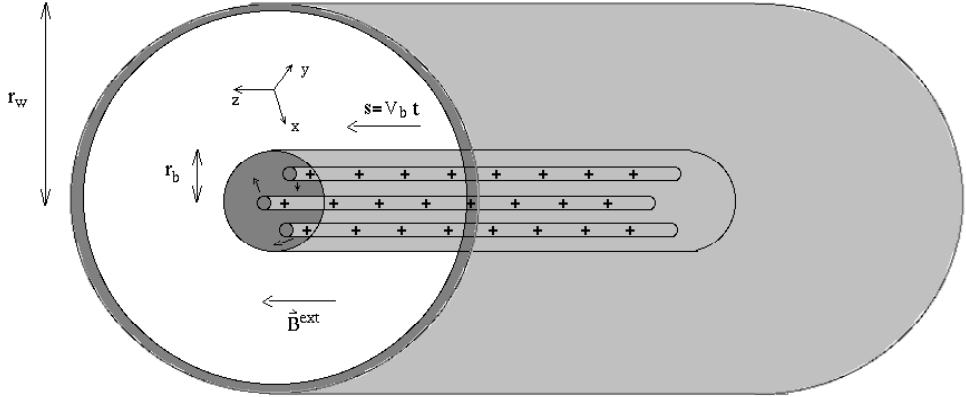}%[width=12cm,height=5cm]{feixedinamics}
\caption{Charged particle beam of characteristic radius $r_b$ propagating along the longitudinal direction $\hat z$ with constant velocity $V_b$. The particles are confined
by a magnetic field ${\bf B}^{ext}=B_0\hat z$, and the beam is isolated from the external environment by a conducting cylindrical wall located at $r_w$.\label{fig:feixeparticula}}
\end{center}
\end{figure}
%%%%%%%%%%%%%%%%%%%%%%%%%%%%%%%%%%%%%%
We assume that the beam has axial symmetry and that the motion along the $\hat z$ direction is uniform. 
Consequently, we consider that the relevant dynamics takes place only in the transverse plane ``$\bot $''\footnote{We approximate $\nabla^2 \approx \nabla^2_{\bot}$ since the variation of the potential along the longitudinal direction is negligible compared to the variations in the transverse plane. Therefore, in this section, $\nabla$ will be understood to represent $\nabla_{\bot}$.}. 
Under these conditions, the time $t$ can be replaced by the longitudinal coordinate $s$, by means of a canonical transformation of the original Hamiltonian, 
where $s=V_b t$ and $V_b = \beta_b c $, $ c $ being the speed of light in vacuum, as illustrated in Fig \ref{fig:feixeparticula}. 

The charge of the beam particle is $Z_ie$, where $Z_i$ is the valence  and $e$ is the electron charge. Furthermore, assuming that the transverse velocity of the beam particles is much lower than the longitudinal velocity, the dynamics along the transverse plane may be considered non-relativistic. This set of conditions, known as the paraxial approximation, is sufficient to study narrow and intense charged-particle beams \cite{DavQin2001}. 

The electric ${\bf E}^s$ and magnetic $ {\bf B}^s$ fields satisfy Maxwell's equations \cite{DavQin2001} and the electric potential the Poisson equation,
%%%%%%%%%%%%%%%%
\begin{equation}
\nabla_{\bot}^2\phi^s=\frac{1}{r}\frac{\partial{}}{\partial{r}}\left(r\frac{\partial{}}{\partial{r}}
\right)\phi^s(r,s)=-4\pi Z_ien_b \label{poisson0}
\end{equation} 
%%%%%%%%%%%%%%%%
with boundary conditions  $\phi(r_w) = 0$ and $\phi'(0) = 0$, where $n_b$ is the number density of the particles. The electric potential is always zero outside the conductive wall, located at $r_w$.
The vector potential, $ \hat z A_z^s(r,s)$, produced by the current of charges $Z_ien_bV_{zb}$ --- the longitudinal velocity of the beam, $V_{zb}(r,s)$, 
is approximated by $V_b$  ---  satisfies
%%%%%%%%%%%%%%%% 
\begin{equation}
\nabla_{\bot}^2A_z^s(r,s)=-4\pi Z_ien_b\beta_{b}. \label{potencialvetor}
\end{equation}
%%%%%%%%%%%%%%%%
Comparing equations \eqref{poisson0} and \eqref{potencialvetor}, we see that the electric and vector potentials are related by 
%%%%%%%%%%%%%%%%
\begin{equation}
A_z^s=\beta_b\phi^s \;.
\end{equation}
%%%%%%%%%%%%%%%%
Thus, solving the Poisson equation \eqref{poisson0}, we find the electromagnetic field
acting on each particle, 
%%%%%%%%%%%%%%%%
\begin{eqnarray}
{\bf E}^s=-\nabla \phi^s(r,s), \\
{\bf B}={\bf B}^{ext}+\beta_b\nabla\phi^s(r,s)\ \times\ \hat z\;.
\end{eqnarray}
%%%%%%%%%%%%%%%%

%%%%%%%%%%%%%%%%%%%%%%%%%%%%%%%%%%%%%%
\begin{figure}[h]
\vspace{5mm}
\begin{center}
\includegraphics[width=0.55\textwidth]{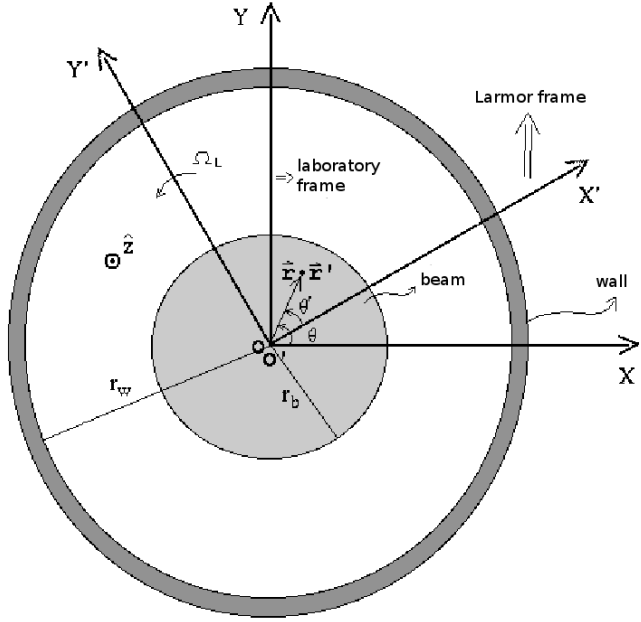}%[width=7.5cm,height=7cm]{feixereference}
\caption{Change of reference frames: $``O"$ represents the laboratory frame and $``O'"$ the Larmor frame.\label{fig:feixereference}}
\end{center}
\end{figure}
%%%%%%%%%%%%%%%%%%%%%%%%%%%%%%%%%%%%%%

As a matter of convenience, \cite{DavQin2001,Rei1991}, we study the system in the Larmor frame which rotates in relation to the laboratory 
with a constant angular velocity
$\Omega_L=-Z_ieB_0/2\gamma_bm\beta_bc^2$, where $\beta_b=V_b/c$, $\gamma_b=(1-\beta_b^2)^{-1/2}$ and
$m$ is the mass of a particle, see Fig \ref{fig:feixereference}. 
We define the dimensionless potential as 
%%%%%%%%%%%%%%%%
\begin{equation}
\psi^{b}(r,s)=\left(Z_ie/\gamma_b^3m\beta_b^2c^2\right)\phi^s(r,s)\;.
\end{equation}
%%%%%%%%%%%%%%%%
In the Larmor frame, the focusing due to magnetic field ${\bf B}^{ext}$, 
results in a radial confining force.
The change to the Larmor frame is accomplished by a change of coordinates $(r,\theta)\rightarrow(r',\theta ')$, where
%%%%%%%%%%%%%%%%
\begin{eqnarray}
r'=r, \nonumber \\
\theta'=\theta-\Omega_L\ s,
\end{eqnarray}
%%%%%%%%%%%%%%%%
as shown in Fig \ref{fig:feixereference}.
The evolution of the distribution function $f({\bf r},{\bf v},s)$ in the Larmor frame satisfies the Poisson-Vlasov systems of equations \cite{DavQin2001},
%%%%%%%%%%%%%%%%
\begin{eqnarray}
\frac{\partial f}{\partial s} + {\bf v} \cdot \nabla f +\left[-\kappa_z^2 {\bf r}
-\nabla \psi^{b}(r)\right] \cdot \nabla_{\bf v} f = 0, \label{ap11} \\
\nabla ^2 \psi^{b}(r) = - {2 \pi K }n ({\bf r}, s), \label{ap12}
\end{eqnarray}
%%%%%%%%%%%%%%%%
where $n ({\bf r}, s)= \int f \ d{\bf v} $ is the density profile of the beam, 
$\kappa_z^2 =|\Omega_L|^2/c^2 $ is the focusing field parameter, and 
$K=2 Z_i^2 e^2 N_b / \gamma_b^3 \beta_b^2mc^2$ is the perveance which
measures the intensity of the beam. The number of particles 
per unit axial length is $N_b$, ${\bf r}$ is the position vector in the transverse plane, 
and $ {\bf v} \equiv d{\bf r} /ds $ is the dimensionless transverse ``velocity''. 
The problem then reduces to studying the dynamics of 2D  pseudo-particles of charge $q=\sqrt{\frac{K}{N_b}}$ confined by an external parabolic potential $U=\kappa_z r^2/2$. The interaction potential between the particles is $V_{b}({\bf r, r'})=q^2 G_{b}({\bf r, r'})$ where
$G_{b}({\bf r}, {\bf r'})$
is the Green's function of the two-dimensional Poisson equation.  
For conducting boundary conditions at $r_w$, the Green's function can be calculated
using Kelvin's inversion theorem \cite{Jac1998,ArfWeb2001}. 
%We will use the dimensionless variables for distances and velocities by scaling them with
%$\sqrt{\varepsilon_0}$ and $\sqrt{\varepsilon_0\kappa_z }$, respectively, where $\varepsilon_0$
%is the initial beam emittance, see Eq. \eqref{emit}. The energy of the system will be scaled with 
%$\varepsilon_0 \kappa_z$~\footnote{We set $m=1$, so that $v_i$ represents the moment $p_i$.}. 
The Hamiltonian for the effective 2D system is then
%%%%%%%%%%%%%%%%
\begin{equation}
 {\cal H}^{b}(r_i,\theta_i,v_{r_i},v_{\theta_{i}})=\sum_{i=1}^{N_b}\left(\frac{v_{r_i}^2}{2}+
\frac{v_{\theta_i}^2}{2r_i^2}\right)-\frac{q^2}{2}\sum_{i,j=1}^{N_b} G_{b}({\bf r}, {\bf r'})+\frac{\kappa_z^2 r_i^2}{2} \,.
\label{hamicolip}
\end{equation}
%%%%%%%%%%%%%%%% 
Starting from an arbitrary initial distribution, the system of particles can now be simulated to obtain the
final qSS. 

If the system has azimuthal symmetry, the simulations can be simplified further. 
In the thermodynamic
limit the Vlasov mean-field description becomes exact, so that each particle moves under the
action of the mean-electromagnetic potential produced by all the other particles. To approach the mean-field
limit with a finite number of particles we can uniformly smear the charge of each particle over a circle
or radius $r_i$ corresponding to its position.  This is the same approximation that
was used to efficiently simulate 2D and 3D gravitational systems.  Using Gauss's law, the 
equation of motion for the radial coordinate of a particle
$i$ becomes 
%%%%%%%%%%%%%%%%
\begin{eqnarray}
 \ddot{r}_{eff}(r_i)=\frac{v_{\theta_i}^2}{r_i^3}+\frac{K}{N_b}\frac{n_{eff}(r_i)}{r_i}- \kappa_z^2 r_i\;,\\
\nonumber\\
  n_{eff}(r_i)=\sum_{j=1}^{N_b}\Theta(r_i-r_j)\;,
\end{eqnarray}
%%%%%%%%%%%%%%%%
where $n_{eff}$ is the number of particles with  $r<r_i$ and $v_{\theta_i}=r_i^2\dot{\theta}_i$.  Since the force acting on each particle is radially
symmetric, $v_{\theta_i}$ is a conserved quantity determined from the initial condition,
$v_{\theta_i}(t)=v_{\theta_i}(0)$.
The effective Hamiltonian in the mean-field limit can then be written as
%%%%%%%%%%%%%%%%
\begin{equation}
\label{hamileffp2d}
 {\cal H}_{eff}^{b}(r_i,\theta_i,v_{r_i},v_{\theta_{i}})=\sum_{i=1}^{N_b}\left(\frac{v_{r_i}^2}{2}+
\frac{v_{\theta_i}^2}{2r_i^2}-\frac{K}{N_b}n_{eff}(r_i)\ln \left(\frac{r_i}{r_w}\right)+\frac{\kappa_z^2 r_i^2}{2}\right)
\end{equation}
%%%%%%%%%%%%%%%%

\subsection{The envelope equation}

We define the beam envelope as $r_b \equiv \left[2\langle r^2 \rangle \right]^{1/2} $. 
Differentiating twice with respect to $s$ 
gives us the beam envelope equation, 
%%%%%%%%%%%%%%%%
\begin{equation}\label{eq:rb2}
{\ddot r}_b+ \kappa_z^2 r_b-\frac{K}{r_b}-\frac{\varepsilon^2(t)}{r_b^3}=0,
\end{equation}
%%%%%%%%%%%%%%%%
where $\varepsilon(t)$ is the emittance, Eq. \eqref{emit}. 
This equation is exact; however, the dynamics of $\varepsilon(t)$ is unknown. 
 For short
times we will set it equal to the initial 
emittance $\varepsilon(t)=\varepsilon(0) \equiv \varepsilon_0$.

The beam envelope will not oscillate if ${\ddot r}_b = 0$. This defines the matched beam radius,
%%%%%%%%%%%%%%%%
\begin{equation}
r_{b}^*=\left\{\frac{K}{2 \kappa_z^2}+\left[\frac{K^2}{4 \kappa_z^4} + \frac{\varepsilon_0^2}{\kappa_z^2} \right]^{1/2}\right\}^{1/2},
\label{rb*}
\end{equation}
%%%%%%%%%%%%%%%%
which is equivalent to the virial condition, Eq \eqref{eq:virialcondition}.

If the initial beam is launched with the radius $r_b=r_b^*$, it will not develop significant oscillations 
and will not suffer emittance growth. 
However, in practice it is virtually impossible to launch a beam precisely at this radius. 
We, therefore,   define the virial parameter as
%%%%%%%%%%%%%%%%
\begin{equation}
\label{mu}
\mu(t)\equiv r_b(t)/r_b^*,
\end{equation}
%%%%%%%%%%%%%%%%
which measures how far the initial beam deviates from the virial condition.

\subsection{Initial conditions}

At $t=0$ 
the $N_b$ particles are distributed uniformly in phase space with $r_i \in [0,r_m]$ 
and velocities $v_i \in [0,v_m]$, 
%%%%%%%%%%%%%%%%
\begin{equation}\label{f0p}
f_0(r_m,v_m)=\eta \Theta(r_m-r)\Theta(v_m-v).
\end{equation}
%%%%%%%%%%%%%%%%
It is convenient to measure all length in units $\sqrt{\varepsilon_0/\kappa_z}$ and ``time'' (longitudinal length) $s$
in units of $1/\kappa_z$.  The transverse velocities will then be measured in units $\sqrt{\varepsilon_0 \kappa_z}$. 
In these dimensionless units the matched beam radius becomes
%%%%%%%%%%%%%%%%
\begin{equation}
r_{b}^*=\left\{\frac{K^*}{2}+\left[\frac{K^{*2}}{4} + 1 \right]^{1/2}\right\}^{1/2},
\label{rb2*}
\end{equation}
%%%%%%%%%%%%%%%%
where $K^*=K/\varepsilon_0 \kappa_z$.
Unlike for self-gravitating systems, for which only the virial number determined the dynamical
evolution, in the case of beams we have two dimensionless parameters, 
$K^*$ and $\mu_0=\mu(0)$.

In the reduced units, $\varepsilon_0=1$ and
%%%%%%%%%%%%%%%%
\begin{equation}\label{vm}
v_m=1/r_m \;,
\end{equation}
%%%%%%%%%%%%%%%%
where $r_m=r_b(0)$ and the emittance growth is $\varepsilon_{qSS}$.
%%%%%%%%%%%%%%%% figure %%%%%%%%%%%%%%%%%%%%%
%\begin{figure}[!ht]
%\begin{center}
%\begin{minipage}{0.45\textwidth}
%\hspace{-4.cm}
%\begin{center}
%\includegraphics[width=7cm]{distiniplas}
%\end{center}
%\end{minipage}
%\end{center}
%\caption{(a) Initial distribution in phase space, normalized by $\eta=\pi^{-2}$.\label{distiniplas}}
%\end{figure}
%%%%%%%%%%%%% end of figure %%%%%%%%%%%%%%%%%

The potential $\psi_{wb}^{b}$ associated with the initial distribution given by equation \eqref{f0p} can 
be obtained by solving the Poisson equation \eqref{ap12},
%%%%%%%%%%%%%%%%
\begin{eqnarray}
 \frac{d^2\psi_{wb}^{b}(r)}{dr^2}+\frac{1}{r}\frac{d\psi_{wb}^{b}(r)}{dr}&=&
\left\{
\begin{array}{l}
-2K^*/r_m^2 \>\ \text {for} \>\ r \le r_m\;, \\
\\
0 \>\ \text {for} \>\ r_m < r \le r_w\;,
\end{array}
\right.
\end{eqnarray}
%%%%%%%%%%%%%%%
with the boundary conditions $\psi_{wb}^{b}(r_w)=0$ and ${\psi'}_{wb}^{b}(0)=0$. The solution is 
%%%%%%%%%%%%%%%
\begin{eqnarray}
\label{potpwb}
\psi_{wb}^{b}(r)&=&\left\{
\begin{array}{l}
 -K^* \left[\frac{(r^2-r_m^2)}{2r_m^2} + \ln(r_m/r_w)\right] \>\ \text {for} \>\ r \le r_m\;, \\
\\
 -K^* \ln(r/r_w) \>\ \text {for} \>\ r_m \le r \le r_w\;.
\end{array}
\right.
\end{eqnarray}
%%%%%%%%%%%%%%
For the initial waterbag distribution \eqref{f0p}, the initial energy of the system is
%%%%%%%%%%%%%%%%
\begin{equation}
\label{e0}
{\cal E}_0(K^*,r_w;\mu_0)=\frac{v_m^2}{4} + \frac{r_m^2}{4} + \frac{K^*}{8} - \frac{K^*}{2}\ln(\frac{r_m}{r_w})\,,
\end{equation}
%%%%%%%%%%%%%%%%
with $r_m$ and $v_m$ defined by Eqs \eqref{vm} and \eqref{mu}, respectively\footnote{If the initial distribution is nonuniform, the functional dependence between $v_m$ and $r_m$ will change.}.

\subsection{Lynden-Bell theory for a charged particle beam}

\begin{figure}[!ht]
\vspace{5mm}
\begin{center}
\includegraphics[width=0.85\textwidth]{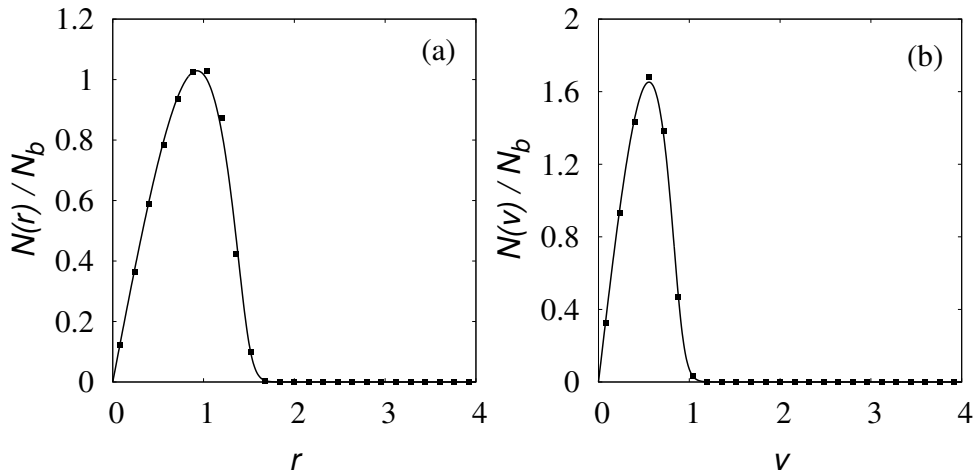}%{figplyn.eps}
\end{center}
\psfrag{r}{$r/\sqrt{/\kappa_z}$}
\caption{Number density of particles in (a) position and (b) velocity for a system initially in a waterbag distribution with $\mu_0=1$, where $K^*=1$ and $r_w=4$. 
The solid line corresponds to the distribution obtained using LB theory, Eq \eqref{eq:flb}, and the points are results of MD simulation with $N_b=50000$
particles, averaged over $100$ dynamical times after the system reached a qSS. Error bars in the distributions are comparable to the symbol size.\label{figplyn}}
\end{figure}

We will first analyze the situation in which the beam envelope at $t=0$ is matched, i.e. satisfies the virial
condition $\mu_0=1$. From our experience with self-gravitating systems, we expect that in this case LB statistics should work reasonably well. 
The electromagnetic potential should then satisfy the Poisson equation \eqref{ap12}, 
with the charge density obtained by integrating the distribution function, Eq \eqref{eq:flb}, over velocities, 
\begin{equation}
\label{eqLBp}
\frac{d^2\psi_{lb}^{b}(r)}{dr^2}+\frac{1}{r}\frac{d\psi_{lb}^{b}(r)}{dr}=
-\frac{4 \pi^2 K^*}{\beta}\ln \left[1+e^{-\beta \left(\psi_{lb}^{b}(r)+\frac{r^2}{2}-\alpha\right)}\right] \,.
\end{equation}
The Lagrange multipliers $\alpha$ and $\beta$ are determined 
using energy and norm conservation. 
The solution to this equation is obtained numerically and 
the resulting marginal distributions  
\begin{equation}
\label{eqnrplb}
N(r)=2 \pi N_b r \int{\rm d^2{\rm v}} f_{lb}({\bf r},{\bf v})%=\frac{4 \pi^2 C}{}%=\frac{2N\lambda^2r}{(\lambda^2+r^2)^2}
\end{equation}
and  
\begin{equation}
\label{eqnvplb}
N(v)=2 \pi N_b v \int{\rm d^2{\rm r}} f_{lb}({\bf r},{\bf v})%=4Nve^{-2v^2}\,.
\end{equation}
are compared with the results of MD simulations in Fig ~\ref{figplyn}, showing a very good agreement.
%%%%%%%%%%%%%%% figure %%%%%%%%%%%%%%%%%%%%%
%%%%%%%%%%%%% end of figure %%%%%%%%%%%%%%%%%%%

\subsection{The test particle model}

In practice, it is very difficult to launch a perfectly matched beam. In most case $\mu_0\neq 1$ and parametric  resonances
will be excited.  To study these, we once again appeal to the model of non-interacting test particles 
moving in an oscillating potential $\psi_{e}(r_b(t))$. We consider $15$ test particles initially distributed uniformly with positions $r_i \in [0,r_m]$ and 
velocities $v_i \in [0,v_m]$. The equation of motion for the particle $i$ is
%%%%%%%%%%%%%%%%%%
\begin{eqnarray}
\label{eqpoinc}
\ddot{r}_i(t)-\frac{{v_{\theta_i}}^2}{{r_i}^3(t)}+r_i(t)&=&
\left\{
\begin{array}{l}
K^*\frac{r_i(t)}{r_b^2(t)}\>\, \text{ for }\>\,r_i(t)\le r_b(t)
\\
\\
K^*\frac{1}{r_i(t)}\>\, \text{ for }\>\,r_i(t)\ge r_b(t)\;,
\end{array}
\right.
\end{eqnarray}
%%%%%%%%%%%%%%%%%%
where $r_b(t)$ evolves according to \eqref{eq:rb2} with $\varepsilon(t)=\varepsilon_0$. 

\begin{figure}[!ht]
\vspace{5mm}
\begin{center}
\includegraphics[width=0.8\textwidth]{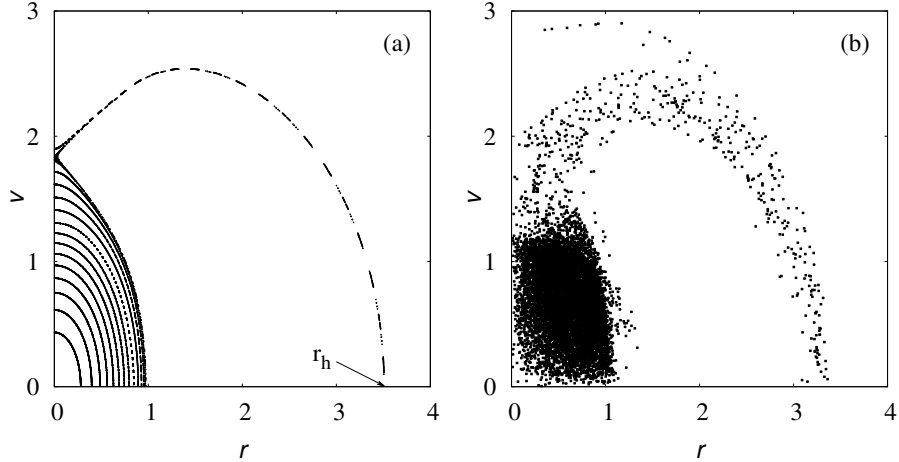}%{figpoincnbp}
\caption{Poincar\'e section of the test particles (a) and phase space of the $N$-body MD simulation (b) using
the Hamiltonian \eqref{hamileffp2d} at $t=200$, for an initial distribution with $\mu_0=1.5$ and $K^*=1$.
The test particle dynamics allows us to determine the maximum position $r_h$ reached and, consequently, the maximum energy $\epsilon_h$ that a particle may attain, $\epsilon_h=\frac{r_h^2}{2}-\ln\frac{r_h}{r_w}$.\label{figpoincnpb}}
\end{center}
\end{figure}

Comparing the result of the test particle dynamics with the full $N$-body MD simulation, shown in Fig \ref{figpoincnpb}, we see that the
reduced test-particle model predicts accurately the location of the resonant orbit.  This allows
us to calculate the maximum energy $\epsilon_h$ that a particle can gain from
the parametric resonance, $\epsilon_h=\frac{r_h^2}{2}-\ln\frac{r_h}{r_w}$, 
where $r_h$ is the maximum distance from the origin reached by a test particle of the initial distribution, see Fig. \ref{figpoincnpb} (a).
Phenomenologically it has been 
found~\cite{AllCha2002} that for beams with 
large space charge $K^*$, $r_h$ is
simply related to the virial parameter and the matched envelope radius,
$r_h=2r_b^*(1+\ln(\mu_0))$. 
%****************************************************************************
%****************************************************************************

\subsection{The core-halo distribution} 

\begin{figure}[!ht]
\vspace{5mm}
\begin{center}
\begin{minipage}{0.45\textwidth}
\includegraphics[width=7cm]{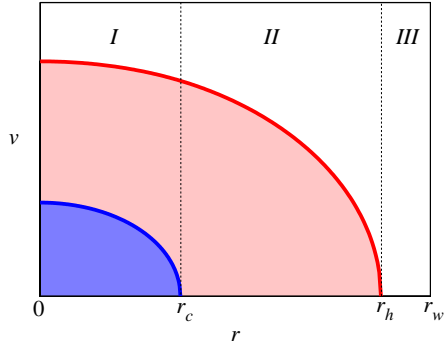}%{distplas3regn.eps}
\end{minipage}
\end{center}
\caption{Regions of phase space used in the solution of equation \eqref{ap12}.\label{distplas3reg}}
\end{figure}

\begin{figure}[!ht]
\begin{center}
\includegraphics[width=0.6\textwidth]{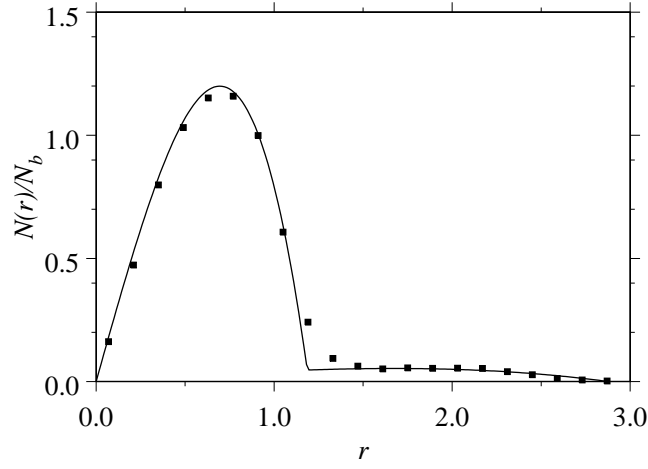}%{distfinalplas1.eps}
\end{center}
\caption{Particle distribution for a mismatched beam, with $\mu_0=1.5$ and $K^*=1$. Points
are results of MD simulation, averaged over $100$ dynamical times in the qSS, and the line shows the prediction obtained using the core-halo distribution, Eq \eqref{eqplasmafrch}. Error bars in the distributions are comparable to the symbol size.
\label{distfinalplas}}
\end{figure}

For mismatched beams ($\mu_0 \neq 1$), we expect that the qSS distribution function will, once again, be of the 
core-halo type, 
%%%%%%%%%%%%%%%
\begin{equation}
f_{ch}({\bf r},{\bf v})=\frac{1}{\pi ^2}\left [\Theta(\epsilon_F-\epsilon(\mathbf{r},\mathbf{v}))
+\chi \Theta(\epsilon_h-\epsilon(\mathbf{r},\mathbf{v})) \Theta(\epsilon(\mathbf{r},\mathbf{v})-\epsilon_F)\right ]\;.
\label{fs}
\end{equation}
%%%%%%%%%%%%%%%
It is convenient to divide phase space into three regions, $I$, $II$, and $III$ (Fig. \ref{distplas3reg}),
corresponding respectively to $r<r_c$, $r_c<r<r_h$, and $r_h<r<r_w$, where $r_c$ is the core radius. 
%%%%%%%%%%%%%%% figure %%%%%%%%%%%%%%%%%%%%%
%%%%%%%%%%%%% end of figure %%%%%%%%%%%%%%%%%
The particle density 
%%%%%%%%%%%%
\begin{equation}\label{eqplasmafrch}
n(r)=\int{f_{ch}({\bf r}, {\bf v})\rm d^2{\rm v}}
\end{equation}
%%%%%%%%%%%
in the three regions can be written as 
%%%%%%%%%%%%%%%%
\begin{equation}
n_I(r)=\frac{2}{\pi}\left[ 
\epsilon_F+\chi(\epsilon_h-\epsilon_F)-V_I(r)
\right],
\label{ni}
\end{equation}
\begin{equation}
n_{II} (r)=\frac{2 \chi}{\pi}\left[\epsilon_h - V_{II}(r)\right ],
\label{nii}
\end{equation}
%%%%%%%%%%%%%%%%
and $n_{III}(r)=0$, where $V_i(r)\equiv {\psi_{ch}}_i(r)+r^2/2$, $i=I,II,III$  
is the total potential that takes into account the effects of the interaction between 
particles as well as the contribution of the external field. The parameter $r_c$ is determined by the condition $V(r_c)=\epsilon_F$. The maximum halo extent $r_h$ is calculated using test particle dynamics, see Fig \ref{figpoincnpb}a. Both ${\psi_{ch}}_i(r)$ and $V_i(r)$ 
and their first derivatives must be continuous at $r=r_c$ and $r=r_h$. 
These conditions, together with the Poisson equation \eqref{ap12}, provide a closed set of equations for the potential in different regions. The equations can be solved analytically, allowing us to calculate the distribution function in the qSS \cite{TelPak2010}. A good agreement between theory and MD simulation is shown in Fig \ref{distfinalplas}.

%%%%%%%%%%%%%%%% figure %%%%%%%%%%%%%%%%%%%%%
%%%%%%%%%%%%% end of figure %%%%%%%%%%%%%%%%%
%To compare with full $N$-body molecular dynamics simulations, besides the initial uniform distribution \eqref{f0p}, we also consider an initial Gaussian distribution, 
%%%%%%%%%%%%%%%%%
%\begin{equation}
%\label{f0g}
% f_{0_g}=\frac{4}{\pi^2r_m^2v_m^2}e^{-2(v^2/v_m^2+r^2/r_m^2)}
%\end{equation}
%%%%%%%%%%%%%%%%%
%and a semi-Gaussian,
%%%%%%%%%%%%%%%%
%\begin{equation}
%\label{f0sg}
%  f_{0_{sg}}=\frac{2}{\pi^2v_m^2r_m^2}e^{-2v^2/v_m^2}\Theta(r_m-r)
%\end{equation}
%%%%%%%%%%%%%%%%
%where $r_m$ and $v_m$ obey eq \eqref{vm}, so that $\varepsilon_0=1$ (see Fig \ref{distfinalplas}).
%\emph{TAKE OUT GAUSSIAN AND SEMIGAUSSIAN PART}

\begin{figure}[!ht]
\begin{center}
%\begin{minipage}{0.45\textwidth}
%\hspace{-4.cm}
%\begin{center}
\includegraphics[width=0.6\textwidth]{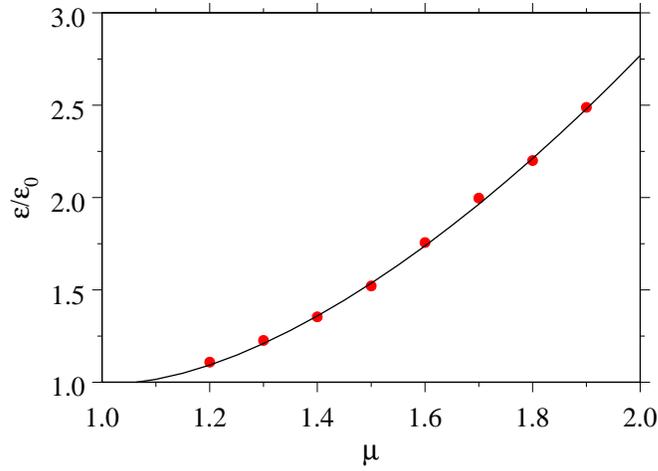}%{emirenato}
%\end{center}
%\end{minipage}
\end{center}
\caption{Emittance growth, $\varepsilon/\varepsilon_0$, as a function of the initial virial parameter $\mu_0$
predicted by the core-halo theory (solid line) and compared with the MD simulations (points)
for $K^*=1$. \label{emigrown}}
\end{figure}

The theory also allows us to predict the emittance
growth, 
a quantity which is of primary
importance for beam physics. Comparing the predictions of the present theory with the results of  MD simulations, an excellent agreement between the two is observed, Fig.~\ref{emigrown}. The theory is also in excellent agreement with the experimental measurements \cite{AllCha2002}.  
%The predictions of the theory are also compared with experimental
%measurements in Fig.~\ref{emigrownwithexp}. 
%%%%%%%%%%%%%%%% figure %%%%%%%%%%%%%%%%%%%%%
%%%%%%%%%%%%% end of figure %%%%%%%%%%%%%%%%%

\begin{figure}[!ht]
\begin{center}
%\begin{minipage}{0.45\textwidth}
\includegraphics[width=0.6\textwidth]{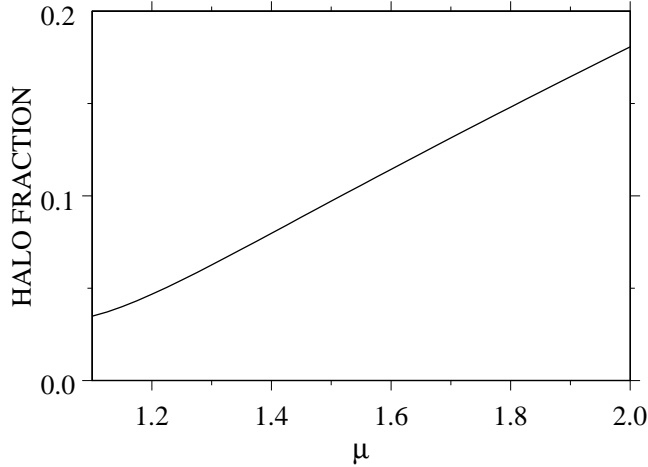}%{haloren.eps}
%\end{minipage}
\end{center}
\caption{Fraction of particles occupying the halo, Eq \eqref{fh}, as a function of the initial
mismatch $\mu_0$, for $K^*=1$.\label{halogrown}}
\end{figure}

The fraction of particles that escape from the core region to form a high energy 
halo can be obtained by integrating the distribution function between the
energies  $\epsilon_F $ and $\epsilon_h$,   
 ${\cal F}_h=(\chi/\pi ^2) \int \Theta(\epsilon_h-\epsilon) \Theta(\epsilon-\epsilon_F) d^2{ r}d^2{\rm v}$ (Fig.~\ref{halogrown}). 
We find 
%%%%%%%%%%%%%%%%
\begin{equation}
{\cal F}_h=1-{2Ar_c^2I_2(\alpha _c r_c)},
\label{fh}
\end{equation} 
%%%%%%%%%%%%%%%%
where $I_n(z)$ is the modified Bessel function of the first kind of order $n$.
%%%%%%%%%%%%%%%% figure %%%%%%%%%%%%%%%%%%%%%
%%%%%%%%%%%%% end of figure %%%%%%%%%%%%%%%%%

\subsection{Relaxation time}

\begin{figure}[!ht]
\begin{center}
%\begin{minipage}{0.45\textwidth}
\hspace{-0.5\textwidth}%-6.cm}
%\includegraphics[width=7cm]{figcolitime}
%\end{minipage}
\includegraphics[width=0.5\textwidth]{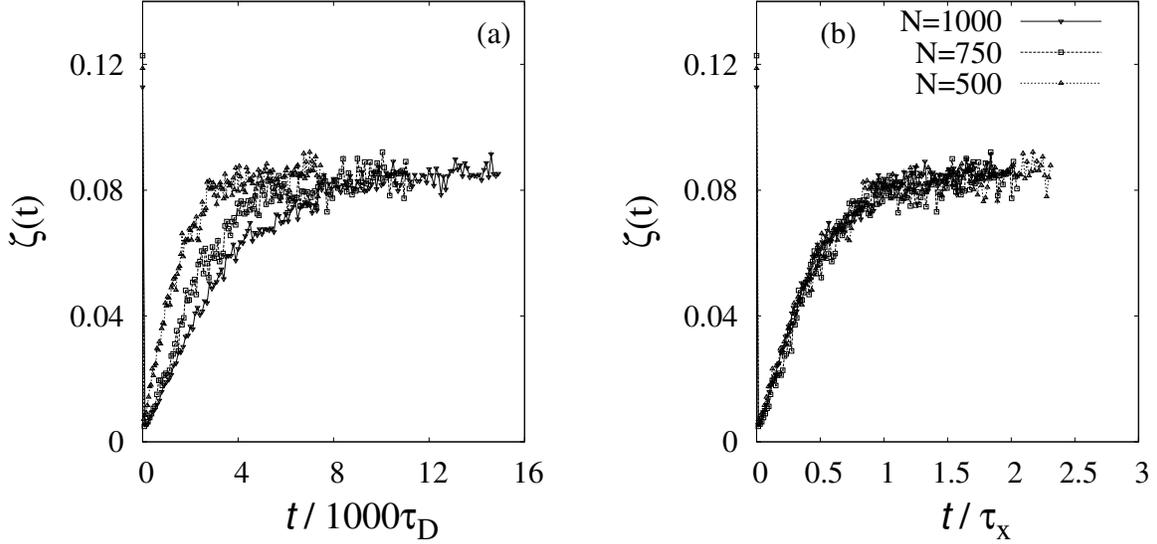}
\end{center}
\caption{(a) $\zeta(t)$ for different numbers of particles in the system. When the dynamical time $\tau_D$ is
rescaled by $\tau_\times$, all points in (a) converge to a universal curve (b). In this case, the asymptotic value of $\zeta$ is $\zeta_{eq}\approx 0.08$.  The simulations were performed with explicit particles 
with initial distribution satisfying the virial condition, $\mu_0=K^*=1$.
\label{figtime2dp}}
\end{figure}

Since plasmas contain astronomical numbers of charged particles, relaxation to Boltzmann-Gibbs thermodynamic
equilibrium will not happen on laboratory time scale.   From the purely theoretical stand point,
however, it is interesting to
study what would happen if the number of particles can be reduced.  This can be easily
achieved on computer, if not in practice.   We thus define a crossover parameter
%%%%%%%%%%%%%%%%
\begin{equation}
\zeta(t)=\frac{1}{N^2}\int_0^{\infty}{[N(v,t)-N_{lb}(v)]^2{\rm d}v}
\end{equation}
%%%%%%%%%%%%%%%%
where $N(v,t)$ is the number of particles with velocity in the interval $[v, v+{\rm d}v]$ at 
simulation time $t$, and $N_{lb}(v)$ is given by Eq \eqref{eqnvplb}. 
The LB distribution was used in the definition of $\zeta(t)$ because we consider cases where the virial condition was initially satisfied. The value of $\zeta(t)$ should tend towards its asymptotic value, $\zeta_{eq}$, as the system approaches thermodynamic
equilibrium. This value is given by
%%%%%%%%%%%%%%%%
\begin{equation}\label{eq:plaszetaeq}
\zeta_{eq}=\frac{1}{N^2}\int_0^{\infty}{[N_{eq}(v)-N_{lb}(v)]^2{\rm d}v},
\end{equation}
%%%%%%%%%%%%%%%%
where $N_{eq}(v)=2\pi N v\int f_{mb}(\mathbf{r},v)\,\mathrm{d}\mathbf{r}$ and $f_{mb}(\mathbf{r},v)$ is the
equilibrium distribution function.
The dynamic time scale is set to $\tau_D = \kappa_z$. If the simulation time 
is scaled with $\tau_\times=N^{\gamma}\tau_D$, where $\gamma=1.3$, all curves fall on the same  
universal curve.   This show that in thermodynamic limit the crossover time diverges as $N^{1.3}\tau_D$(Fig ~\ref{figtime2dp}). 
The result is very similar to the one found in self-gravitating systems. Recently 
a theoretical model based on the  Chandrasekhar collisional mechanism has 
been proposed to account for such large crossover time.  The theory 
predicts that the most important factor in determining the exponent $\gamma$
is the system dimensionality ~\cite{GabJoy2010a,GabJoy2010b}. 
%%%%%%%%%%%%%%%% figure %%%%%%%%%%%%%%%%%%%%%
%%%%%%%%%%%%% end of figure %%%%%%%%%%%%%%%%% 

\subsection{Thermodynamic equilibrium} 

\begin{figure}[!ht]
\begin{center}
\includegraphics[width=0.8\textwidth]{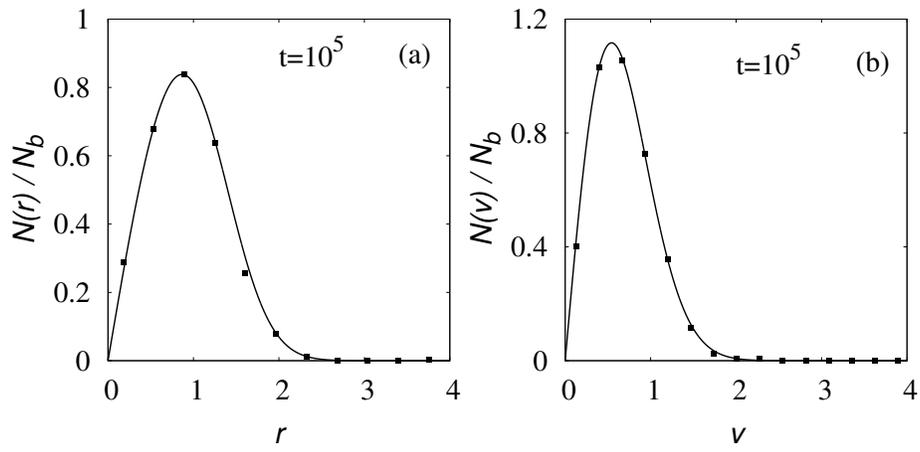}%{figpcoli}
\end{center}
\psfrag{r}{$r/\sqrt{/\kappa_z}$}
\caption{Distribution in (a) position and (b) velocity for a system with ${\cal E}_0=1.597$. The solid line represents
the equilibrium results $N(r)$ and $N(v)$, obtained using the Maxwell-Boltzmann distribution, and the points are the results
of molecular dynamics simulations with $N=1000$ particles. A fourth-order symplectic integrator with constant step size of $dt=10^{-2}$ was used for the molecular dynamics ~\cite{HaiLub2006}.\label{figcolip}}
\end{figure}

After the crossover time $\tau_\times(N)$, during which plasma remains 
trapped in an out of equilibrium qSS, 
it should relax to the thermodynamic equilibrium in which the particle density and velocity
distributions should
be given by the usual Boltzmann-Gibbs statistical mechanics 
%%%%%%%%%%%%%%%%
\begin{equation}\label{eqMBp}
n({\bf r})=Ce^{-\beta \left[\omega({\bf r})+\frac{r^2}{2}\right]} 
\end{equation}
%%%%%%%%%%%%%%%%
and
%%%%%%%%%%%%%%%%
\begin{equation}\label{eqMBpv}
n({\bf v})=\frac{\beta}{2\pi}e^{-\frac{\beta |\mathbf{v}|^2}{2}} \,,
\end{equation}
%%%%%%%%%%%%%%%%
where $C$ is the normalization constant, $\beta=1/T$ is the Lagrange multiplier for conservation of energy, and $\omega({\bf r})$ is the potential of mean force \cite{Lev2002}. 
For large number of particles, the correlations become unimportant and 
$\omega({\bf r}) \approx \psi({\bf r})$. 
The potential $\psi_{eq}$ must then satisfy the Poisson-Boltzmann equation,
%%%%%%%%%%%%%%%%
\begin{equation}\label{eqMB}
\frac{d^2\psi_{eq}(r)}{dr^2}+\frac{1}{r}\frac{d\psi_{eq}(r)}{dr}=-\frac{4\pi^2 K^*C}{\beta}e^{-\beta \left[\psi_{eq}(r)+\frac{r^2}{2}\right]}
\end{equation}
%%%%%%%%%%%%%%%%
with the boundary conditions $\psi_{eq}(r_w)=0$ and ${\psi'}_{eq}(0)=0$.
The solution to this equation can be obtained 
numerically.
In Fig \ref{figcolip} we compare the predictions of the Boltzmann-Gibbs statistical mechanics with the results of 
MD simulations.  The computer runs were performed with not too many particles to allow the system
to relax to equilibrium within reasonable CPU time. Fig \ref{figcolip} shows the 
marginal distributions  $N(r)=2 \pi r n(r)$, and $N(v)=2 \pi v n(v)$ with $n(r)$ and $n(v)$ given by Eqs \eqref{eqMBp} and \eqref{eqMBpv}. As expected, after
a sufficiently long time the system relaxes to the thermodynamic equilibrium.
%%%%%%%%%%%%%%% figure %%%%%%%%%%%%%%%%%%%%%
%%%%%%%%%%%%% end of figure %%%%%%%%%%%%%%%%%%%

\section{The Hamiltonian Mean Field model}

The gravitational and plasma systems studied up to now are of great practical importance.  From the perspective
of statistical mechanics, however, they have a serious drawback --- they do not exhibit a phase transition.  
In the last two sections of this review we will consider two systems with long-range forces which
do show a spontaneous symmetry breaking.  
In particular, we are interested to explore how the phase transitions
between the  qSSs differ from the usual equilibrium phase transitions.
 
The first system that we will study is the Hamiltonian Mean Field (HMF) model.  
 The HMF is a
mean-field version of the $XY$-model, in which all spins interact with each other \cite{KonKan1992,AntRuf1995}. It has become a paradigm of a system with  long-range interaction \cite{LatRap1999,PluLat2004,AntFan2007,BacCha2008}, and is especially interesting due to its phase transition. 
For one-dimensional systems with short-range forces the Mermin-Wagner theorem prohibits spontaneous symmetry breaking in 1D \cite{MerWag1966}. 
The phase transition in the HMF
is only possible  because of the infinite range interaction between the spins \cite{TamAnt2000,DeLeo2013}. 
The HMF model can 
also be considered a simplified representation of a one-dimensional self-gravitating \cite{Val2006a} or Coulomb system \cite{Daw1962} on a ring, and has some similarity with the Colson-Bonifacio model of a single-pass free electron laser \cite{Col1976,BonPel1984,BarBou2005,AntEls2006}.

\subsection{The model}\label{subsec:hmfmodel}

The HMF model can be interpreted in terms of interacting spins or as particles confined to move
on a circle of radius one.  The particle interpretation is more convenient for studying the dynamics of this model, so we will adopt it for most of 
our discussion.
The dynamics of $N$ particles of the HMF  is governed by the Hamiltonian \cite{AntRuf1995}
%%%%%%%%%%%%%%%%%%%%%%%%%%%%%%%%%%%%%%%%%%%%%%%%%% %%
\begin{equation}
\label{eq:hmfHamiltonian}
\mathcal{H} = \sum_{i = 1}^{N}\frac{p_i^2}{2}+\frac{\gamma}{2N}\sum_{i,j=1}^{N}
\left[1 - \cos(\theta_i-\theta_j) \right],
\end{equation}
%%%%%%%%%%%%%%%%%%%%%%%%%%%%%%%%%%%%%%%%%%%%%%%%%% %%
where $\theta_i$ is the coordinate  and $p_i$ the conjugate momentum of the $i$-th particle, and $\gamma$ is a parameter that controls the intensity of the interaction.  The sign of $\gamma $ determines the type of coupling between the particles: if $\gamma > 0 $, the interaction is attractive and the coupling is ferromagnetic; if $\gamma < 0 $, the interaction is repulsive and the coupling is antiferromagnetic.  

The Hamiltonian \eqref{eq:hmfHamiltonian} is a simplification of a one-dimensional gravitational or a Coulomb system with periodic boundary conditions and a neutralizing background. For example, consider a system formed by $ N $ particles distributed along a ring of unit radius, i.e. with position $\theta \in [-\pi, \, \pi] $. The Poisson equation is
%%%%%%%%%%%%%%%%%%%%%%%%%%%%%%%%%%%%%%%%%%%%%%%%%% %%
\begin{equation}
\nabla^2\psi(\theta) = \xi \sum_{i = 1}^N \left[\delta(\theta-\theta_i) - \frac{1}{2 \pi}\right]
\end{equation}
%%%%%%%%%%%%%%%%%%%%%%%%%%%%%%%%%%%%%%%%%%%%%%%%%% %%
where $ \xi $ depends on the system under consideration, and $\psi(-\pi)=\psi(\pi)$, ${\psi'}(-\pi)={\psi'}(\pi)=0$ if $\theta_i=0,\,\forall\, i$. In the gravitational case $ \xi = 4 \pi G m $, where $ G $ is the gravitational constant, $ m = M / N $ is the particle mass and $ M $ the total mass. For the Coulomb case, $ \xi = -q / \varepsilon_0 $, where $q = Q / N $ is the charge density, $Q$ the total charge and $\varepsilon_0$ the vacuum permittivity. The term $ 1/2\pi$  represents the uniform neutralizing background which
is necessary both for Coulomb and gravitational systems with periodic boundary conditions.

Expressing the Dirac delta in its Fourier representation, $ \delta(\theta-\theta_i) = \sum_{n} \exp[\, \mathrm{i} \, n(\theta-\theta_i)] / 2\pi $ and integrating the Poisson equation, the potential produced by $N$ particles is found to be
%%%%%%%%%%%%%%%%%%%%%%%%%%%%%%%%%%%%%%%%%%%%%%%%%%%%
\begin{equation}
\psi(\theta)=\xi\sum_{i=1}^N\sum_{n=1}^{\infty}\left[\frac{1-\cos(n(\theta-\theta_i))}{\pi n^2}\right].
\end{equation}
%%%%%%%%%%%%%%%%%%%%%%%%%%%%%%%%%%%%%%%%%%%%%%%%%%%%
The potential is normalized so that $\psi(0)=0$ when $\theta_i=0, \, \forall i $. 
Truncating the series at $ n = 1 $ and taking $\gamma/N=\xi/\pi$, we recover the potential of the HMF model.

We will consider the ferromagnetic HMF model.  Rescaling time, we can set $ \gamma = 1 $. 
The Hamiltonian \eqref{eq:hmfHamiltonian} can then be written as
%%%%%%%%%%%%%%%%%%%%%%%%%%%%%%%%%%%%%%%%%%%%%%%%%% %%
\begin{equation}
\mathcal{H} = \sum_{i = 1}^N \frac{p_i^2}{2} + \frac{1}{2N} \sum_{i, j = 1}^N
(1 - \cos \theta_i \cos \theta_j-\sin \theta_i \sin \theta_j) \,,
\end{equation}
%%%%%%%%%%%%%%%%%%%%%%%%%%%%%%%%%%%%%%%%%%%%%%%%%%
or
%%%%%%%%%%%%%%%%%%%%%%%%%%%%%%%%%%%%%%%%%%%%%%%%%%
\begin{equation}
\label{eq:hmfHamiltonian2}
\mathcal{H} = \sum_{i = 1}^N \frac{p_i^2}{2} + \frac{1}{2} - \frac{1}{2N}\left(\sum_{i = 1}^N \cos \theta_i \right)^2
- \frac{1}{2N}\left(\sum_{i = 1}^N \sin \theta_i \right)^2.
\end{equation}
%%%%%%%%%%%%%%%%%%%%%%%%%%%%%%%%%%%%%%%%%%%%%%%%%%

The order parameter of the system is the magnetization per particle, $\mathbf{M} = (M_x, M_y) $, which measures
how ``bunched'' is the particle distribution.  If $M=0$ the particles are uniformly distributed over the ring. 
The components of the magnetization are
%%%%%%%%%%%%%%%%%%%%%%%%%%%%%%%%%%%%%%%%%%%%%%%%%%
\begin{align}
\label{eq:hmfmagnetizationx}
M_x = & \langle \cos \theta \rangle = \, \frac{1}{N} \sum_{i = 1}^N \cos \theta_i
\end{align}
and
\begin{align}
\label{eq:hmfmagnetizationy}
M_y = & \langle \sin \theta \rangle = \, \frac{1}{N} \sum_{i = 1}^N \sin \theta_i.
\end{align}
%%%%%%%%%%%%%%%%%%%%%%%%%%%%%%%%%%%%%%%%%%%%%%%%%%
The energy per particle, $ \mathcal{E} = \mathcal{H} / N $, can be written as
%%%%%%%%%%%%%%%%%%%%%%%%%%%%%%%%%%%%%%%%%%%%%%%%%%
%%%%%%%%%%%%%%%%%%%%%%%%%%%%%%%%%%%%%%%%%%%%%%%%%%
\begin{equation}\label{eq:hmfmeanenergy}
\mathcal{E}=\frac{\langle p^2 \rangle}{2} + \frac{1-M_x^2-M_y^2}{2},
\end{equation}
%%%%%%%%%%%%%%%%%%%%%%%%%%%%%%%%%%%%%%%%%%%%%%%%%%
and the one particle energy is
%%%%%%%%%%%%%%%%%%%%%%%%%%%%%%%%%%%%%%%%%%%%%%%%%%
\begin{equation}
\label{eq:hmfepsilon}
\epsilon(\theta_i, p_i) = \frac{p_i^2}{2}+1-M_x \cos(\theta_i)-M_y\sin(\theta_i).
\end{equation}
%%%%%%%%%%%%%%%%%%%%%%%%%%%%%%%%%%%%%%%%%%%%%%%%%% %%
If the initial distribution is 
symmetric in $\theta$, then $ M_y = 0 $,  and in the thermodynamic limit, 
it will remain so throughout the evolution \cite{PakLev2011}.
For now we will only consider symmetric distributions and set $M_y(t)=0$.

\subsection{Thermodynamic equilibrium}\label{subsec:hmfeq}

Classical statistical mechanics provides a prediction for the thermodynamic equilibrium of the HMF model \cite{AntRuf1995}. In this subsection, we shall briefly describe the results in the microcanonical ensemble. A more extensive treatment of the equilibrium state of the HMF model can be found in ref. \cite{CamDau2009}.

The microcanonical ensemble is defined by the surface of constant energy $ E $ in the $2Nd$-dimensional configuration space, $ d $ being the number of degrees of freedom of each particle ($ d = 1 $ for the HMF),
%%%%%%%%%%%%%%%%%%%%%%%%%%%%%%%%%%%%%%%%%%%%%%%%%%%%
\begin{equation}\label{eq:hmfomega}
\Omega(E,N)=\int_{-\pi}^{\pi}d\boldsymbol\theta\int_{-\infty}^{\infty}d\mathbf{p}\,\,
\delta(H(\mathbf{p},\boldsymbol\theta)-E),
\end{equation}
%%%%%%%%%%%%%%%%%%%%%%%%%%%%%%%%%%%%%%%%%%%%%%%%%%%%
where $\boldsymbol\theta$ and $\mathbf{p}$ are $N$-dimensional vectors representing the positions and velocities of all $N$ particles that compose the system: $\boldsymbol\theta=(\theta_1,\,\theta_2,\,\hdots,\,\theta_N)$ and $\mathbf{p}=(p_1,\,p_2,\,\hdots,\,p_N)$. Thus, we also write $d\boldsymbol\theta=\prod_{i=1}^N d\theta_i$ and $d\mathbf{p}=\prod_{i=1}^N d\mathrm{p}_i$. 

\begin{figure}
\begin{center}
\vspace{5mm}
\includegraphics[width=0.65\textwidth]{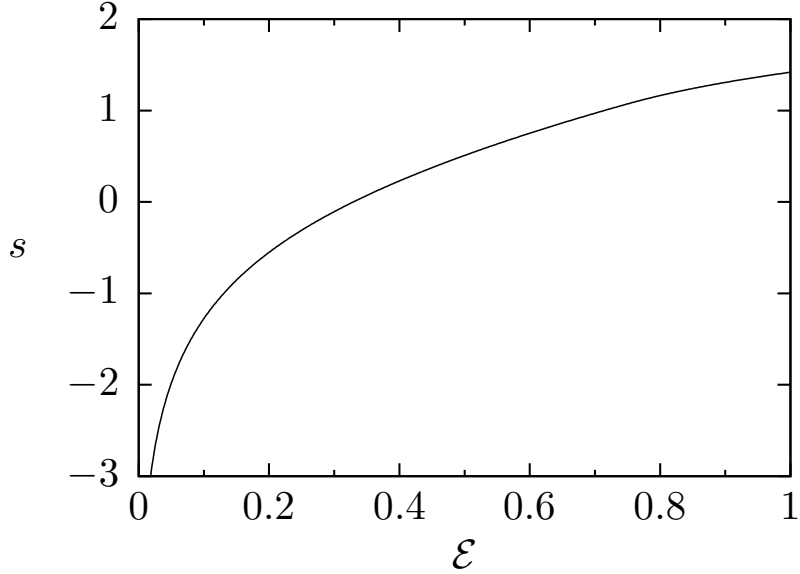}
\caption{Microcanonical entropy as a function of the mean energy for the HMF model.\label{fig:su}}
\end{center}
\end{figure}

\begin{figure}
\begin{center}
\vspace{5mm}
\includegraphics[width=0.65\textwidth]{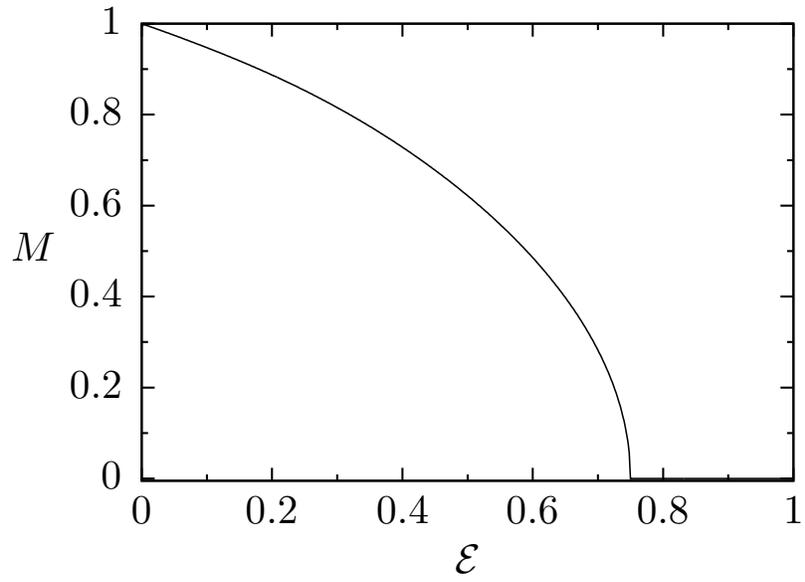}
\caption{Equilibrium magnetization as a function of the mean energy $\mathcal{E}$ for the HMF.\label{fig:mageq}}
\end{center}
\end{figure}

\begin{figure}
\begin{center}
\vspace{5mm}
\includegraphics[width=0.65\textwidth]{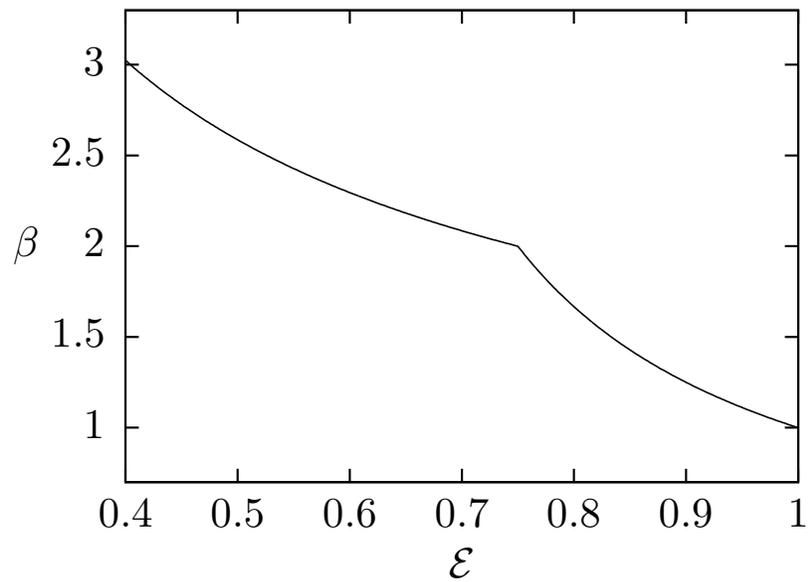}
\caption{The inverse temperature $\beta=1/(2\mathcal{E}-1+M^2)$ as a function of the mean energy $\mathcal{E}$ for the HMF. The sharp corner at $\mathcal{E}=0.75$ indicates a second-order phase transition.\label{fig:hmfbeta}}
\end{center}
\end{figure}

The Boltzmann entropy per particle is $s=\frac{1}{N}\ln\Omega$ which is calculated to be \cite{BarBou2002a,CamDau2009}
%%%%%%%%%%%%%%%%%%%%%%%%%%%%%%%%%%%%%%%%%%%%%%%%%%%%
\begin{equation}
\label{eq:hmfmcentropy}
s(\mathcal{E})=\frac{1}{2}(\ln 4\pi+1)+\sup_{M}\left[\frac{1}{2}\ln\left(\mathcal{E}-\frac{1-M^2}{2}\right)
-\frac{M^2}{2\mathcal{E}-1+M^2}+\ln I_0\left(\frac{M}{2\mathcal{E}-1+M^2}\right)\right].
\end{equation}
%%%%%%%%%%%%%%%%%%%%%%%%%%%%%%%%%%%%%%%%%%%%%%%%%%%%
where $I_n(z)=\int d\theta \cos n\theta \exp(z\cos\theta)$ is the modified Bessel function of the first kind. The curve $s(\mathcal{E})$ is shown on Fig \ref{fig:su}. The equilibrium magnetization is obtained by solving the equation
%%%%%%%%%%%%%%%%%%%%%%%%%%%%%%%%%%%%%%%%%%%%%%%%%%%%
\begin{equation}\label{eq:hmfmageq}
\frac{I_1\left(\frac{M}{2\mathcal{E}-1+M^2}\right)}{I_0\left(\frac{M}{2\mathcal{E}-1+M^2}\right)}=M,
\end{equation}
%%%%%%%%%%%%%%%%%%%%%%%%%%%%%%%%%%%%%%%%%%%%%%%%%%%%
and is plotted as a function of $\mathcal{E}$ in Fig \ref{fig:mageq}. Finally, Fig \ref{fig:hmfbeta} shows the inverse temperature $\beta=1/T$ as a function of $\mathcal{E}$. These figures indicate a second-order phase transition between ferromagnetic and paramagnetic states at $\mathcal{E}_c=0.75$.
%***************************************************
%***************************************************
%****************************************************
%****************************************************

\subsection{Out of equilibrium, quasistationary states}

The results shown in subsection \ref{subsec:hmfeq} are valid if 
the HMF  is able to relax to
thermodynamic equilibrium. However, as we have seen throughout this report, in thermodynamic limit
systems with long-range interactions do not reach the equilibrium, but become trapped in a qSS, the lifetime
of which diverges with the number of particles \cite{YamBar2004}. Thus, in practice the equilibrium state will never
be achieved by the HMF model with a large enough number of particles.  To explore the properties of the qSS and the possible phase transitions between the different nonequilibrium states, we use MD simulations. In this report we focus on simulations with initial distributions of the one-level waterbag type --- Eq \eqref{eq:hmff0}; for results of studies of the qSSs of the HMF model using other types of initial distributions, see for example Refs. \cite{MorKan2006a,PakLev2013a,CamCha2013}.

At $t=0$ the particles are distributed in accordance with the one-level waterbag distribution, 
%%%%%%%%%%%%%%%%%%%%%%%%%%%%%%%%%%%%%%%%%%%%%%%%%%%%
\begin{equation}\label{eq:hmff0}
f_0(\theta,p)=\eta\Theta(|\theta|-\theta_m)\Theta(|p|-p_m),
\end{equation}
%%%%%%%%%%%%%%%%%%%%%%%%%%%%%%%%%%%%%%%%%%%%%%%%%%%%
where $\Theta(x)$ is the Heaviside step function. The constants $\eta$ (density), $\theta_m$ (maximum value of $\theta$) and $p_m$ (maximum value of $p$) are determined by the normalization of the distribution, initial magnetization ($M_0$) and mean energy ($\mathcal{E}$), respectively, 
%%%%%%%%%%%%%%%%%%%%%%%%%%%%%%%%%%%%%%%%%%%%%%%%%%%%
\begin{align}
1&=\int_{-\pi}^{\pi}\mathrm{d}\theta\int_{-\infty}^{\infty}\mathrm{d} p\,f_0(\theta,p),\\
M_0&=\int_{-\pi}^{\pi}\mathrm{d}\theta\int_{-\infty}^{\infty}\mathrm{d} p\,f_0(\theta,p)\cos\theta,
\end{align}
%%%%%%%%%%%%%%%%%%%%%%%%%%%%%%%%%%%%%%%%%%%%%%%%%%%%
and
%%%%%%%%%%%%%%%%%%%%%%%%%%%%%%%%%%%%%%%%%%%%%%%%%%%%
\begin{equation}
\mathcal{E}=\int_{-\pi}^{\pi}\mathrm{d}\theta\int_{-\infty}^{\infty}\mathrm{d} p\,f_0(\theta,p)\frac{p^2}{2}+\frac{1-M_0^2}{2} \,.
\end{equation}
%%%%%%%%%%%%%%%%%%%%%%%%%%%%%%%%%%%%%%%%%%%%%%%%%%%%
These lead to
%%%%%%%%%%%%%%%%%%%%%%%%%%%%%%%%%%%%%%%%%%%%%%%%%%%%
\begin{gather}
\eta=\frac{1}{4\theta_m\,p_m},\label{eq:hmfeta0}\\
M_0=\frac{\sin\theta_m}{\theta_m}\label{eq:hmfthetam} \,,
\end{gather}
%%%%%%%%%%%%%%%%%%%%%%%%%%%%%%%%%%%%%%%%%%%%%%%%%%%%
and
%%%%%%%%%%%%%%%%%%%%%%%%%%%%%%%%%%%%%%%%%%%%%%%%%%%%
\begin{equation}\label{eq:hmfpm} 
p_m=\sqrt{3\,(\,2\mathcal{E}-1+M_0^2)}\,.
\end{equation}
%%%%%%%%%%%%%%%%%%%%%%%%%%%%%%%%%%%%%%%%%%%%%%%%%%%%

To simulate a system composed of $N$ particles, we use two vectors of dimension $N/2$, where the $i$-th component of the first vector represents the angle $\theta_i$ of the $i$-th particle, and similarly the $i$-th component of the second vector is the momentum $p_i$ of the respective particle. As the initial condition, each $\theta_i$ and $p_i$ take a random value between $[-\theta_m,\theta_m]$ and $[-p_m, p_m]$, respectively. For each of these particles, we consider that there exists a particle in a symmetrical position in phase space: $\theta_{i+N/2}=-\theta_{i}$ and $p_{i+N/2}=-p_i$, which ensures that $M_y(t)=0\,\forall\, t$ and increases the simulation speed --- since the dynamics is symmetric, we only need to integrate the motion of half of the particles.

The trajectory of each particle is governed by the equation of motion $\ddot{\theta}_i=\dot{p}_i=-\partial H/\partial\theta_i$, or 
%%%%%%%%%%%%%%%%%%%%%%%%%%%%%%%%%%%%%%%%%%%%%%%%%%%%
\begin{align}\label{eq:hmfforce}
\ddot{\theta}_i &= -\frac{1}{N}\sin\theta_i\sum_{j=1}^N\cos\theta_j+
\frac{1}{N}\cos\theta_i\sum_{j=1}^N\sin\theta_j \nonumber\\
 &=-M_x \sin\theta_i+M_y\cos\theta_i\nonumber \\
&=-M \sin\theta_i \,.
\end{align}
%%%%%%%%%%%%%%%%%%%%%%%%%%%%%%%%%%%%%%%%%%%%%%%%%%%%
The numerical integration is implemented using a fourth-order symplectic integrator \cite{Yos1990}, available online from E. Hairer \cite{Hai2004}. To control the numerical precision, the error in conservation of 
energy per particle $\mathcal{E}$, given by equation \eqref{eq:hmfmeanenergy}, was kept at approximately $10^{-8}$. 

\begin{figure}
\begin{center}
\vspace{5mm}
\includegraphics[width=0.7\textwidth]{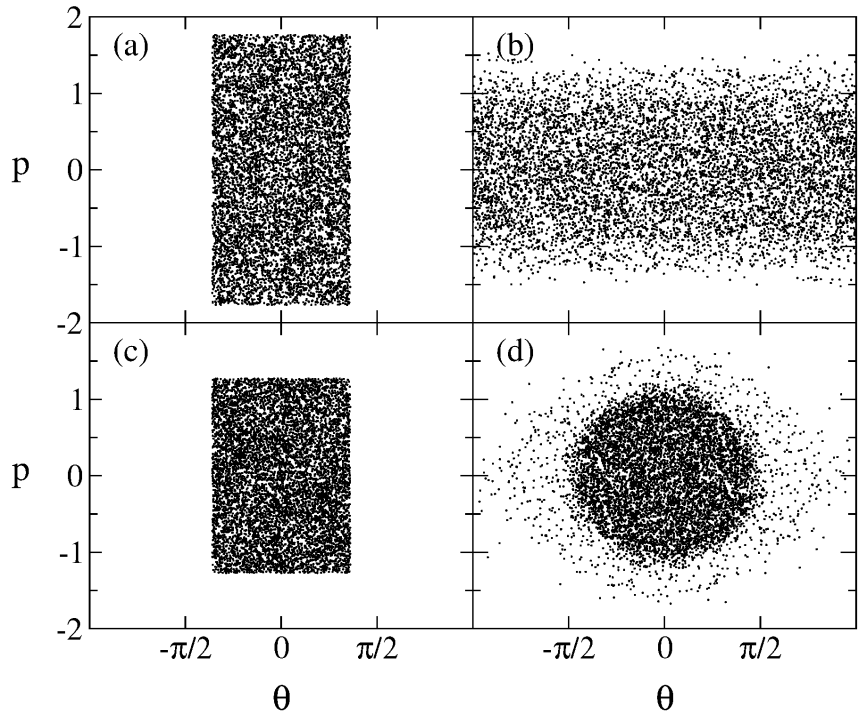}%{m08phases2.eps}
\caption{Phase space of molecular dynamics with $N=2\times 10^5$. The left column shows the initial distributions with (a) $\mathcal{E}=0.7$  and (c) $\mathcal{E}=0.45$. The initial magnetization is the same for both cases, $M_0=0.8$ . The right column shows the two final qSS to which the system relaxes: (b) paramagnetic  and (d) ferromagnetic.  The simulation time is $t=5000$. \label{fig:hmfm08phases}}
\end{center}
\end{figure}

Figure \ref{fig:hmfm08phases} shows examples of two initial phase space distributions, panels (a) and (c), and the respective distributions after a qSS have been achieved, panels (b) and (d). The simulations were performed with $N=2\times 10^5$ particles. The initial magnetization was the same in both simulations, $M_0=0.8$ --- both 
initial waterbags had the same $\theta_m$. The $p_m$'s for the two distributions were different 
corresponding to energies (a) $\mathcal{E} = 0.7 $ and (c) $\mathcal{E} = 0.45 $. The
two initial conditions lead to different phases: the higher energy configuration leads to a paramagnetic (homogeneous) distribution, panel (b),  while the system with lower energy remains magnetized, panel (d). 
%**************************************************
%**************************************************
\begin{figure}
\begin{center}
\vspace{5mm}
\includegraphics[width=0.6\textwidth]{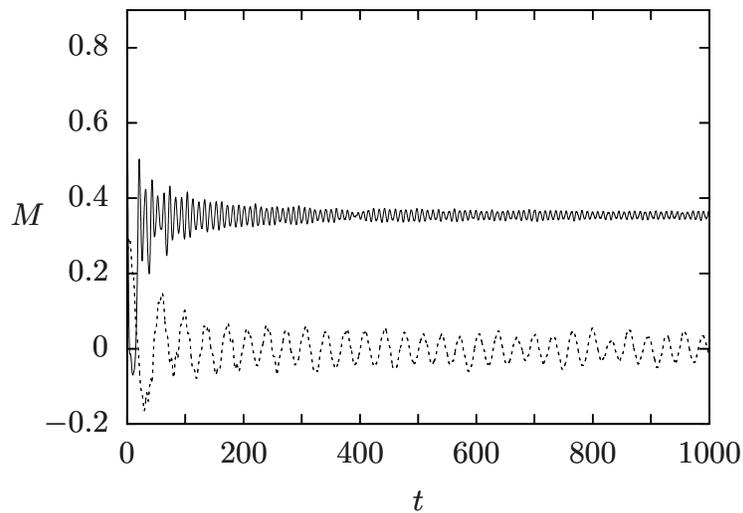}%{mag_u062.eps}
\caption{Magnetization as a function of time obtained using MD simulation with $N=10^6$ particles. 
For the same energy $\mathcal{E}=0.62$,
different initial magnetizations result in distinct qSS: for $M_0=0.8$ ferromagnetic (solid line) and  
for $M_0=0.2$ paramagnetic (dashed line).\label{fig:hmfmagu062}}
\end{center}
\end{figure}
%***************************************************

The final qSS state depends both on the initial magnetization $M_0$ and energy $\mathcal{E}$. This is very
different from the state of thermodynamic equilibrium which depends only on $\mathcal{E}$.   The evolution of $M$
for two systems with the same
energy $\mathcal{E}=0.62$ and different values of $M_0$ is shown in  Fig \ref{fig:hmfmagu062}. 
A system with an initial magnetization $M_0=0.2$ quickly relaxes to a paramagnetic state in 
which its magnetization oscillates around $M=0$.  
On the other hand, a system with $M_0=0.8$ remains magnetized. 
In both cases, the magnetization $M(t)$ oscillates
around its quasi-stationary value $M_s$, given by the temporal average  of $M(t)$  \cite{MorKan2006a}. 
However, while the
oscillations inside the ferromagnetic state are clearly damped, the amplitude of oscillations in the
paramagnetic state remains finite.  The difference between the two states is that inside the ferromagnetic
phase the particles experience a finite mean-field potential produced by $M(t)$ while in the paramagnetic phase the average potential 
is zero.  
This means that inside the ferromagnetic state
some particles can enter in resonance with the oscillations of the potential and gain energy from the
collective motion.  This, in turn, will result in Landau damping of the magnetization
and the relaxation to qSS.  In the paramagnetic phase, $M(t)$ oscillates around zero, so there is no
resonant mechanism to dampen the oscillations.   

\begin{figure}
\begin{center}
\vspace{5mm}
\includegraphics[width=0.6\textwidth]{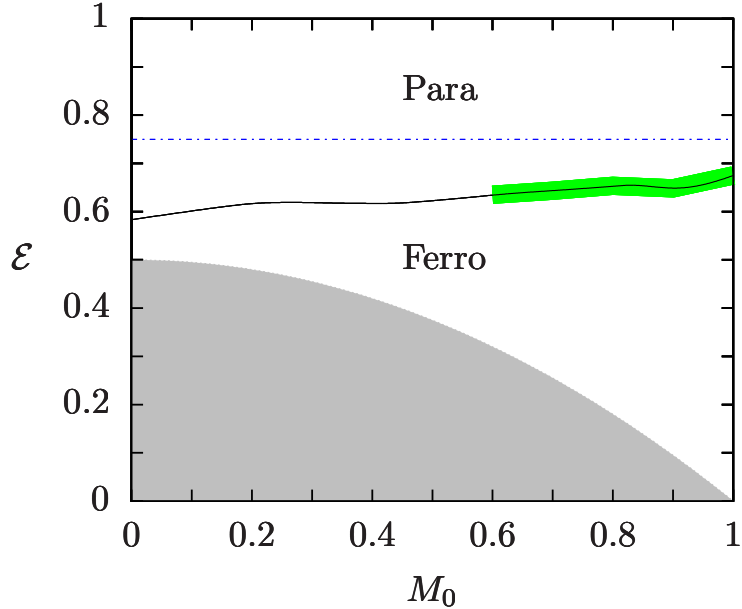}%{phasediag_1}
\caption{Phase diagram of the HMF model. The solid line shows the nonequilibrium transition, obtained using MD simulations. Around this line, for $M_0>0.6$, approximately, the green line shows a region in which the transition is not
very well defined, where ``reentrances'', small ferromagnetic regions exist above the critical line, inside the paramagnetic region. The equilibrium transition, at $\mathcal{E}=0.75$, is represented by the blue dash-dotted line. The gray area represents forbidden initial conditions, delimited by the minimum energy necessary for a given $M_0$.\label{fig:hmfphasediag1}}
\end{center}
\end{figure}

The location of the phase transition can be determined by performing simulations for different initial conditions, varying $\mathcal{E}$ for a fixed initial magnetization and calculating $M_s$. The resulting nonequilibrium phase diagram for the HMF model is shown in Fig \ref{fig:hmfphasediag1}. The results are fairly similar to the nonequilibrium phase diagram found using the Lynden-Bell entropy \cite{AntFan2007a}, yet has some differences, primarily as to the order of the phase transition in some regions, as will be seen further on in this chapter, and in the location of the transition for higher initial magnetizations.
%Performing simulations for different initial conditions, we construct a nonequilibrium phase diagram for the HMF model, shown in Fig \ref{fig:hmfphasediag1}.

It is interesting to compare the nonequilibrium phase diagram with the one found
for the equilibrium of the HMF model.  In equilibrium, the critical energy $\mathcal{E}_e=0.75$  separates the
paramagnetic ($\mathcal{E}>\mathcal{E}_e$) from the ferromagnetic phase ($\mathcal{E}<\mathcal{E}_e$) and is independent of the initial magnetization, 
as is shown by the  
dashed-dotted line of zero slope in the phase diagram, Fig \ref{fig:hmfphasediag1}. 
On the other hand, the transition between the nonequilibrium ferromagnetic and paramagnetic phases occurs at different values of $\mathcal{E}$, depending on the initial magnetization. This transition is represented by a solid line.
The shaded region is the forbidden zone  --- since the minimum kinetic energy is zero, 
$M_0$  determines the minimum allowed 
energy per particle $\mathcal{E}_{\mathrm{min}}=(1-M_0^2)/2$. 
The diagram also shows a region in which the nonequilibrium order-disorder 
transition is not well defined: the wide, shaded line around the critical line for $M_0>0.6$, approximately.
For these values of $M_0$, there are regions where the average energy $\mathcal{E}$ is above the critical line, yet in which the system remains magnetized. Similar regions, or reentrances, have also been observed in studies of the HMF model using numerical resolution of the Vlasov equation \cite{RocAma2012} and Lynden-Bell statistics \cite{StaCha2009}.  Finally, while the equilibrium phase transition between the ferromagnetic and paramagnetic phases is of second order \cite{CamDau2009}, the nonequilibrium phase transition is of first order. 
%***************************************************
%***************************************************

In our studies of  self-gravitating systems and plasmas, we saw the importance
of the virial theorem to determine when strong collective oscillations will occur. However, since 
the potential of the HMF is not a homogeneous function of the separation between the particles, 
we can not directly apply
the results of Section \ref{subsec:virial} to determine the virial condition. To 
discover under what conditions the magnetization of the HMF model will remain constant, 
so that the parametric resonances will not be excited, we need to derive 
a Generalized Virial Condition (GVC).
%***************************************************
\begin{figure}[ht]
\begin{center}
\includegraphics[width=0.6\textwidth]{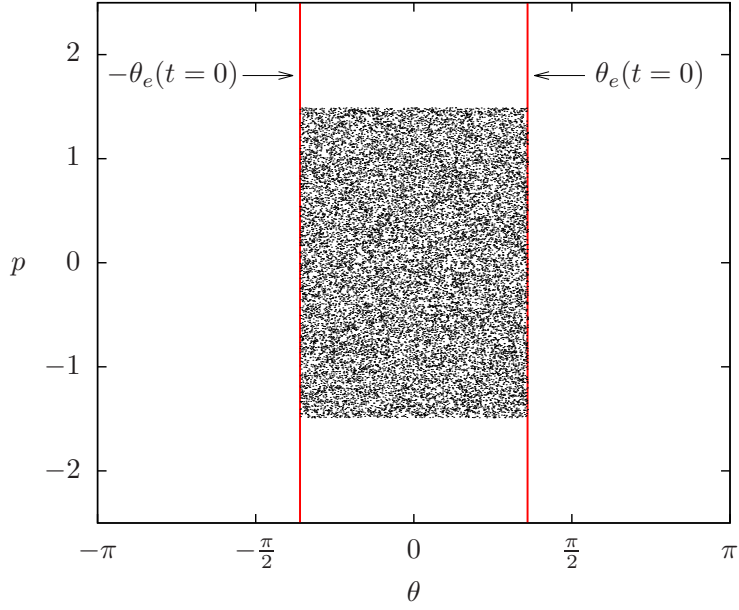}%{env_example}
\caption{Example of the envelope $\theta_e$ at $t=0$ (red lines) in comparison with an initial waterbag distribution of particles (dots).\label{fig:hmfenvexample}}
\end{center}
\end{figure}
%***************************************************
To do this we define the envelope of the particle distribution of the HMF as \cite{BenTel2012}
%%%%%%%%%%%%%%%%%%%%%%%%%%%%%%%%%%%%%%%%%%%%%%%%%%%%
\begin{equation}\label{eq:hmfenvdefinition}
\theta_e(t) = \sqrt{3\langle\theta^2(t)\rangle}.
\end{equation}
%%%%%%%%%%%%%%%%%%%%%%%%%%%%%%%%%%%%%%%%%%%%%%%%%%%%
Note that at $t=0$, the envelope coincides with the maximum $\theta$ of the initial waterbag distribution, $\theta_m$, see Fig \ref{fig:hmfenvexample}. Differentiating equation \eqref{eq:hmfenvdefinition} twice with respect to time, we find
%%%%%%%%%%%%%%%%%%%%%%%%%%%%%%%%%%%%%%%%%%%%%%%%%%%%
\begin{equation}\label{eq:hmfenvelope1}
\ddot{\theta}_e(t)=\frac{3\langle\dot{\theta}^2(t)\rangle}{\theta_e(t)}+\frac{3\langle\theta\ddot{\theta}(t)\rangle}{\theta_e(t)}-\frac{9\langle\theta(t)\dot{\theta}(t)\rangle^2}{\theta_e^3(t)}.
\end{equation}
%%%%%%%%%%%%%%%%%%%%%%%%%%%%%%%%%%%%%%%%%%%%%%%%%%%%
As the result of the conservation of energy, see equation \eqref{eq:hmfmeanenergy}, in the first term, the mean square velocity $\langle\dot{\theta}^2(t)\rangle$ is $2\mathcal{E}-1+M^2(t)$ . To calculate the other averages, we assume the marginal distribution in $\theta$ remains uniform in the interval $[-\theta_e(t),\theta_e(t)]$ and zero outside.
Using this approximation, the second term of equation \eqref{eq:hmfenvelope1} reduces to
%%%%%%%%%%%%%%%%%%%%%%%%%%%%%%%%%%%%%%%%%%%%%%%%%%%%
\begin{align*}
\langle\theta\ddot{\theta}(t)\rangle & =\frac{-M(t)}{2\theta_e(t)} \int_{-\theta_e(t)}^{\theta_e(t)} \theta\sin\theta d\theta \\
 & =\frac{M(t)}{2\theta_e(t)}\left[\theta\cos\theta-\sin\theta\right]^{\theta_e(t)}_{-\theta_e(t)}  \\
 & =\frac{M(t)}{2\theta_e(t)}\left[2\theta_e(t)\cos\theta_e(t)-2\sin\theta_e(t)\right] \\
 & =M(t)\cos\theta_e-M(t)\frac{\sin\theta_e(t)}{\theta_e(t)}.
\end{align*}
%%%%%%%%%%%%%%%%%%%%%%%%%%%%%%%%%%%%%%%%%%%%%%%%%%%%
The last term of equation \eqref{eq:hmfenvelope1} may be neglected by disregarding the correlations between $\theta$ and $p$. The resulting envelope equation is
%%%%%%%%%%%%%%%%%%%%%%%%%%%%%%%%%%%%%%%%%%%%%%%%%%%%
\begin{equation}\label{eq:hmfenvelope2}
\ddot{\theta}_e(t)=\frac{3}{\theta_e}\left(2\mathcal{E}+M_e(t)\cos\theta_e-1\right),
\end{equation}
%%%%%%%%%%%%%%%%%%%%%%%%%%%%%%%%%%%%%%%%%%%%%%%%%%%%
where we have used
%%%%%%%%%%%%%%%%%%%%%%%%%%%%%%%%%%%%%%%%%%%%%%%%%%%%
\begin{align}\label{eq:hmfmagenvelope}
M_e(t)& =\frac{1}{2\theta_e(t)}\int_{-\theta_e(t)}^{\theta_e(t)}\cos\theta d\theta \nonumber\\
 & =\frac{\sin\theta_e(t)}{\theta_e(t)}.
\end{align}
%%%%%%%%%%%%%%%%%%%%%%%%%%%%%%%%%%%%%%%%%%%%%%%%%%%%
Fig \ref{fig:hmfmagenvelope} compares the evolution of magnetization $M_e(t)$, predicted by the equations \eqref{eq:hmfenvelope2} and \eqref{eq:hmfmagenvelope}, with the magnetization obtained using the full $N$-body MD simulation. We see an excellent agreement between the theory and simulation, especially at short times.  
For longer times, the amplitude of the magnetization observed in the simulations is damped, while the envelope oscillations do not. This occurs because the envelope equation is conservative, while in the simulation the
parametric resonances transfer the energy from the collective oscillations to the individual particles.  
%***************************************************
\begin{figure}
\begin{center}
\vspace{5mm}
\includegraphics[width=0.6\textwidth]{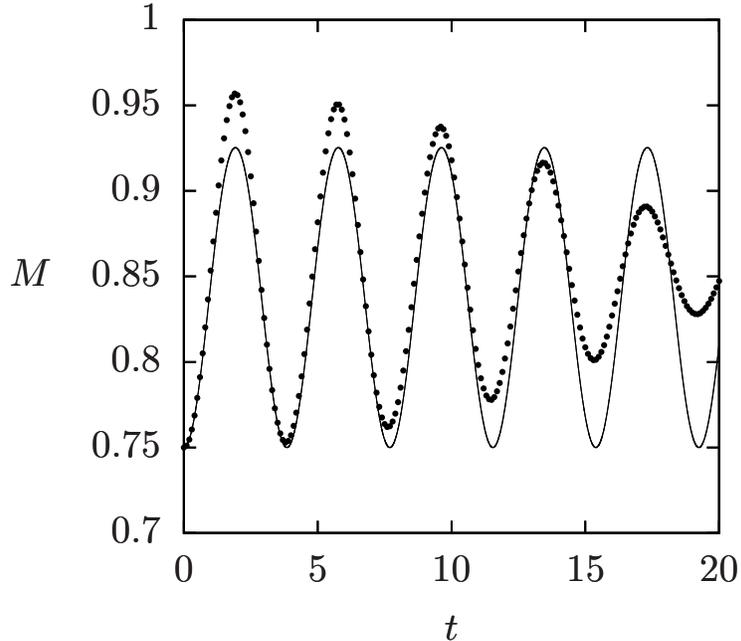}%{env_m075e025}
\caption{Comparison of the magnetization $M(t)=\sum_i\cos\theta_i/N$ of molecular dynamics (dots) and of the envelope magnetization $M_e(t)=\sin\theta_e(t)/\theta_e(t)$ (line). The initial condition is $(M_0=0.75,\mathcal{E}=0.25)$, off the generalized virial curve.\label{fig:hmfmagenvelope}}
\end{center}
\end{figure}
%****************************************************

\begin{figure}[ht]
\begin{center}
\includegraphics[width=0.6\textwidth]{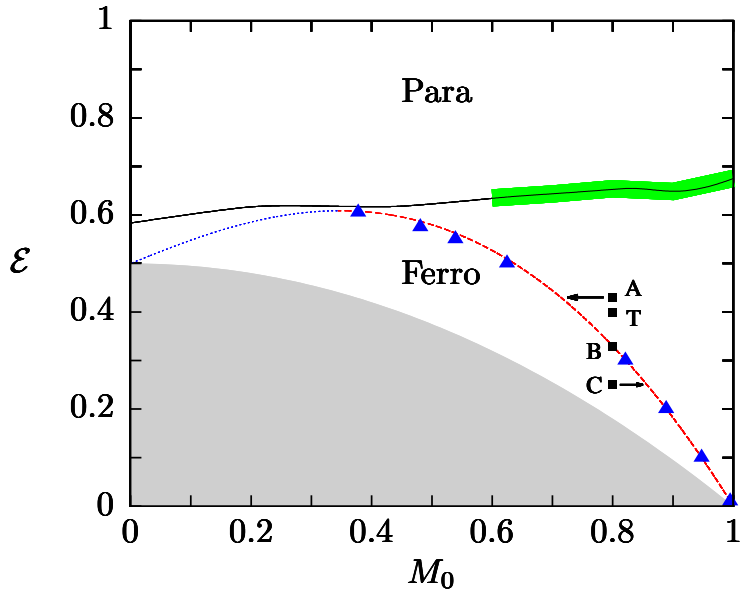}%{phasediag_2}
\caption{Phase diagram of the HMF model, exhibiting the generalized virial condition (red dashed line). Triangles represent the stationary magnetization $M_s$, determined using MD simulations, of systems with initial conditions on the generalized virial condition. Points $A$, $B$ and $C$ show the initial state of the systems corresponding to Fig \ref{fig:hmfdist}, and $T$ and $B$ to Fig \ref{fig:hmfpoincare}. The arrows next to $A$ and $C$ indicate that the stationary magnetization corresponding to these initial conditions is close to the magnetization of the GVC curve for the same energy $\mathcal{E}$. The continuation of the GVC curve, the blue dotted line for $M_0<0.343$, shows an unstable region. The lower gray area represents inaccessible initial conditions. The black solid line shows the phase transition, and the thick green line represents the region of reentrances, where some small ferromagnetic regions exist above the critical line.\label{fig:hmfphasediagram}}
\end{center}
\end{figure}

The GVC corresponds to the initial condition for which the envelope does not oscillate, so that $M_e(t)=M_0$.  This happens when \cite{BenTel2012},
%%%%%%%%%%%%%%%%%%%%%%%%%%%%%%%%%%%%%%%%%%%%%%%%%%%%
\begin{equation}\label{eq:hmfgvc}
2\mathcal{E}+M_0\cos\theta_m-1=0,
\end{equation}
%%%%%%%%%%%%%%%%%%%%%%%%%%%%%%%%%%%%%%%%%%%%%%%%%%%%
so that $\ddot{\theta_e}(t)=0$.  Eq. \eqref{eq:hmfgvc} defines the GVC condition which is plotted by the dashed line in the nonequilibrium phase diagram of Fig \ref{fig:hmfphasediagram}. 

To test the GVC we perform  MD simulations starting with initial waterbag distributions which lie 
directly on top of the GVC curve \eqref{eq:hmfgvc}.  We then plot with triangles in 
Fig \ref{fig:hmfphasediagram}
the final magnetization to which the system relaxes (note that for both the initial and the final state the energy is the same).  We see that the final stationary magnetizations 
$M_s$ are almost exactly  the same as the initial magnetizations $M_0$.
Furthermore, for systems with initial conditions {\it off} the GVC curve, the magnetization quickly changes and begins to oscillate around the stationary value corresponding to $M_s$ on the GVC curve with the same $\mathcal{E}$. For example, points $A$ and $C$ of Fig \ref{fig:hmfphasediagram} each represent initial conditions {\it off} the GVC. Let us call the coordinates of these points $(M_0^A,\mathcal{E}^A)$ and $(M_0^C,\mathcal{E}^C)$, respectively. The stationary values obtained using the  MD simulations correspond to  $(M_s^A,\mathcal{E}^A)$ and $(M_s^C,\mathcal{E}^C)$. The arrows next to points $A$ and $C$ indicate the values of $M_s^A$ and $M_s^C$ on the GVC curve
to which the system relaxes. This result is quite surprising, since the distribution functions for the initial and the final state are very different for systems that do not satisfy the GVC \cite{BenTel2012}.  It is not clear at this moment why the approximate GVC derived using the waterbag distribution works so well to predict
the final magnetizations for systems which initially are very far from their qSS. 

Eq. \eqref{eq:hmfgvc} has an unstable branch, represented by the blue dotted line in Fig \ref{fig:hmfphasediagram}.  
If the initial conditions place the system exactly on this branch, 
the magnetization will remain the same, however, any perturbation will make the system evolve 
from the line of unstable fixed points toward the line of stable ones, represented by the red dashed curve.
%***************************************************
%***************************************************

\subsection{Lynden-Bell theory for the HMF model}

The LB theory has been extensively applied to the HMF model, in some cases showing reasonable agreement with the
results of MD simulations \cite{BarBou2002,Cha2006,AntCal2007,AssFan2012}.  From the examples of gravity and plasma, however, 
we expect that LB theory should only work when the initial distribution satisfies 
the GVC. For non-virial initial conditions, resonances should drive the 
HMF into a qSS with 
a core-halo particle distribution \cite{PakLev2011,BenTel2012}.

The LB distribution for the HMF model is given by \cite{AntCal2007}
%%%%%%%%%%%%%%%%%
\begin{equation}
\label{eq:hmflbdistribution}
\bar{f}_{lb}(\theta,p)=\eta\frac{e^{-\beta(p^2/2-M[\bar{f}_{lb}]\cos\theta-\mu)}}{1+e^{-\beta(p^2/2-M[\bar{f}_{lb}]\cos\theta-\mu)}},
\end{equation}
%%%%%%%%%%%%%%%%%
where $M(\bar{f}_{lb})=\int\bar{f}_{lb}\cos\theta dp d\theta$. The phase space density $\eta$ is determined by the initial 
distribution \eqref{eq:hmfeta0}, while $\beta$ and $\mu$ are the Lagrange multipliers used to preserve the 
norm and the energy.   Solving the system of equations
%%%%%%%%%%%%%%%%%%%%%%%%%%%%%%%%%%%%%%%%%%%%%%%%%%%%
\begin{align}
\mathcal{E}&=\frac{\eta}{2}\int p^2\left[1+\exp(\beta p^2/2-\beta M(\bar{f})\cos\theta+\beta \mu)\right]^{-1}d pd\theta
+\frac{1-M(\bar{f})^2}{2},\label{eq:hmflbencons}\\
1&=\eta\int\left[1+\exp(\beta p^2/2-\beta M(\bar{f})\cos\theta+\beta \mu)\right]^{-1}d pd\theta\label{eq:hmflbncons}
\end{align}
%%%%%%%%%%%%%%%%%%%%%%%%%%%%%%%%%%%%%%%%%%%%%%%%%%%%
and
%%%%%%%%%%%%%%%%%%%%%%%%%%%%%%%%%%%%%%%%%%%%%%%%%%%%
\begin{equation}
M=\eta\int\cos\theta\left[1+\exp(\beta p^2/2-\beta M(\bar{f})\cos\theta+ \beta \mu)\right]^{-1}d pd\theta.\label{eq:hmflbmag}
\end{equation}
%%%%%%%%%%%%%%%%%%%%%%%%%%%%%%%%%%%%%%%%%%%%%%%%%%%%
we can calculate $\beta$, $\mu$ and $M$ and obtain the particle distribution predicted by LB for the qSS.  

\begin{figure}
\begin{center}
\vspace{5mm}
\includegraphics[width=0.6\textwidth]{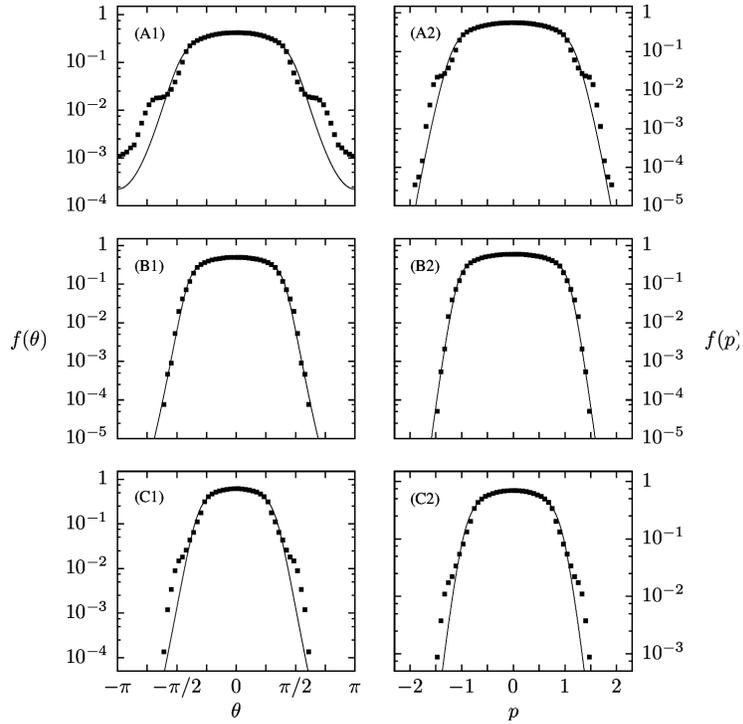}%{histograms_lb_m08}
\caption{Distributions in angle (left column) and momentum (right column) of the stationary states calculated using molecular dynamics (squares) and LB theory (lines) for three different initial conditions --- top row (point $A$ of Fig \ref{fig:hmfphasediagram}): $M_0=0.8,\mathcal{E}=0.43$ (off the generalized virial curve); middle row (point $B$ of Fig \ref{fig:hmfphasediagram}): $M_0=0.8,\mathcal{E}=0.3297$ (on the generalized virial curve); bottom row (point $C$ of Fig \ref{fig:hmfphasediagram}): $M_0=0.8, \mathcal{E}=0.25$ (off the GVC). In the MD simulations were used $N=10^5$ and the corresponding distributions were averaged between times $t=15000$ and $t=17000$. Error bars in the distributions are smaller than the symbol size.\label{fig:hmfdist}}
\end{center}
\end{figure}

In Fig \ref{fig:hmfdist}, we show the marginal distributions,  in angle and momentum, obtained using MD simulations, and compare them with the predictions of LB theory.  
Three different initial conditions are  shown in Fig \ref{fig:hmfphasediagram}: panels $A$ and $C$ 
correspond to non-virial initial conditions, while panel $B$ shows the
initial condition that lies on the GVC. For the non-virial initial conditions, the distribution
functions show a significant deviation from the LB theory.  
On the other hand, the initial distribution that satisfies
the GVC is found to relax to the qSS which is well described by LB theory, panel $B$ of Fig.\ref{fig:hmfphasediagram}.  
%***************************************************
%***************************************************

\subsection{The test particle model}

The discrepancies between the results of MD simulations and the LB theory, for initial distributions which
do not satisfy the GVC, are a consequence of the parametric resonances which transfer the energy from the collective
motion to the individual particles \cite{BenTel2012}. To study these resonances we, once again, appeal to the test particle model.  
%*******************************************
\begin{figure}
\begin{center}
\vspace{5mm}
\includegraphics[width=0.7\textwidth]{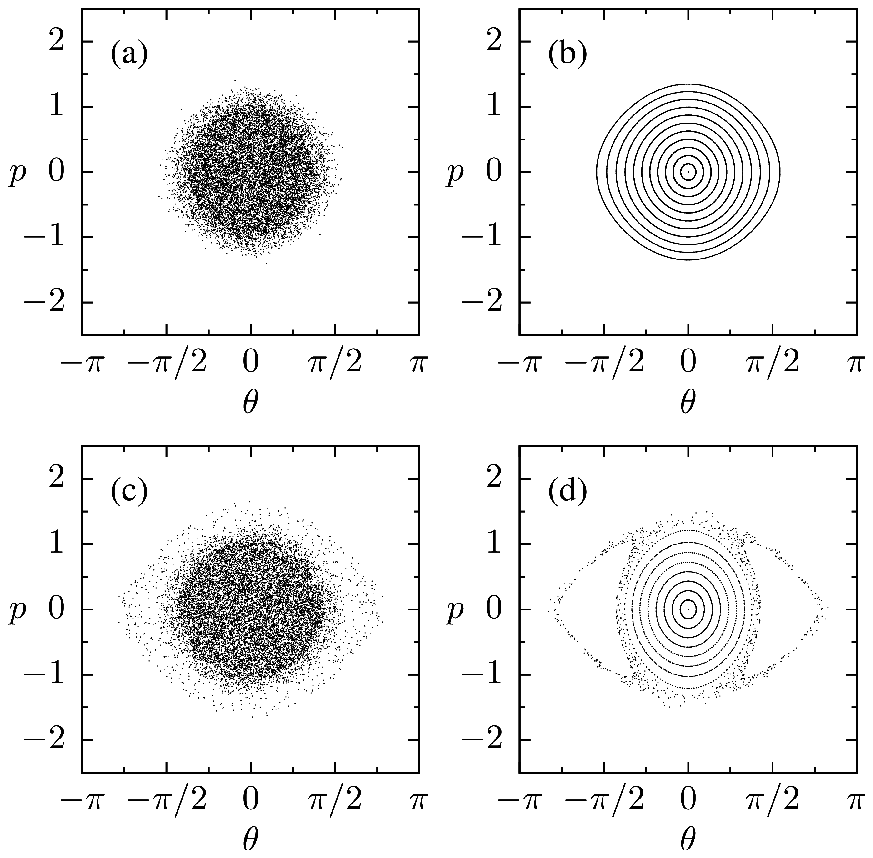}%{poin_xv_m08e04e03297}
\caption{Right column: Poincar\'e sections of test particle dynamics. Left column: phase space of molecular dynamics of $N=10^5$ particles. Top row: initial conditions on the generalized virial curve ($M_0=0.8,\mathcal{E}=0.3297$) -- point $B$ of Fig \ref{fig:hmfphasediagram}. Bottom row: initial conditions off the generalized virial curve ($M_0=0.8,\mathcal{E}=0.4$) -- point $T$ of Fig \ref{fig:hmfphasediagram}.\label{fig:hmfpoincare}}
\end{center}
\end{figure}
%*******************************************
The test particles obey the equation of motion \eqref{eq:hmfforce}, with the magnetization determined by the envelope equation, $M_e(t)$. Fig \ref{fig:hmfpoincare} shows the Poincar\'e sections of test particle dynamics --- the phase space of the test particles plotted when $M_e(t)$ is at its minimum --- compared with the phase space of the HMF, obtained using MD simulation. Two cases are shown:  top panels correspond to the initial conditions that obey the GVC (point $B$ of Fig \ref{fig:hmfphasediagram}), while the bottom panels correspond to the initial conditions slightly off the GVC (point $T$ of Fig \ref{fig:hmfphasediagram}). For the initial distribution satisfying the GVC, the test particle dynamics is regular and no halo is formed. On the other hand, for the non-virial initial distributions (off the GVC), we see resonances which lead to the halo formation in the HMF.

The mechanism of core-halo formation in the HMF is the same as was discussed for gravitational and plasma systems. 
The parametric resonances transfer the energy from the collective motion
to the individual particles.  This, in turn, dampens the collective 
oscillations, forcing the core particles into low energy orbits.  
Once the oscillations die out completely, the dynamics of all the particles becomes integrable, and the 
ergodicity is irreversibly broken. The high energy particles become trapped inside a halo, 
while the low energy particles form a degenerate core.  The LB theory, which relies on the assumptions of ergodicity and
efficient mixing \cite{SakGou1991}, is not able to describe such qSSs \cite{BenTel2012}.

\subsection{The core-halo distribution}

The core-halo distribution for the HMF model is \cite{PakLev2011}
\begin{equation}\label{eq:chhmf}
\bar{f}_{ch}(\theta,p)=\eta\Theta(\epsilon_F-\epsilon(\theta,p))+\chi\Theta(\epsilon(\theta,p)-\epsilon_F)\Theta(\epsilon_h-\epsilon(\theta,p)) \,,
\end{equation}
with the one-particle energy given by $\epsilon(\theta, p) = \frac{p^2}{2}+1-M \cos(\theta)$.  To calculate this distribution we need to determine 
$\epsilon_h$, $\epsilon_F$, $M_s$ and $\chi$. 
The parameters $\epsilon_F$ and $\chi$ are calculated using the conservation of energy and norm, respectively,
%%%%%%%%%%%%%%%%%
\begin{equation}
\mathcal{E}=\frac{1}{2}\int p^2 f_{ch}(\theta, p)d pd \theta +\frac{1}{2}(1-M_s^2),
\end{equation}

\begin{equation}
1=\int f_{ch}(\theta, p)d p d \theta,
\end{equation}
%%%%%%%%%%%%%%%%%
and $M_s$ is given by
%%%%%%%%%%%%%%%%%%%%%%%%%%%%%%%%%%%%%%%%%%%%%%%%%%%%
\begin{equation}
M_s=\int\cos\theta f_{ch}(\theta, p)d \theta d p.
\end{equation}
%%%%%%%%%%%%%%%%%%%%%%%%%%%%%%%%%%%%%%%%%%%%%%%%%%%%

\begin{figure}
\begin{center}
\vspace{5mm}
\includegraphics[width=0.6\textwidth]{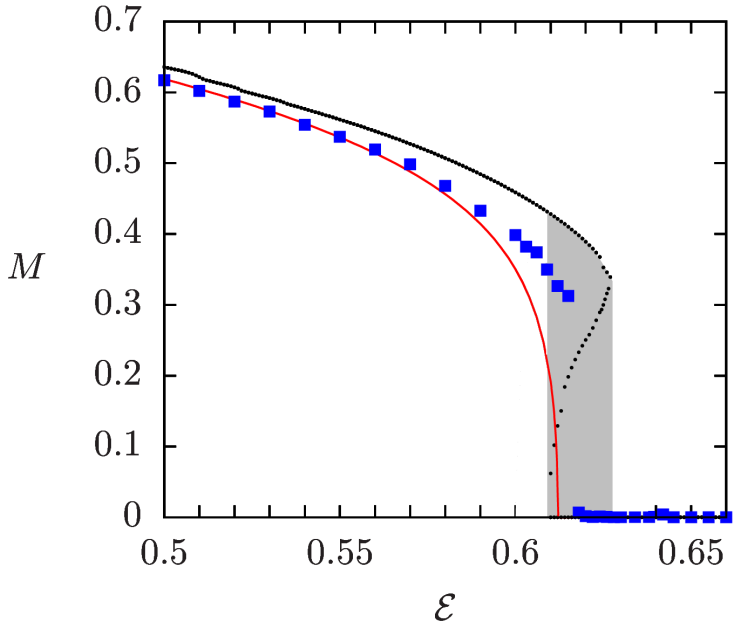}%{m04_transition}
\caption{qSS magnetization according to the core-halo theory (black dots), LB theory (red line), and as determined by MD simulations with $N=2\times 10^6$ particles (blue squares), averaged over $200$ dynamical times in the qSS. For the core-halo theory, $\epsilon_h$ was determined by test particle dynamics. The shaded region shows where the first order transition predicted by the core-halo theory will occur. Error bars of the MD simulation results are comparable to the symbol size.\label{fig:hmfnhlbtrans}}
\end{center}
\end{figure}

To calculate $\epsilon_h$ for gravitational systems and plasmas we have used the test particle dynamics to locate precisely the resonant orbit. However, there is an inherent difficulty in using this approach for the HMF model. The interaction potential for HMF particles is bounded from above. Depending on the initial conditions, some particles can gain enough energy to completely escape the confining potential, and start moving in rotating orbits.
This  makes it difficult to pinpoint the highest possible energy of the resonant particle. In this sense, the HMF model is similar to 3d self-gravitating systems, for which particles can escape the gravitational potential of the cluster. Another difficulty with the test particle dynamics is that while the system is spatially periodic, the envelope equation \eqref{eq:hmfenvelope2} is not. The oscillations of the envelope may be so large that the envelope surpasses $\theta_e=\pi$, in which case an artificial periodicity must be introduced into the test particle dynamics. In spite of these difficulties, we can still attempt to use the core-halo distribution with the approximate values of  $\epsilon_h$
to locate the order-disorder transition in this model. Figure \ref{fig:hmfnhlbtrans} shows the qSS magnetization $M_s$ as determined by the core-halo theory and the test particle dynamics for various values of $\mathcal{E}$ at fixed initial magnetization $M_0=0.4$. 
The core-halo theory predicts a first order phase transition between the paramagnetic and ferromagnetic phases.  
In the same figure we also plot
the prediction of LB theory.  Although the distribution functions of LB theory deviate significantly from the results of MD simulations in the tails, far from the transition point the theory accounts quite accurately for the values of $M_s$.  
LB theory, however, incorrectly predicts that the phase transition between the qSSs for  $M_0=0.4$ is of second order \cite{AntFan2007a}, 
while the simulations find it to be of first order, Fig \ref{fig:hmfnhlbtrans}.
Numerical resolution of the Vlasov equation, which may be used to study the dynamics of the HMF model \cite{Fil2013}, also shows only first-order transitions in the HMF \cite{RocAma2012} . 
%************************************
%************************************

\begin{figure}
\begin{center}
\vspace{5mm}
\includegraphics[width=0.7\textwidth]{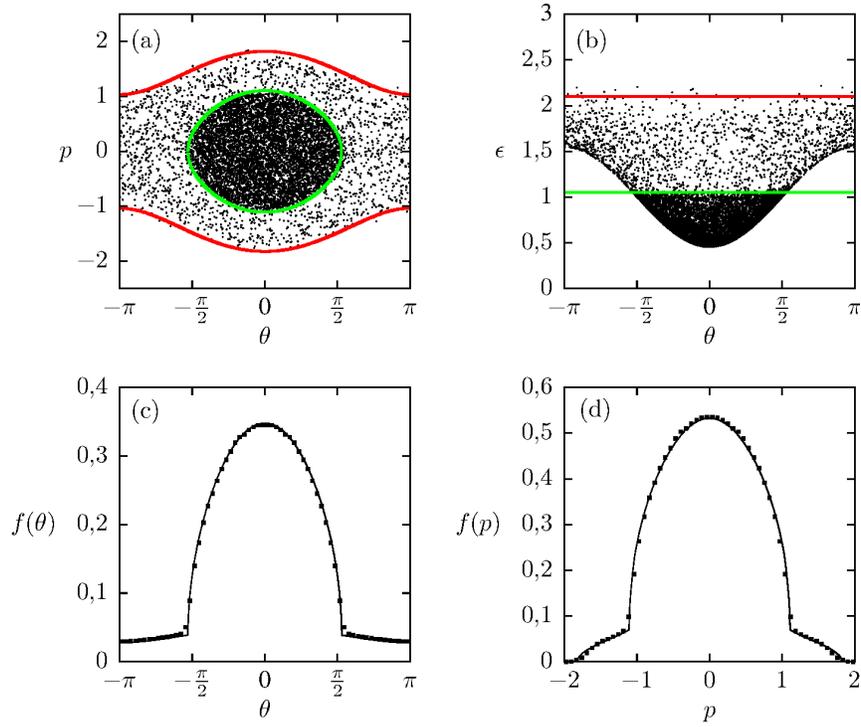}%{testpart_m08e055}
\caption{Comparison of MD simulation with $N=8\times 10^5$ particles and predictions of the core-halo theory for $(M_0=0.8, \mathcal{E}=0.55)$. Panel (a) shows the phase space at $t=10000$ (black dots) and the curves $\epsilon(\theta,p)=\epsilon_h$ (red line) and $\epsilon(\theta,p)=\epsilon_F$ (green line). Panel (b) shows the one-particle energy $\epsilon(\theta,p)$ (black dots) and the energies $\epsilon_h$ (red line) and $\epsilon_F$ (green line). Panels (c) and (d) show the distributions in $\theta$ and $p$, respectively, of molecular dynamics (squares) and core-halo theory (lines). The halo energy $\epsilon_h$ was determined using a short MD simulation with $N=1000$. The distributions of MD simulations are averaged over $100$ dynamical times in the qSS, and error bars are comparable to symbol size.\label{fig:hmfcorehalo}}
\end{center}
\end{figure}

At the moment, we lack a general  method to calculate  the halo energy $\epsilon_h$ for arbitrary values of $M_0$ and $\mathcal{E}$.
The envelope equation and the test particles dynamics allow us to make accurate predictions of $\epsilon_h$ 
for distributions close to the GVC. To predict  the final particle distributions in 
the qSS which are far from the GVC, we can use a short MD simulation 
of the full HMF model with not too many particles. 
Since the formation
of resonances is a fast process, the  $\epsilon_h$ can be defined as the highest energy achieved by any particle after a few oscillations of $M(t)$.  Fig \ref{fig:hmfcorehalo} shows that this procedure leads to an excellent description of the final qSS.
%*************************************
%**************************************

\subsection{Relaxation to equilibrium}

\begin{figure}
\begin{center}
\vspace{5mm}
\includegraphics[width=0.98\textwidth]{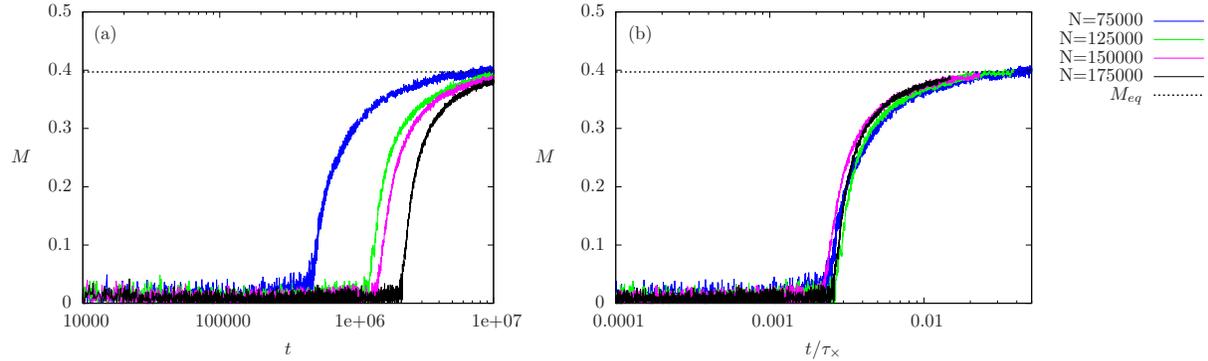}%{fighmfm04e065_gamma17}
\caption{Magnetization as a function of time for different values of $N$: $N=75\times 10^3$ (blue), $N=125\times 10^3$ (green), $N=150\times 10^3$ (magenta) e $N=175\times 10^3$ (black). The results are from MD simulations with initial magnetization $M_0=0.4$, and mean energy $\mathcal{E}=0.65$. For this energy, the equilibrium state is ferromagnetic, while the qSS is paramagnetic. The black dotted line represents the equilibrium magnetization, $M_{eq}=0.397$, corresponding to this energy.\label{fig:hmfmagrelax}}
\end{center}
\end{figure}

For finite $ N $, the lifespan of the qSS is finite, and eventually a crossover to thermodynamic equilibrium occurs \cite{BouGup2010}. In equilibrium the 
particle distribution has the usual Maxwell-Boltzmann form, with the magnetization given by the solution of equation \eqref{eq:hmfmageq} \cite{CamDau2009}. The relaxation to equilibrium is shown in Fig \ref{fig:hmfmagrelax}, which demonstrates the evolution of $M$ for different values of $N$.  The 
initial condition ($M_0 = 0.4$ and $ \mathcal{E} = 0.65 $) is such that the qSS is paramagnetic, while the equilibrium state is ferromagnetic. For this energy, the equilibrium magnetization is $ M_{eq} =0.397$, represented by the black dotted line in Fig \ref{fig:hmfmagrelax}. As the figure shows, the fewer particles in
the system, the faster the magnetization relaxes to the equilibrium value. Rescaling time with $N^{\gamma}$, with $\gamma \approx 1.7$, all the curves collapse onto one universal curve.  The lifespan of the qSS therefore scales with $\tau_{\times}\sim N^{\gamma}$. The exponent $\gamma  \approx 1.7$  is the same as the value found in other studies of the HMF model \cite{YamBar2004}. However, recent large-scale MD simulations show that for large $N$, the exponent $\gamma$ crosses over to $\gamma=2$. This is consistent with the arguments based on the Balescu-Lenard equation, which suggest that the crossover time from a paramagnetic (homogeneous) qSS to a ferromagnetic
equilibrium state should scale as $N^2$ \cite{MarcianoCondMat,FigRoc2013,Cha2012}.
%******************
%*****************

\section{The Generalized Hamiltonian Mean Field model}

From the perspective of statistical mechanics, the HMF model is significantly richer than self-gravitating or plasma systems.  Unlike these systems, the
HMF possesses a genuine nonequilibrium phase transition between qSSs.  
The structure of the phase diagram of the HMF, however, is still relatively simple, since only paramagnetic and ferromagnetic phases exist.  
To explore further the differences between equilibrium and nonequilibrium phase transitions, we introduce 
a Generalized Hamiltonian Mean Field (GHMF) model.  In addition to
paramagnetic and ferromagnetic phases, this model also has a nematic phase. 
In this section we will compare the equilibrium and nonequilibrium 
phase diagrams of the GHMF and show that in the new qSS nematic phase, particles
are once again distributed in accordance with the core-halo distribution.

\subsection{The model}

The Hamiltonian of the GHMF model is given by
%%%%%%%%%%%%%%%%%%%%%%%%%%%%%%%%%%%%%%%%%%%%%%%%%%%%%%%%%%%%%%%%%%%%%%%%%%%%%%%%%%%%%%%%%%
\begin{equation}
H= \sum_{i=1}^{N}\frac{p_i^2}{2}+\frac{1}{2N}\sum_{i,j=1}^{N}\left[1-\Delta\cos(\theta_i-\theta_j)-(1-\Delta)\cos(q \theta_i -q \theta_j)\right],\label{eq:ghmfham}
\end{equation}
%%%%%%%%%%%%%%%%%%%%%%%%%%%%%%%%%%%%%%%%%%%%%%%%%%%%%%%%%%%%%%%%%%%%%%%%%%%%%%%%%%%%%%%%%%
where $q \in \mathbb{N} $ and $ \Delta \in [0,1] $ \cite{TelBen2012}. This model is a long-range version of the models studied in references \cite{LeeGri1985, PodAre2011}. Considering the particles as a collection of spins, the generalized nematic coupling $\cos(q\theta_i-q\theta_j)$ favors either 
alignment or misalignment of spins.  For example, for $q = 2 $, it favors either parallel or antiparallel spins. 
From the perspective of the particle dynamics, either homogeneous or bunched states are possible, with the number of bunches controlled by
the parameter $q$. 

The order parameters for the GHMF model are the generalized magnetizations
%%%%%%%%%%%%%%%%%%%%%%%%%%%%%%%%%%%%%%%%%%%%%%%%%%%%%%%%%%%%%%%%%%%%%%%%%%%%%%%%%%%%%%%%%%
\begin{equation}
\label{eq:ghmfmag1}
M_1=\frac{1}{N}\sum_{i=1}^N \cos\theta
\end{equation}
and
\begin{equation}
\label{eq:ghmfmagq}
M_q=\frac{1}{N}\sum_{i=1}^N \cos (q\theta).
\end{equation}
%%%%%%%%%%%%%%%%%%%%%%%%%%%%%%%%%%%%%%%%%%%%%%%%%%%%%%%%%%%%%%%%%%%%%%%%%%%%%%%%%%%%%%%%%%
Note that the full definition of the magnetizations should include $\langle \sin\theta \rangle$ and $\langle \sin q \theta\rangle$, analogous to the HMF model; however, we neglect these terms because only initial distributions symmetric in $\theta$ will be considered.

The GHMF Hamiltonian  \eqref{eq:ghmfham} can be rewritten as
%%%%%%%%%%%%%%%%%%%%%%%%%%%%%%%%%%%%%%%%%%%%%%%%%%%
\begin{equation*}
\label{eq:ghmfham2}
H =\sum_{i=1}^{N}\frac{p_i^2}{2}+\frac{1}{2}-\frac{1}{2N}\Delta\left(\sum_{i=1}^N\cos\theta_i\right)^2
-\frac{1}{2N}(1-\Delta)\left(\sum_{i=1}^N\cos(q\theta_i)\right)^2,
\end{equation*}
%%%%%%%%%%%%%%%%%%%%%%%%%%%%%%%%%%%%%%%%%%%%%%%%%%%%
The average energy per particle is
%%%%%%%%%%%%%%%%%%%%%%%%%%%%%%%%%%%%%%%%%%%%%%%%%%%%
\begin{equation}
\label{eq:ghmfmeanenergy}
\mathcal{E} = \frac{\langle p^2\rangle}{2}+\frac{1-\Delta M_1^2-(1-\Delta)M_q^2}{2}.
\end{equation}
%%%%%%%%%%%%%%%%%%%%%%%%%%%%%%%%%%%%%%%%%%%%%%%%%%%%
and the one-particle energy is
%%%%%%%%%%%%%%%%%%%%%%%%%%%%%%%%%%%%%%%%%%%%%%%%%%%%
\begin{equation}
\label{eq:ghmfepsilon}
\epsilon(\theta,p) = \frac{p^2}{2}+1-\Delta M_1 \cos\theta-(1-\Delta)M_q\cos(q\theta).
\end{equation}
%%%%%%%%%%%%%%%%%%%%%%%%%%%%%%%%%%%%%%%%%%%%%%%%%%%%

\subsection{Thermodynamic equilibrium}

The procedure for obtaining the equilibrium values of $M_1$ and $M_q$ 
is the same as used for the HMF model. 
Here we present only the final results; more details can be found in the reference \cite{TelBen2012}. The microcanonical entropy is given by
%%%%%%%%%%%%%%%%%%%%%%%%%%%%%%%%%%%%%%%%%%%%%%%%%%%%%%%%%%%%%%%%%%%%%%%%%%%%%%%%%%%%%%%%%%%%%%%%%%%
\begin{align}
\label{eq:ghmfmcentropyfinal}
s(\mathcal{E})=\frac{1}{2}&\ln 2\pi +\frac{1}{2}+\sup_{M_1,M_q}\left[\frac{1}{2}\ln\left(2\mathcal{E} -1+\Delta M_1^2+(1-\Delta) M_q^2\right)-M_1 a(M_1,M_q)\right. \nonumber \\
&\left.-M_q\, b(M_1,M_q)+\ln \left(\int d\theta \exp[a(M_1,M_q)\cos\theta +b(M_1,M_q)\cos q\theta]\right)\right].
\end{align}
%%%%%%%%%%%%%%%%%%%%%%%%%%%%%%%%%%%%%%%%%%%%%%%%%%%%%%%%%%%%%%%%%%%%%%%%%%%%%%%%%%%%%%%%%%%%%%%%%%%
The equilibrium magnetizations correspond to the maximum of the entropy \eqref{eq:ghmfmcentropyfinal} 
and must satisfy the coupled equations 
%%%%%%%%%%%%%%%%%%%%%%%%%%%%%%%%%%%%%%%%%%%%%%%%%%%%%%%%%%%%%%%%%%%%%%%%%%%%%%%%%%%%%%%%%%%%%%%%%%%
\begin{equation}
M_1=\frac{\int d\theta \cos\theta \exp\left[a\cos\theta +b\cos q\theta\right]}{\int d\theta \exp\left[a\cos\theta +b\cos q\theta\right]}\label{eq:xcondition}
\end{equation}
and
\begin{equation}
M_q=\frac{\int d\theta \cos q\theta \exp\left[a\cos\theta +b\cos q\theta\right]}{\int d\theta \exp\left[a\cos\theta +b\cos q\theta\right]},\label{eq:ycondition}
\end{equation}
%%%%%%%%%%%%%%%%%%%%%%%%%%%%%%%%%%%%%%%%%%%%%%%%%%%%%%%%%%%%%%%%%%%%%%%%%%%%%%%%%%%%%%%%%%%%%%%%%%%
where 
%%%%%%%%%%%%%%%%%%%%%%%%%%%%%%%%%%%%%%%%%%%%%%%%%%%%%%%%%%%%%%%%%%%%%%%%%%%%%%%%%%%%%%%%%%%%%%%%%%%
\begin{equation}
\label{eq:ghmfmcm1solution}
a(M_1,M_q)=\frac{\Delta M_1}{2\mathcal{E} -1+\Delta M_1^{ 2}+(1-\Delta)M_q^{ 2}},
\end{equation}
\begin{equation}
\label{eq:ghmfmcmqsolution}
b(M_1,M_q)=\frac{(1-\Delta)M_q}{2\mathcal{E} -1+\Delta M_1^{ 2}+(1-\Delta) M_q^{ 2}}
\end{equation}
%%%%%%%%%%%%%%%%%%%%%%%%%%%%%%%%%%%%%%%%%%%%%%%%%%%%%%%%%%%%%%%%%%%%%%%%%%%%%%%%%%%%%%%%%%%%%%%%%%%

The roots of equations \eqref{eq:xcondition}, \eqref{eq:ycondition}, \eqref{eq:ghmfmcm1solution} and \eqref{eq:ghmfmcmqsolution} determine the equilibrium magnetizations for a given $\mathcal{E} $, $q$ and $\Delta$. 
%************************
%\begin{figure}[ht]
%\begin{center}
%\includegraphics[width=0.95\textwidth]{phasesq2_hor}
%\caption{Equilibrium (Maxwell-Boltzmann) angular distributions of the three phases of the GHMF model with $q=2$: (a) ferromagnetic ($|M_1|>|M_2|\geq 0$), (b) nematic ($|M_2|>|M_1|\geq 0$) and (c) paramagnetic ($M_1=M_2=0$).\label{fig:ghmfphasesq2}}
%\end{center}
%\end{figure}
%************************
\begin{figure}
\vspace{5mm}
\begin{center}
\includegraphics[width=0.75\textwidth]{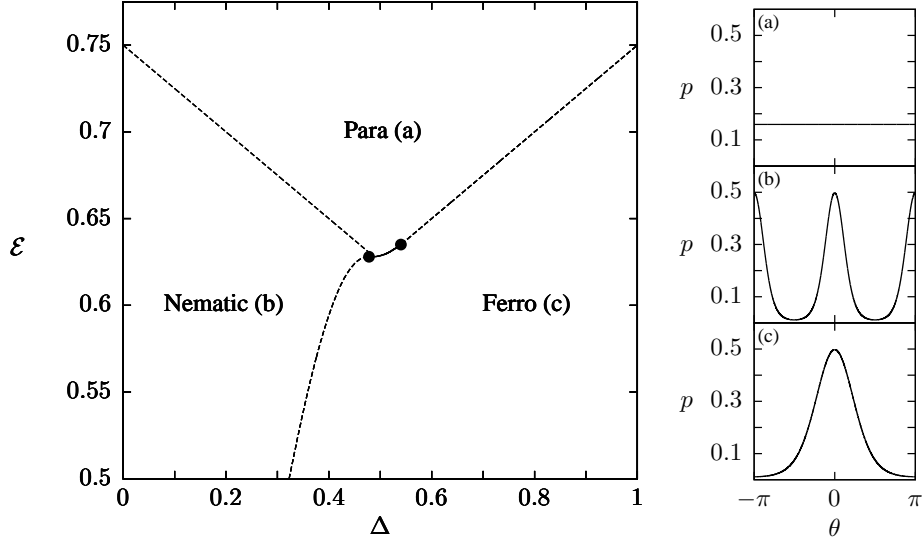}%{eqphase}
\caption{Equilibrium phase diagram (microcanonical ensemble) for $q=2$. The transitions are second order (dashed lines), with the exception of a small region in the center, between two tricritical points (solid circles), in which the transition is first order (solid line). On the right, the three panels show the equilibrium (MB) angular distributions $f(\theta)$ for each phase: paramagnetic (a), nematic (b), and ferromagnetic (c).\label{fig:ghmfmcphaseq2}}
\end{center}
\end{figure}
%************************
Figure \ref{fig:ghmfmcphaseq2} shows the phase diagram for $q=2$ \cite{TelBen2012}. Most transitions are of second order (dashed lines), except for a small region near $\Delta = 0.5$, where the transition is of first order (solid line). 
The equilibrium distribution functions $f(\theta)$ 
for the three phases are illustrated in the right-hand panels of Fig \ref{fig:ghmfmcphaseq2} :
\begin{enumerate}[(a)] 
\item the paramagnetic phase ($M_1=M_2=0$), \\
\item the nematic phase ($|M_2|>|M_1|\geq 0$) and \\ 
\item the ferromagnetic phase ($|M_1|>0, |M_2|\geq 0$).
\end{enumerate}

\begin{figure}
\vspace{5mm}
\begin{center}
\includegraphics[width=.9\textwidth]{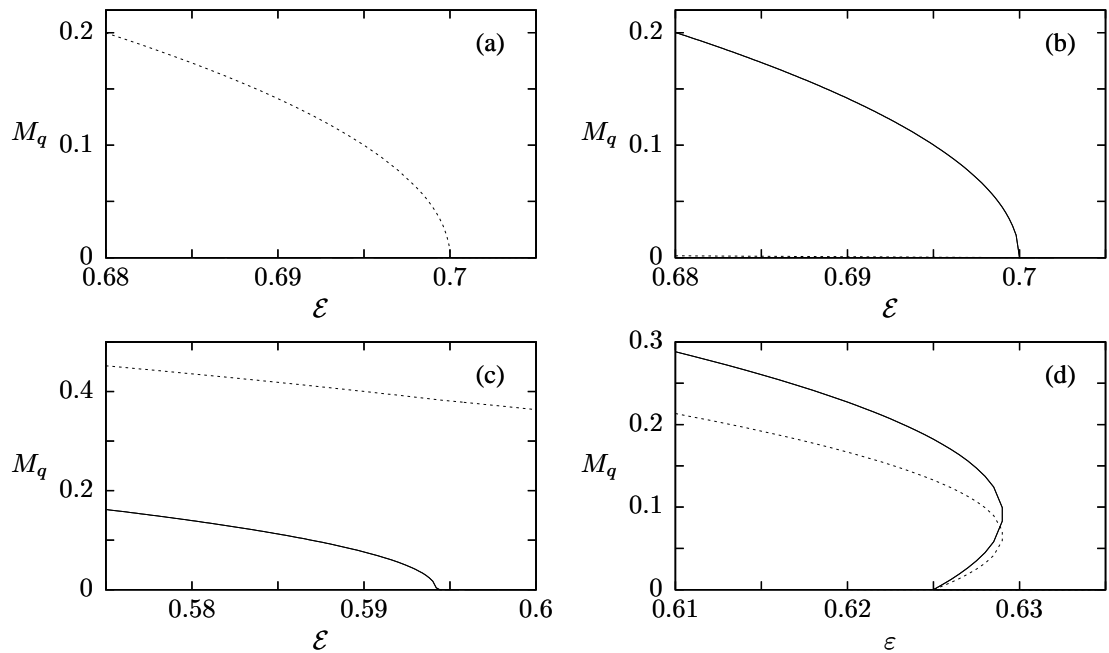}%{eq_transitions}
\caption{Equilibrium solutions of $M_1$ (solid line) and $M_2$ (dotted line) as a function of the mean energy $\mathcal{E}$, exhibiting the (a) nematic-paramagnetic, (b) ferromagnetic-paramagnetic, (c) ferromagnetic-nematic, and (d) 
ferromagnetic-paramagnetic phase transitions, at $\Delta=0.2$, $0.8$, $0.4$ and $0.5$, respectively. The transitions shown in (a), (b) and (c) are second order, and the transition in (d) is first order.\label{fig:ghmfeqtrans}}
\end{center}
\end{figure}

The generalized magnetizations $M_1$ (solid line) and $M_2$ (dotted line) as a function of energy, for four values of $\Delta$, are shown in Fig \ref{fig:ghmfeqtrans}: panels (a), (b) and (c) show second order
transitions (nematic-paramagnetic, ferromagnetic-paramagnetic, and ferromagnetic-nematic, respectively), and panel (d) shows a first order ferromagnetic-paramagnetic transition. In the latter case, the critical energy is the energy for which the entropies of the ferromagnetic and paramagnetic phases are equal.
%************************
%************************

\subsection{Nonequilibrium quasi-stationary states}

Unlike the equilibrium states of the GHMF, which only depends on the initial energy, the qSSs depend explicitly on the initial particle distribution.
In this Report we will explore how the ordered ferromagnetic and nematic phases arise from the initially homogeneous particle distribution of the waterbag form,
%%%%%%%%%%%%%%%%%%%%%%%%%%%%%%%%%%%%%%%%%%%%%%%%%%%%
\begin{equation}\label{eq:ghmfwaterbag}
f_0(\theta,p)=\frac{1}{4\pi p_m}\Theta(\pi-|\theta|)\Theta(p_m-|p|).
\end{equation}
%%%%%%%%%%%%%%%%%%%%%%%%%%%%%%%%%%%%%%%%%%%%%%%%%%%%

In MD simulations, $N$ particles are distributed so that $(-\pi,-p_m)\leq(\theta_i,p_i)\leq(\pi,p_m) $, where $ (\theta_i, p_i) $ is the position and momentum of the $i$-th particle. The average energy per particle is $\mathcal{E}=p_m^2/6$. 
The equation of motion for the $i$-th particle is given by
%%%%%%%%%%%%%%%%%%%%%%%%%%%%%%%%%%%%%%%%%%%%%%%%%%%%
\begin{equation}\label{eq:ghmfforce}
\ddot{\theta_i}=-\frac{\partial H}{\partial \theta_i}=-\Delta M_1(t) \sin\theta_i-2(1-\Delta)M_2(t)\sin(2\theta_i).
\end{equation}
%%%%%%%%%%%%%%%%%%%%%%%%%%%%%%%%%%%%%%%%%%%%%%%%%%%%

In simulations we observe that the system quickly relaxes into a qSS in which $M_1(t)$ and $M_2(t)$ oscillate slightly around their average values ($M_1$ and $M_2$), which 
depend on  $\mathcal{E}$ and $\Delta$. Phase transitions are located by performing a series of simulations varying $\Delta$, for a given value of $\mathcal{E}$, and calculating the average value of $M_1(t)$ and $M_2(t)$ over a time interval during the QSS. The transitions are found to be of first order.

\subsection{Stability of the homogeneous state}

The distribution given by Eq \eqref{eq:ghmfwaterbag} is a stationary solution of the Vlasov equation. Therefore, a transition between a homogeneous state and a non-homogeneous state, either ferromagnetic or nematic, can occur only as  a result of a dynamical instability. Therefore, by studying 
the stability of the homogeneous solution, we should be able to gain an insight into the structure of the phase diagram of the GHMF model. A similar approach has also been used to study the HMF model in an external magnetic field \cite{PakLev2013} and
was shown to agree with the predictions of the linear response theory \cite{PatGup2012}.

To explore the stability of the distribution function Eq. \eqref{eq:ghmfwaterbag}, 
we perturb the upper momentum limit, $p_m$, as
%%%%%%%%%%%%%%%%%%%%%%%%%%%%%%%%%%%%%%%%%%%%%%%%%%%%
\begin{equation}\label{eq:ghmfpm}
p_m(t)=p_0+\sum_{k=1}^{\infty}A_k(t)\cos(k\theta).
\end{equation}
%%%%%%%%%%%%%%%%%%%%%%%%%%%%%%%%%%%%%%%%%%%%%%%%%%%%
We define the generalized magnetizations $M_n$ as
%%%%%%%%%%%%%%%%%%%%%%%%%%%%%%%%%%%%%%%%%%%%%%%%%%%%
\begin{align}
M_n(t)&=\eta\int_{-\infty}^{\infty}d p\int_{-\pi}^{\pi}d \theta\cos(n\theta) \Theta(p_m(t)-|p|)\Theta(\pi-|\theta|)\nonumber \\
 &=2\eta\int_{-\pi}^{\pi}d \theta p_m(t)\cos(n\theta)\nonumber \\
 &=2\eta\int_{-\pi}^{\pi}d \theta p_0\cos(n\theta)+2\eta\sum_{k=1}^{\infty}\int_{-\pi}^{\pi}d \theta A_k(t)\cos(k\theta)\cos(n\theta)\nonumber \\
 &=2\pi\eta A_n(t) \nonumber \\
 &=\frac{A_n(t)}{2p_0},\label{eq:ghmfmagpert}
\end{align}
%%%%%%%%%%%%%%%%%%%%%%%%%%%%%%%%%%%%%%%%%%%%%%%%%%%%
where  $\eta = 1/4 \pi p_0$. Differentiating the term $\langle\cos(n\theta)\rangle $ twice with respect to time, we find the equation of motion
%%%%%%%%%%%%%%%%%%%%%%%%%%%%%%%%%%%%%%%%%%%%%%%%%%%%
\begin{equation}
\ddot{M}_n(t)=-n\langle F(\theta)\sin(n\theta)\rangle-n^2\langle p^2\cos(n\theta)\rangle.
\end{equation}
%%%%%%%%%%%%%%%%%%%%%%%%%%%%%%%%%%%%%%%%%%%%%%%%%%%%
The average values are calculated using the distribution function $f(\theta,p,t)=\eta\,\Theta(p_m(t)-|p|)\Theta
(\pi-|\theta|)$. Thus, the integral above involves an infinite series of cosines. For our analysis, we consider the series up to $k=4$, which will prove to be sufficient to locate and determine the order of the phase transitions. Performing the averages, we obtain a system of differential equations for
the generalized magnetizations,
%%%%%%%%%%%%%%%%%%%%%%%%%%%%%%
\begin{align}
\ddot M_1+\left( \frac{12 \mathcal{E}-6-\Delta}{2} \right )M_1 = f_1(M_1,M_2,M_3,M_4) \label{eq:ghmfop1} \\
\ddot M_2+2\left( 12 \mathcal{E}+\Delta-7 \right )M_2 = f_2(M_1,M_2,M_3,M_4) \label{eq:ghmfop2}\\
\ddot M_3+ 27 (2 \mathcal{E}-1) \, M_3 = f_3(M_1,M_2,M_3,M_4) \label{eq:ghmfop3}\\
\ddot M_4+ 48 (2 \mathcal{E}-1) \, M_4 = f_4(M_1,M_2,M_3,M_4), \label{eq:ghmfop4}
\end{align}
%%%%%%%%%%%%%%%%%%%%%%%%%%%%%%
where
%%%%%%%%%%%%%%%%%%%%%%%%%%%%%%
\begin{align}
f_1\,&=\,{M_1}\,M_2 \left( 1 - \frac{3\,\Delta}{2}\right)+(\Delta-1) \,{M_2}\,
   {M_3}- 3( 2\,\mathcal{E} -1) \,
 \{ {{M_1^3}} + {{M_1^2}}\,{M_3}\, +\nonumber\\
&{M_3}\,[ {M_2}\,( 2 + {M_2} )  + 
        2\,( 1 + {M_2} ) \,{M_4} ]  + 
     2\,{M_1}\,[ {M_2} + {{M_2^2}} + {{M_3^2}} + 
        {M_2}\,{M_4} + {{M_4^2}}\, ]\,\},\label{eq:ghmffunctions1}\\
 f_2\,&=\, \Delta \,
( {{M_1^2}} - {M_1}\,{M_3} + 
     2\,{M_2}\,{M_4} )-2\,{M_2}\,{M_4}-12 \,( 2\,\mathcal{E}-1 ) \,
   [ \,{{M_2^3}} + {{M_3^2}}\,{M_4}\, +\nonumber\\
&2\,{M_1}\,{M_3}\,
      ( 1 + {M_2} + {M_4} )  + 
     {{M_1^2}}\,( 1 + 2\,{M_2} + {M_4} )  + 
     2\,{M_2}\,( {{M_3^2}} + {M_4} + {{M_4}^2}\,
       )  ],\label{eq:ghmffunctions2}\\
  f_3 \,&=\, \frac{3\,{M_1}}{2}\,[\,( 2 - \Delta  ) \,
        {M_2} - \Delta \,{M_4}\, ] -9\,( 2\,\mathcal{E} -1)\,
\{ {{M_1^3}} + 6\,{{M_1^2}}\,{M_3}\,+\nonumber\\
&3\,{M_1}\,[\, {M_2}\,
( 2 + {M_2} )  + 
        2\,( 1 + {M_2}) \,{M_4}\, ]  + 
     3\,{M_3}\,[ {{M_3^2}} + 
        2\,( {{M_2^2}} + {M_2}\,{M_4} + 
           {{M_4^2}}\, )\, ]\,\}\label{eq:ghmffunctions3}
\end{align}
and
\begin{align}
  f_4 \,&=\, 2 \Delta \, M_1M_3-4\,(\Delta -1) M_2^2- 
48\, (2 \mathcal{E} -1)\, [\, 2\, {M_1}\,( 1 + {M_2})\,  {M_3} + 
    {M_2}\, ( {M_2} + {{M_3^2}})\,+\nonumber\\  & 
    2\,( {{M_2^2}} + {{M_3^2}}) \, {M_4} + 
    {{M_4^3}} + {{M_1^2}} \,
     ( {M_2} + 2 {M_4})\,].\label{eq:ghmffunctions4}
\end{align}
%%%%%%%%%%%%%%%%%%%%%%%%%%%%%%%%%%%%%%%%%%%%%%%%%%
Equations \eqref{eq:ghmfop1}--\eqref{eq:ghmfop4} have been written so as to separate linear terms on the left hand side and the nonlinear terms on the right hand side of the equality. To calculate the paramagnetic-ferromagnetic and paramagnetic-nematic phase boundaries, we analyze the linear stability of $M_1(t)$ and $M_2(t)$. Neglecting the nonlinear terms \eqref{eq:ghmffunctions1}--\eqref{eq:ghmffunctions4}, equations \eqref{eq:ghmfop1} and \eqref{eq:ghmfop2} take the form $\ddot{M}_{1, 2} = -\kappa_{1, 2} M_{1, 2}$, whose solutions are $\exp(\pm i \sqrt{\kappa_{1, 2}} t)$.  Thus, the
magnetizations  will remain stable only if  $\kappa_{1, 2} \geq 0$. If $\kappa_{1, 2} < 0$, the exponents will 
become real and any
infinitesimal  fluctuation will experience an exponential growth, destabilizing the paramagnetic phase. The phase 
boundary that separates the paramagnetic phase from the ferromagnetic and nematic phases is, therefore, determined by the conditions $\kappa_1=0$ and $\kappa_2=0$, respectively. According to the equations  \eqref{eq:ghmfop1} and \eqref{eq:ghmfop2}, $\kappa_1=(12\mathcal{E}-6-\Delta)/2$ and $\kappa_2=2(12\mathcal{E}+\Delta-7)=0$ and we find the phase boundaries to be
%%%%%%%%%%%%%%%%%%%%%%%%%%%%%%%%%%%%%%%%%%%%%%%%%%
\begin{equation}\label{eq:ghmfecpf}
\mathcal{E}_c^{pf}(\Delta)=\frac{6+\Delta}{12}
\end{equation}
and
\begin{equation}\label{eq:ghmfecpn}
\mathcal{E}_c^{pn}(\Delta)=\frac{7-\Delta}{12},
\end{equation}
%%%%%%%%%%%%%%%%%%%%%%%%%%%%%%%%%%%%%%%%%%%%%%%%%%
where $\mathcal{E}_c^{pf}$ and $\mathcal{E}_c^{pn}$ are the  boundaries for the paramagnetic-ferromagnetic and paramagnetic-nematic transitions, respectively.

To determine the order of the phase transitions, we study the fixed points of the system of equations \eqref{eq:ghmfop1}--\eqref{eq:ghmfop4}, including the nonlinear terms \eqref{eq:ghmffunctions1}-\eqref{eq:ghmffunctions4}. 
Although the equations are conservative, we expect that in the full GHMF, the 
Landau damping will provide dissipation which will drive the
system towards the qSS.  The dissipation can be included 
by adding  terms proportional to $\dot M_n$ into Eqs. \eqref{eq:ghmfop1}--\eqref{eq:ghmfop4}.
This will make the system relax to the stable fixed points of equations \eqref{eq:ghmfop1}--\eqref{eq:ghmfop4}, 
which will then correspond to the generalized
magnetizations in the final qSS.
We find that once the paramagnetic-nematic boundary is crossed, the value of $M_2$ jumps discontinuously from zero to approximately $0.459$, while $M_1$ remains zero.  The jump in  $M_2$ is 
very close to the value observed in MD simulation, $0.450$, independent of $\Delta$. 
For the paramagnetic-ferromagnetic transition, the two 
magnetizations jump from zero to finite values which depend on $\Delta$. 
In this case the theory is again consistent with the simulations predicting that when crossing the phase transition 
boundary, $M_2$ is always negative, while $M_1$ may be positive or negative. 
%although the predicted quantitative values are not so close (for example, the equations
%determine $(M_1,M_2)$ as $(0.51,-0.19)$, while molecular dynamics results in $(0.41,-0.10)$, for $\Delta=0.5$).

The ferromagnetic-nematic phase boundary should be determined by the two growth rates ($\sqrt{\kappa_{1,2}}$) of $M_1(t)$ and $M_2(t)$. If $M_1$ grows faster than $M_2$, the system will reach the ferromagnetic fixed point prior to reaching the nematic one,
and vice versa.  Therefore, we expect that the ferromagnetic-nematic phase boundary should 
be close to the curve $\kappa_1=\kappa_2$, 
%%%%%%%%%%%%%%%%%%%%%%%%%%%%%%%%%%%%%%%%%%%%%%%%%%
\begin{equation}\label{eq:ghmfecnf}
\mathcal{E}_c^{nf}=(22-5\Delta)/36.
\end{equation}
%%%%%%%%%%%%%%%%%%%%%%%%%%%%%%%%%%%%%%%%%%%%%%%%%%
%****************
\begin{figure}
\vspace{5mm}
\begin{center}
\includegraphics[width=0.75\textwidth]{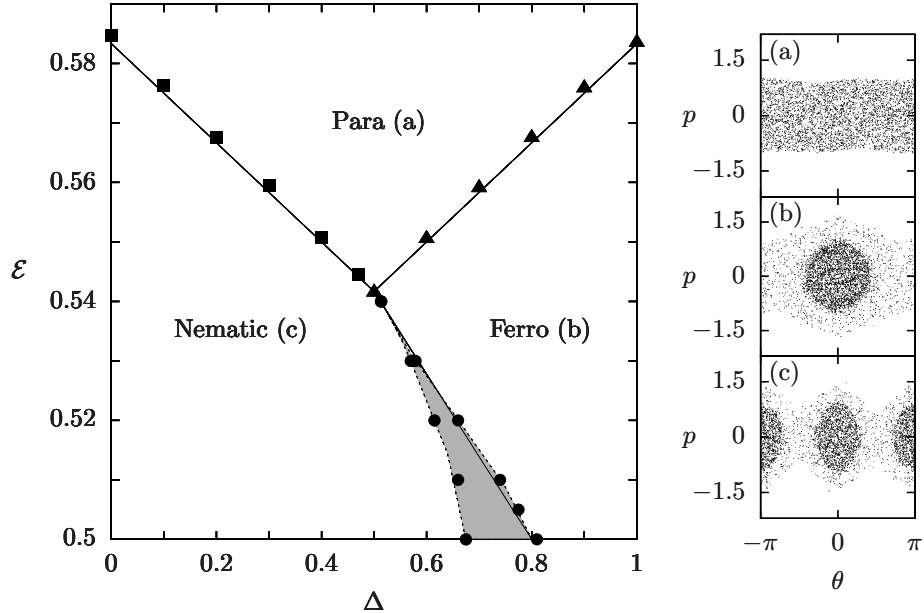}%{neq_phase}
\caption{The nonequilibrium phase diagram of the GHMF model ($q=2$). Lines are the phase transitions predicted by the linear stability analysis. Squares and triangles are the results of MD simulations and represent the paramagnetic-nematic and the paramagnetic-ferromagnetic phase boundaries, respectively. Solid circles show the limits of the nematic-ferromagnetic transition region. Error bars are smaller than the size of the symbols. The gray area, between the circles, is an unstable region where MD simulations find both nematic and ferromagnetic phases, with almost equal probability, see Fig. \ref{fig:nf_un}. The right hand panels show examples of the phase space distributions obtained using the MD simulations for each of the three phases: (a) paramagnetic, (b) ferromagnetic, and (c) nematic.\label{fig:ghmfphasefull}}

\end{center}
\end{figure}
%*****************

%****************
\begin{figure}
\vspace{5mm}
\begin{center}
\includegraphics[width=0.75\textwidth]{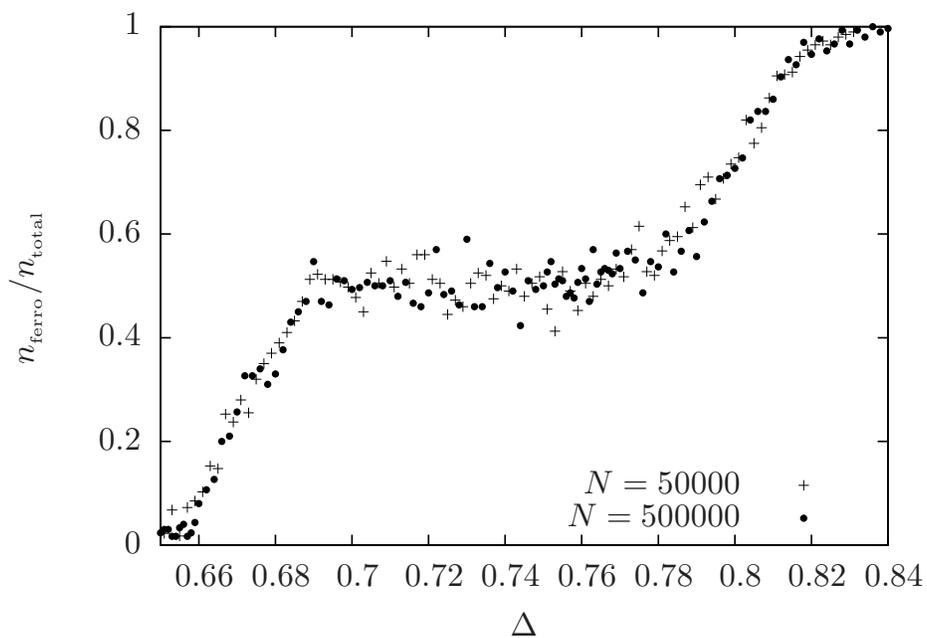}%{neq_phase}
\caption{The probability of finding  a ferromagnetic phase, within
the instability region of Fig. \ref{fig:ghmfphasefull}, at energy $\mathcal{E}=0.5$ for various values of $\Delta$.
To calculate the probability for $N=50000$, we have used $n_{total}=100$ different initial conditions drawn from the same
waterbag distribution, Eq. \ref{eq:ghmfwaterbag}, and observed how many of these ($n_{ferro}$) evolved into a ferromagnetic phase.  For $N=500000$, we have used $n_{total}=300$ different initial conditions
for each value of $\Delta$. \label{fig:nf_un}}

\end{center}
\end{figure}
%*****************
Figure \ref{fig:ghmfphasefull} show the nonequilibrium phase diagram for the GHMF model for an initially homogeneous particle  distribution. The theoretically calculated phase boundaries  obtained using equations \eqref{eq:ghmfecpf}, \eqref{eq:ghmfecpn} and \eqref{eq:ghmfecnf} are shown as the solid lines.  The results of MD simulations are shown as symbols. The paramagnetic-nematic and the paramagnetic-ferromagnetic phase boundaries predicted by the theory are in perfect agreement with the results of MD simulations. For the ferromagnetic-nematic transition the simulations find an instability region in which either
phase can occur with equal probability, Fig. \ref{fig:nf_un}.  The theoretically predicted phase boundary for the ferromagnetic-nematic transition Eq. \eqref{eq:ghmfecnf} passes through the instability region.

\subsection{The core-halo distribution}

\begin{figure}
\vspace{5mm}
\begin{center}
\includegraphics[width=0.75\textwidth]{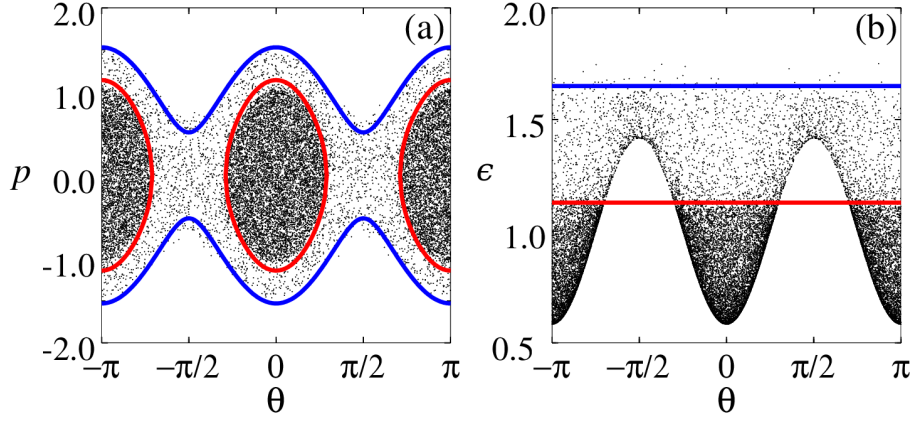}%{nematicophase.eps}
\caption{(a) Phase space particle distribution and (b) one-particle energy obtained 
using MD simulation for GHMF with $\Delta=0.2$ and $N=10^5$ particles.  
In panel (a) the blue line shows the orbit corresponding to 
energy $\epsilon_h$ and the red line to the orbit with energy  $\epsilon_F$.  In panel (b)
the same color lines show the halo and Fermi energies. The initial distribution was homogeneous (paramagnetic) waterbag of energy $\mathcal{E}=0.55$\label{fig:nempha}}
\end{center}
\end{figure}

The particle distributions in the ferromagnetic and nematic phases are, once again,
of the core-halo form, Eq. \eqref{eq:chhmf}, with the one-particle energy given
by Eq. \eqref{eq:ghmfepsilon}.  In Fig. \ref{fig:nempha} we plot a snapshot of the
phase space of the GHMF and the energy
of each particle once the system has relaxed into a nematic qSS.  In both panels of Fig. \ref{fig:nempha} a core-halo structure can be clearly seen.  In the nematic 
phase it actually 
appears that there are two cores.  This happens because $M_1=0$ and
the one-particle energy has two minimums at $\theta=0$ and $\theta=\pi$.  Both cores, however, 
appear in the core-halo distribution function, given by
%%%%%%%%%%%%%%%
\begin{equation}\label{eq:ghmffch}
f_{ch}(\theta,p)=\eta\Theta(\epsilon_F-\epsilon(\theta,p))+\chi\Theta(\epsilon_h-\epsilon(\theta,p))\Theta(\epsilon(\theta,p)-\epsilon_F),
\end{equation}
%%%%%%%%%%%%%%%
where $\eta$ and $\chi$ are the phase space densities of the core and halo, respectively; $\epsilon_F$ and $\epsilon_h$ are the maximum energies of the core and halo, respectively; and the one-particle energy $\epsilon(\theta,p)$ is given by Eq \eqref{eq:ghmfepsilon}. 
%****************
%*****************

\begin{figure}
\vspace{5mm}
\begin{center}
\includegraphics[width=0.75\textwidth]{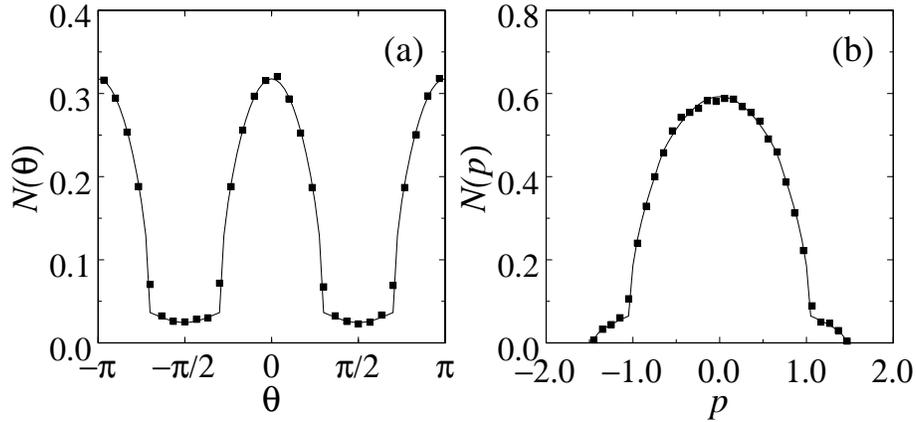}%{nematicodist.eps}
\caption{Marginal distributions $N(\theta)$, Eq \eqref{eq:ghmfntch}, and $N(p)$, Eq \eqref{eq:ghmfnpch}, for a nematic qSS of the GHMF model.  
All the parameters are the same as in Fig. \ref{fig:nempha}. \label{fig:nemdis}}
\end{center}
\end{figure}

In Fig. \ref{fig:nemdis} we plot the marginal distributions calculated
using the core-halo theory,
%%%%%%%%%%%%%
\begin{equation}\label{eq:ghmfntch}
N(\theta)=\int f_{ch}(\theta,p)\,dp
\end{equation}
and
\begin{equation}\label{eq:ghmfnpch}
N(p)=\int f_{ch}(\theta,p)\,d\theta
\end{equation}
%%%%%%%%%%%%%
with $f_{ch}(\theta,p)$ given by Eq \eqref{eq:ghmffch}, and compare them with the results of MD simulations.  
The halo energy $\epsilon_h$ (blue line in Fig \ref{fig:nempha}(b)) was obtained
using a short simulation with $N=1000$ particles, which ran for 
only $10$ dynamical times. The 
Fermi energy $\epsilon_F$ and the halo phase space density $\chi$ were
calculated using the conservation of energy and of norm.  
The predicted value for the Fermi energy $\epsilon_F$ is the red
line in Fig. \ref{fig:nempha}(b).  In panel (a) of the same figure we show the 
orbit of a particle with energy equal to $\epsilon_F$ (red line).  This orbit
perfectly encloses the core. In the same panel, the blue line represents an orbit of a particle
with energy $\epsilon_h$.

%****************
%*****************

As with other long-range systems, eventually the GHMF will relax to thermodynamic equilibrium described
by the Boltzmann-Gibbs statistical mechanics. The resultant phase diagram will then change to the one shown in Fig \ref{fig:ghmfmcphaseq2}. In the thermodynamic limit $N\to\infty$, this relaxation, however, will never occur and the system will
remain trapped forever in one of the qSSs.  

\section{Conclusions and Perspectives}

In this Review we have explored statistical mechanics of 
systems with long-range interactions.  
A number of different examples have been considered, ranging from
plasmas and self-gravitating systems to the kinetic spin models.  In the thermodynamic limit,  these 
systems do not relax to the Boltzmann-Gibbs 
equilibrium, but become trapped in the qSSs, the life 
time of which diverges
with the number of particles $N$.  If $N$
is small, after staying in the qSS for a time of approximately  
$\tau_\times \sim N^\gamma$, where $\gamma$  is usually larger or equal to one,  
a system relaxes to  the thermodynamic equilibrium
described by the usual Boltzmann-Gibbs statistical mechanics.  
This is  what has been observed for all the models studied so far --- 
after a time $\tau_\times$, they all (with the exception of 3D gravity, 
which always remains out of equilibrium) relaxed to 
thermodynamic equilibrium.
In this respect, speculations
that long-ranged systems should be described by the
non-extensive Tsallis statistics are unfounded \cite{Tsa2009}.  

In the case of plasmas and elliptical galaxies, the number of ``particles'' is 
so large that the state of thermodynamic equilibrium can not be reached within the life time
of the universe.  Furthermore, for 3D gravity, we saw that there is an 
additional problem related to the bounded (from above) nature of Newton's gravitational 
potential and the resulting
flux of evaporating particles. For 1D and 2D gravitational systems, on the other hand,
there is no problem with particle evaporation.  
After a short time, these systems relax to qSSs which have a characteristic 
core-halo structure.  The distribution function that describes qSSs of self-gravitating systems
is the same as the one that describes the qSS of magnetically confined plasmas and of spin systems.  
The ubiquity of core-halo
distributions, observed in so many different contexts, suggests that there is
a significant degree of universality to the process of collisionless relaxation.
The core-halo distribution appears to be a 
universal attractor  --- in a coarse-grained sense --- analogous to the 
Maxwell-Boltzmann distribution for 
systems with short-range forces. 

\begin{figure}[!htb]
\begin{center}
\includegraphics[width=0.8\textwidth]{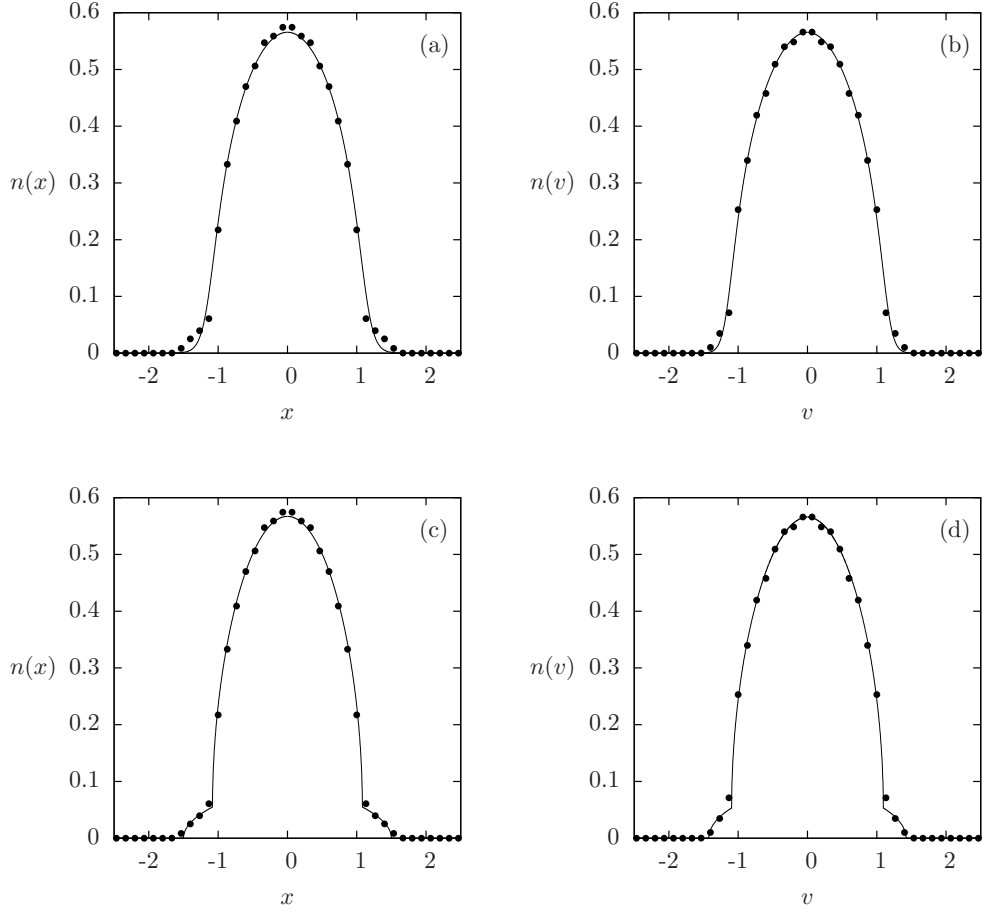}%{figconcg1dchdist}
\end{center}
\caption{Comparison of theoretical (lines) and $N$-body MD simulation (dots) results for the 1D self-gravitating system that
initially satisfies the virial condition (${\cal R}_0=1$). In panels (a) and (b), the theoretical distribution corresponds to LB theory, and in panels (c) and (d) corresponding to core-halo theory. \label{fig:concchdist}}
\end{figure}

A qSS reached by a long-range interacting system depends explicitly on the
initial particle distribution. In this Report we have considered only the initial conditions of the
waterbag form.  In the future, it will be important to extend the theory to more complex initial conditions.
Preliminary work in this direction indicates that multilevel distributions lead to significantly
more complex qSSs, with very interesting topological structure which, nevertheless, preserves 
some of the 
core-halo characteristics \cite{PakLev2013a}. Curiously, for such initial distributions, the
LB theory
fails to describe the qSSs, 
even when initial conditions satisfy the virial theorem. 
This indicates that for multilevel distributions 
mixing is even poorer than it is for one level waterbags. Furthermore, even
for one-level waterbag distributions satisfying the virial condition, there are small deviations between
the results of simulations and the LB theory, and some halo formation may be observed.  This suggests that  that
the core-halo distribution may also be relevant for predicting the qSS of initially virialized waterbag distributions.  Since for ${\cal R}_0=1$ the parametric resonances are not be excited,
the halo energy in this case should be the same as the energy of the most energetic particle of the initial distribution.   
In Fig \ref{fig:concchdist} we compare the predictions of the core-halo and the LB theories with the
results of MD simulations for 1D 
self-gravitating system with ${\cal R}_0=1$.  It appears that even in this case the 
core-halo theory agrees better with the results of simulations than does the LB approach. This suggests that
mixing and ergodicity are not perfect even for initially virialized distributions. This, however, should be tested 
for other models discussed in this Review. 
%%%%%%%%%%%%%%%% Figure%%%%%%%%%%%%%%%%%%%%%
%%%%%%%%%%%%% End of figure%%%%%%%%%%%%%%%%%

A trapping of a system in a qSS is a consequence of the ergodicity breaking.  The 
process of Landau damping decreases the amplitude of collective oscillations which are responsible for the 
energy transfer between the particles.  For long-range systems, there are no collisions (correlations) 
between  the particles, and the only mechanism of energy transfer
is the wave-particle interaction.  Therefore, once the oscillations have completely died out, 
each particles will move in a static mean-field potential and the ergodity of the system will be broken.  
All the systems that have been considered 
so far
had either spherical (in 3D) or polar (in 2D) symmetry.  
The equations of motion for a particle inside such potentials are integrable.  
This, in general, is not true
for asymmetric potentials for which  particle trajectories can become 
chaotic.  
It should be of great interest to explore if the chaotic dynamics 
in the qSS can lead to a faster relaxation to the Boltzmann-Gibbs equilibrium and a shorter lifetime  
of a qSS.

There are a number of outstanding open question which remain to be addressed.  Can the 
core-halo theory developed above be extended to study 3D self-gravitating systems?  For such systems 
the halo will extend all the way to infinity.  At the moment we do not have an understanding of the structure
of such halos.  Furthermore, both 2D and 3D gravitational systems are susceptible to symmetry breaking
instabilities \cite{AguMer1990}.  The simulation methods used in the present work, which primarily 
relied on the Gauss's law,
do not allow us to study such instabilities. The
theoretical understanding of the symmetry breaking mechanism that leads to asymmetric QSS 
is still lacking and it is not clear
how to extend the core-halo theory to describe the asymmetric stationary states.  Finally, 
in the future it will be 
important
to move beyond the waterbag initial distributions.  As discussed above, multilevel initial distributions
appear to exhibit ergodicity breaking and poor mixing even when they are virialized.  
This makes the study of such initial conditions very challenging~\cite{PakLev2013a}. 
Nevertheless, it has been observed that even 
such complex initial distributions
also relax to core-halo QSS, with the particle distribution in the core well
fitted by polytropic distributions \cite{CamCha2013}.

In spite of their ubiquity, long-range interacting systems are still
poorly understood.  They are the unexplored frontier of statistical physics.
We hope that the present Review helps to attract attention of the statistical mechanics
community to this fascinating field.  

Y.L. would like to thank Michael Fisher, without whose insistence and encouragement this review 
would not have been written.
This work was partially supported by the CNPq, FAPERGS, INCT-FCx, and by the US-AFOSR under the grant FA9550-12-1-0438.

\bibliographystyle{elsarticle-num}
\bibliography{root}

\begin{thebibliography}{100}
\expandafter\ifx\csname url\endcsname\relax
  \def\url#1{\texttt{#1}}\fi
\expandafter\ifx\csname urlprefix\endcsname\relax\def\urlprefix{URL }\fi
\expandafter\ifx\csname href\endcsname\relax
  \def\href#1#2{#2} \def\path#1{#1}\fi

\bibitem{Kle1967}
M.~J. Klein, {T}hermodynamics in {E}instein's {T}hought, {S}cience 157~(3788)
  (1967) 509.
\newblock \href {http://dx.doi.org/10.1126/science.157.3788.509}
  {\path{doi:10.1126/science.157.3788.509}}.

\bibitem{Fis1964}
M.~E. Fisher, {T}he free energy of a macroscopic system, {A}rchive for
  {R}ational {M}echanics and {A}nalysis 17~(5) (1964) 377.

\bibitem{Rue1970}
D.~Ruelle, {S}uperstable interactions in classical statistical mechanics,
  {C}ommunications in {M}athematical {P}hysics 18~(2) (1970) 127.

\bibitem{Kac1959}
M.~Kac, \href{http://link.aip.org/link/?PFLDAS/2/8/1}{{On the Partition
  Function of a One-Dimensional Gas}}, {P}hysics of {F}luids 2~(1) (1959)
  8--12.
\newline\urlprefix\url{http://link.aip.org/link/?PFLDAS/2/8/1}

\bibitem{KacUhl1963}
M.~Kac, G.~E. Uhlenbeck, P.~C. Hemmer, {O}n the van der {W}aals {T}heory of the
  {V}apor-{L}iquid {E}quilibrium. {I}. {D}iscussion of a {O}ne-{D}imensional
  {M}odel, {J}ournal of {M}athematical {P}hysics 4~(2) (1963) 216.
\newblock \href {http://dx.doi.org/10.1063/1.1703946}
  {\path{doi:10.1063/1.1703946}}.

\bibitem{Bak1963}
G.~A. Baker, {I}sing {M}odel with a {L}ong-{R}ange {I}nteraction in the
  {P}resence of {R}esidual {S}hort-{R}ange {I}nteractions, {P}hysical {R}eview
  130~(4) (1963) 1406.
\newblock \href {http://dx.doi.org/10.1103/PhysRev.130.1406}
  {\path{doi:10.1103/PhysRev.130.1406}}.

\bibitem{CamDau2009}
A.~Campa, T.~Dauxois, S.~Ruffo, {Statistical mechanics and dynamics of solvable
  models with long-range interactions}, {P}hysics {R}eports 480~(3-6) (2009)
  57--159.
\newblock \href {http://dx.doi.org/10.1016/j.physrep.2009.07.001}
  {\path{doi:10.1016/j.physrep.2009.07.001}}.

\bibitem{BarMuk2001}
J.~Barr\'{e}, D.~Mukamel, S.~Ruffo, {Inequivalence of Ensembles in a System
  with Long-Range Interactions}, {P}hysical {R}eview {L}etters 87~(3) (2001)
  30601.
\newblock \href {http://dx.doi.org/10.1103/PhysRevLett.87.030601}
  {\path{doi:10.1103/PhysRevLett.87.030601}}.

\bibitem{Thi1970}
W.~Thirring, {S}ystems with {N}egative {S}pecific {H}eat, {Z}eitschrift f\"ur
  {P}hysik 235~(4) (1970) 339.

\bibitem{LynLyn1977}
D.~Lynden-Bell, R.~M. Lynden-Bell, {O}n the negative specific heat paradox,
  {M}onthly {N}otices of the {R}oyal {A}stronomical {S}ociety 181 (1977) 405.

\bibitem{El-1998}
A.~El-Zant, {A}pproach to equilibrium in {N}-body gravitational systems,
  {P}hysical {R}eview {E} 58~(4) (1998) 4152.
\newblock \href {http://dx.doi.org/10.1103/PhysRevE.58.4152}
  {\path{doi:10.1103/PhysRevE.58.4152}}.

\bibitem{Lyn1999}
D.~Lynden-Bell, {N}egative specific heat in astronomy, physics and chemistry,
  {P}hysica {A}: {S}tatistical {M}echanics and its {A}pplications 263 (1999)
  293.
\newblock \href {http://dx.doi.org/10.1016/S0378-4371(98)00518-4}
  {\path{doi:10.1016/S0378-4371(98)00518-4}}.

\bibitem{KieNeu2003}
M.~Kiessling, T.~Neukirch,
  \href{http://www.pnas.org/content/100/4/1510.short}{{Negative specific heat
  of a magnetically self-confined plasma torus}}, {P}roceedings of the
  {N}ational {A}cademy of {S}ciences of the {U}nited {S}tates of {A}merica
  100~(4) (2003) 1510.
\newline\urlprefix\url{http://www.pnas.org/content/100/4/1510.short}

\bibitem{ThiNar2003}
W.~Thirring, H.~Narnhofer, H.~Posch, {N}egative {S}pecific {H}eat, the
  {T}hermodynamic {L}imit, and {E}rgodicity, {P}hysical {R}eview {L}etters
  91~(13) (2003) 130601.
\newblock \href {http://dx.doi.org/10.1103/PhysRevLett.91.130601}
  {\path{doi:10.1103/PhysRevLett.91.130601}}.

\bibitem{Rei1998}
L.~Reichl, {A} {M}odern {C}ourse {I}n {S}tatistical {P}hysics, 2nd Edition,
  Wiley-Interscience, 1998.

\bibitem{TsuKon1994}
T.~Tsuchiya, T.~Konishi, N.~Gouda, {Q}uasiequilibria in one-dimensional
  self-gravitating many-body systems, {P}hysical {R}eview {E} 50~(4) (1994)
  2607.
\newblock \href {http://dx.doi.org/10.1103/PhysRevE.50.2607}
  {\path{doi:10.1103/PhysRevE.50.2607}}.

\bibitem{BorCel2004}
F.~Borgonovi, G.~L. Celardo, M.~Maianti, E.~Pedersoli, {B}roken {E}rgodicity in
  {C}lassically {C}haotic {S}pin {S}ystems, {J}ournal of {S}tatistical
  {P}hysics 116~(5--6) (2004) 1435.
\newblock \href {http://dx.doi.org/10.1023/B:JOSS.0000041745.62340.00}
  {\path{doi:10.1023/B:JOSS.0000041745.62340.00}}.

\bibitem{CheCro1996}
P.~Chen, M.~Cross, {Mixing and thermal equilibrium in the dynamical relaxation
  of a vortex ring}, {P}hysical {R}eview {L}etters 77~(20) (1996) 4174--4177.
\newblock \href {http://dx.doi.org/10.1103/PhysRevLett.77.4174}
  {\path{doi:10.1103/PhysRevLett.77.4174}}.

\bibitem{BenTel2012}
F.~{P. da C. Benetti}, T.~N. Teles, R.~Pakter, Y.~Levin, {E}rgodicity
  {B}reaking and {P}arametric {R}esonances in {S}ystems with {L}ong-{R}ange
  {I}nteractions, {P}hysical {R}eview {L}etters 108 (2012) 140601.
\newblock \href {http://dx.doi.org/10.1103/PhysRevLett.108.140601}
  {\path{doi:10.1103/PhysRevLett.108.140601}}.

\bibitem{BraHep1977}
W.~Braun, K.~Hepp, {The Vlasov dynamics and its fluctuations in the 1/N limit
  of interacting classical particles}, {C}ommunications in {M}athematical
  {P}hysics 56~(2) (1977) 101--113.
\newblock \href {http://dx.doi.org/10.1007/BF01611497}
  {\path{doi:10.1007/BF01611497}}.

\bibitem{GabJoy2010}
A.~Gabrielli, M.~Joyce, {Gravitational force in an infinite one-dimensional
  Poisson distribution}, {P}hysical {R}eview {E} 81~(2) (2010) 1--9.
\newblock \href {http://dx.doi.org/10.1103/PhysRevE.81.021102}
  {\path{doi:10.1103/PhysRevE.81.021102}}.

\bibitem{LuwSev1984}
M.~Luwel, G.~Severne, P.~Rousseeuw, {N}umerical study of the relaxation of
  one-dimensional gravitational systems, {A}strophysics and {S}pace {S}cience
  100~(1--2) (1984) 261.
\newblock \href {http://dx.doi.org/10.1007/BF00651601}
  {\path{doi:10.1007/BF00651601}}.

\bibitem{LevPak2008a}
Y.~Levin, R.~Pakter, T.~Teles, {Collisionless Relaxation in Non-Neutral
  Plasmas}, {P}hysical {R}eview {L}etters 100~(4) (2008) 040604.
\newblock \href {http://dx.doi.org/10.1103/PhysRevLett.100.040604}
  {\path{doi:10.1103/PhysRevLett.100.040604}}.

\bibitem{TelPak2009}
T.~N. Teles, R.~Pakter, Y.~Levin, {Relaxation and emittance growth of a thermal
  charged-particle beam}, {A}pplied {P}hysics {L}etters 95~(17) (2009) 173501.
\newblock \href {http://dx.doi.org/10.1063/1.3254245}
  {\path{doi:10.1063/1.3254245}}.

\bibitem{PakLev2011}
R.~Pakter, Y.~Levin, {C}ore-{H}alo {D}istribution in the {H}amiltonian
  {M}ean-{F}ield {M}odel, {P}hysical {R}eview {L}etters 106 (2011) 200603.
\newblock \href {http://dx.doi.org/10.1103/PhysRevLett.106.200603}
  {\path{doi:10.1103/PhysRevLett.106.200603}}.

\bibitem{Hen1964}
M.~H\'enon,
  \href{http://adsabs.harvard.edu/full/1964AnAp...27...83H}{{L'\'{e}volution
  initiale d'un amas sph\'{e}rique}}, {A}nnales d'{A}strophysique 27 (1964) 83.
\newline\urlprefix\url{http://adsabs.harvard.edu/full/1964AnAp...27...83H}

\bibitem{Sal1965}
A.~M. Salzberg, {Exact statistical thermodynamics of gravitational interactions
  in one and two dimensions}, {J}ournal of {M}athematical {P}hysics 6~(1)
  (1965) 158.
\newblock \href {http://dx.doi.org/10.1063/1.1704254}
  {\path{doi:10.1063/1.1704254}}.

\bibitem{Lec1966}
M.~Lecar, \href{http://adsabs.harvard.edu/full/1966IAUS...25...46L}{{A}
  one-dimensional self-gravitating stellar gas}, in: G.~I. Kontopoulos (Ed.),
  {T}he {T}heory of {O}rbits in the {S}olar {S}ystem and in {S}tellar
  {S}ystems. {P}roceedings from {S}ymposium no. 25 held in {T}hessaloniki,
  1964., International Astronomical Union, Academic Press, 1966, p.~46.
\newline\urlprefix\url{http://adsabs.harvard.edu/full/1966IAUS...25...46L}

\bibitem{HohCam1968}
F.~Hohl, J.~W. Campbell,
  \href{http://adsabs.harvard.edu/full/1968AJ.....73..611H}{{Statistical
  Mechanics of a Collisionless Self-Gravitating System}}, {T}he {A}stronomical
  {J}ournal 73~(7) (1968) 611.
\newline\urlprefix\url{http://adsabs.harvard.edu/full/1968AJ.....73..611H}

\bibitem{Mil1971}
R.~H. Miller, {N}umerical experiments in collisionless systems, {A}strophysics
  and {S}pace {S}cience 14~(1) (1971) 73.
\newblock \href {http://dx.doi.org/10.1007/BF00649196}
  {\path{doi:10.1007/BF00649196}}.

\bibitem{WriMil1982}
H.~Wright, B.~Miller, W.~Stein, {The relaxation time of a one-dimensional
  self-gravitating system}, {A}strophysics and {S}pace {S}cience 84~(2) (1982)
  421.
\newblock \href {http://dx.doi.org/10.1007/BF00651321}
  {\path{doi:10.1007/BF00651321}}.

\bibitem{LuwSev1985}
M.~Luwel, G.~Severne, {C}ollisionless mixing in 1-dimensional gravitational
  systems initially in a stationary waterbag configuration, {A}stronomy \&
  {A}strophysics 152 (1985) 305.

\bibitem{Car1986}
R.~Carlberg, {The phase space density in elliptical galaxies}, {T}he
  {A}strophysical {J}ournal 310 (1986) 593--596.
\newblock \href {http://dx.doi.org/10.1086/164711} {\path{doi:10.1086/164711}}.

\bibitem{ReiMil1987}
C.~J. Reidl~Jr., B.~N. Miller, {G}ravity in one dimension: selective
  relaxation?, {T}he {A}strophysical {J}ournal 318 (1987) 248.
\newblock \href {http://dx.doi.org/10.1086/165364} {\path{doi:10.1086/165364}}.

\bibitem{YawMil2003}
K.~R. Yawn, B.~N. Miller, {I}ncomplete relaxation in a two-mass one-dimensional
  self-gravitating system, {P}hysical {R}eview {E} 68~(5) (2003) 1--17.
\newblock \href {http://dx.doi.org/10.1103/PhysRevE.68.056120}
  {\path{doi:10.1103/PhysRevE.68.056120}}.

\bibitem{DeWu2004}
R.~De~Simone, X.~Wu, S.~Tremaine, {T}he stellar velocity distribution in the
  solar neighbourhood, {M}onthly {N}otices of the {R}oyal {A}stronomical
  {S}ociety 350~(1) (2004) 627.
\newblock \href {http://dx.doi.org/10.1111/j.1365-2966.2004.07675.x}
  {\path{doi:10.1111/j.1365-2966.2004.07675.x}}.

\bibitem{KliMil2004}
P.~Klinko, B.~N. Miller, {D}ynamical study of a first order gravitational phase
  transition, {P}hysics {L}etters {A} 333 (2004) 187.

\bibitem{LevPak2008}
Y.~Levin, R.~Pakter, F.~Rizzato, {Collisionless relaxation in gravitational
  systems: From violent relaxation to gravothermal collapse}, {P}hysical
  {R}eview {E} 78~(2) (2008) 021130.
\newblock \href {http://dx.doi.org/10.1103/PhysRevE.78.021130}
  {\path{doi:10.1103/PhysRevE.78.021130}}.

\bibitem{TelLev2010}
T.~N. Teles, Y.~Levin, R.~Pakter, F.~B. Rizzato, {Statistical mechanics of
  unbound two-dimensional self-gravitating systems}, {J}ournal of {S}tatistical
  {M}echanics: {T}heory and {E}xperiment 2010~(05) (2010) P05007.
\newblock \href {http://dx.doi.org/10.1088/1742-5468/2010/05/P05007}
  {\path{doi:10.1088/1742-5468/2010/05/P05007}}.

\bibitem{JalTre2012}
M.~A. Jalali, S.~Tremaine, {D}ensity waves in debris discs and galactic nuclei,
  {M}onthly {N}otices of the {R}oyal {A}stronomical {S}ociety 421~(3) (2012)
  2368.
\newblock \href {http://dx.doi.org/10.1111/j.1365-2966.2012.20469.x}
  {\path{doi:10.1111/j.1365-2966.2012.20469.x}}.

\bibitem{Len1961}
A.~Lenard, \href{http://link.aip.org/link/doi/10.1063/1.1703757}{{Exact
  Statistical Mechanics of a One-Dimensional System with Coulomb Forces}},
  {J}ournal of {M}athematical {P}hysics 2~(5) (1961) 682.
\newblock \href {http://dx.doi.org/10.1063/1.1703757}
  {\path{doi:10.1063/1.1703757}}.
\newline\urlprefix\url{http://link.aip.org/link/doi/10.1063/1.1703757}

\bibitem{EdwLen1962}
S.~F. Edwards, A.~Lenard, {E}xact {S}tatistical {M}echanics of a
  {O}ne-{D}imensional {S}ystem with {C}oulomb {F}orces. {II}. {T}he {M}ethod of
  {F}unctional {I}ntegration, {J}ournal of {M}athematical {P}hysics 3~(4)
  (1962) 778.
\newblock \href {http://dx.doi.org/10.1063/1.1724281}
  {\path{doi:10.1063/1.1724281}}.

\bibitem{EldFei1963}
O.~Eldridge, M.~Feix,
  \href{http://link.aip.org/link/?PFLDAS/6/398/1}{{Numerical experiments with a
  plasma model}}, {P}hysics of {F}luids 6~(3) (1963) 398.
\newline\urlprefix\url{http://link.aip.org/link/?PFLDAS/6/398/1}

\bibitem{Ste1963}
M.~J. Stephen,
  \href{http://prola.aps.org/abstract/PR/v129/i3/p997_1}{{Oscillations of a
  Plasma in a Magnetic Field}}, {P}hysical {R}eview 129~(3) (1963) 997--1004.
\newline\urlprefix\url{http://prola.aps.org/abstract/PR/v129/i3/p997_1}

\bibitem{Ran1965}
S.~Rand, \href{http://link.aip.org/link/?PFLDAS/8/143/1}{{Collision Damping of
  Electron Plasma Waves}}, {P}hysics of {F}luids 8~(1) (1965) 143.
\newline\urlprefix\url{http://link.aip.org/link/?PFLDAS/8/143/1}

\bibitem{KadPog1970}
B.~B. Kadomtsev, O.~P. Pogutse, {C}ollisionles {R}elaxation in {S}ystems with
  {C}oulomb {I}nteractions, {P}hysical {R}eview {L}etters 25 (1970) 1155.
\newblock \href {http://dx.doi.org/10.1103/PhysRevLett.25.1155}
  {\path{doi:10.1103/PhysRevLett.25.1155}}.

\bibitem{Rei1991}
M.~Reiser, {F}ree energy and emittance growth in nonstationary charged particle
  beams, {J}ournal of {A}pplied {P}hysics 70~(4) (1991) 1919.
\newblock \href {http://dx.doi.org/10.1063/1.349474}
  {\path{doi:10.1063/1.349474}}.

\bibitem{HuaDri1994}
X.-P. Huang, C.~F. Driscoll, {R}elaxation of 2{D} turbulence to a
  metaequilibrium near the minimum enstrophy state, {P}hysical {R}eview
  {L}etters 72~(14) (1994) 2187.
\newblock \href {http://dx.doi.org/10.1103/PhysRevLett.72.2187}
  {\path{doi:10.1103/PhysRevLett.72.2187}}.

\bibitem{LunBar1995}
S.~M. Lund, J.~J. Barnard, J.~M. Miller, {O}n the relaxation of semi-{G}aussian
  and {K}-{V} beams to thermal equilibrium, in: {P}roceedings of the 1995
  {P}article {A}ccelerator {C}onference, Vol.~5, 1995, p. 3278.
\newblock \href {http://dx.doi.org/10.1109/PAC1995.505854}
  {\path{doi:10.1109/PAC1995.505854}}.

\bibitem{AllCha2002}
C.~Allen, K.~Chan, P.~Colestock, K.~Crandall, R.~Garnett, J.~Gilpatrick,
  W.~Lysenko, J.~Qiang, J.~Schneider, M.~Schulze, R.~Sheffield, H.~Smith,
  T.~Wangler, {Beam-Halo Measurements in High-Current Proton Beams}, {P}hysical
  {R}eview {L}etters 89~(21) (2002) 214802.
\newblock \href {http://dx.doi.org/10.1103/PhysRevLett.89.214802}
  {\path{doi:10.1103/PhysRevLett.89.214802}}.

\bibitem{OkaIke1997}
H.~Okamoto, M.~Ikegami, {S}imulation study of halo formation in breathing round
  beams, {P}hysical {R}eview {E} 55~(4) (1997) 4694.
\newblock \href {http://dx.doi.org/10.1103/PhysRevE.55.4694}
  {\path{doi:10.1103/PhysRevE.55.4694}}.

\bibitem{WanCra1998}
T.~P. Wangler, K.~R. Crandall, R.~Ryne, T.~S. Wang, {Particle-core model for
  transverse dynamics of beam halo}, {P}hysical {R}eview {S}pecial {T}opics --
  {A}ccelerators and {B}eams 1~(8) (1998) 84201.
\newblock \href {http://dx.doi.org/10.1103/PhysRevSTAB.1.084201}
  {\path{doi:10.1103/PhysRevSTAB.1.084201}}.

\bibitem{TelLev2011}
T.~Teles, Y.~Levin, R.~Pakter, {S}tatistical mechanics of 1{D} self-gravitating
  systems: the core-halo distribution, {M}onthly {N}otices of the {R}oyal
  {A}stronomical {S}ociety 417 (2011) L21--L25.
\newblock \href {http://dx.doi.org/10.1111/j.1745-3933.2011.01112.x}
  {\path{doi:10.1111/j.1745-3933.2011.01112.x}}.

\bibitem{BarBou2002}
J.~Barr\'e, F.~Bouchet, T.~Dauxois, S.~Ruffo, {O}ut-of-{E}quilibrium {S}tates
  as {S}tatistical {E}quilibria of an {E}ffective {D}ynamics in a {S}ystem with
  {L}ong-{R}ange {I}nteractions, {P}hys. {R}ev. {L}ett. 89 (2002) 110601.
\newblock \href {http://dx.doi.org/10.1103/PhysRevLett.89.110601}
  {\path{doi:10.1103/PhysRevLett.89.110601}}.

\bibitem{AntCal2007}
A.~Antoniazzi, F.~Califano, D.~Fanelli, S.~Ruffo, {Exploring the Thermodynamic
  Limit of Hamiltonian Models: Convergence to the Vlasov Equation}, {P}hysical
  {R}eview {L}etters 98~(15) (2007) 150602.
\newblock \href {http://dx.doi.org/10.1103/PhysRevLett.98.150602}
  {\path{doi:10.1103/PhysRevLett.98.150602}}.

\bibitem{Kap1922}
J.~Kapteyn,
  \href{http://articles.adsabs.harvard.edu/full/1922ApJ....55..302K}{{First
  Attempt at a Theory of the Arrangement and Motion of the Sidereal System}},
  {A}strophysical {J}ournal 55 (1922) 302.
\newline\urlprefix\url{http://articles.adsabs.harvard.edu/full/1922ApJ....55..%
302K}

\bibitem{Jea1922}
J.~Jeans, \href{http://adsabs.harvard.edu/full/1922MNRAS..82..122J}{{The
  motions of stars in a Kapteyn universe}}, {M}onthly {N}otices of the {R}oyal
  {A}stronomical {S}ociety 82 (1922) 122--132.
\newline\urlprefix\url{http://adsabs.harvard.edu/full/1922MNRAS..82..122J}

\bibitem{Oor1932}
F.~H. Oort, {T}he force exerted by stellar system in the direction
  perpendicular to the galactic plane and some related problems, {B}ulletin
  {O}f {T}he {A}stronomical {I}nstitutes of the {N}etherlands 6~(17) (1932)
  238.

\bibitem{Cam1950}
G.~Camm,
  \href{http://adsabs.harvard.edu/full/1950MNRAS.110..305C}{{Self-gravitating
  star systems}}, {M}onthly {N}otices of the {R}oyal {A}stronomical {S}ociety
  110 (1950) 305.
\newline\urlprefix\url{http://adsabs.harvard.edu/full/1950MNRAS.110..305C}

\bibitem{Ogo1957}
K.~Ogorodnikov,
  \href{http://adsabs.harvard.edu/full/1957SvA.....1..748O}{{Statistical
  Mechanics of the Simplest Types of Galaxies.}}, {S}oviet {A}stronomy 1 (1957)
  748.
\newline\urlprefix\url{http://adsabs.harvard.edu/full/1957SvA.....1..748O}

\bibitem{Pad1990}
T.~Padmanabhan, {S}tatistical mechanics of gravitating systems, {P}hysics
  {R}eports 188~(5) (1990) 285--362.
\newblock \href {http://dx.doi.org/10.1016/0370-1573(90)90051-3}
  {\path{doi:10.1016/0370-1573(90)90051-3}}.

\bibitem{TreOst1999}
S.~Tremaine, J.~P. Ostriker, {R}elaxation in stellar systems, and the shape and
  rotation of the inner dark halo, {M}onthly {N}otices of the {R}oyal
  {A}stronomical {S}ociety 306~(3) (1999) 662.
\newblock \href {http://dx.doi.org/10.1046/j.1365-8711.1999.02558.x}
  {\path{doi:10.1046/j.1365-8711.1999.02558.x}}.

\bibitem{Tre1999}
S.~Tremaine, {T}he geometry of phase mixing, {M}onthly {N}otices of the {R}oyal
  {A}stronomical {S}ociety 307~(4) (1999) 877.
\newblock \href {http://dx.doi.org/10.1046/j.1365-8711.1999.02690.x}
  {\path{doi:10.1046/j.1365-8711.1999.02690.x}}.

\bibitem{BinTre2009}
J.~Binney, S.~Tremaine, {G}alactic {D}ynamics, 2nd Edition, Princeton
  University Press, 2009.

\bibitem{ChaSom1996}
P.~H. Chavanis, J.~Sommeria, R.~Robert, {S}tatistical {M}echanics of
  {T}wo-{D}imensional {V}ortices and {C}ollisionless {S}tellar {S}ystems, {T}he
  {A}strophysical {J}ournal 471~(1) (1996) 385.
\newblock \href {http://dx.doi.org/10.1086/177977} {\path{doi:10.1086/177977}}.

\bibitem{Cha2000}
P.-H. Chavanis, {Quasilinear theory of the 2D Euler equation}, {P}hysical
  {R}eview {L}etters 84~(24) (2000) 5512--5515.
\newblock \href {http://dx.doi.org/10.1103/PhysRevLett.84.5512}
  {\path{doi:10.1103/PhysRevLett.84.5512}}.

\bibitem{AndLim2007}
T.~Andersen, C.~Lim, {Negative Specific Heat in a Quasi-2D Generalized
  Vorticity Model}, {P}hysical {R}eview {L}etters 99~(16) (2007) 165001.
\newblock \href {http://dx.doi.org/10.1103/PhysRevLett.99.165001}
  {\path{doi:10.1103/PhysRevLett.99.165001}}.

\bibitem{BouBar2008}
F.~Bouchet, J.~Barr\'{e}, A.~Venaille, A.~Campa, A.~Giansanti, G.~Morigi, F.~S.
  Labini, {E}quilibrium and out of equilibrium phase transitions in systems
  with long range interactions and in 2{D} flows, {AIP} {C}onference
  {P}roceedings (2008) 117--152\href {http://dx.doi.org/10.1063/1.2839113}
  {\path{doi:10.1063/1.2839113}}.

\bibitem{VenBou2009}
A.~Venaille, F.~Bouchet, {S}tatistical {E}nsemble {I}nequivalence and
  {B}icritical {P}oints for {T}wo-{D}imensional {F}lows and {G}eophysical
  {F}lows, {P}hysical {R}eview {L}etters 102 (2009) 104501.
\newblock \href {http://dx.doi.org/10.1103/PhysRevLett.102.104501}
  {\path{doi:10.1103/PhysRevLett.102.104501}}.

\bibitem{BouVen2012}
F.~Bouchet, A.~Venaille, {S}tatistical mechanics of two-dimensional and
  geophysical flows, {P}hysics {R}eports 515 (2012) 227.
\newblock \href {http://dx.doi.org/10.1016/j.physrep.2012.02.001}
  {\path{doi:10.1016/j.physrep.2012.02.001}}.

\bibitem{Kas2011}
M.~Kastner, {Diverging Equilibration Times in Long-Range Quantum Spin Models},
  {P}hysical {R}eview {L}etters 106~(13) (2011) 1--4.
\newblock \href {http://dx.doi.org/10.1103/PhysRevLett.106.130601}
  {\path{doi:10.1103/PhysRevLett.106.130601}}.

\bibitem{BerKez2012}
O.~L. Berman, R.~Y. Kezerashvili, G.~V. Kolmakov, Y.~E. Lozovik, {T}urbulence
  in a {B}ose-{E}instein condensate of dipolar excitons in coupled quantum
  wells, {P}hysical {R}eview {B}: {C}ondensed {M}atter and {M}aterials
  {P}hysics 86~(4) (2012) 045108.
\newblock \href {http://dx.doi.org/10.1103/PhysRevB.86.045108}
  {\path{doi:10.1103/PhysRevB.86.045108}}.

\bibitem{SlaKre2008}
S.~Slama, G.~Krenz, S.~Bux, C.~Zimmermann, P.~W. Courteille, {C}ollective
  {A}tomic {R}ecoil {L}asing and {S}uperradiant {R}ayleigh {S}cattering in a
  high-{Q} ring cavity, in: A.~Campa, A.~Giansanti, G.~Morigi, F.~S. Labini
  (Eds.), {D}ynamics and {T}hermodynamics of {S}ystems with {L}ong-range
  {I}nteraction: {T}heory and {E}xperiments, Vol. CP 970 of {M}athematical and
  {S}tatistical {P}hysics, American Institute of Physics, 2008, p. 319.
\newblock \href {http://dx.doi.org/10.1063/1.2839129}
  {\path{doi:10.1063/1.2839129}}.

\bibitem{CheDav1994}
C.~Chen, R.~C. Davidson, {N}onlinear properties of the
  {K}apchinskij-{V}ladimirskij equilibrium and envelope equation for an intense
  charged-particle beam in a periodic focusing field, {P}hysical {R}eview {E}
  49~(6) (1994) 5679.
\newblock \href {http://dx.doi.org/10.1103/PhysRevE.49.5679}
  {\path{doi:10.1103/PhysRevE.49.5679}}.

\bibitem{DavQin2001}
R.~C. Davidson, H.~Qin, {P}hysics of {I}ntense {C}harged {P}article {B}eams in
  {H}igh {E}nergy {A}ccelerators, 1st Edition, World Scientific, 2001.

\bibitem{KagDav2010}
I.~D. Kaganovich, R.~C. Davidson, M.~a. Dorf, E.~a. Startsev, a.~B. Sefkow,
  E.~P. Lee, A.~Friedman, {Physics of neutralization of intense high-energy ion
  beam pulses by electrons}, {P}hysics of {P}lasmas 17~(5) (2010) 056703.
\newblock \href {http://dx.doi.org/10.1063/1.3335766}
  {\path{doi:10.1063/1.3335766}}.

\bibitem{Cal1985}
H.~B. Callen, {T}hermodynamics and an {I}ntroduction to {T}hermostatics, 2nd
  Edition, John Wiley \& Sons, 1985.

\bibitem{FisRue1966}
M.~E. Fisher, D.~Ruelle, {T}he {S}tability of {M}any-{P}article {S}ystems,
  {J}ournal of {M}athematical {P}hysics 7 (1966) 260.
\newblock \href {http://dx.doi.org/10.1063/1.1704928}
  {\path{doi:10.1063/1.1704928}}.

\bibitem{Hoh1978}
F.~Hohl,
  \href{http://adsabs.harvard.edu/full/1978AJ.....83..768H}{{Three-dimensional
  galaxy simulations}}, {T}he {A}stronomical {J}ournal 83~(7) (1978) 768--778.
\newline\urlprefix\url{http://adsabs.harvard.edu/full/1978AJ.....83..768H}

\bibitem{YawMil1997}
K.~R. Yawn, B.~N. Miller, {E}quipartition and {M}ass {S}egregation in a
  {O}ne-{D}imensional {S}elf-{G}ravitating {S}ystem, {P}hysical {R}eview
  {L}etters 79~(19) (1997) 3561.
\newblock \href {http://dx.doi.org/10.1103/PhysRevLett.79.3561}
  {\path{doi:10.1103/PhysRevLett.79.3561}}.

\bibitem{HerThi1971}
P.~Hertel, W.~Thirring, {A} soluble model for a system with negative specific
  heat, {A}nnals of {P}hysics 63~(2) (1971) 520.
\newblock \href {http://dx.doi.org/10.1016/0003-4916(71)90025-X}
  {\path{doi:10.1016/0003-4916(71)90025-X}}.

\bibitem{BouBar2005}
F.~Bouchet, J.~Barr\'e, {C}lassification of {P}hase {T}ransitions and
  {E}nsemble {I}nequivalence, in {S}ystems with {L}ong {R}ange {I}nteractions,
  {J}ournal of {S}tatistical {P}hysics 118~(516) (2005) 1073.
\newblock \href {http://dx.doi.org/10.1007/s10955-004-2059-0}
  {\path{doi:10.1007/s10955-004-2059-0}}.

\bibitem{Cha2006a}
P.~H. Chavanis, {P}hase transitions in self-gravitating systems,
  {I}nternational {J}ournal of {M}odern {P}hysics {B} 20~(22) (2006) 3113.
\newblock \href {http://dx.doi.org/10.1142/S0217979206035400}
  {\path{doi:10.1142/S0217979206035400}}.

\bibitem{FilAma2009}
T.~M.~R. Filho, M.~A. Amato, A.~Figueiredo, {A novel approach to the
  determination of equilibrium properties of classical Hamiltonian systems with
  long-range interactions}, {J}ournal of {P}hysics {A}: {M}athematical and
  {T}heoretical 42~(16) (2009) 165001.
\newblock \href {http://dx.doi.org/10.1088/1751-8113/42/16/165001}
  {\path{doi:10.1088/1751-8113/42/16/165001}}.

\bibitem{CohMuk2012}
O.~Cohen, D.~Mukamel, {E}nsemble inequivalence: {L}andau theory and the {ABC}
  model, {J}ournal of {S}tatistical {M}echanics: {T}heory and {E}xperiment 2012
  (2012) P12017.
\newblock \href {http://dx.doi.org/10.1088/1742-5468/2012/12/P12017}
  {\path{doi:10.1088/1742-5468/2012/12/P12017}}.

\bibitem{RamLar2008}
A.~Ram\'{i}rez-Hern\'{a}ndez, H.~Larralde, F.~Leyvraz, {V}iolation of the
  {Z}eroth {L}aw of {T}hermodynamics in {S}ystems with {N}egative {S}pecific
  {H}eat, {P}hysical {R}eview {L}etters 100~(12) (2008) 120601.
\newblock \href {http://dx.doi.org/10.1103/PhysRevLett.100.120601}
  {\path{doi:10.1103/PhysRevLett.100.120601}}.

\bibitem{RamLar2008a}
A.~Ram\'{i}rez-Hern\'{a}ndez, H.~Larralde, F.~Leyvraz, {S}ystems with negative
  specific heat in thermal contact: {V}iolation of the zeroth law, {P}hysical
  {R}eview {E} 78~(6) (2008) 1--8.
\newblock \href {http://dx.doi.org/10.1103/PhysRevE.78.061133}
  {\path{doi:10.1103/PhysRevE.78.061133}}.

\bibitem{MicSan2009}
K.~Michaelian, I.~Santamar\'{i}a-Holek, A.~P\'{e}rez-Madrid, {C}omment on
  ``{V}iolation of the {Z}eroth {L}aw of {T}hermodynamics in {S}ystems with
  {N}egative {S}pecific {H}eat'', {P}hysical {R}eview {L}etters 102~(13) (2009)
  138901.
\newblock \href {http://dx.doi.org/10.1103/PhysRevLett.102.138901}
  {\path{doi:10.1103/PhysRevLett.102.138901}}.

\bibitem{RamLar2009}
A.~Ram\'{i}rez-Hern\'{a}ndez, H.~Larralde, F.~Leyvraz,
  {R}am\'{i}rez-{H}ern\'{a}ndez, {L}arralde, and {L}eyvraz {R}eply:, {P}hysical
  {R}eview {L}etters 102~(13) (2009) 138902.
\newblock \href {http://dx.doi.org/10.1103/PhysRevLett.102.138902}
  {\path{doi:10.1103/PhysRevLett.102.138902}}.

\bibitem{Pen1979}
O.~Penrose,
  \href{http://iopscience.iop.org/0034-4885/42/12/002/pdf/0034-4885_42_12_002.%
pdf}{{Foundations of statistical mechanics}}, {R}eports on {P}rogress in
  {P}hysics 42~(12) (1979) 1937--2007.
\newline\urlprefix\url{http://iopscience.iop.org/0034-4885/42/12/002/pdf/0034-%
4885_42_12_002.pdf}

\bibitem{LebPen1973}
J.~L. Lebowitz, O.~Penrose,
  \href{http://ergodic.ugr.es/FisicaEstadistica/copialibros/libros/LeboPen
  PT1973.pdf}{{Modern Ergodic Theory}}, {P}hysics {T}oday 26~(2) (1973)
  155--175.
\newline\urlprefix\url{http://ergodic.ugr.es/FisicaEstadistica/copialibros/lib%
ros/LeboPen PT1973.pdf}

\bibitem{Leb1999}
J.~L. Lebowitz,
  \href{http://linkinghub.elsevier.com/retrieve/pii/S0378437198005147}{{Micros%
copic origins of irreversible macroscopic behavior}}, {P}hysica {A}:
  {S}tatistical {M}echanics and its {A}pplications 263~(1-4) (1999) 516--527.
\newblock \href {http://dx.doi.org/10.1016/S0378-4371(98)00514-7}
  {\path{doi:10.1016/S0378-4371(98)00514-7}}.
\newline\urlprefix\url{http://linkinghub.elsevier.com/retrieve/pii/S0378437198%
005147}

\bibitem{MukRuf2005}
D.~Mukamel, S.~Ruffo, N.~Schreiber, {B}reaking of {E}rgodicity and {L}ong
  {R}elaxation {T}imes in {S}ystems with {L}ong-{R}ange {I}nteractions, {P}hys.
  {R}ev. {L}ett. 95 (2005) 240604.
\newblock \href {http://dx.doi.org/10.1103/PhysRevLett.95.240604}
  {\path{doi:10.1103/PhysRevLett.95.240604}}.

\bibitem{Bal1997}
R.~Balescu, {S}tatistical {D}ynamics: {M}atter {O}ut of {E}quilibrium, World
  Scientific, 1997.

\bibitem{Hua1987}
K.~Huang, {S}tatistical {M}echanics, 2nd Edition, John Wiley \& Sons, 1987.

\bibitem{Gib1928}
J.~W. {G}ibbs, {C}ollected {W}orks, {L}ongmans, {G}reen and {C}o., 1928.

\bibitem{Lev2002}
Y.~Levin, {E}lectrostatic correlations: from plasma to biology, {R}eports on
  {P}rogress in {P}hysics 65~(11) (2002) 1577.
\newblock \href {http://dx.doi.org/10.1088/0034-4885/65/11/201}
  {\path{doi:10.1088/0034-4885/65/11/201}}.

\bibitem{YamBar2004}
Y.~Yamaguchi, J.~Barr\'{e}, F.~Bouchet, T.~Dauxois, S.~Ruffo, {Stability
  criteria of the Vlasov equation and quasi-stationary states of the HMF
  model}, {P}hysica {A}: {S}tatistical and {T}heoretical {P}hysics 337~(1-2)
  (2004) 36--66.
\newblock \href {http://dx.doi.org/10.1016/j.physa.2004.01.041}
  {\path{doi:10.1016/j.physa.2004.01.041}}.

\bibitem{JaiBou2007}
K.~Jain, F.~Bouchet, D.~Mukamel, {Relaxation times of unstable states in
  systems with long range interactions}, {J}ournal of {S}tatistical
  {M}echanics: {T}heory and {E}xperiment 2007~(11) (2007) P11008--P11008.
\newblock \href {http://dx.doi.org/10.1088/1742-5468/2007/11/P11008}
  {\path{doi:10.1088/1742-5468/2007/11/P11008}}.

\bibitem{SakGou1991}
M.-a. Sakagami, N.~Gouda,
  \href{http://adsabs.harvard.edu/full/1991MNRAS.249..241S}{{On the collective
  relaxation in self-gravitating stellar systems}}, {M}onthly {N}otices of the
  {R}oyal {A}stronomical {S}ociety 249 (1991) 241.
\newline\urlprefix\url{http://adsabs.harvard.edu/full/1991MNRAS.249..241S}

\bibitem{ChaBou2005}
P.~Chavanis, F.~Bouchet, {On the coarse-grained evolution of collisionless
  stellar systems}, {A}stronomy and {A}strophysics 430~(3) (2005) 771--778.
\newblock \href {http://dx.doi.org/10.1051/0004-6361:20041462}
  {\path{doi:10.1051/0004-6361:20041462}}.

\bibitem{FilFig2005}
T.~M.~R. Filho, A.~Figueiredo, M.~Amato, {Entropy of Classical Systems with
  Long-Range Interactions}, {P}hysical {R}eview {L}etters 95~(19) (2005)
  190601.
\newblock \href {http://dx.doi.org/10.1103/PhysRevLett.95.190601}
  {\path{doi:10.1103/PhysRevLett.95.190601}}.

\bibitem{TreHen1986}
S.~Tremaine, M.~H\'enon, D.~Lynden-Bell,
  \href{http://adsabs.harvard.edu/full/1986MNRAS.219..285T}{{H}-functions and
  mixing in violent relaxation}, {M}onthly {N}otices of the {R}oyal
  {A}stronomical {S}ociety 219 (1986) 285.
\newline\urlprefix\url{http://adsabs.harvard.edu/full/1986MNRAS.219..285T}

\bibitem{Lyn1967}
D.~Lynden-Bell,
  \href{http://adsabs.harvard.edu/full/1967MNRAS.136..101L}{{Statistical
  mechanics of violent relaxation in stellar systems}}, {M}onthly {N}otices of
  the {R}oyal {A}stronomical {S}ociety 136 (1967) 101--121.
\newline\urlprefix\url{http://adsabs.harvard.edu/full/1967MNRAS.136..101L}

\bibitem{Shu1978}
F.~H. Shu, {O}n the statistical mechanics of violent relaxation,
  {A}strophysical {J}ournal 225 (1978) 83.
\newblock \href {http://dx.doi.org/10.1086/156470} {\path{doi:10.1086/156470}}.

\bibitem{Nak2000}
T.~K. Nakamura, {S}tatistical {M}echanics of a {C}ollisionless {S}ystem {B}ased
  on the {M}aximum {E}ntropy {P}rinciple, {T}he {A}strophysical {J}ournal
  531~(2) (2000) 739.
\newblock \href {http://dx.doi.org/10.1086/308484} {\path{doi:10.1086/308484}}.

\bibitem{AraLyn2005}
I.~Arad, D.~Lynden-Bell, {Inconsistency in theories of violent relaxation},
  {M}onthly {N}otices of the {R}oyal {A}stronomical {S}ociety 361~(2) (2005)
  385--395.
\newblock \href {http://dx.doi.org/10.1111/j.1365-2966.2005.09133.x}
  {\path{doi:10.1111/j.1365-2966.2005.09133.x}}.

\bibitem{BinSec2008}
D.~Bindoni, L.~Secco, {Violent relaxation in phase-space}, {N}ew {A}stronomy
  {R}eviews 52~(1) (2008) 1--18.
\newblock \href {http://dx.doi.org/10.1016/j.newar.2007.11.001}
  {\path{doi:10.1016/j.newar.2007.11.001}}.

\bibitem{MilPre1968}
R.~H. Miller, K.~H. Prendergast,
  \href{http://adsabs.harvard.edu/full/1968ApJ...151..699M}{{S}tellar
  {D}ynamics in a {D}iscrete {P}hase {S}pace}, {A}strophysical {J}ournal 151
  (1968) 699.
\newline\urlprefix\url{http://adsabs.harvard.edu/full/1968ApJ...151..699M}

\bibitem{MilPre1970}
R.~H. Miller, K.~H. Prendergast, W.~J. Quirk,
  \href{http://adsabs.harvard.edu/abs/1970ApJ...161..903M}{{N}umerical
  experiments on spiral structure}, {A}strophysical {J}ournal 161 (1970) 903.
\newline\urlprefix\url{http://adsabs.harvard.edu/abs/1970ApJ...161..903M}

\bibitem{SevLuw1984}
G.~Severne, M.~Luwel, P.~Rousseeuw, {Equipartition and mass segregation in
  1-dimensional gravitational systems}, {A}stronomy and {A}strophysics 138~(2)
  (1984) 365.

\bibitem{WriMil1984}
H.~L. Wright, B.~N. Miller, {G}ravity in one dimension: {A} dynamical and
  statistical study, {P}hysical {R}eview {A} 29~(3) (1984) 1411.
\newblock \href {http://dx.doi.org/10.1103/PhysRevA.29.1411}
  {\path{doi:10.1103/PhysRevA.29.1411}}.

\bibitem{SevLuw1986}
G.~Severne, M.~Luwel, {V}iolent relaxation and mixing in non-uniform
  one-dimensional gravitational systems, {A}strophysics and {S}pace {S}cience
  122~(2) (1986) 299.
\newblock \href {http://dx.doi.org/10.1007/BF00650198}
  {\path{doi:10.1007/BF00650198}}.

\bibitem{YanGou1998}
T.~Yano, N.~Gouda, {E}volution of the {P}ower {S}pectrum and
  {S}elf-{S}imilarity in the {E}xpanding {O}ne-dimensional {U}niverse, {T}he
  {A}strophysical {J}ournal {S}upplement {S}eries 118~(2) (1998) 267.
\newblock \href {http://dx.doi.org/10.1086/313142} {\path{doi:10.1086/313142}}.

\bibitem{YouMil2000}
V.~P. Youngkins, B.~N. Miller, {G}ravitational phase transitions in a
  one-dimensional spherical system, {P}hysical {R}eview {E} 62 (2000) 4583.
\newblock \href {http://dx.doi.org/10.1103/PhysRevE.62.4583}
  {\path{doi:10.1103/PhysRevE.62.4583}}.

\bibitem{MilRou2002}
B.~N. Miller, J.~L. Rouet, {I}nfluence of expansion on hierarchical structure,
  {P}hysical {R}eview {E} 65~(5) (2002) 056121.
\newblock \href {http://dx.doi.org/10.1103/PhysRevE.65.056121}
  {\path{doi:10.1103/PhysRevE.65.056121}}.

\bibitem{Val2006}
P.~Valageas, {Relaxation of a one-dimensional gravitational system}, {P}hysical
  {R}eview {E} 74~(1) (2006) 1.
\newblock \href {http://dx.doi.org/10.1103/PhysRevE.74.016606}
  {\path{doi:10.1103/PhysRevE.74.016606}}.

\bibitem{JoySic2011}
M.~Joyce, F.~Sicard, {N}on-linear gravitational clustering of cold matter in an
  expanding universe: indications from 1{D} toy models, {M}onthly {N}otices of
  the {R}oyal {A}stronomical {S}ociety 413~(2) (2011) 1439.
\newblock \href {http://dx.doi.org/10.1111/j.1365-2966.2011.18225.x}
  {\path{doi:10.1111/j.1365-2966.2011.18225.x}}.

\bibitem{HohFei1967}
F.~Hohl, M.~R. Feix, {N}umerical experiments with a one-dimensional model for a
  self-gravitating star system, {A}strophysical {J}ournal 147 (1967) 1164.

\bibitem{Ryb1971}
G.~B. Rybicki, {Exact statistical mechanics of a one-dimensional
  self-gravitating system}, {A}strophysics and {S}pace {S}cience 14~(1) (1971)
  56--72.
\newblock \href {http://dx.doi.org/10.1007/BF00649195}
  {\path{doi:10.1007/BF00649195}}.

\bibitem{Mat1990}
S.~D. Mathur,
  \href{http://adsabs.harvard.edu/full/1990MNRAS.243..529M}{{Existence of
  oscillation modes in collisionless gravitating systems}}, {M}onthly {N}otices
  of the {R}oyal {A}stronomical {S}ociety 243 (1990) 529--536.
\newline\urlprefix\url{http://adsabs.harvard.edu/full/1990MNRAS.243..529M}

\bibitem{JoyWor2010}
M.~Joyce, T.~Worrakitpoonpon, {Relaxation to thermal equilibrium in the
  self-gravitating sheet model}, {J}ournal of {S}tatistical {M}echanics:
  {T}heory and {E}xperiment 2010~(10) (2010) P10012.
\newblock \href {http://dx.doi.org/10.1088/1742-5468/2010/10/P10012}
  {\path{doi:10.1088/1742-5468/2010/10/P10012}}.

\bibitem{JoyWor2011}
M.~Joyce, T.~Worrakitpoonpon, {Quasi-stationary states in the self-gravitating
  sheet model}, {P}hysical {R}eview {E} 84~(1) (2011) 011139.
\newblock \href {http://dx.doi.org/10.1103/PhysRevE.84.011139}
  {\path{doi:10.1103/PhysRevE.84.011139}}.

\bibitem{MilRou2006}
B.~N. Miller, J.-L. Rouet, {Development of fractal geometry in a
  one-dimensional gravitational system}, {C}omptes {R}endus {P}hysique 7~(3-4)
  (2006) 383--390.
\newblock \href {http://dx.doi.org/10.1016/j.crhy.2006.02.005}
  {\path{doi:10.1016/j.crhy.2006.02.005}}.

\bibitem{MilRou2007}
B.~Miller, J.-L. Rouet, E.~{Le Guirriec}, {Fractal geometry in an expanding,
  one-dimensional, Newtonian universe}, {P}hysical {R}eview {E} 76~(3) (2007)
  1--14.
\newblock \href {http://dx.doi.org/10.1103/PhysRevE.76.036705}
  {\path{doi:10.1103/PhysRevE.76.036705}}.

\bibitem{MilRou2010}
B.~N. Miller, J.-L. Rouet, {Cosmology in one dimension: fractal geometry, power
  spectra and correlation}, {J}ournal of {S}tatistical {M}echanics: {T}heory
  and {E}xperiment 2010~(12) (2010) P12028.
\newblock \href {http://dx.doi.org/10.1088/1742-5468/2010/12/P12028}
  {\path{doi:10.1088/1742-5468/2010/12/P12028}}.

\bibitem{SchDeh2013}
A.~E. Schulz, W.~Dehnen, G.~Jungman, S.~Tremaine, {G}ravitational collapse in
  one dimension, {M}onthly {N}otices of the {R}oyal {A}stronomical {S}ociety
  431~(1) (2013) 49.
\newblock \href {http://dx.doi.org/10.1093/mnras/stt073}
  {\path{doi:10.1093/mnras/stt073}}.

\bibitem{NouFan2003}
A.~Noullez, D.~Fanelli, E.~Aurell, {A} heap-based algorithm for the study of
  one-dimensional particle systems, {J}ournal of {C}omputational {P}hysics
  186~(2) (2003) 697.
\newblock \href {http://dx.doi.org/10.1016/S0021-9991(03)00048-2}
  {\path{doi:10.1016/S0021-9991(03)00048-2}}.

\bibitem{PreTeu1992}
W.~H. Press, S.~A. Teukolsky, W.~T. Vetterling, B.~P. Flannery, {F}ortran
  {N}umerical {R}ecipes, 2nd Edition, Vol.~1, Cambridge University Press, 1992.

\bibitem{CupGol1969}
S.~Cuperman, S.~Goldstein, M.~Lecar, {N}umerical experimental check of
  {L}ynden-{B}ell statistics-{II}. {T}he core-halo structure and the role of
  the violent relaxation, {M}onthly {N}otices of the {R}oyal {A}stronomical
  {S}ociety 146 (1969) 161--169.

\bibitem{GolCup1969}
S.~Goldstein, S.~Cuperman, M.~Lecar, {N}umerical {E}xperimental {C}heck of
  {L}ynden-{B}ell {S}tatistics for a {C}ollisionless {O}ne-{D}imensional
  {S}tellar {S}ystem, {M}onthly {N}otices of the {R}oyal {A}stronomical
  {S}ociety 143 (1969) 209.

\bibitem{LecCoh1971}
M.~Lecar, L.~Cohen, {N}umerical experiments on {L}ynden-{B}ell's statistics,
  {A}strophysics and {S}pace {S}cience 13~(2) (1971) 397.
\newblock \href {http://dx.doi.org/10.1007/BF00649169}
  {\path{doi:10.1007/BF00649169}}.

\bibitem{AarLec1975}
S.~J. Aarseth, M.~Lecar,
  \href{http://adsabs.harvard.edu/full/1975ARA%26A..13....1A}{{C}omputer
  simulations of stellar systems}, {A}nnual {R}eview of {A}stronomy and
  {A}strophysics 13 (1975) 1.
\newline\urlprefix\url{http://adsabs.harvard.edu/full/1975ARA%26A..13....1A}

\bibitem{MinFei1990}
P.~Mineau, M.~Feix, J.~Rouet, {N}umerical simulations of violent relaxation and
  formation of phase space holes in gravitational systems, {A}stronomy and
  {A}strophysics 228~(2) (1990) 344.

\bibitem{Yam2008}
Y.~Y. Yamaguchi, {One-dimensional self-gravitating sheet model and Lynden-Bell
  statistics}, {P}hysical {R}eview {E} 78~(4) (2008) 1.
\newblock \href {http://dx.doi.org/10.1103/PhysRevE.78.041114}
  {\path{doi:10.1103/PhysRevE.78.041114}}.

\bibitem{Duf2001}
D.~G. Duffy, {Green's Functions with Applications}, {S}tudies in {A}dvanced
  {M}athematics, CRC Press, 2001.

\bibitem{Wu1962}
C.-S. Wu, {Landau Damping and Resonant Energy Absorption}, {P}hysical {R}eview
  127~(5) (1962) 1419.
\newblock \href {http://dx.doi.org/10.1103/PhysRev.127.1419}
  {\path{doi:10.1103/PhysRev.127.1419}}.

\bibitem{Sag1994}
D.~Sagan, {On the physics of Landau damping}, {A}merican {J}ournal of {P}hysics
  62~(5) (1994) 450.
\newblock \href {http://dx.doi.org/10.1119/1.17547}
  {\path{doi:10.1119/1.17547}}.

\bibitem{Lan1946}
L.~D. Landau, {O}n the vibrations of the electronic plasma, {J}.
  {P}hys.({USSR}).

\bibitem{Glu1994}
R.~Gluckstern, {Analytic Model for Halo Formation in High Current Ion Linacs},
  {P}hysical {R}eview {L}etters 73~(9) (1994) 1247--1250.
\newblock \href {http://dx.doi.org/10.1103/PhysRevLett.73.1247}
  {\path{doi:10.1103/PhysRevLett.73.1247}}.

\bibitem{GolPoo2001}
H.~Goldstein, C.~P. Poole, J.~L. Safko, {C}lassical {M}echanics, 3rd Edition,
  Addison Wesley, 2001.

\bibitem{ChaSir2006}
P.-H. Chavanis, C.~Sire, {V}irial theorem and dynamical evolution of
  self-gravitating {B}rownian particles in an unbounded domain: {II}.
  {I}nertial models, {P}hysical {R}eview {E}: {S}tatistical, {N}onlinear, and
  {S}oft {M}atter {P}hysics 73~(6) (2006) 066104.
\newblock \href {http://dx.doi.org/10.1103/PhysRevE.73.066104}
  {\path{doi:10.1103/PhysRevE.73.066104}}.

\bibitem{ArfWeb2001}
G.~B. Arfken, H.~J. Weber, F.~Harris, {M}athematical methods for physicists,
  5th Edition, Academic Press, 2001.

\bibitem{SimRiz2006}
W.~Simeoni, F.~B. Rizzato, R.~Pakter, {Nonlinear coupling between breathing and
  quadrupole-like oscillations in the transport of mismatched beams in
  continuous magnetic focusing fields}, {P}hysics of {P}lasmas 13~(6) (2006)
  063104.
\newblock \href {http://dx.doi.org/10.1063/1.2208293}
  {\path{doi:10.1063/1.2208293}}.

\bibitem{RizPak2007}
F.~B. Rizzato, R.~Pakter, Y.~Levin, {W}ave breaking and particle jets in
  intense inhomogeneous charged beams, {P}hysics of {P}lasmas 14~(11) (2007)
  110701.
\newblock \href {http://dx.doi.org//10.1063/1.2802072}
  {\path{doi:/10.1063/1.2802072}}.

\bibitem{RizPak2009}
F.~B. Rizzato, R.~Pakter, Y.~Levin, {D}riven one-component plasmas, {P}hysical
  {R}eview {E}: {S}tatistical, {N}onlinear, and {S}oft {M}atter {P}hysics 80
  (2009) 021109.
\newblock \href {http://dx.doi.org/10.1103/PhysRevE.80.021109}
  {\path{doi:10.1103/PhysRevE.80.021109}}.

\bibitem{HohBro1967}
F.~Hohl, D.~T. Broaddus, {T}hermalization effects in a one-dimensional
  self-gravitating system, {P}hysics {L}etters {A} 25~(10) (1967) 713.
\newblock \href {http://dx.doi.org/10.1016/0375-9601(67)90956-5}
  {\path{doi:10.1016/0375-9601(67)90956-5}}.

\bibitem{TsuGou1996}
T.~Tsuchiya, N.~Gouda, T.~Konishi,
  \href{http://pre.aps.org/abstract/PRE/v53/i3/p2210_1}{{Relaxation processes
  in one-dimensional self-gravitating many-body systems}}, {P}hysical {R}eview
  {E} 53~(3) (1996) 2210--2216.
\newline\urlprefix\url{http://pre.aps.org/abstract/PRE/v53/i3/p2210_1}

\bibitem{Hoh1971}
F.~Hohl, {N}umerical {E}xperiments with a {D}isk of {S}tars, {A}strophysical
  {J}ournal 168 (1971) 343.
\newblock \href {http://dx.doi.org/10.1086/151091} {\path{doi:10.1086/151091}}.

\bibitem{DorKot1980}
A.~Doroshkevich, E.~Kotok, I.~Novikov, A.~Poliudov, S.~Shandarin, Y.~S. Sigov,
  {T}wo-dimensional simulation of the gravitational system dynamics and
  formation of the large-scale structure of the universe, {M}onthly {N}otices
  of the {R}oyal {A}stronomical {S}ociety 192 (1980) 321.

\bibitem{Aly1994}
J.~Aly, {Thermodynamics of a two-dimensional self-gravitating system},
  {P}hysical {R}eview {E} 49~(5) (1994) 3771.
\newblock \href {http://dx.doi.org/10.1103/PhysRevE.49.3771}
  {\path{doi:10.1103/PhysRevE.49.3771}}.

\bibitem{AlyPer1999}
J.-J. Aly, J.~Perez, {Thermodynamics of a two-dimensional unbounded
  self-gravitating system}, {P}hysical {R}eview {E} 60~(5) (1999) 5185.
\newblock \href {http://dx.doi.org/10.1103/PhysRevE.60.5185}
  {\path{doi:10.1103/PhysRevE.60.5185}}.

\bibitem{Mar2012}
B.~{Marcos}, {Collisional relaxation of two-dimensional gravitational systems},
  {A}r{X}iv e-prints{ }\href {http://arxiv.org/abs/1212.0959}
  {\path{arXiv:1212.0959}}.

\bibitem{Bin2004}
J.~Binney, {D}iscreteness effects in cosmological ${N}$-body simulations,
  {M}onthly {N}otices of the {R}oyal {A}stronomical {S}ociety 350~(3) (2004)
  939.
\newblock \href {http://dx.doi.org/10.1111/j.1365-2966.2004.07699.x}
  {\path{doi:10.1111/j.1365-2966.2004.07699.x}}.

\bibitem{Hen2006}
R.~N. Henriksen, {I}solated and non-isolated dark matter haloes and the
  {N}avarro, {F}renk and {W}hite profile, {M}onthly {N}otices of the {R}oyal
  {A}stronomical {S}ociety 366~(2) (2006) 697.
\newblock \href {http://dx.doi.org/10.1111/j.1365-2966.2005.09915.x}
  {\path{doi:10.1111/j.1365-2966.2005.09915.x}}.

\bibitem{Sax2013}
C.~J. Saxton, {G}alaxy stability within a self-interacting dark matter halo,
  {M}onthly {N}otices of the {R}oyal {A}stronomical {S}ociety 430~(3) (2013)
  1578.
\newblock \href {http://dx.doi.org/10.1093/mnras/sts689}
  {\path{doi:10.1093/mnras/sts689}}.

\bibitem{Vau1948}
G.~de~Vaucouleurs, {R}echerches sur les {N}\'{e}buleuses {E}xtragalactiques,
  {A}nnales d'{A}strophysique 11 (1948) 247.

\bibitem{Ser1963}
J.~L. S\'{e}rsic, {I}nfluence of the atmospheric and instrumental dispersion on
  the brightness distribution in a galaxy, {B}oletin de la {A}sociacion
  {A}rgentina de {A}stronomia {L}a {P}lata {A}rgentina 6 (1963) 41.

\bibitem{HjoMad1995}
J.~Hjorth, J.~Madsen, {S}mall deviations from the ${R}^{1/4}$ law, the
  fundamental plane, and phase densities of elliptical galaxies,
  {A}strophysical {J}ournal 445 (1995) 55.
\newblock \href {http://dx.doi.org/10.1086/175672} {\path{doi:10.1086/175672}}.

\bibitem{NavFre1996}
J.~F. Navarro, C.~S. Frenk, S.~D.~M. White, {T}he {S}tructure of {C}old {D}ark
  {M}atter {H}alos, {T}he {A}strophysical {J}ournal 462 (1996) 563.
\newblock \href {http://dx.doi.org/10.1086/177173} {\path{doi:10.1086/177173}}.

\bibitem{NavFre1997}
J.~F. Navarro, C.~S. Frenk, S.~D. White, {A} {U}niversal density profile from
  hierarchical clustering, {T}he {A}strophysical {J}ournal 490~(2) (1997) 493.
\newblock \href {http://dx.doi.org/10.1086/304888} {\path{doi:10.1086/304888}}.

\bibitem{WilHjo2010a}
L.~L.~R. Williams, J.~Hjorth, {S}tatistical mechanics of collisionless orbits.
  {II}. {S}tructure of halos, {A}strophysical {J}ournal 722~(1) (2010) 856.
\newblock \href {http://dx.doi.org/10.1088/0004-637X/722/1/856}
  {\path{doi:10.1088/0004-637X/722/1/856}}.

\bibitem{WilHjo2010b}
L.~L.~R. Williams, J.~Hjorth, R.~Wojtak, {S}tatistical mechanics of
  collisionless orbits. {III}. {C}omparison with {N}-body simulations, {T}he
  {A}strophysical {J}ournal 725~(1) (2010) 282.
\newblock \href {http://dx.doi.org/10.1088/0004-637X/725/1/282}
  {\path{doi:10.1088/0004-637X/725/1/282}}.

\bibitem{ChaSom1998}
P.-H. Chavanis, J.~Sommeria, {D}egenerate equilibrium states of collisionless
  stellar systems, {M}onthly {N}otices of the {R}oyal {A}stronomical {S}ociety
  296~(3) (1998) 569.
\newblock \href {http://dx.doi.org/10.1046/j.1365-8711.1998.01414.x}
  {\path{doi:10.1046/j.1365-8711.1998.01414.x}}.

\bibitem{DubGil1962}
D.~DuBois, V.~Gilinsky, M.~Kivelson, {Collision Damping of Plasma
  Oscillations}, {P}hysical {R}eview {L}etters 8~(11) (1962) 419--421.
\newblock \href {http://dx.doi.org/10.1103/PhysRevLett.8.419}
  {\path{doi:10.1103/PhysRevLett.8.419}}.

\bibitem{Bal1975}
R.~Balescu, {E}quilibrium and nonequilibrium statistical mechanics, {NASA}
  {STI}/{R}econ {T}echnical {R}eport {A} 76 (1975) 32809.

\bibitem{NunPak2007}
R.~P. Nunes, R.~Pakter, F.~B. Rizzato, {S}implified self-consistent model for
  emittance growth in charged beams with mismatched envelopes, {P}hysics of
  {P}lasmas 14~(2) (2007) 023104.
\newblock \href {http://dx.doi.org/10.1063/1.2472294}
  {\path{doi:10.1063/1.2472294}}.

\bibitem{BanSch2002}
S.~Banna, L.~Sch\"achter, {A}nalytic method for evaluation of the field of a
  charge traversing a geometric discontinuity, {A}pplied {P}hysics {L}etters 80
  (2002) 2842.
\newblock \href {http://dx.doi.org/10.1063/1.1472477}
  {\path{doi:10.1063/1.1472477}}.

\bibitem{CheGon2005}
Y.~Chekh, A.~Goncharov, I.~Protsenko, I.~G. Brown, {E}ffect of the
  electrostatic plasma lens on the emittance of a high-current heavy ion beam,
  {A}pplied {P}hysics {L}etters 86 (2005) 041502.
\newblock \href {http://dx.doi.org/10.1063/1.1855428}
  {\path{doi:10.1063/1.1855428}}.

\bibitem{MugBlu2008}
P.~Muggli, B.~Blue, C.~Clayton, F.~Decker, M.~Hogan, C.~Huang, C.~Joshi,
  T.~Katsouleas, W.~Lu, W.~Mori, C.~O'Connell, R.~Siemann, D.~Walz, M.~Zhou,
  {Halo Formation and Emittance Growth of Positron Beams in Plasmas},
  {P}hysical {R}eview {L}etters 101~(5) (2008) 1--4.
\newblock \href {http://dx.doi.org/10.1103/PhysRevLett.101.055001}
  {\path{doi:10.1103/PhysRevLett.101.055001}}.

\bibitem{ChePak2000}
C.~Chen, R.~Pakter, {M}echanisms and control of beam halo formation in intense
  microwave sources and accelerators, {P}hysics of {P}lasmas 7~(5) (2000) 2203.
\newblock \href {http://dx.doi.org/10.1063/1.874042}
  {\path{doi:10.1063/1.874042}}.

\bibitem{HesChe2000}
M.~Hess, C.~Chen, {C}onfinement criterion for a highly bunched beam, {P}hysics
  of {P}lasmas 7~(12) (2000) 5206.
\newblock \href {http://dx.doi.org/10.1063/1.1319639}
  {\path{doi:10.1063/1.1319639}}.

\bibitem{MorPak2004}
J.~S. Moraes, R.~Pakter, F.~B. Rizzato, {E}quilibrium and {S}tability of
  {O}ff-{A}xis {P}eriodically {F}ocused {P}article {B}eams, {P}hysical {R}eview
  {L}etters 93~(24) (2004) 244801.
\newblock \href {http://dx.doi.org/10.1103/PhysRevLett.93.244801}
  {\path{doi:10.1103/PhysRevLett.93.244801}}.

\bibitem{MorPak2005}
J.~S. Moraes, R.~Pakter, F.~B. Rizzato, {C}entroid motion in periodically
  focused beams, {P}hysics of {P}lasmas 12~(2) (2005) 023104.
\newblock \href {http://dx.doi.org/10.1063/1.1848546}
  {\path{doi:10.1063/1.1848546}}.

\bibitem{Hes2008}
M.~Hess, {O}ff-{A}xis {S}pace-{C}harge {L}imit for a {B}unched {E}lectron
  {B}eam in a {C}oaxial {C}onducting {S}tructure, {P}lasma {S}cience, {IEEE}
  {T}ransactions on 36~(3) (2008) 729.
\newblock \href {http://dx.doi.org/10.1109/TPS.2008.917163}
  {\path{doi:10.1109/TPS.2008.917163}}.

\bibitem{MarRiz2009}
L.~C. Martins, F.~B. Rizzato, R.~Pakter, {O}ff-axis stability of intense
  continuous relativistic beams, {J}ournal of {A}pplied {P}hysics 106~(4)
  (2009) 043305.
\newblock \href {http://dx.doi.org/10.1063/1.3204972}
  {\path{doi:10.1063/1.3204972}}.

\bibitem{BabGos2012}
P.~S. Babu, A.~Goswami, V.~S. Pandit, {A} {V}lasov equilibrium for space charge
  dominated beam in a misaligned solenoidal channel, {P}hysics of {P}lasmas
  19~(8) (2012) 080702.
\newblock \href {http://dx.doi.org/10.1063/1.4747694}
  {\path{doi:10.1063/1.4747694}}.

\bibitem{BerKis1999}
S.~Bernal, R.~A. Kishek, M.~Reiser, I.~Haber, {O}bservations and {S}imulations
  of {T}ransverse {D}ensity {W}aves in a {C}ollimated {S}pace-{C}harge
  {D}ominated {E}lectron {B}eam, {P}hysical {R}eview {L}etters 82~(20) (1999)
  4002.
\newblock \href {http://dx.doi.org/10.1103/PhysRevLett.82.4002}
  {\path{doi:10.1103/PhysRevLett.82.4002}}.

\bibitem{AndRos2000}
S.~G. Anderson, J.~B. Rosenzweig, {N}onequilibrium transverse motion and
  emittance growth in ultrarelativistic space-charge dominated beams,
  {P}hysical {R}eview {S}pecial {T}opics - {A}ccelerators and {B}eams 3 (2000)
  094201.
\newblock \href {http://dx.doi.org/10.1103/PhysRevSTAB.3.094201}
  {\path{doi:10.1103/PhysRevSTAB.3.094201}}.

\bibitem{LunGro2005}
S.~M. Lund, D.~P. Grote, R.~C. Davidson, {S}imulations of beam emittance growth
  from the collective relaxation of space-charge nonuniformities, {N}uclear
  {I}nstruments \& {M}ethods in {P}hysics {R}esearch, {S}ection {A}:
  {A}ccelerators, {S}pectrometers, {D}etectors, and {A}ssociated {E}quipment
  544~(1--2) (2005) 472.
\newblock \href {http://dx.doi.org/10.1016/j.nima.2005.01.280}
  {\path{doi:10.1016/j.nima.2005.01.280}}.

\bibitem{QiaZho2003}
B.~L. Qian, J.~Zhou, C.~Chen, {I}mage-charge effects on the envelope dynamics
  of an unbunched intense charged-particle beam, {P}hysical {R}eview {S}pecial
  {T}opics - {A}ccelerators and {B}eams 6~(1) (2003) 014201.
\newblock \href {http://dx.doi.org/10.1103/PhysRevSTAB.6.014201}
  {\path{doi:10.1103/PhysRevSTAB.6.014201}}.

\bibitem{ZhoQia2003}
J.~Zhou, B.~L. Qian, C.~Chen, {C}haotic particle motion and beam halo formation
  induced by image-charge effects in a small-aperture alternating-gradient
  focusing system, {P}hysics of {P}lasmas 10~(11) (2003) 4203.
\newblock \href {http://dx.doi.org/10.1063/1.1622388}
  {\path{doi:10.1063/1.1622388}}.

\bibitem{PakLev2007}
R.~Pakter, Y.~Levin, F.~B. Rizzato, {I}mage effects on the transport of intense
  nonaxisymmetric charged beams, {A}pplied {P}hysics {L}etters 91~(25) (2007)
  251503.
\newblock \href {http://dx.doi.org/10.1063/1.2827580}
  {\path{doi:10.1063/1.2827580}}.

\bibitem{OCWan1993}
J.~S. O'Connell, T.~P. Wangler, R.~S. Mills, K.~R. Crandall, {B}eam halo
  formation from space-charge dominated beams in uniform focusing channels, in:
  {P}roceedings of the 1993 {P}article {A}ccelerator {C}onference, Vol.~5,
  1993, p. 3657.
\newblock \href {http://dx.doi.org/10.1109/PAC.1993.309749}
  {\path{doi:10.1109/PAC.1993.309749}}.

\bibitem{Jac1998}
J.~D. Jackson, {C}lassical electrodynamics, 3rd Edition, Wiley, 1998.

\bibitem{TelPak2010}
T.~Teles, R.~Pakter, Y.~Levin, {Emittance growth and halo formation in the
  relaxation of mismatched beams}, {P}hysical {R}eview {S}pecial {T}opics -
  {A}ccelerators and {B}eams 13~(11) (2010) 1--8.
\newblock \href {http://dx.doi.org/10.1103/PhysRevSTAB.13.114202}
  {\path{doi:10.1103/PhysRevSTAB.13.114202}}.

\bibitem{GabJoy2010a}
A.~Gabrielli, M.~Joyce, B.~Marcos, {Quasistationary States and the Range of
  Pair Interactions}, {P}hysical {R}eview {L}etters 105~(21) (2010) 1--4.
\newblock \href {http://dx.doi.org/10.1103/PhysRevLett.105.210602}
  {\path{doi:10.1103/PhysRevLett.105.210602}}.

\bibitem{GabJoy2010b}
A.~Gabrielli, M.~Joyce, B.~Marcos, F.~Sicard, {A Dynamical Classification of
  the Range of Pair Interactions}, {J}ournal of {S}tatistical {P}hysics 141~(6)
  (2010) 970--989.
\newblock \href {http://dx.doi.org/10.1007/s10955-010-0090-x}
  {\path{doi:10.1007/s10955-010-0090-x}}.

\bibitem{HaiLub2006}
E.~Hairer, C.~Lubich, G.~Wanner, {G}eometric {N}umerical {I}ntegration:
  {S}tructure-{P}reserving {A}lgorithms for {O}rdinary {D}ifferential
  {E}quations, 2nd Edition, Springer, 2006.

\bibitem{KonKan1992}
T.~Konishi, K.~Kaneko, {C}lustered motion in symplectic coupled map systems,
  {J}ournal of {P}hysics {A}: {M}athematical and {G}eneral 25~(23) (1992) 6283.
\newblock \href {http://dx.doi.org/10.1088/0305-4470/25/23/023}
  {\path{doi:10.1088/0305-4470/25/23/023}}.

\bibitem{AntRuf1995}
M.~Antoni, S.~Ruffo, {Clustering and relaxation in Hamiltonian long-range
  dynamics}, {P}hysical {R}eview {E} 52~(3) (1995) 2361--2374.
\newblock \href {http://dx.doi.org/10.1103/PhysRevE.52.2361}
  {\path{doi:10.1103/PhysRevE.52.2361}}.

\bibitem{LatRap1999}
V.~Latora, A.~Rapisarda, S.~Ruffo, {S}uperdiffusion and out-of-equilibrium
  chaotic dynamics with many degrees of freedoms, {P}hysical {R}eview {L}etters
  83~(11) (1999) 2104.
\newblock \href {http://dx.doi.org/10.1103/PhysRevLett.83.2104}
  {\path{doi:10.1103/PhysRevLett.83.2104}}.

\bibitem{PluLat2004}
A.~Pluchino, V.~Latora, A.~Rapisarda, {G}lassy phase in the {H}amiltonian
  mean-field model, {P}hysical {R}eview {E} 69 (2004) 056113.
\newblock \href {http://dx.doi.org/10.1103/PhysRevE.69.056113}
  {\path{doi:10.1103/PhysRevE.69.056113}}.

\bibitem{AntFan2007}
A.~Antoniazzi, D.~Fanelli, J.~Barr\'e, P.-H. Chavanis, T.~Dauxois, S.~Ruffo,
  {M}aximum entropy principle explains the quasistationary states in systems
  with long-range interactions: {T}he example of the {H}amiltonian mean-field
  model, {P}hysical {R}eview {E} 75 (2007) 011112.
\newblock \href {http://dx.doi.org/10.1103/PhysRevE.75/011112}
  {\path{doi:10.1103/PhysRevE.75/011112}}.

\bibitem{BacCha2008}
R.~Bachelard, C.~Chandre, D.~Fanelli, X.~Leoncini, S.~Ruffo, {A}bundance of
  {R}egular {O}rbits and {N}onequilibrium {P}hase {T}ransitions in the
  {T}hermodynamic {L}imit for {L}ong-{R}ange {S}ystems, {P}hysical {R}eview
  {L}etters 101~(26) (2008) 260603--1.
\newblock \href {http://dx.doi.org/10.1103/PhysRevLett.101.260603}
  {\path{doi:10.1103/PhysRevLett.101.260603}}.

\bibitem{MerWag1966}
N.~D. Mermin, H.~Wagner, {A}bsence of {F}erromagnetic or {A}ntiferromagnetism
  in {O}ne- or {T}wo-{D}imensional {I}sotropic {H}eisenberg {M}odels,
  {P}hysical {R}eview {L}etters 17~(22) (1966) 1133.
\newblock \href {http://dx.doi.org/10.1103/PhysRevLett.17.1133}
  {\path{doi:10.1103/PhysRevLett.17.1133}}.

\bibitem{TamAnt2000}
F.~Tamarit, C.~Anteneodo, {R}otators with {L}ong-{R}ange {I}nteractions:
  {C}onnection with the {M}ean-{F}ield {A}pproximation, {P}hysical {R}eview
  {L}etters 84~(2) (2000) 208.
\newblock \href {http://dx.doi.org/10.1103/PhysRevLett.84.208}
  {\path{doi:10.1103/PhysRevLett.84.208}}.

\bibitem{DeLeo2013}
S.~{De Nigris}, X.~Leoncini, {E}mergence of a non-trivial fluctuating phase in
  the {XY}-rotors model on regular networks, {EPL} ({E}urophysics {L}etters)
  101~(1) (2013) 10002.
\newblock \href {http://dx.doi.org/10.1209/0295-5075/101/10002}
  {\path{doi:10.1209/0295-5075/101/10002}}.

\bibitem{Val2006a}
P.~Valageas, {T}hermodynamics and dynamics of a 1-{D} gravitational system,
  {A}stronomy \& {A}strophysics 450~(2) (2006) 445.
\newblock \href {http://dx.doi.org/10.1051/0004-6361:20054472}
  {\path{doi:10.1051/0004-6361:20054472}}.

\bibitem{Daw1962}
J.~Dawson, {O}ne-{D}imensional {P}lasma {M}odel, {P}hysics of {F}luids 5~(4)
  (1962) 445.

\bibitem{Col1976}
W.~B. Colson, {T}heory of a free electron laser, {P}hysics {L}etters {A} 59~(3)
  (1976) 187.
\newblock \href {http://dx.doi.org/10.1016/0375-9601(76)90561-2}
  {\path{doi:10.1016/0375-9601(76)90561-2}}.

\bibitem{BonPel1984}
R.~Bonifacio, C.~Pellegrini, L.~M. Narducci, {C}ollective instabilities and
  high-gain regimei in a free electron laser, {O}ptics {C}ommunications 50~(6)
  (1984) 373.
\newblock \href {http://dx.doi.org/10.1016/0030-4018(84)90105-6}
  {\path{doi:10.1016/0030-4018(84)90105-6}}.

\bibitem{BarBou2005}
J.~Barr\'{e}, F.~Bouchet, T.~Dauxois, S.~Ruffo, {Large deviation techniques
  applied to systems with long-range interactions}, {J}ournal of {S}tatistical
  {P}hysics 119~(3) (2005) 677--713.
\newblock \href {http://dx.doi.org/10.1007/s10955-005-3768-8}
  {\path{doi:10.1007/s10955-005-3768-8}}.

\bibitem{AntEls2006}
A.~Antoniazzi, Y.~Elskens, D.~Fanelli, S.~Ruffo, {S}tatistical mechanics and
  {V}lasov equation allow for a simplified {H}amiltonian description of
  {S}ingle-{P}ass {F}ree {E}lectron {L}aser saturated dynamics, {T}he
  {E}uropean {P}hysical {J}ournal {B} 50~(4) (2006) 603--611.
\newblock \href {http://dx.doi.org/10.1140/epjb/e2006-00175-0}
  {\path{doi:10.1140/epjb/e2006-00175-0}}.

\bibitem{BarBou2002a}
J.~Barr\'e, F.~Bouchet, T.~Dauxois, S.~Ruffo, {B}irth and long-time
  stabilization of out-of-equilibrium coherent structures, {E}uropean
  {P}hysical {J}ournal {B}: {C}ondensed {M}atter {P}hysics 29~(4) (2002) 577.
\newblock \href {http://dx.doi.org/10.1140/epjb/e2002-00342-3}
  {\path{doi:10.1140/epjb/e2002-00342-3}}.

\bibitem{MorKan2006a}
H.~Morita, K.~Kaneko, {C}ollective {O}scillation in a {H}amiltonian {S}ystem,
  {P}hysical {R}eview {L}etters 96 (2006) 050602.
\newblock \href {http://dx.doi.org/10.1103/PhysRevLett.96.050602}
  {\path{doi:10.1103/PhysRevLett.96.050602}}.

\bibitem{PakLev2013a}
R.~Pakter, Y.~Levin, {T}opology of {C}ollisionless {R}elaxation, {P}hysical
  {R}eview {L}etters 110~(14) (2013) 140601.
\newblock \href {http://dx.doi.org/10.1103/PhysRevLett.110.140601}
  {\path{doi:10.1103/PhysRevLett.110.140601}}.

\bibitem{CamCha2013}
A.~Campa, P.-H. Chavanis, {C}aloric curves fitted by polytropic distributions
  in the {HMF} model, {T}he {E}uropean {P}hysical {J}ournal {B} 86 (2013) 170.
\newblock \href {http://dx.doi.org/10.1140/epjb/e2013-30947-0}
  {\path{doi:10.1140/epjb/e2013-30947-0}}.

\bibitem{Yos1990}
H.~Yoshida, {C}onstruction of higher order symplectic integrators, {P}hysics
  {L}etters {A} 150~(5-7) (1990) 262.
\newblock \href {http://dx.doi.org/10.1016/0375-9601(90)90092-3}
  {\path{doi:10.1016/0375-9601(90)90092-3}}.

\bibitem{Hai2004}
E.~Hairer, \href{http://www.unige.ch/~hairer/software.html}{{F}ortran and
  {M}atlab {C}odes}, Website (2004).
\newline\urlprefix\url{http://www.unige.ch/~hairer/software.html}

\bibitem{AntFan2007a}
A.~Antoniazzi, D.~Fanelli, S.~Ruffo, Y.~Y. Yamaguchi, {Nonequilibrium
  Tricritical Point in a System with Long-Range Interactions}, {P}hysical
  {R}eview {L}etters 99~(4) (2007) 2--5.
\newblock \href {http://dx.doi.org/10.1103/PhysRevLett.99.040601}
  {\path{doi:10.1103/PhysRevLett.99.040601}}.

\bibitem{RocAma2012}
T.~M.~R. Filho, M.~A. Amato, A.~Figueiredo, {N}onequilibrium phase transitions
  and violent relaxation in the {H}amiltonian mean-field model, {P}hysical
  {R}eview {E} 85~(6) (2012) 062103.
\newblock \href {http://dx.doi.org/10.1103/PhysRevE.85.062103}
  {\path{doi:10.1103/PhysRevE.85.062103}}.

\bibitem{StaCha2009}
F.~Staniscia, P.~Chavanis, G.~{De Ninno}, D.~Fanelli, {O}ut-of-equilibrium
  phase re-entrance(s) in long-range interacting systems, {P}hysical {R}eview
  {E} 80~(2) (2009) 1.
\newblock \href {http://dx.doi.org/10.1103/PhysRevE.80.021138}
  {\path{doi:10.1103/PhysRevE.80.021138}}.

\bibitem{Cha2006}
P.-H. Chavanis, {L}ynden-{B}ell and {T}sallis distributions for the {HMF}
  model, {E}uropean {P}hysical {J}ournal {B}: {C}ondensed {M}atter {P}hysics 53
  (2006) 487.
\newblock \href {http://dx.doi.org/10.1140/epjb/e2006-00405-5}
  {\path{doi:10.1140/epjb/e2006-00405-5}}.

\bibitem{AssFan2012}
M.~Assllani, D.~Fanelli, A.~Turchi, T.~Carletti, X.~Leoncini, {S}tatistical
  theory of quasistationary states beyond the single water-bag case study,
  {P}hysical {R}eview {E} 85~(2) (2012) 021148.
\newblock \href {http://dx.doi.org/10.1103/PhysRevE.85.021148}
  {\path{doi:10.1103/PhysRevE.85.021148}}.

\bibitem{Fil2013}
T.~M.~R. Filho, {S}olving the {V}lasov equation for one-dimensional models with
  long range interactions on a {GPU}, {C}omputer {P}hysics {C}ommunications
  184~(1) (2013) 34.
\newblock \href {http://dx.doi.org/10.1016/j.cpc.2012.08.005}
  {\path{doi:10.1016/j.cpc.2012.08.005}}.

\bibitem{BouGup2010}
F.~Bouchet, S.~Gupta, D.~Mukamel, {T}hermodynamics and dynamics of systems with
  long-range interactions, {P}hysica {A}: {S}tatistical {M}echanics and its
  {A}pplications 389 (2010) 4389.
\newblock \href {http://dx.doi.org/10.1016/j.physa.2010.02.024}
  {\path{doi:10.1016/j.physa.2010.02.024}}.

\bibitem{MarcianoCondMat}
T.~M. {Rocha Filho}, A.~E. {Santana}, J.~R.~S. {Moura}, M.~A. {Amato},
  A.~{Figueiredo}, {Dynamics and physical interpretation of quasi-stationary
  states in systems with long-range interactions}, {A}r{X}iv e-prints{ }\href
  {http://arxiv.org/abs/1305.2903} {\path{arXiv:1305.2903}}.

\bibitem{FigRoc2013}
A.~Figueiredo, T.~M. {Rocha Filho}, A.~E. Santana, M.~A. Amato, {S}caling of
  the dynamics of homogeneous states of one-dimensional long-range interacting
  systems, {A}r{X}iv e-prints{}\href {http://arxiv.org/abs/1305.4417}
  {\path{arXiv:1305.4417}}.

\bibitem{Cha2012}
P.~H. Chavanis, {K}inetic theory of spatially homogeneous systems with
  long-range interactions: {I}. {G}eneral results, {T}he {E}uropean {P}hysical
  {J}ournal {P}lus 127 (2012) 19.
\newblock \href {http://dx.doi.org/10.1140/epjp/i2012-12019-9}
  {\path{doi:10.1140/epjp/i2012-12019-9}}.

\bibitem{TelBen2012}
T.~N. Teles, F.~{P. da C. Benetti}, R.~Pakter, Y.~Levin, {N}onequilibrium phase
  transitions in systems with long-range interactions, {P}hysical {R}eview
  {L}etters 109 (2012) 230601.
\newblock \href {http://dx.doi.org/10.1103/PhysRevLett.109.230601}
  {\path{doi:10.1103/PhysRevLett.109.230601}}.

\bibitem{LeeGri1985}
D.~Lee, G.~Grinstein, {S}trings in two-dimensional classical {XY} models,
  {P}hysical {R}eview {L}etters 55~(5) (1985) 541--544.
\newblock \href {http://dx.doi.org/10.1103/PhysRevLett.55.541}
  {\path{doi:10.1103/PhysRevLett.55.541}}.

\bibitem{PodAre2011}
F.~C. Poderoso, J.~J. Arenzon, Y.~Levin, {N}ew {O}rdered {P}hases in a {C}lass
  of {G}eneralized {XY} {M}odels, {P}hysical {R}eview {L}etters 106~(6) (2011)
  067202.
\newblock \href {http://dx.doi.org/10.1103/PhysRevLett.106.067202}
  {\path{doi:10.1103/PhysRevLett.106.067202}}.

\bibitem{PakLev2013}
R.~Pakter, Y.~Levin, {N}on-equilibrium {D}ynamics of an {I}nfinite {R}ange
  ${XY}$ {M}odel in an {E}xternal {F}ield, {J}ournal of {S}tatistical {P}hysics
  150~(3) (2013) 531.
\newblock \href {http://dx.doi.org/10.1007/s10955-012-0576-9}
  {\path{doi:10.1007/s10955-012-0576-9}}.

\bibitem{PatGup2012}
A.~Patelli, S.~Gupta, C.~Nardini, S.~Ruffo, {L}inear response theory for
  long-range interacting systems in quasistationary states, {P}hysical {R}eview
  {E} 85~(2) (2012) 021133.
\newblock \href {http://dx.doi.org/10.1103/PhysRevE.85.021133}
  {\path{doi:10.1103/PhysRevE.85.021133}}.

\bibitem{Tsa2009}
C.~Tsallis, {I}ntroduction to {N}onextensive {S}tatistical {M}echanics,
  Springer, 2009.

\bibitem{AguMer1990}
L.~A. Aguilar, D.~Merritt, {T}he structure and dynamics of galaxies formed by
  cold dissipationless collapse, {T}he {A}strophysical {J}ournal 354 (1990) 33.
\newblock \href {http://dx.doi.org/10.1086/168665} {\path{doi:10.1086/168665}}.

\end{thebibliography}

\end{document}